\documentclass[a4paper,10pt]{article}

\usepackage{epsfig}

\newtheorem{thm}{Theorem}
\newtheorem{lem}{Lemma}
\newtheorem{tab}{Table}
\newtheorem{fig}{Figure}

\def\leurre{\noindent\leftskip0pt\small\baselineskip 10pt}
\def\grostrait{\ligne{\vrule height 1pt depth 1pt width \hsize}}
\def\demitrait{\ligne{\vrule height 0.5pt depth 0.5pt width \hsize}}

\def\encadre#1#2{%
\setbox100=\hbox{\kern#1{#2}\kern#1}
\dimen100=\ht100 \advance \dimen100 by #1
\dimen101=\dp100 \advance \dimen101 by #1
\setbox100=\hbox{\vrule height \dimen100 depth \dimen101\box100\vrule}
\setbox100=\vbox{\hrule\box100\hrule}
\advance \dimen100 by .4pt \ht100=\dimen100
\advance \dimen101 by .4pt \dp100=\dimen101
\box100
\relax
}

\def\ligne#1{\hbox to \hsize{#1}}
\def\PlacerEn#1 #2 #3 {\rlap{\kern#1\raise#2\hbox{#3}}}

\font\rmx=cmr10
\font\rmxii=cmr12
\font\rmviii=cmr8
\font\itx=cmti10
\font\ttx=cmtt10
\font\ttxii=cmtt12
\font\ttv=cmtt5
\font\ttvi=cmtt6
\font\rmix=cmr9 \font\mmix=cmmi9 \font\symix=cmsy9
\def\mathix{\textfont0=\rmix \textfont1=\mmix \textfont2=\symix}

\title{A new universal cellular automaton on the ternary heptagrid
\vskip 15pt
\rmxii
\ligne{\hfill Maurice Margenstern\hfill} 
\vskip 15pt
\rmx\baselineskip=12pt
\ligne{\hfill
Laboratoire d'Informatique Th\'eorique et Appliqu\'ee, EA 3097,\hfill}
\ligne{\hfill Universit\'e de Metz, I.U.T. de Metz,\hfill}
\ligne{\hfill D\'epartement d'Informatique,\hfill}
\ligne{\hfill \^Ile du Saulcy,\hfill}
\ligne{\hfill 57045 Metz Cedex, France,\hfill}
\ligne{\hfill {\itx email:} {\ttx margens@univ-metz.fr}\hfill}
}
\begin{document}
\maketitle

\vskip 10pt
\begin{abstract}
In this paper, we construct a new weakly universal cellular automaton
on the ternary heptagrid. The previous result, obtained by the same author 
and Y. Song required 
six states only. This time, the number of states is four. This is the best result
up to date for cellular automata in the hyperbolic plane.
\end{abstract}
\vskip 10pt

\def\cqfd{\hbox{\kern 2pt\vrule height 6pt depth 2pt width 8pt\kern 1pt}}
\def\Hii{$I\!\!H^2$}
\def\Hiii{$I\!\!H^3$}
\def\Hiv{$I\!\!H^4$}
\def\norm{\hbox{$\vert\vert$}}
\section{Introduction}

   As indicated in the abstract, this paper is a significant improvement of the first 
result about a universal cellular automaton on the ternary heptagrid which was obtained
by the same author and Y. Song, see \cite{mmsyENTCS}. This time we have a weakly universal 
cellular automaton on the ternary heptagrid which is the smallest universal cellular 
automaton obtained in the hyperbolic plane, up to date. As noticed in the quoted
paper, the translation of the present result to the pentagrid is not straightforward and
would require at least one more state with the same pattern of simulation. We remind
that for the pentagrid, the best result was obtained also by the authors of~\cite{mmsyENTCS}
in~\cite{mmsyBristol,mmsyPPL}, also see~\cite{mmbook2}. The latter result was a significant 
improvement of the first result established in the pentagrid, see~\cite{fhmmTCS}. For 
the pentagrid, papers~\cite{mmsyBristol,mmsyPPL,mmbook2} reduce the number of states 
from~22 down to~9. In \cite{mmsyENTCS}, we proved that there is a weakly universal cellular
automaton on the heptagrid with six states. 

   In this paper, we reduce the number of states to four ones as
indicated in the following:

\begin{thm}\label{universal} {\rm(Margenstern)} $-$ 
There is a cellular automaton on the ternary heptagrid which is weakly 
universal and which has four states. Moreover, the rules of the cellular 
automaton are rotation invariant. The cellular automaton has an infinite 
initial configuration which
is ultimately periodic along two different rays of mid-points~$r_1$ and~$r_2$ 
of the 
ternary heptagrid
and finite in the complement of the parts attached to~$r_1$ and~$r_2$.
\end{thm}

   Our present cellular automaton also simulates a railway circuit, as this
used in papers~\cite{fhmmTCS,mmsyBristol,mmsyPPL,mmbook2,mmsyENTCS}. In order to make 
this paper self-contained, Section~2 reminds the principles of this simulation.
In Section~3, we remind the reader about hyperbolic geometry and cellular automata
on the ternary heptagrid. Still in Section~3, we give the general features of
the implementation of a railway circuit in the ternary heptagrid and
in Section~4, we precisely define the implementation in the heptagrid. In Section~5, 
we give the format of the rules and the transition table of the automaton whose action 
is described in Section~4. We also indicate how a computer program contributed to the
construction of the table.

\section{The railway circuit}

   As initially devised in~\cite{stewart} and then mentioned
in~\cite{mmCSJMtrain,fhmmTCS,mmsyBristol,mmsyENTCS,mmbook2},
the circuit uses tracks represented by lines and quarters of circles and switches.
There are three kinds of switches: the {\bf fixed}, the {\bf memory} and the
{\bf flip-flop} switches. They are represented by the schemes given in
Fig.~\ref{aiguillages}.

\vskip 10pt
\vtop{
\setbox110=\hbox{\epsfig{file=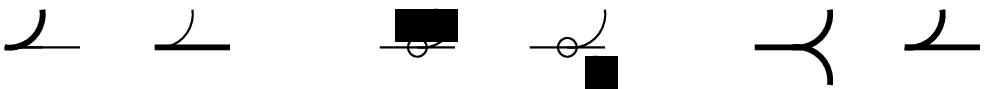,width=280pt}}
\ligne{\hfill
\PlacerEn {-305pt} {0pt} \box110
}
\vskip-15pt
\begin{fig}
\label{aiguillages}
\leurre
The three kinds of switches. From left to right: fixed, flip-flop and memory switches.
\end{fig}
}
\vskip 10pt

   Note that a switch is an oriented structure: on one side, it has a single
track~$u$ and, on the the other side, it has two tracks~$a$ and~$b$. This 
defines two ways of crossing a switch. Call the way from~$u$ to~$a$ or~$b$
{\bf active}. Call the other way, from~$a$ or~$b$ to~$u$ {\bf passive}. The 
names comes from the fact that in a passive way, the switch plays no role on 
the trajectory of the locomotive. On the contrary, in an active
crossing, the switch indicates which track between~$a$ and~$b$ will be followed by
the locomotive after running on~$u$: the new track is called the {\bf selected}
track.

   With the help of these three kind of switches, we define an 
{\bf elementary circuit} as in~\cite{stewart}, which exactly contains one bit of 
information. The circuit is illustrated by Fig.~\ref{element}, above.
It can be remarked that the working of the circuit strongly depends on how
the locomotive enters it. If the locomotive enters the circuit through~$E$,
it leaves the circuit through~$O_1$ or~$O_2$, depending on the selected track
of the memory switch which stands near~$E$. If the locomotive enters through~$U$,
the application of the given definitions shows that the selected track at the
switches near~$E$ and~$U$ are both changed: the switch at~$U$ is a flip-flop which
is changed by the very active passage of the locomotive and the switch at~$E$
is a memory one which is changed because it is passively crossed by the 
locomotive and through the non-selected track. The just described actions of
the locomotive correspond to a {\bf read} and a {\bf write} operation on the 
bit contained by the circuit which consists of the configurations of the 
switches at~$E$ and at~$U$. It is assumed that the write operation is triggered
when we know that we have to change the bit which we wish to rewrite.

\vskip 10pt
\vtop{
\setbox110=\hbox{\epsfig{file=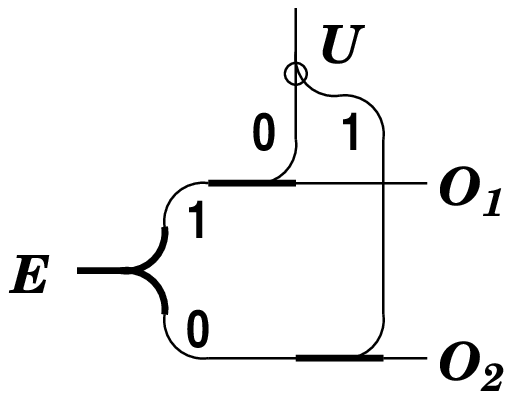,width=160pt}}
\ligne{\hfill
\PlacerEn {-285pt} {0pt} \box110
}
\vskip-15pt
\begin{fig}
\label{element}
\leurre
The elementary circuit.
\end{fig}
}
\vskip 10pt

   From this element, it is easy to devise circuits which represent different
parts of a register machine. As an example, Fig.~\ref{unit} illustrates
an implementation of a unit of a register.
\vskip 10pt
\vtop{
\setbox110=\hbox{\epsfig{file=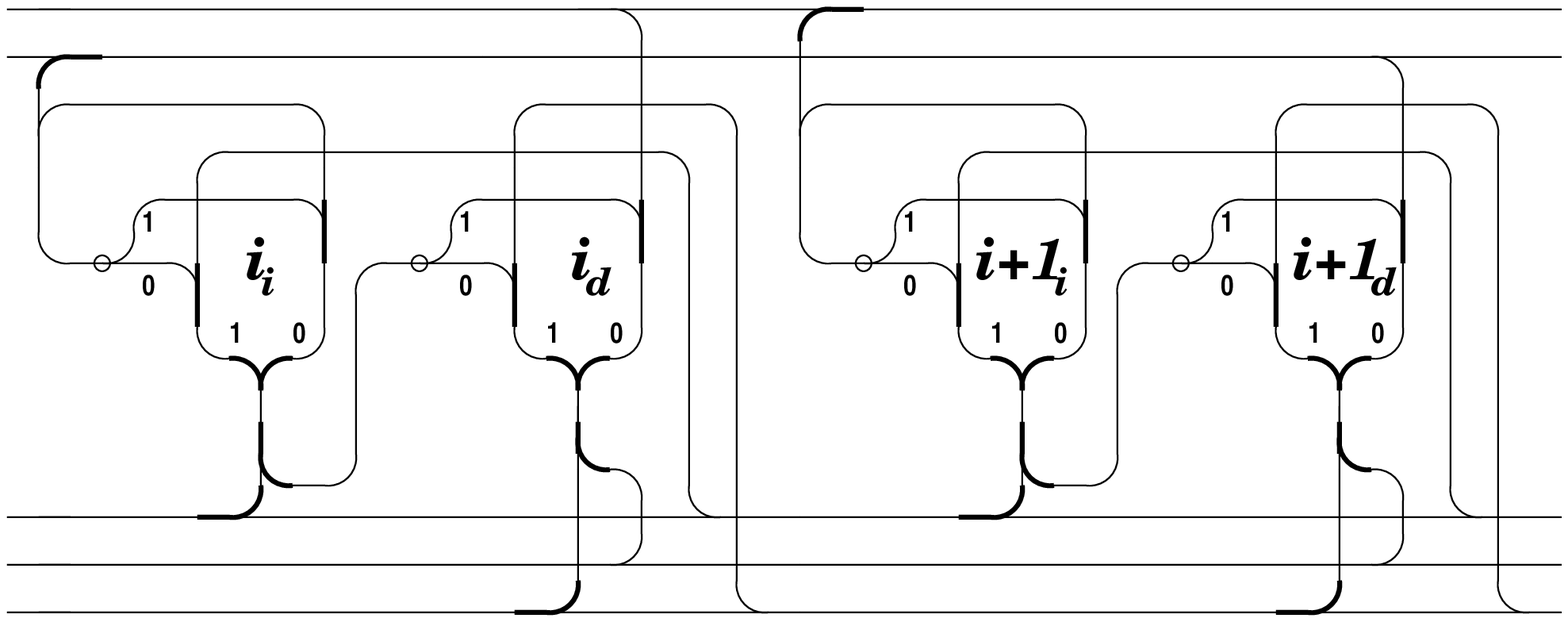,width=320pt}}
\ligne{\hfill
\PlacerEn {-335pt} {0pt} \box110
\PlacerEn {-328pt} {24pt} {\mathix$i$}
\PlacerEn {-328pt} {14.5pt} {\mathix$d$}
\PlacerEn {-328pt} {5pt} {\mathix$r$}
\PlacerEn {-328pt} {127pt} {\mathix$j_1$}
\PlacerEn {-328pt} {117pt} {\mathix$j_2$}
}
\vskip-15pt
\begin{fig}
\label{unit}
\leurre
Here, we have two consecutive units of a register. A register contains 
infinitely many copies of units. Note the tracks $i$, $d$, $r$, $j_1$ and~$j_2$.
For incrementing, the locomotive arrives at a unit through~$i$ and it leaves the
unit through~$r$. For decrementing, it arrives though~$d$ and it leaves 
also through~$r$ if decrementing the register was possible, otherwise, it leaves
through~$j_1$ or~$j_2$.
\end{fig}
}
\vskip 10pt
   As indicated by its name, the {\bf fixed switch} is left unchanged by the 
passage of the locomotive. It always remains in the same position: when
actively crossed by the locomotive, the switch always sends it onto the same 
track. The flip-flop switch is assumed to be crossed actively only. Now,
after each crossing by the locomotive, it changes the selected track. 
The memory switch can be crossed by the locomotive actively and passively.
In an active passage, the locomotive is sent onto the selected track. Now, the
selected track is defined by the track of the last passive crossing by the 
locomotive. Of course, at initial time, the selected track is fixed.

   Other parts of the needed circuitry are described 
in~\cite{mmCSJMtrain,fhmmTCS}. The main idea in these different parts is
to organize the circuit in possibly visiting several elementary circuits
which represent the bits of a configuration which allow the whole system
to remember the last visit  of the locomotive. The use of this technique is
needed for the following two operations.

   When the locomotive arrives to a register~$R$, it arrives either to 
increment~$R$ or to decrement it. As can be seen on Fig.~\ref{unit}, when the
instruction is performed, the locomotive goes back from the register by the
same track. Accordingly, we need somewhere to keep track of the fact whether
the locomotive incremented~$R$ or it decremented~$R$. This is one type of control.
The other control comes from the fact that several instructions usually apply
to the same register. Again, when the locomotive goes back from~$R$,
in general it goes back to perform a new instruction which depends on the one
it has just performed on~$R$. Again this can be controlled by what we called
the {\bf selector} in~\cite{mmCSJMtrain,fhmmTCS}. 

   At last, the dispatching of the locomotive on the right track for the next 
instruction is performed by the {\bf sequencer}, a circuit whose main structure
looks like its implementation in the classical models of cellular automata such 
as the game of life or the billiard ball model. The reader is referred to the 
already quoted papers for full details on the circuit. Remember that this 
implementation is performed in the Euclidean plane, as clear from 
Fig.~\ref{example} which illustrates the case of a few lines of a program of 
a register machine.

\vskip 10pt
\vtop{
\setbox110=\hbox{\epsfig{file=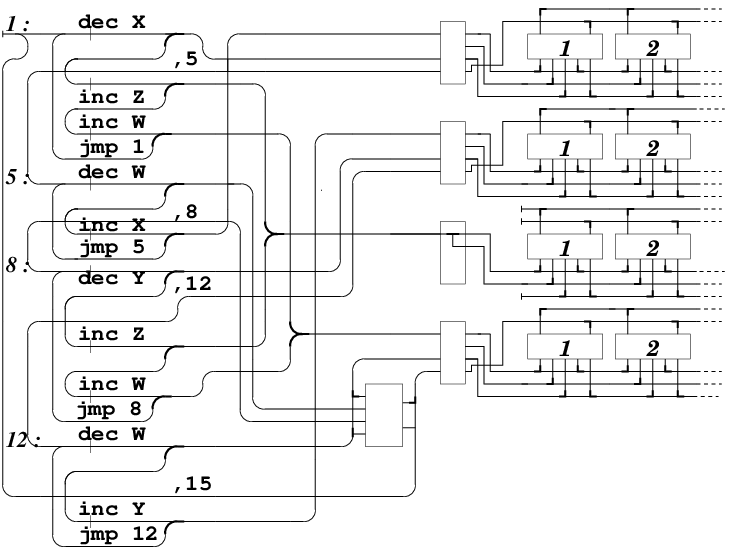,width=320pt}}
\ligne{\hfill
\PlacerEn {-335pt} {0pt} \box110
}
\vskip-15pt
\begin{fig}
\label{example}
\leurre
An example of the implementation of a small program of a register machine.
On the left-hand side of the figure, the part of the sequencer. It can be noticed
how the tracks are attached to each instruction of the program. Note that there 
are four decrementing instructions for~$W$: this is why a selector gathers 
the arriving tracks before sending the locomotive to the
control of the register. On the way back, the locomotive is sent on the right
track. 
\end{fig}
}
\vskip 10pt
   Now, we turn to the implementation in the hyperbolic plane, which first
requires some features of hyperbolic geometry.

\section{Implementation in the hyperbolic plane}

   Hyperbolic geometry appeared in the first half of the 19$^{\rm th}$ century,
in the last attempts to prove the famous parallel axiom of Euclid's {\it Elements} 
from the remaining ones. Independently, Lobachevsky and Bolyai discovered a new geometry
by assuming that in the plane, from a point out of a given line, there are at
least two lines which are parallel to the given line. Later, models of the new
geometry were found, in particular Poincar\'e's model, which is the frame of
all this study.

\vskip 7pt
   In this model, the hyperbolic plane is the set of points
which lie in the open unit disc of the Euclidean plane whose border is the
unit circle. The lines of the hyperbolic plane in Poincar\'e's disc
model are either the trace of diametral lines or the trace of circles
which are orthogonal to the unit circle, see Fig.~\ref{model}.
We say that the considered lines or circles {\bf support} the hyperbolic
line, simply {\bf line} for short, when there is no ambiguity, $h$-{\bf line}
when it is needed to avoid it. Fig.~\ref{model}
illustrates the notion of {\bf parallel} and {\bf non-secant} lines in
this setting.

   The angle between two $h$-lines are defined as the Euclidean angle between
the tangents to their support. The reason for choosing the Poincar\'e's model
is that hyperbolic angles between $h$-lines are, in a natural way, the 
Euclidean angle between the corresponding supports. In particular, orthogonal circles 
support perpendicular $h$-lines.

\vskip 14pt
\setbox110=\hbox{\epsfig{file=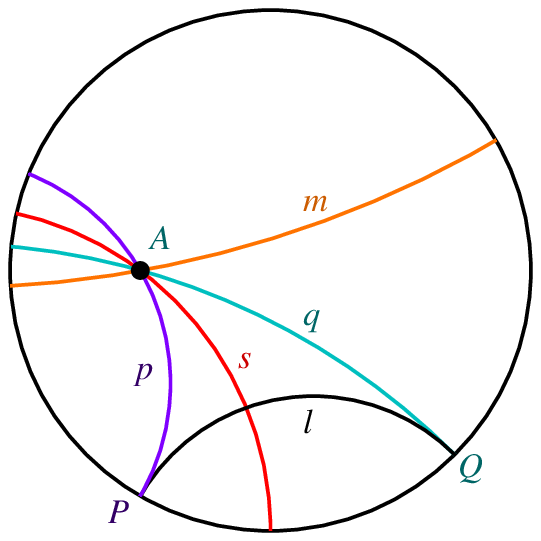,width=200pt}}
\vtop{
\ligne{\hfill
\PlacerEn {-110pt} {0pt} \box110
\hfill}
\vskip 0pt
\begin{fig}\label{model}
\leurre
The lines $p$ and $q$ are {\bf parallel} to the line~$\ell$, with points at
infinity~$P$ and~$Q$, on the border of the unit disc. The $h$-line $m$ is
{\bf non-secant} with $\ell$: it can be seen that there are infinitely 
many such lines.
\end{fig}
}

\subsection{The heptagrid}

   Remember that in the Euclidean plane and up to similarities,
there are only three kinds of tilings based on the recursive replication of 
a regular polygon by reflection in its sides and of the images in their sides. 
In the hyperbolic plane, where the notion of similarity is meaningless,
there are infinitely many such tilings. In this paper, we consider the 
smallest regular polygon defined by the property that three copies of it can 
be put around a vertex in order to cover a neighbourhood of the vertex 
with no overlapping. This tiling is called the {\bf ternary heptagrid}, 
see Fig.~\ref{hepta} and~\ref{eclate_73} for an illustrative representation.
Later on, we shall simply say the {\bf heptagrid}. Here, we give a rough explanation 
of these objects, referring to~\cite{mmbook1} and to~\cite{mmDMTCS} for more details 
and references.

   The left-hand side of Fig.~\ref{hepta} illustrates the heptagrid.
But, besides the occurrence of a lot of symmetries, nothing can be grasped on the
structure of the tiling from this mere picture. The right-hand side picture of
Fig.~\ref{hepta} illustrates the main tool to make the structure visible.
There, we can see two lines which we call {\bf mid-point lines} as they
join mid-points of edges of heptagons belonging to the tiling. On the figure,
a half of each line is drawn with a thicker stroke. It is a {\bf ray} issued
from the common point of these lines: here, a mid-point of an edge of the central
heptagon of the figure. We shall say a {\bf ray of mid-points}. These two rays 
define an angle, and the set of tiles
whose all mid-points of the edges fall inside the angle is called a {\bf sector}.

   Fig.~\ref{hepta} and~\ref{eclate_73} sketchily remember that the tiling 
is spanned by a generating tree. In fact, as can be noticed on both the right-hand
side of Fig.~\ref{hepta} and the left-hand side of Fig.~\ref{eclate_73}, the
set of tiles constituting a sector is spanned by a Fibonacci tree, 
see~\cite{mmbook1,mmDMTCS} for references. The name of the tree comes from
the fact that the number of nodes on a given level~$n$ is $f_{2n+1}$, where
$\{f_n\}$ denotes the Fibonacci sequence with $f_1=1$, $f_2=2$.

Now, as indicated in Fig.~\ref{eclate_73},
seven sectors around a central tile allow us to exactly cover the
hyperbolic plane with the heptagrid which is the tessellation
obtained from the regular heptagon described above and easily seen on the figures.

\vskip 7pt
\vskip 14pt
\setbox110=\hbox{\epsfig{file=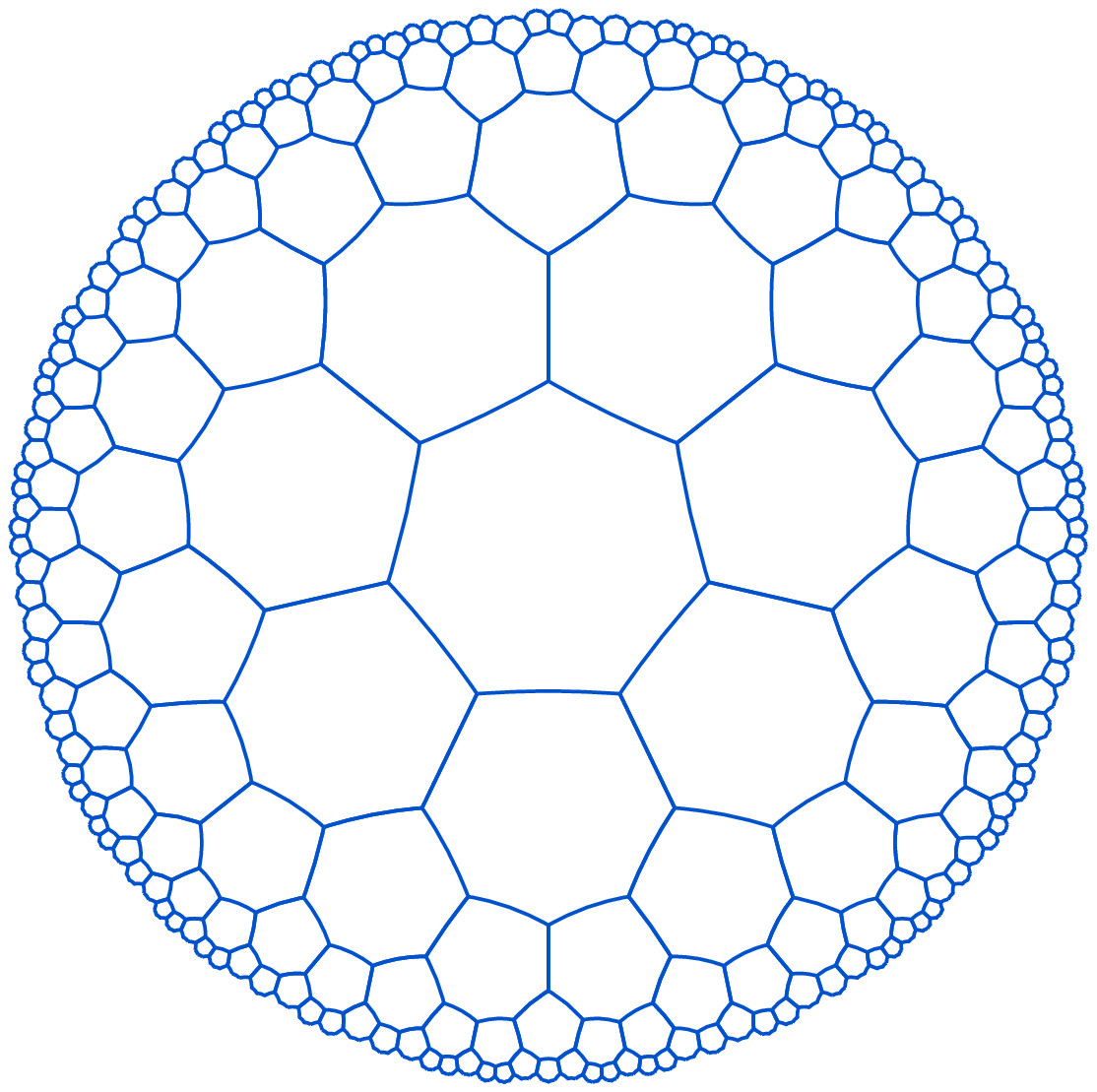,width=150pt}}
\setbox112=\hbox{\epsfig{file=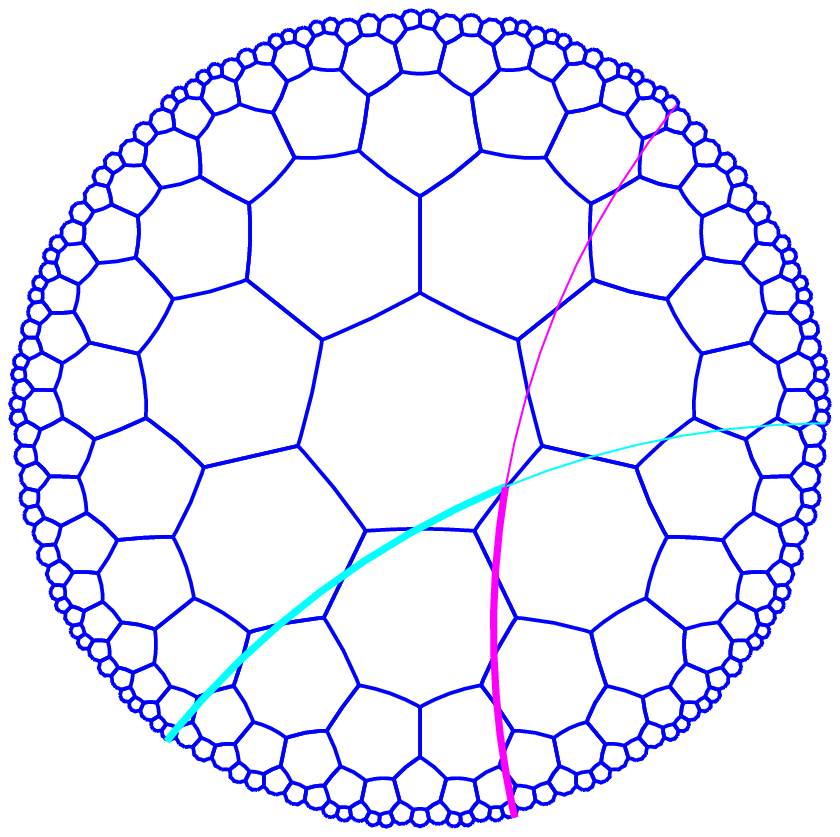,width=150pt}}
\vtop{
\ligne{\hfill
\PlacerEn {-335pt} {0pt} \box110
\PlacerEn {-180pt} {0pt} \box112
}
\vskip-15pt
\begin{fig}\label{hepta}
\leurre
On the left: the tiling; on the right: the delimitation of the sectors
which are spanned by a tree. Note the rays of mid-points. They are issued
from the same point: a mid-point of an edge of the central cell of the figure.
\end{fig}
}
\vskip 10pt

   In the left-hand side picture of Fig.~\ref{eclate_73},
we represent the sectors in terms of tiles. The tiles are in bijection 
with the tree which is represented on the right-hand side part of the figure.
This allows to define the coordinates in a sector of the heptagrid, 
see \cite{mmbook1}. We number
the nodes of the tree, starting from the root and going on, level
by level and, on each level, from the left to the right. Then, we
represent each number in the basis defined by the quoted Fibonacci sequence,
taking the maximal representation, see\cite{mmJUCSii,mmbook1}.

\vskip 10pt
\setbox110=\hbox{\epsfig{file=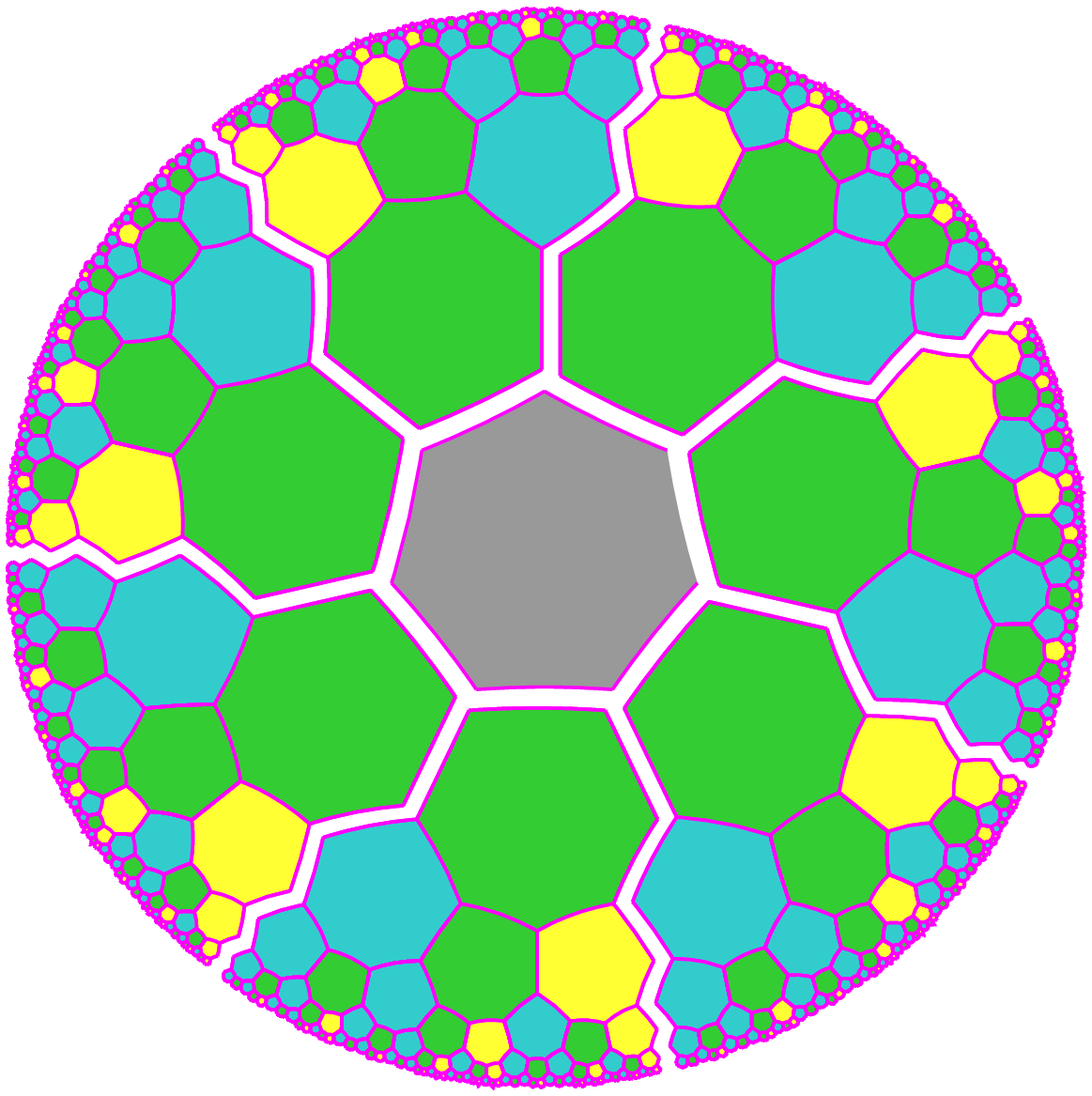,width=140pt}}
\setbox112=\hbox{\epsfig{file=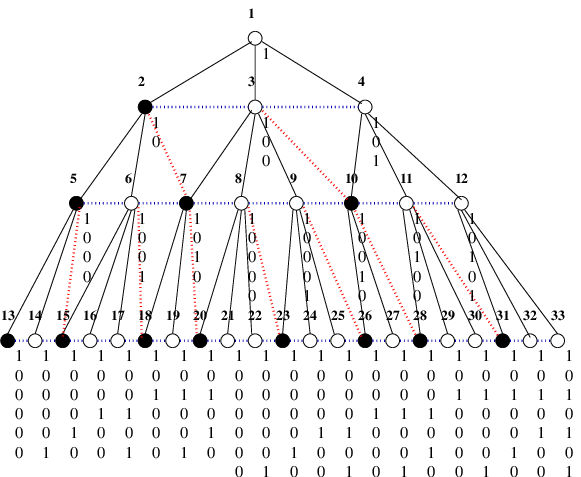,width=170pt}}
\vtop{
\ligne{\hfill
\PlacerEn {-335pt} {10pt} \box110
\PlacerEn {-185pt} {0pt} \box112
}
\begin{fig}\label{eclate_73}
\leurre
On the left: seven sector around a central tile;
on the right: the representations of the numbers attached to the
nodes of the Fibonacci tree.
\end{fig}
}
\vskip 10pt

  One of the reasons to use this system of coordinates
is that from any cell, we can find out the coordinates of its
neighbours in linear time with respect to the coordinate of the cell.
Also in linear time from the coordinate of the cell, we can compute
the path which goes from the central cell to the cell. These properties
are established in \cite{mmASTC03,mmbook1} and they rely on a particular property
of the coordinates in the tree which allow to compute the coordinate of the 
father of a node in constant time from the coordinate of the node. In the paper,
the coordinate of a cell is of the form $\nu(\sigma)$ where $\sigma$
is the number of the sector where the cell is and $\nu$ is its number
in the Fibonacci tree which spans the sector.
   Now, as the system of coordinates is fixed, we can turn to
the application to the implementation of cellular automata on the
ternary heptagrid, we shall say {\bf heptagrid} for short.

\subsection{Cellular automata on the heptagrid}

   A cellular automaton on the heptagrid is defined by a {\bf local
transition function} which can be put in form of a table. Each
row of the table defines a {\bf rule} and the table has nine columns
numbered from~0 to~8, each entry of the table containing a state of the
automaton. On each row, column~0 contains the state of the cell
to which the rule applies. The rule applies because columns~1 to~7
contain the states of the neighbours of the cell defined in the following
way. For the central cell, its neighbour~1 is fixed once and for all. 
For another cell, its neighbour~1 is its father. In all cases, the other 
neighbours are increasingly numbered from~2 to~7 while counter-clockwise turning 
around the cell starting from side~1. The representation mentioned in 
Subsection~3.1 allows to find the coordinates of
the neighbours from that of the coordinate of the cell in linear time.
The list of states on a row, from column~0 to~7 is called the {\bf context}
of a rule. It is required that two different rules have different contexts.
We say that the cellular automaton is {\bf deterministic}. As there is a
single row to which a rule can be applied to a given cell, the state of
column~8 defines the {\bf new state} of the cell. The local transition function
is the function which transforms the state of a cell into its new one, also
depending on the states of the neighbours as just mentioned.

   An important case in the study of cellular automata is what are called
{\bf rotation invariant} cellular automata. To define this notion,
we consider the following transformation on the rules. Say that
the context of a rule is the {\bf rotated image} of another one if and only
if both contexts have the same state in column~0 and if one context is obtained 
from the other by a {\bf circular} permutation on the contents of columns~1 to~7. 
Now, a cellular automaton is {\bf rotation invariant} if and only if its table of 
transition~$T$ possesses the following properties:

{\leftskip 20pt\parindent 0pt
- for each row~$\rho$ of~$T$, $T$ also contains six rules exactly
whose contexts are the rotated image of that of~$\rho$ and whose new state
is that of~$\rho$;

- if $\rho_1$ and~$\rho_2$ are two rules of~$T$ whose contexts are 
the rotated image of each other, then their column~8 contains the same state.
\par}

   In the rest of the paper, sometimes we shall have to write the rules
of the automaton for a precise situation. The rules can be written according to the 
following format:
\vskip 3pt
\ligne{\hfill 
$\eta_0$, $\eta_1$, $\eta_2$, $\eta_3$, $\eta_4$, $\eta_5$, $\eta_6$,
$\eta_7 \rightarrow \eta_0^1$,\hfill}
\vskip 2pt
\noindent 
where $\eta_0$ is the state of the cell,
$\eta_i$ the state of its neighbour~$i$ and~$\eta_0^1$ is its new state.

   However, in tables and also in order to have a more compact notation, a
rule will be written as a word. The above is rewritten as the following
word: 
$\underline{\eta_0}\eta_1\eta_2\eta_3\eta_4\eta_5\eta_6\eta_7
\underline{\eta_0^1}$, using the same notations.

   The name of rotation invariance comes from the fact that a rotation around
a tile~$T$ leaving the heptagrid globally invariant is characterized by a circular
permutation on the neighbours of~$T$ defined as above.

   Note that the universal cellular automata devised 
in~\cite{fhmmTCS,mmsyBristol,mmsyENTCS} are rotation invariant while the one 
of~\cite{mmkmTCS} is not. For the question of rotation invariance for cellular
automata on the heptagrid, we refer the reader to~\cite{mmACMC07}.
  
   Now, we can turn to the simulation of the railway circuit by a cellular
automaton.

\section{The implementation of the railway circuit}

   In~\cite{fhmmTCS}, the various elements of the circuit mentioned 
in~\cite{mmCSJMtrain} are implemented. In fact, the paper does not give an exact
description of the implementation: it only gives the guidelines, but with
enough details, so that an exact implementation is useless. In this paper,
we take the same model, and we repeat the main lines of implementation mentioned
in~\cite{fhmmTCS,mmsyENTCS}. So that we refer the reader to these papers for more
precise details. Just to help him/her to have a better of view of the overall
configuration, we refer the reader to Fig.~\ref{hypexample}. The figure provides
a simplified illustration of the implementation of the example given by 
Fig.~\ref{example}. 
  
   If the reader carefully looks at the figure, he/she will notice that
the tracks mostly follow branches of a Fibonacci tree and sets of nodes which
are on the same level of the Fibonacci tree. In this implementation, we have
to pay a very precise attention to this situation. We shall tune it a bit
with the help of an intermediary structure. As it will be used for the initial
configuration only, there is no need to translate this structure into the states
of the automaton.

\vskip 10pt
\setbox110=\hbox{\epsfig{file=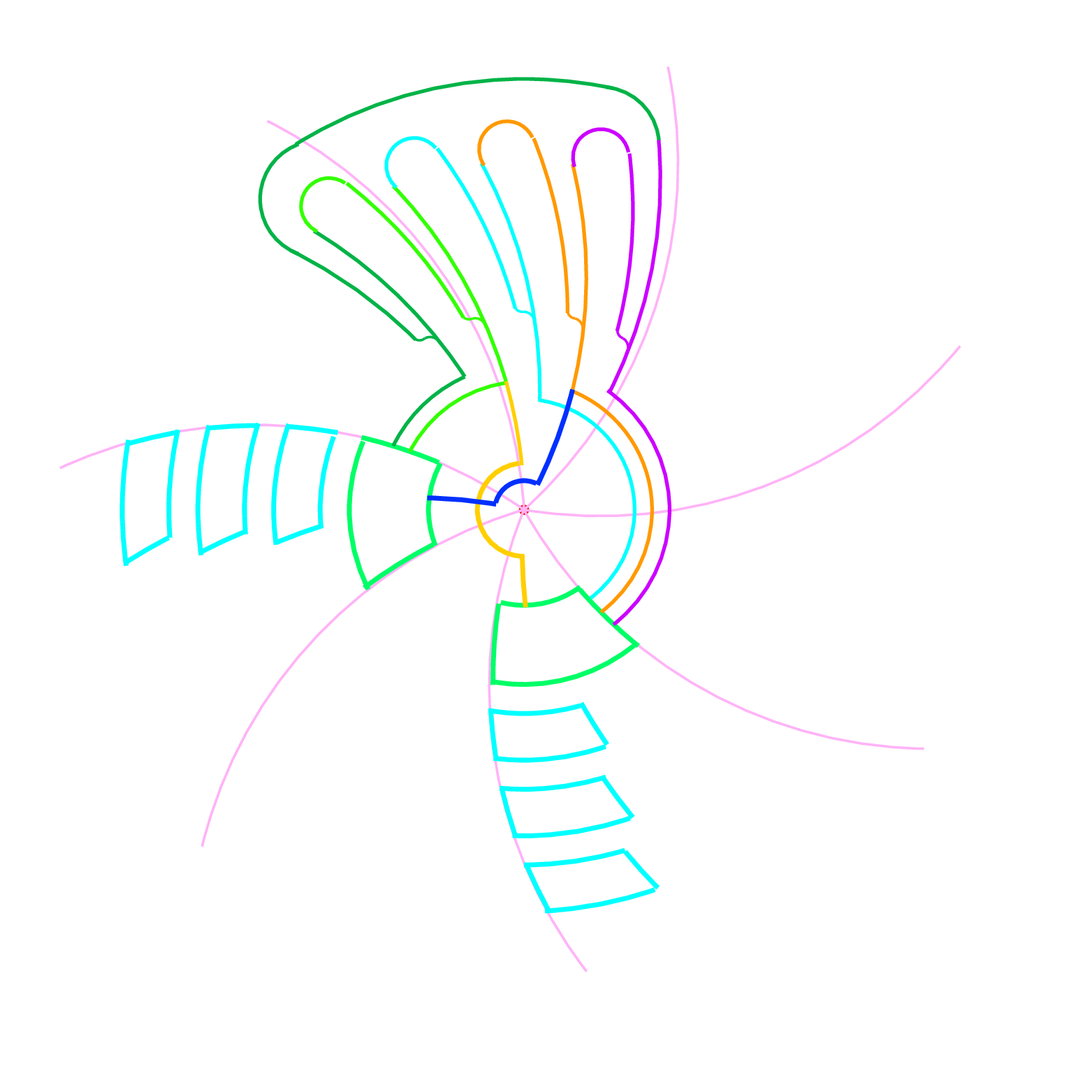,width=240pt}}
\vtop{
\ligne{\hfill
\PlacerEn {-285pt} {10pt} \box110
}
\vspace{-35pt}
\begin{fig}\label{hypexample}
\leurre
The implementation, on the heptagrid, of the example of Fig.~{\rm\ref{example}}.
In sector~$1$, also overlapping onto sector~$2$, the sequencing of the 
instructions of the program of the register machine. In sector~$3$, we can see
the first register and,  in sector~$5$, the second one. For simplicity, the 
figure represents two registers only.
\vskip 0pt
Note the instructions which arrive to the control of the register
through tracks in the shape of an arc of circle. Also note the return from the 
controller of the register when decrementing a register fails, because its content
was zero.
\end{fig}
}

\subsection{Verticals and horizontals}

   The intermediary structure which we shall use consists of a new tiling of the
heptagrid with the help of four colours only, green, blue, yellow and orange,
denoted $G$, $B$, $Y$ and~$O$ respectively.

   The colours allow to implement two kind of Fibonacci trees, studied 
in~\cite{mmJUCSii,mmbook1} with much detail. From this property, we shall derive
a way to precisely define what we later on call {\bf verticals} and {\bf horizontals}. 

   The colours are attached to rules which allow to define a family of uncountably many
tilings. The rules are the following~:

\vskip 5pt
\ligne{\hskip 40pt
$G \rightarrow YBG$,\hfill $Y \rightarrow YBG$,\hfill $O \rightarrow YBO$,
\hfill $B\rightarrow BO$\hskip 40pt}

   If $G$, $Y$ and~$O$ are identified to the white colour in a Fibonacci tree and~$B$
to the black colour, we have the display of a central Fibonacci tree, 
see~\cite{mmJUCSii,mmbook1}. If we identify $B$ and~$O$ with the white colour and
$G$ together with~$Y$ with the black colour, we get a standard Fibonacci tree, 
see~\cite{mmJUCSii,mmbook1}. In order to make the levels more clear in the trees,
we draw arcs on the tiles: a convex arc on the $Y$- and $G$-tiles, a concave arc
on the~$B$- and $O$-tiles. As the arcs join the mid-point of an edge to the mid-point of 
another edge, the arc is continued through the mid-point on the other tile. The arcs define
curves which never intersect and the curves go through any tile of the heptagrid.
We shall call these curves the {\bf isoclines}, following~\cite{mmBEATCS,mmTCShypundec} 
where they where introduced.
\vskip 15pt
\setbox110=\hbox{\epsfig{file=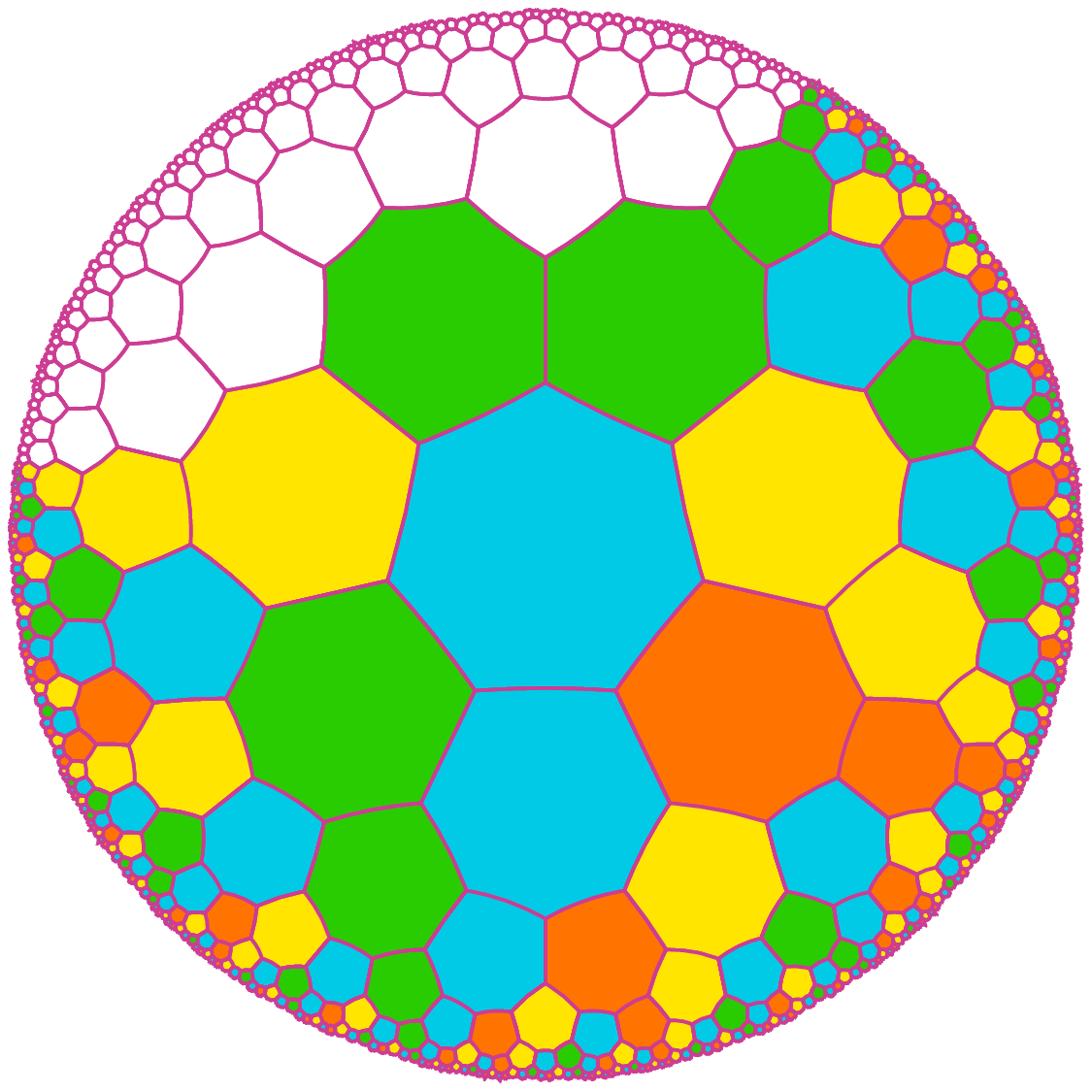,width=150pt}}
\setbox112=\hbox{\epsfig{file=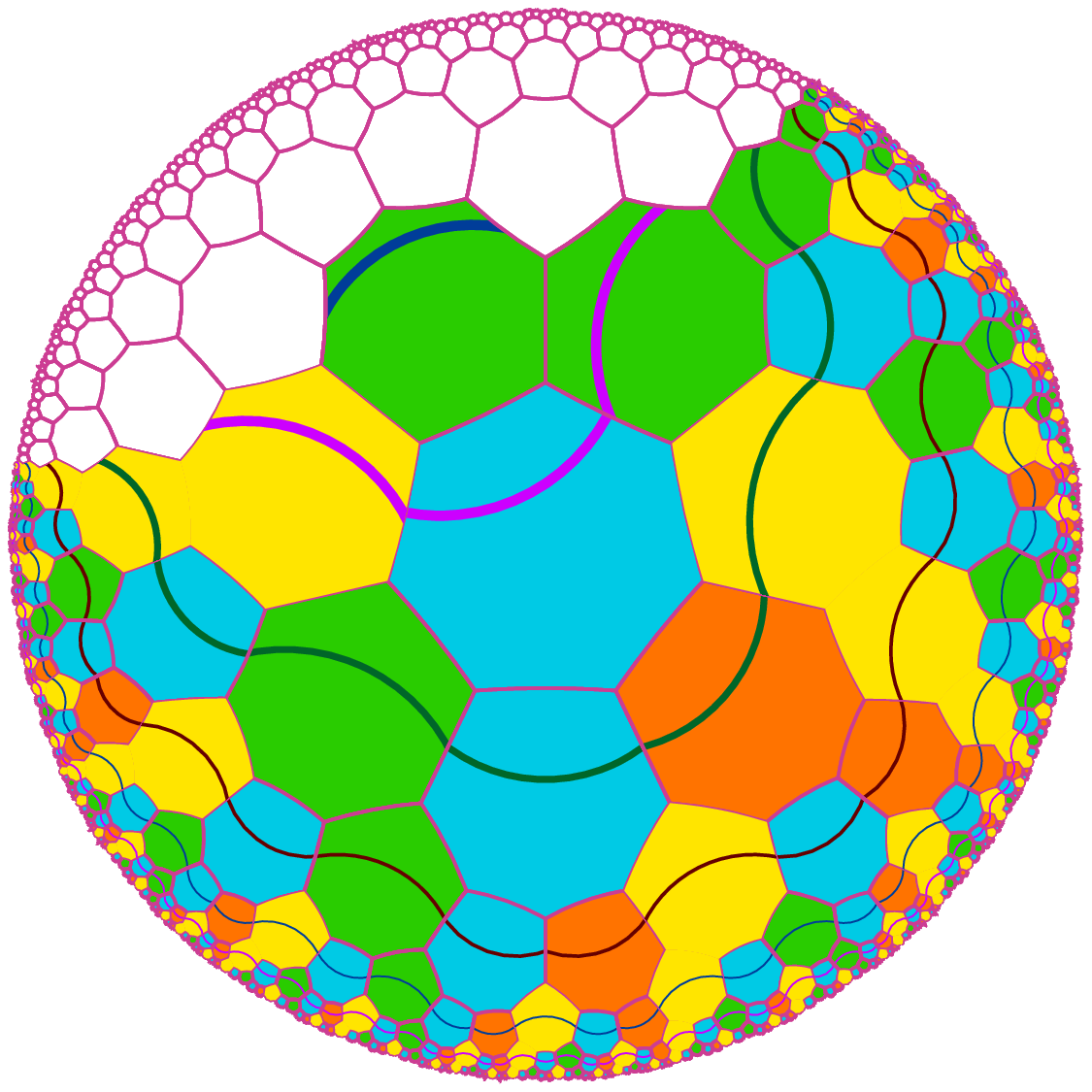,width=150pt}}
\vtop{
\ligne{\hfill
\PlacerEn {-335pt} {0pt} \box110
\PlacerEn {-165pt} {0pt} \box112
}
\begin{fig}\label{hortandvert}
\leurre
The definition of horizontals and verticals:
\vskip 0pt
On the left-hand side, the coloration which allows to define the {\bf isoclines},
which are drawn on the right-hand side.
The {\bf verticals} are represented by rays of yellow tiles. On the right-hand side picture,
note that the common side of adjacent yellow tiles is not drawn: rays of yellow tiles 
appear as solid blocks of tiles. 
\end{fig}
}
\vskip 10pt
   The isoclines have to be seen as the {\bf horizontals} which we need for the
tracks of our circuit. The {\bf verticals} are constituted by the rays of yellow cells.
It can be noticed that these rays intersect all the isoclines. We have to notice that
in general, we do not have a full line of yellow tiles. From Fig.~\ref{hortandvert},
we can see that once the colour a tile~$A$ is fixed, the colour of the tiles in the tree
rooted at~$A$ are also fixed. But for the tiles which belong to a tree which contains
the tree rooted at~$A$, their colour depend on the colour of the root of this new tree.
This is the reason why there is a family of uncountably many different tilings of the 
heptagrid with the colouring defined by the above rules. Now, in some of these tilings,
there may be a unique line of yellow tiles and there are, of course, infinitely many rays
of yellow tiles. But in the other realizations of the heptagrid with this colouring,
there are only rays of yellow tiles. We may assume that we are in a realization
in which there are only rays of yellow tiles.

   It is important to notice that the fact that our verticals are rays only does not
prevent them from constituting a {\bf grid} with the horizontals we defined. We have
the fact that in between two verticals starting from an isocline~$\iota$, new verticals 
appear as we go down from an isocline to the next one, starting from~$\iota$. We may 
ignore these new verticals if we do not need them. And so, these verticals with the piece
of~$\iota$ and a piece of another lower isocline at which we decide to stop constitute
a figure which we may call a {\bf quadrangle}. These quadrangles allow us to implement 
the pieces of circuitry described in Section~2. For the registers, it is enough
to display such quadrangles in such a way that the quadrangles have a side along the
same yellow branch of a tree. This is enough to see that we can consider the setting
of Fig.~\ref{hypexample} as enough for our purpose.

\subsection{The implementation of the tracks}

   From now on, we call tiles {\bf cells}. Most cells are in a quiescent state
which we also call the {\bf blank}. In the following figures of the paper,
it is represented by a light blue colour. In our setting we have three other
colours: blue, green and red. Also, cells are said to be neighbours if and only
if they share a side.

   In the setting of this paper, we have an important difference in the 
implementation of the railway circuit with respect to what was done in the previous 
works of~\cite{fhmmTCS,mmsyBristol,mmsyPPL,mmbook2,mmsyENTCS}. In these quoted papers,
the tracks consists of sets of blue cells which have exactly two blue neighbours.
The intersection of tracks has a specific colour in~\cite{fhmmTCS} which entails a rather
big number of states while in~\cite{mmsyBristol,mmsyPPL,mmbook2,mmsyENTCS}, the intersection
is a blue cell too. In these papers, the locomotive is a set of two neighbouring cells, 
a green one and a red one, which successively replace two cells of the track. 

   Here, we obtain the reduction of the number of states at the price of a more complex
structure of the tracks and, consequently, of the crossings and the switches. The main
reason is that we replace the green cell of the locomotive with a blue one.
As the track make use of blue cells also, the simple structure of the tracks 
in~\cite{mmsyBristol,mmsyPPL,mmbook2,mmsyENTCS} cannot be used here. 

   Call {\bf elementary track} a sequence~$S$ of cells such that any cell belonging
to~$S$ has exactly two neighbours which also belong to~$S$. In our setting, a {\bf path} 
consists of four elementary tracks~$S_1$, $S_2$, $S_3$ and~$S_4$ such that
$S_1$ with~$S_2$, $S_2$ with~$S_3$ and $S_3$ with~$S_4$ have a common border which is
a sequence of sides which belong to both tiles of the considered pair of elementary tracks.
Moreover, $S_1$ is green, $S_2$ is blue and both $S_3$ and~$S_4$ are blank. Now, to
distinguish $S_3$ from~$S_4$, some cells of~$S_4$ are blue, defining what we shall later
call {\bf milestones} and they are more or less regularly dispatched on~$S_4$. Later,
we give the precise rule for placing the milestones. The locomotive moves on~$S_3$ 
and it is represented by two contiguous cells, a blue one, the {\bf front}, and a 
red one, the {\bf rear}.  The motion of the locomotive on~$S_3$ is exactly that of the 
locomotive on a track in~\cite{fhmmTCS,mmsyBristol,mmbook2,mmsyENTCS}. Below, 
Fig.~\ref{mouvement}, Fig.~\ref{pathhoriz} and Fig.~\ref{pathvert} illustrate this motion, 
the motion on~$S_3$ only being illustrated by Fig.~\ref{mouvement}. Later, we shall
call $S_1$, $S_2$, $S_3$ and~$S_4$, the {\bf green}, {\bf blue}, {\bf proper} and 
{\bf safeguard} tracks respectively.

\vskip 10pt
\setbox110=\hbox{\epsfig{file=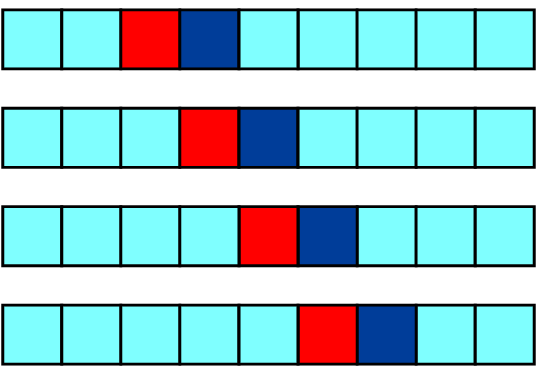,width=120pt}}
\vtop{
\ligne{\hfill
\PlacerEn {-265pt} {0pt} \box110
}
\vskip-15pt
\begin{fig}\label{mouvement}
The motion of the locomotive on~$S_3$: first approximation.
\end{fig}
}
\vskip 10pt
   The tracks of the railway circuit of Section~2 are implemented by portions of 
paths between two switches. 

   In between switches, the path either follows a vertical or a horizontal.

   Fig.~\ref{mouvement} is a space-time diagram of the evolution of the
states of the cells which are along an elementary track~$S_3$. The easy rules are given
here:

\vskip 5pt
\ligne{\hskip 20pt
\ttxii
B\ W\ W\ $\rightarrow$\ B,\hfill
R\ B\ W\ $\rightarrow$\ R,\hfill
W\ R\ B\ $\rightarrow$\ W,\hfill
W\ W\ R\ $\rightarrow$\ W\hskip 20pt
}   
\vskip 5pt
   Note that, as we wish to get rules which are invariant by rotation, the same rules
can be used for the motion on~$S_3$ in the opposite direction.

   In this setting, there are much more rules to describe the motion of the
locomotive than in papers~\cite{fhmmTCS,mmsyBristol,mmbook2,mmsyENTCS}. 
The green cells of~$S_1$ have simple rules: they remain green and they have
two green neighbours. These green neighbours define a partition of the tile by considering
the line~$\ell$ which joins the mid-points of the sides shared with a green neighbour.
On one side of~$\ell$, the neighbours are blank and on the other side, they are blue.
The number of blue cells depends on the shape of~$S_1$ at this tile. 

\vskip 5pt
\setbox110=\hbox{\epsfig{file=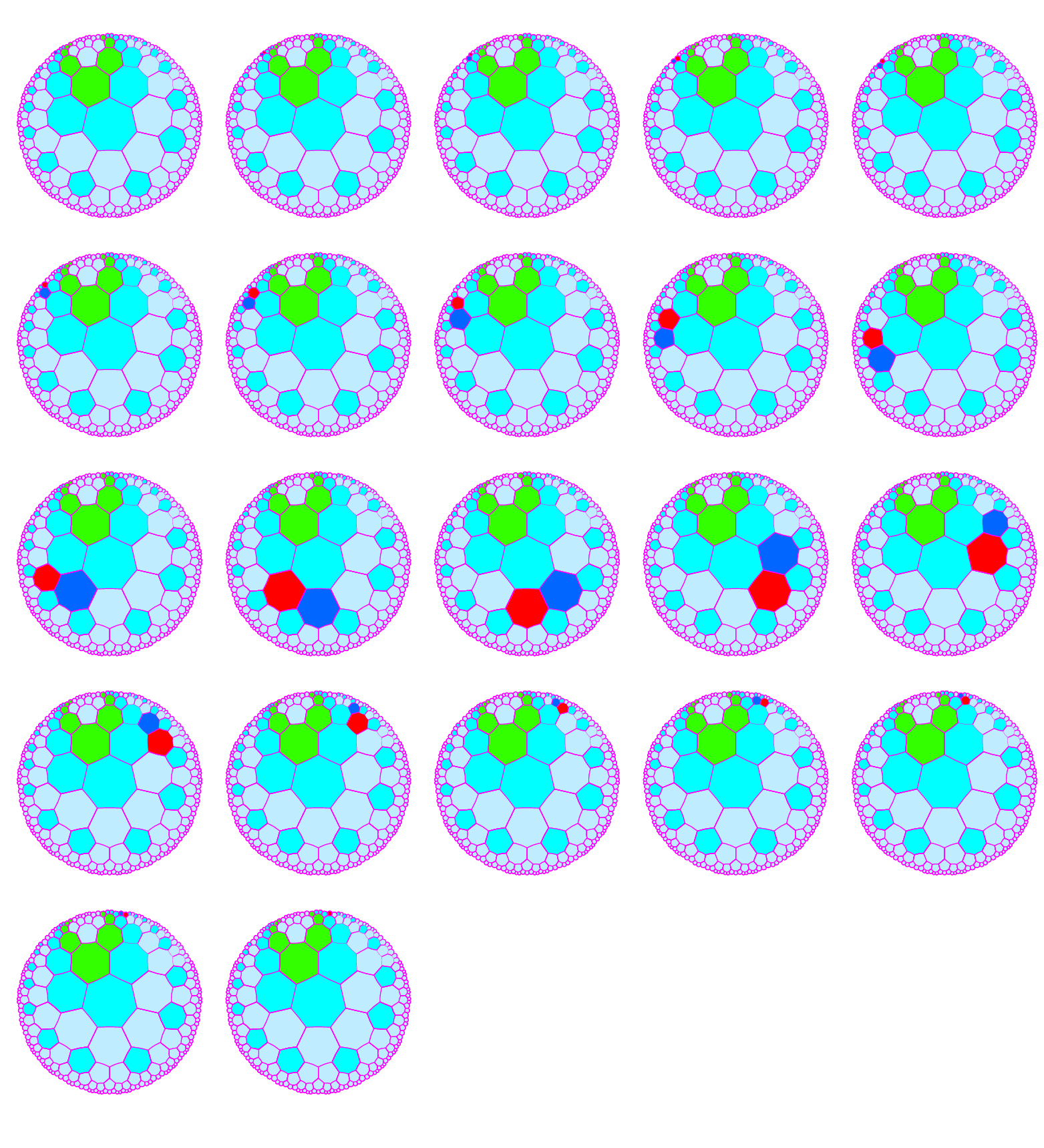,width=350pt}}
\vtop{
\ligne{\hfill
\PlacerEn {-345pt} {0pt} \box110
}
\vspace{-15pt}
\begin{fig}\label{pathhoriz}
The motion of the locomotive on its tracks: here the path follows an isocline.
\end{fig}
}
\vskip 5pt
The blue cells of~$S_2$ have more complex rules. When no locomotive is
in contact with the cell, the rules are similar to those for a green cell of~$S_1$:
they have two blue neighbours and on one side they have green neighbours and blank ones on
the other side. Now, when a locomotive is present in the neighbourhood of the cell,
at least by one cell, the front or the rear, we have new rules which take into account this
situation. This means that we have an additional blue neighbour on the side of the blank ones
and, possibly a red one too. Also, there may be a red neighbour in the side of the blank 
ones. The blank cells of~$S_3$ also have similar rules when the locomotive is not
present. When it is present, the locomotive is in a neighbouring cell, by its front or its
rear and, in this case, a single neighbour of the cell is occupied by the locomotive.
If it is the front, the cell becomes blue at the next time. 

   Fig.~\ref{pathhoriz} shows the structure of a path along an isocline and the 
behaviour of the locomotive on it. We notice that $S_1$, $S_2$, $S_3$ and~$S_4$ follow
consecutive isoclines, the isocline of~$S_i$ being above that of~$S_{i+1}$ for 
$i\in\{1..3\}$. Note that on~$S_4$, the milestones are placed on the tiles which have two
neighbours on~$S_3$. Note that the other cells of~$S_4$ have one neighbour on~$S_3$ 
exactly. Note that in order the locomotive can be better indentified, its front is on 
a darker blue than the colour of the blue track. This is not a new state, this is for the
conveniency of the reader only.
 
\setbox110=\hbox{\epsfig{file=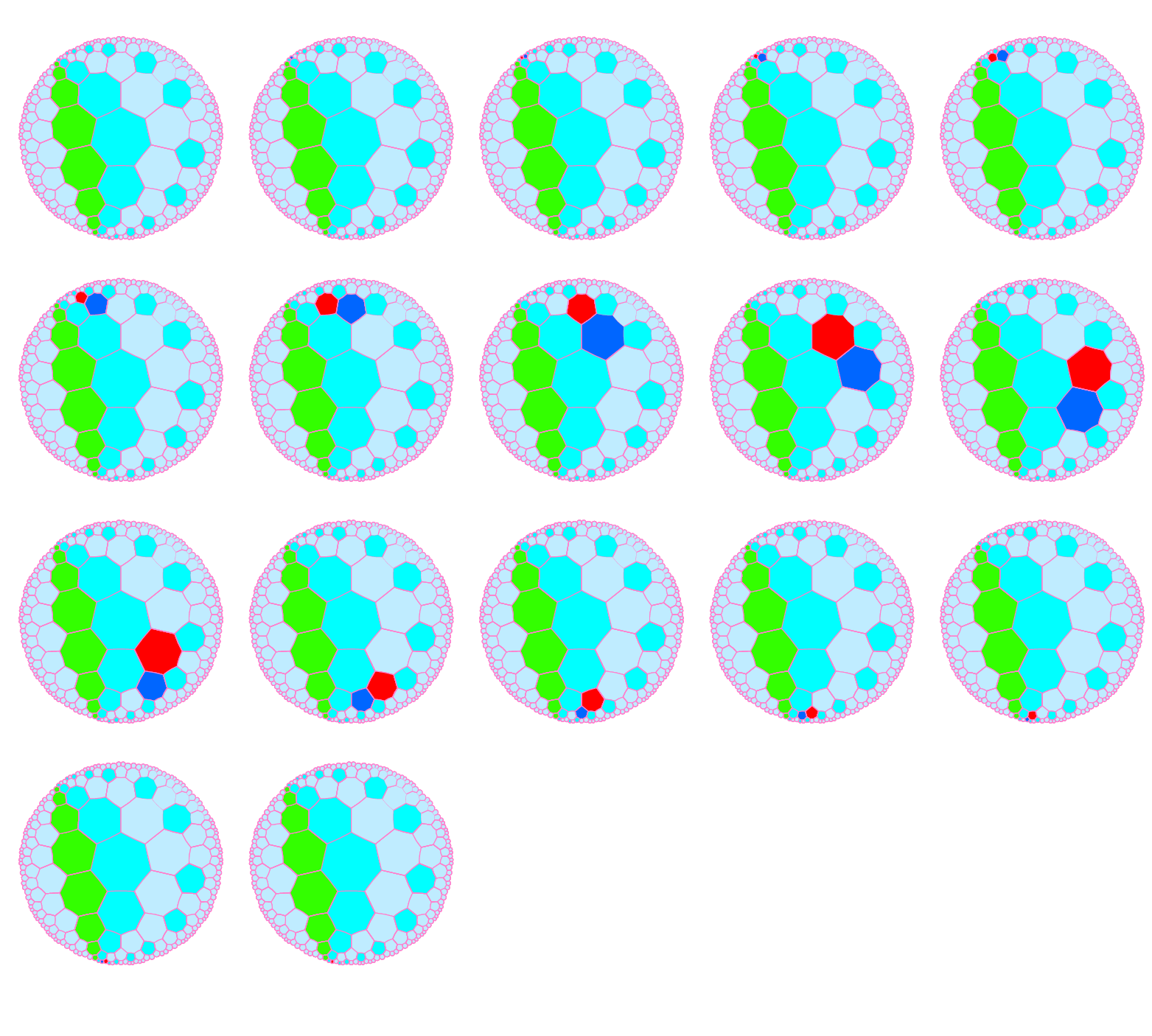,width=350pt}}
\vtop{
\ligne{\hfill
\PlacerEn {-345pt} {0pt} \box110
}
\vspace{-20pt}
\begin{fig}\label{pathvert}
\leurre
The motion of the locomotive along a path which follows a vertical. Note that
the blue track is exactly on the vertical.
\end{fig}
}

   Fig.~\ref{pathvert} illustrates the motion along a vertical. As easily seen from
the pictures of the figure, the blue track of the path follows the vertical. The green
track corresponds to a border of a tree, adjacent to the one whose left-hand side border
supports the blue track. The proper track follows a more complex pattern which is determined
by the edges of the blue track which are not shared by two cells of the track. The way
followed by the safeguard track is still more complex. It has the property that blue cells
of the safeguard track have two neighbours of the proper track and the blank cells of
the safeguard have one neighbours exactly the proper track.

   In this figure as well as in all the further figures of the paper, the front of the
locomotive is represented by a darker colour than it should be, in order the reader could
more easily identify it.

   In the Appendix, Fig.~\ref{turnA}, Fig.~\ref{turnB}, Fig.~\ref{turnC} and Fig.~\ref{turnD}
illustrate how the turn is performed. It makes use of a {\bf slip road} which allows
to go from one kind of path to the other. 

   To understand the structure of the slip road, remember that we called
{\bf flower} a ball of radius~1{} in the heptagrid: a central tile, the {\bf centre}
of the flower, and its seven {\bf petals}: the neighbours of the centre,
see~\cite{mmBEATCS,mmTCShypundec}. In such a flower, we distinguish three tiles 
among the petals. They are the milestones of the path. In the flower, the milestones
define an isosceles triangle. The proper track always crosses the equal legs of
such a triangle, never the basis. The flowers intersect each other in such a way
that the locomotive enter through one short leg of the triangle and exits through the
other short leg. It can be seen that in these conditions, the path is reduced to
three tracks: the proper track which zig-zags around milestones and two safeguard tracks
which are constituted, in each flower, of the two tiles defining the basis of the triangle.
There are no blue nor green tracks in a slip road. This structure of a slip road
is illustrated by Fig.~\ref{explain_sliproad}. 

\setbox110=\hbox{\epsfig{file=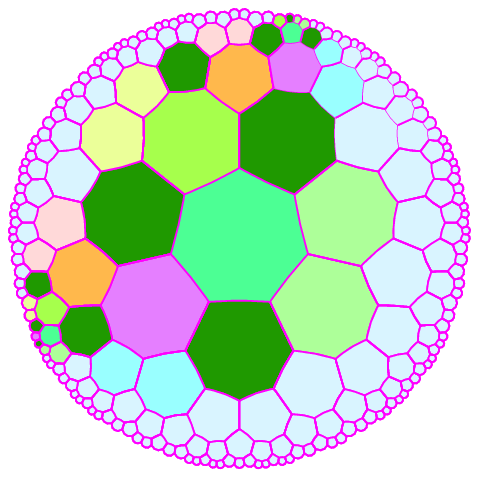,width=200pt}}
\vtop{
\ligne{\hfill
\PlacerEn {-265pt} {0pt} \box110
}
\vspace{-45pt}
\begin{fig}\label{explain_sliproad}
\leurre
The structure of a slip road. The milestones, in dark green on the figure, are
the vertices of isosceles triangles inscribed in a flower. 
\vskip 0pt
The safeguard paths consists of the milestones and the very light coloured tiles. 
The tiles of the proper track are in light colours, a bit darker than those of the
safeguard track, the milestones excepted.
\vskip 0pt
Note how the proper track 
zig-zags around the milestones. Also note that, entering a triangle through a short
leg, the proper path exits through the other short leg.

\end{fig}
}

   It is not difficult to see that the slip road structure could replace the
vertical paths. It would be enough to place the milestones at appropriate places, alternating
on both sides of a vertical. This can be seen on Fig~\ref{sliproad} which illustrates
the motion of a locomotive along such a path. In order to facilitate the interpretation
of the figure, the blue colour of the locomotive is darker than the blue colour of the
milestones. Of course, in the implementation, both colours are the same, as well as
in the computer program used to check the correctness of the rules.

   The high interest of the slip road structure is its flexibility. The same rules
allow to produce the motion of the locomotive along various routes. This makes it possible
to go from a vertical to a horizontal or conversely.

\vskip 10pt 
\setbox110=\hbox{\epsfig{file=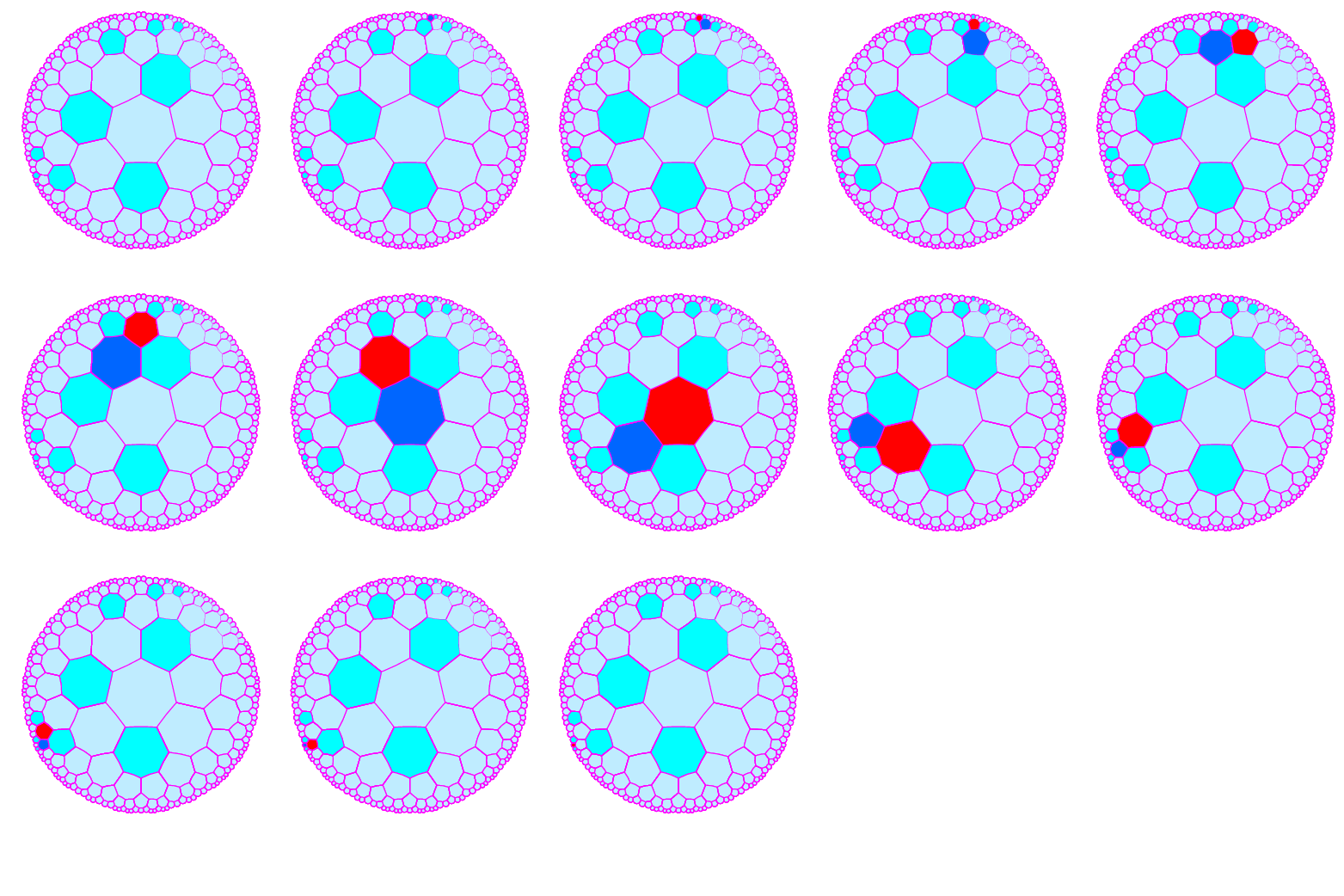,width=350pt}}
\vtop{
\ligne{\hfill
\PlacerEn {-345pt} {0pt} \box110
}
\vspace{-15pt}
\begin{fig}\label{sliproad}
An example of a slip road. 
Note the darker colour of the front of the locomotive, for illustration only.
\end{fig}
}
\vskip 10pt
   With the flexibility of the slip roads at our disposal, we can implement the connection
between a path which follows a vertical to a path following an isocline which intersects
the vertical. This allows the locomotive to perform quadrangular motions.
As the configuration of the intersection of an isocline with a vertical is not symmetric, 
we have four cases, one for each corner of the quadrangle. In each case, there are two 
sub-cases, depending on the direction of the motion along the track. In the Appendix, 
Fig.~\ref{turnA}, Fig.~\ref{turnB}, Fig.~\ref{turnC} and 
Fig.~\ref{turnD} illustrate the four cases in the same direction of the motion.
\vskip 10pt
   Note that in this situation, the motion of the locomotive requires a higher number of
rules than in the previous settings. In particular, the cells of~$S_4$ have to behave
differently. Those which are blue have blank neighbours only when the locomotive is
not present. When it is present they have either one or two non-blank neighbour.
If it has a single one it may be blue or red. If it has two neighbours, there is a blue
and a red ones. In particular, these blue and red neighbours are always contiguous.

   When the locomotive is not present, each blank cell of~$S_4$ has one or two 
blue neighbours which are also on~$S_4$: they are milestones. When the locomotive
is on its proper track, close to a cell of~$S_4$ this makes a difference for the cell
of~$S_4$: it has one or two additional non-blank neighbours. It may be a blue or a red one
if there is a single additional non-blank neighbours. Otherwise, the additional coloured
two neighbours are blue and red. In these situation, the milestones and the blank cell
of~$S_4$ must remain the same. Now, when a blank cell of the proper track has the
front of the locomotive as a neighbour, it has at least three contiguous blue neighbours, 
a situation which cannot occur for a blank cell of the safeguard track.

   There are two series of basic patterns for the rules of the cells which lie on
the proper track, which appear in Table~\ref{basic_rules}.

\def\uneregle #1 #2 #3 #4 #5 #6 #7 #8 #9 {   
\hbox{\ttxii $\underline{\hbox{\ttxii#1}}$#2#3#4#5#6#7#8$\underline{\hbox{\ttxii#9}}$}  
}

\vtop{
\begin{tab}\label{basic_rules}
The basic rules of the motion of the locomotive for the cells of a proper track.
\end{tab}
\vspace{-16pt}
\grostrait
\vskip 5pt
\ligne{\hfill
\uneregle W B B W B W B W W \hfill}
\ligne{\hskip 40pt
\uneregle W B B W B W B B B \hfill
\uneregle B B B W B W B R R \hfill
\uneregle R B B B B W B W W \hfill
\uneregle W B B R B W B W W \hskip 40pt}
\ligne{\hskip 40pt
\uneregle W B B B B W B W B \hfill
\uneregle B B B R B W B W R \hfill
\uneregle R B B W B W B B W \hfill
\uneregle W B B W B W B R W \hskip 40pt}
\vspace{-5pt}
\demitrait
\ligne{\hfill
\uneregle W B W B W W B W W \hfill}
\ligne{\hskip 40pt
\uneregle W B W B W W B B B \hfill
\uneregle B B W B W W B R R \hfill
\uneregle R B B B W W B W W \hfill
\uneregle W B R B W W B W W \hskip 40pt}
\ligne{\hskip 40pt
\uneregle W B B B W W B W B \hfill
\uneregle B B W B W W B R R \hfill
\uneregle R B W B W W B B W \hfill
\uneregle W B W B W W B R W \hskip 40pt}
\vspace{-5pt}
\demitrait
\vskip 5pt
}
\vskip 10pt
   The table has two parts which correspond to the two possible configurations for
a blank cell of the proper track. Such a cell may have one or two neighbours belonging
to the blue track. The upper half of the table deals with the case of two neighbours
while the lower half deals with the case of a single neighbour.

   In each half-table, the first row indicates the rule when the locomotive is
not nearby. We can see that the state of the cell is unchanged. We say that it is
a {\bf conservative} rule. The second row indicates the four rules corresponding to
the passage of the locomotive. The last row indicates the rules corresponding to
a passage in the opposite direction.

   Note that the rules are assumed to be rotation invariant. We can also notice that
the the patterns are different because the number of letters and their respective position
up to a circular permutation are different. It is not difficult to see that these
rules also apply for the proper track of a slip road, whatever the association of the
triangles defined by the milestones of each flower constituting the slip road.

\subsection{The crossings}   
  
   Now, we are in the position to study the crossing of two paths. This makes in fact
four half-paths meeting at the centre of the intersection. 

   Thanks to the slip roads, we always can assume that the paths meet according to the
configuration which is indicated by Fig.~\ref{idle_crossing}.

   In the papers~\cite{mmsyBristol,mmsyPPL,mmsyENTCS}, the colour of the centre 
of the intersection is the same as the colour of the path. Here, as the path has a complex
structure, we define the centre of the intersection as the blank cell which is the 
intersection of all the proper tracks belonging to the meeting paths.

   The two paths are distinguished by a green cell which identifies one of them.
We shall later speak of the green path for the path marked by the green cell. The other
path will be called the blue path.

    Two green cells are neighbours of the centre. The other neighbours are blank cells.
Four of them belong to the arriving proper tracks. The remaining blank cell takes no part 
in the motion of the locomotive. Call this cell the {\bf idle cell}. The blank neighbours 
of the centre which belong to a proper track are called the {\bf first cells} of the 
corresponding half-path.

\vskip-30pt 
\setbox110=\hbox{\epsfig{file=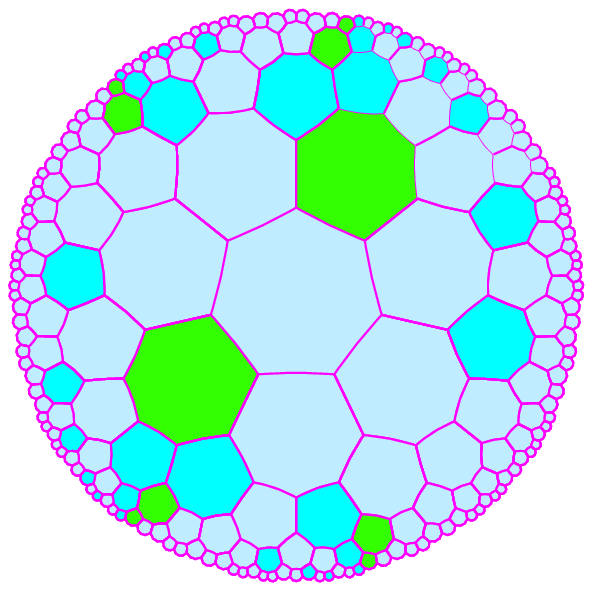,width=220pt}}
\vtop{
\ligne{\hfill
\PlacerEn {-280pt} {0pt} \box110
}
\vspace{-5pt}
\begin{fig}\label{idle_crossing}
\leurre
The idle configuration at the crossing of two paths. The milestones allow to identify each
half-path. The green cells close to the centre of the intersection identify two
half-paths which constitute one path of the locomotive. The other is marked by the
missing of the green cells.
\end{fig}
}

\vskip-15pt
\setbox110=\hbox{\epsfig{file=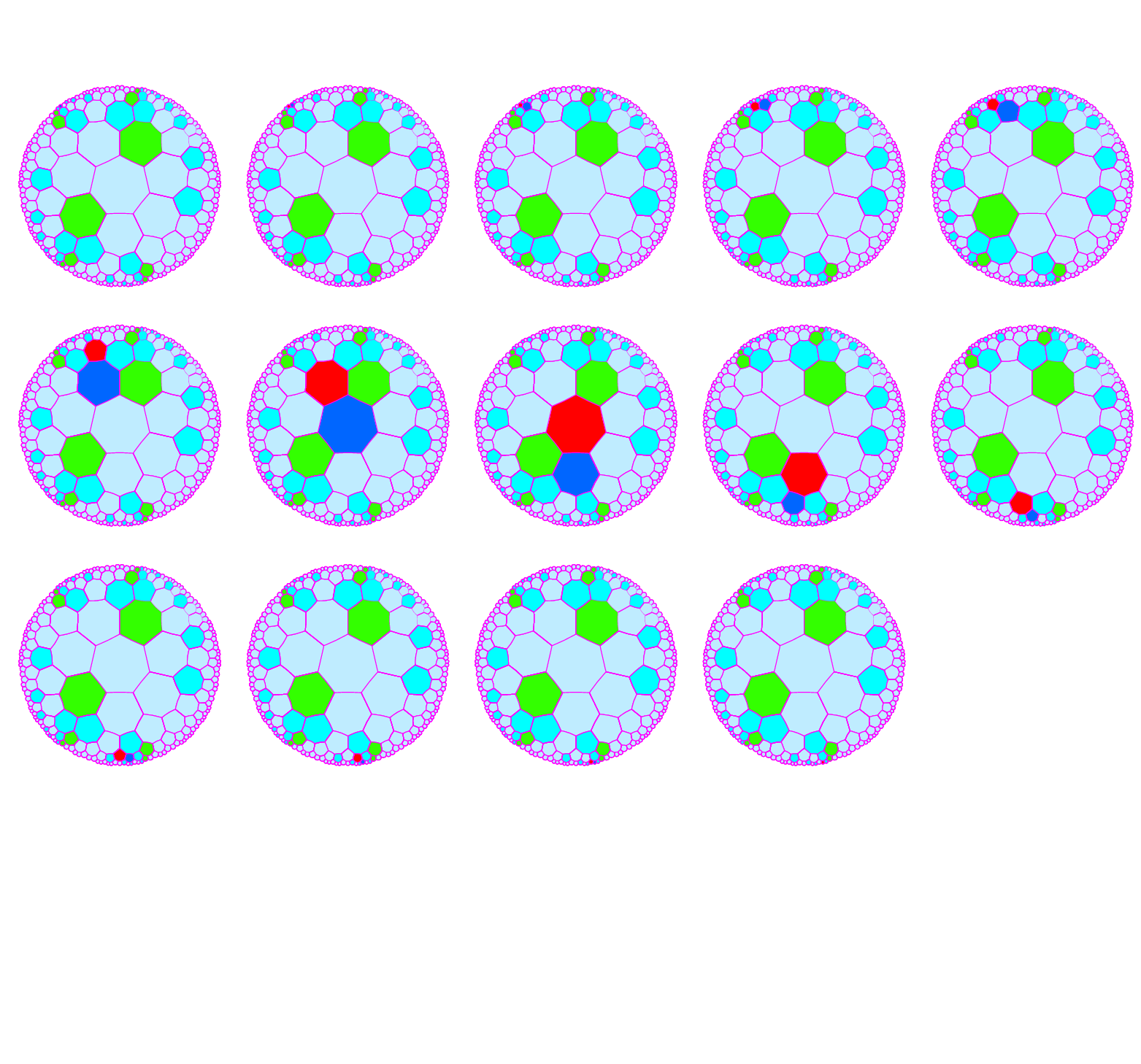,width=340pt}}
\vtop{
\ligne{\hfill
\PlacerEn {-346pt} {0pt} \box110
}
\vspace{-85pt}
\begin{fig}\label{croisement_1}
\leurre
The locomotive goes through a crossing: here from sector~$1$ to sector~$4$.
Here, for illustration only, the front of the locomotive is dark blue.
\end{fig}
}
\vspace{-50pt}

    The motion of the locomotive on the blue path is conformal to the rules of 
Table~\ref{basic_rules}, see Fig.~\ref{croisement_1} which illustrates such a motion. 
Only the rules for the first cells and the centre are different.
Also, the rules for the idle cell are different. Note that the idle cell remains blank
but it witnesses the motion of the locomotive which passes nearby. The three consecutive 
blank neighbours of the idle cell allow to identify it and to prevent the change of its
state. 

The configurations of the first cells and of the centre are also different from
the configurations of cells of a proper track. They all have a green cell among their 
neighbours. Moreover, the centre has two green neighbours exactly.
\vskip 5pt
   The motion of the locomotive along the green path is different from that on the blue
one. The difference lies in the following. When the locomotive sees the green cell for the
first time, it is on the cell~$c$ of the proper path which has the green cell as neighbour.
Adjacent to the green cell, a blue cell of the blue track is also a neighbour of~$c$.
When the locomotive arrives at~$c$, its front becomes green. We call~$c$ the 
{\bf green trigger}. There is another green trigger on the other proper track whose first
cell is a neighbour of the other green cell of the intersection. This allows to adapt the
rules for crossing the centre and to take the proper track identified by the other
green cell. The front of the locomotive remains green until it reaches the other green
trigger. When it leaves it, its front becomes blue again.   

   This particular motion is illustrated by Fig.~\ref{croisement_7}. 

\vskip-15pt
\setbox110=\hbox{\epsfig{file=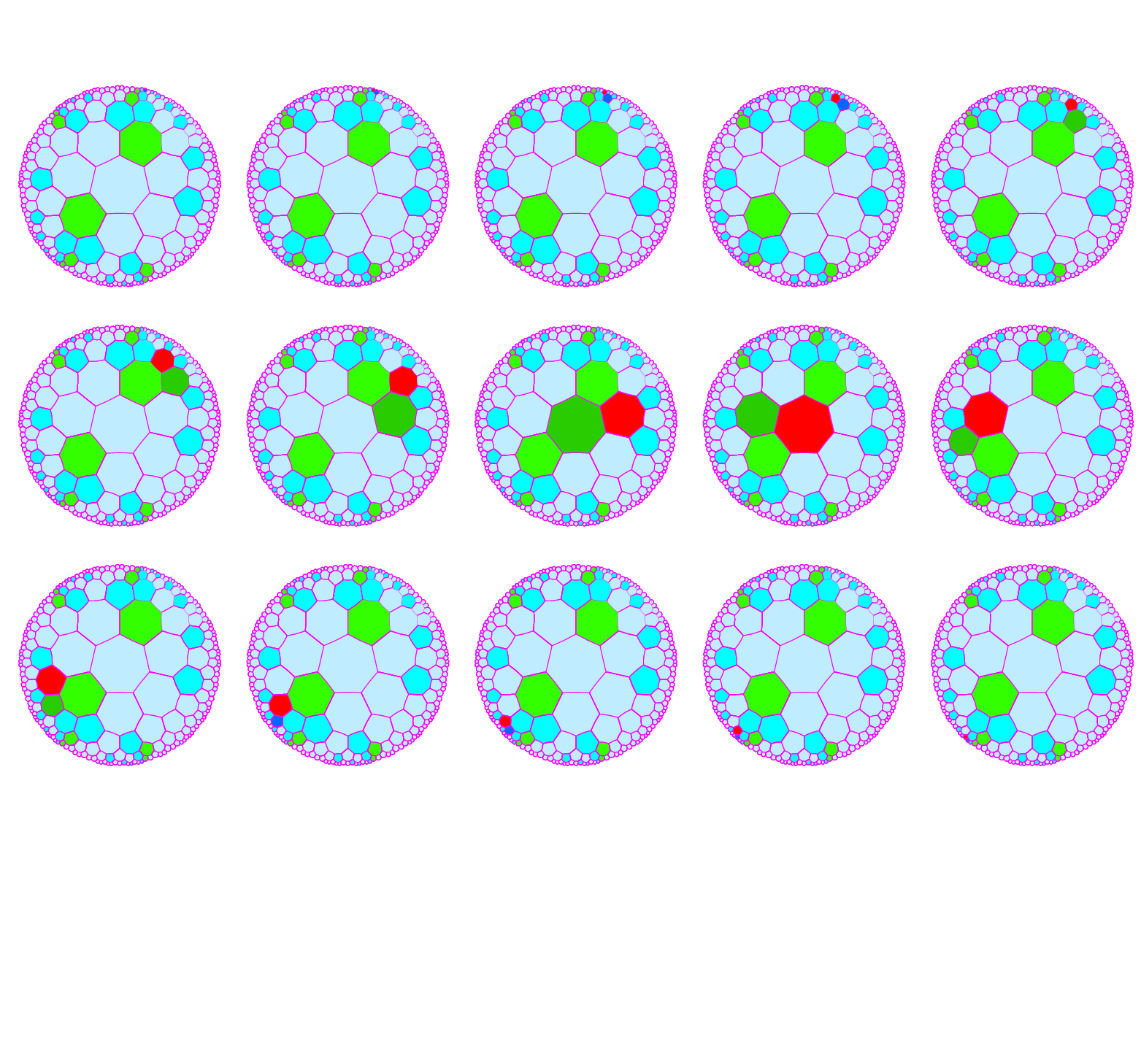,width=340pt}}
\vtop{
\ligne{\hfill
\PlacerEn {-346pt} {0pt} \box110
}
\vspace{-85pt}
\begin{fig}\label{croisement_7}
\leurre
The locomotive goes through a crossing: here from sector~$7$ to sector~$4$: the locomotive
arrives to the crossing from the other path, whose first cells are green. As the previous
figures, for illustration only, the front of the locomotive is darker than the
required colour.
\end{fig}
}
\vspace{-25pt}
   Note that Fig.~\ref{croisement_1} and~\ref{croisement_7} also illustrate the motion
of the locomotive when it arrives from the other half-paths. Indeed, if we except
the green cells, the configuration of each half-path is obtained from the other by a 
rotation by $\displaystyle{{2\pi}\over5}$ around the centre. The occurrence of the idle
cell does not introduce any perturbation on this regard.

\subsection{The switches}

   The switches are characterized by the fact that three paths arrive to a centre.
As mentioned in the introduction, when the locomotive goes to the {\bf arriving path} 
and then goes on through one of the two exiting ones, this is an active crossing of 
the switch. The path taken by the locomotive after the crossing is the {\bf selected path}. 
The other path is the {\bf non-selected} one. A passive crossing occurs when the 
locomotive arrives through the selected or the non-selected path and it always exits 
through the arriving path. As already mentioned, the flip-flop switch must always be crossed 
actively.

The centre of the switch is a blank cell and it is the intersection of the three proper
tracks which arrive to the switch. This is an important difference with the
previous mentioned papers as already noticed for the crossings.

   Below, Fig.~\ref{idle_switches} shows the different configurations when the
locomotive is not nearby. As noticed in~\cite{mmsyENTCS}, we may assume that
the fixed switches always send the locomotive to the left-hand side path. The other
situation can be obtained by an appropriate crossing of the non-selected path with the 
selected path which is put after the crossing of the switch.

   However, for the memory and the flip-flop switches, there is another configuration
corresponding to the other choice of the selected path. 

\vskip-20pt
\setbox112=\hbox{\epsfig{file=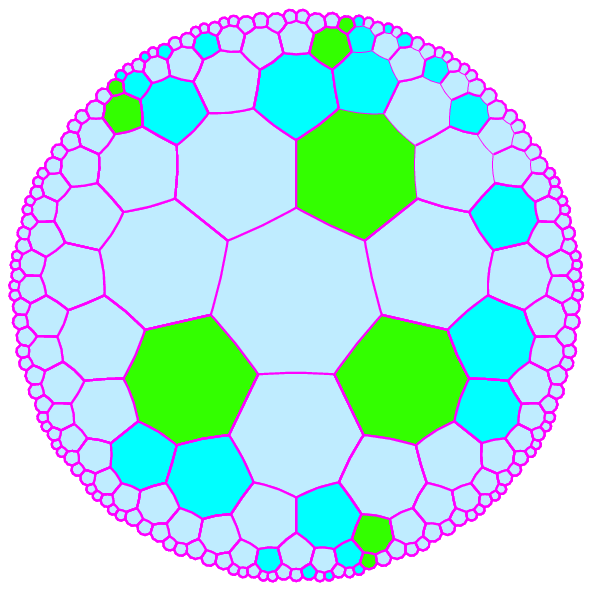,width=140pt}}
\setbox114=\hbox{\epsfig{file=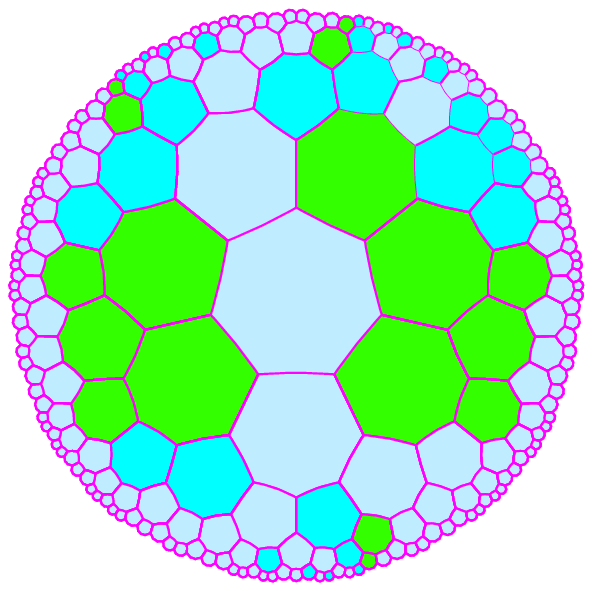,width=140pt}}
\setbox116=\hbox{\epsfig{file=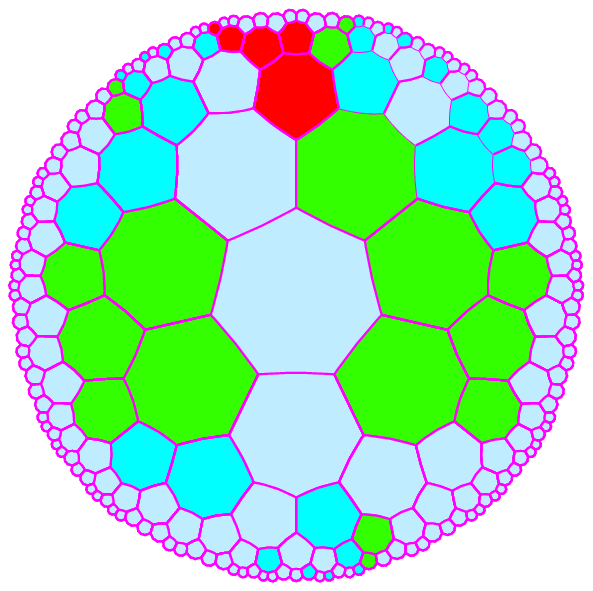,width=140pt}}
\setbox118=\hbox{\epsfig{file=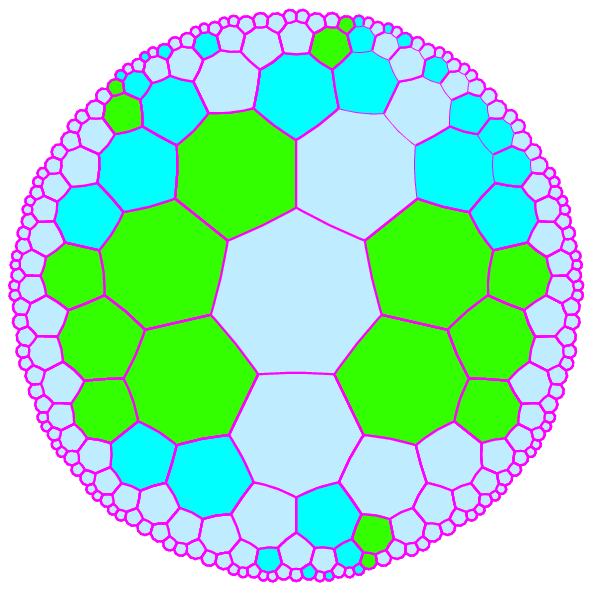,width=140pt}}
\setbox120=\hbox{\epsfig{file=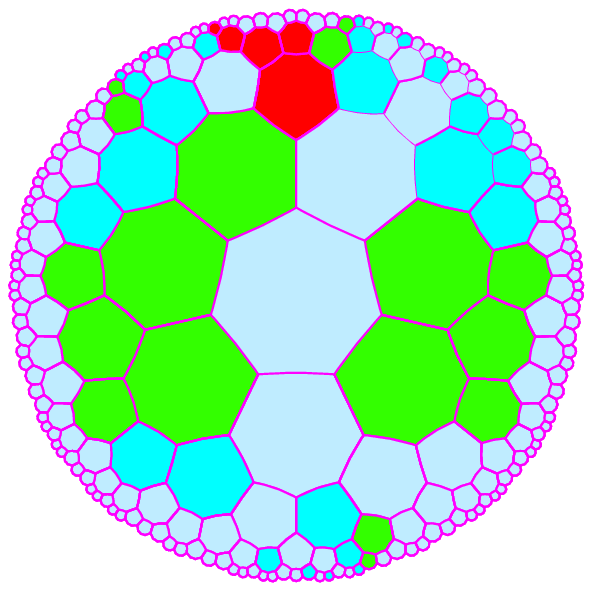,width=140pt}}
\vtop{
\ligne{\hfill
\PlacerEn {-350pt} {0pt} \box112
\PlacerEn {-240pt} {0pt} \box114
\PlacerEn {-130pt} {0pt} \box116
\PlacerEn {-295pt} {-120pt} \box118
\PlacerEn {-185pt} {-120pt} \box120
\PlacerEn {-292pt} {-7pt} {$(a)$}
\PlacerEn {-182pt} {-7pt} {$(b)$}
\PlacerEn {-72pt} {-7pt} {$(c)$}
\PlacerEn {-237pt} {-127pt} {$(d)$}
\PlacerEn {-127pt} {-127pt} {$(e)$}
}
\vskip-15pt
\begin{fig}\label{idle_switches}
\leurre
The idle configurations for switches. Upper row: from left to right,
fixed, memory and flip-flop switches. Lower row, right-hand side versions of
memory and flip-flop switches.
\end{fig}
}
\vskip 10pt
   Now, we shall consider that the locomotive arrives at a switch. Each case is
examined in an appropriate Sub-subsection.

\subsubsection{Fixed switches}

   The working of the switch is mainly illustrated by Fig.~\ref{fixe_7}. In this
figure, the locomotive arrives through the non-selected path. We can see that it
goes out through the arriving path and that the selected path remains the same as before
the passage of the locomotive.

   For completeness, we also give the illustrations for the other crossings:
in Fig.~\ref{fixe_1} for the other passive crossing, in Fig.~\ref{fixe_4} for the
active crossing.

\vskip-15pt
\setbox110=\hbox{\epsfig{file=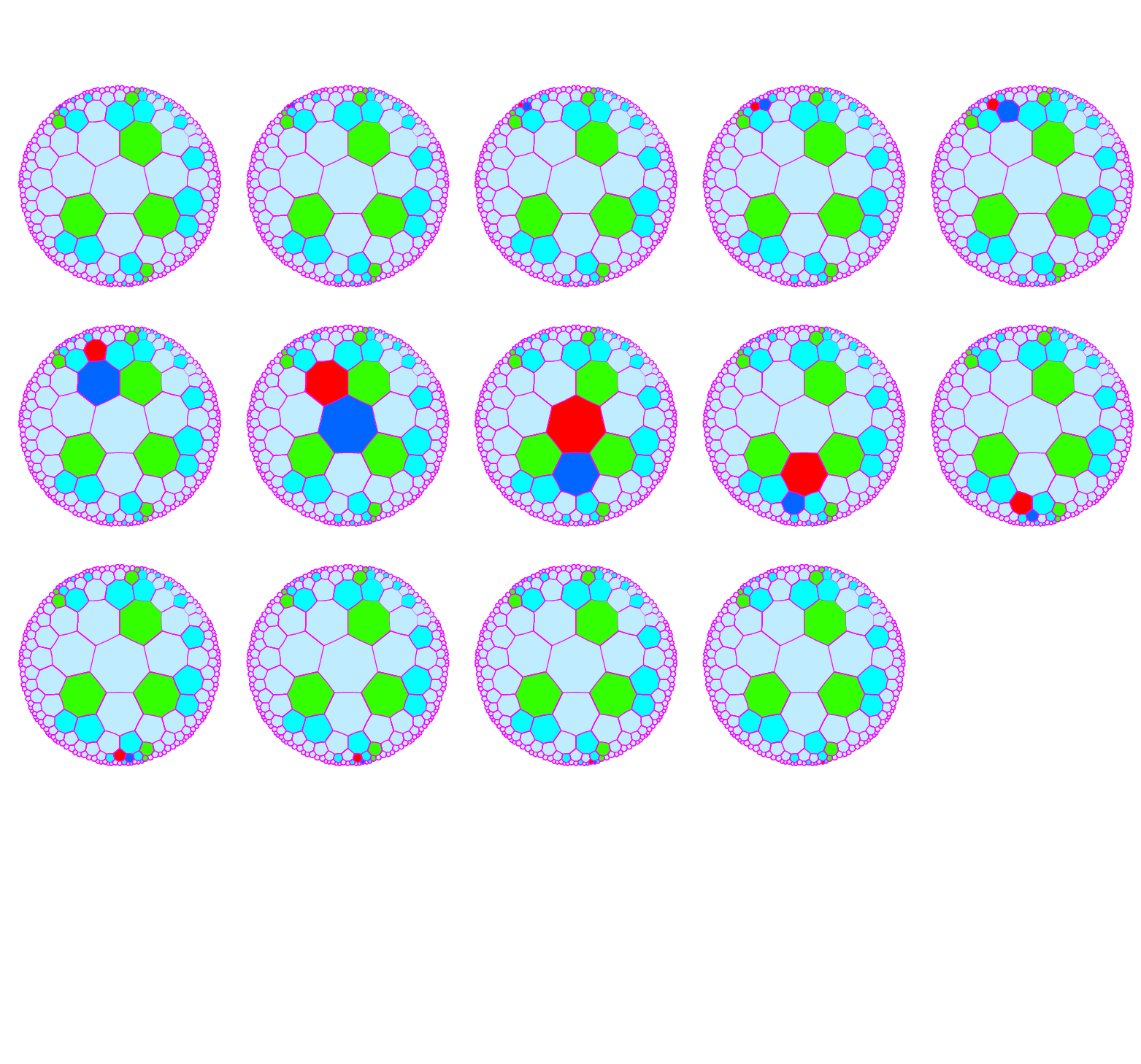,width=340pt}}
\vtop{
\ligne{\hfill
\PlacerEn {-346pt} {0pt} \box110
}
\vspace{-85pt}
\begin{fig}\label{fixe_1}
\leurre
The locomotive passively crosses a fixed switch from the selected track. 
\end{fig}
}

\vskip-30pt
\setbox110=\hbox{\epsfig{file=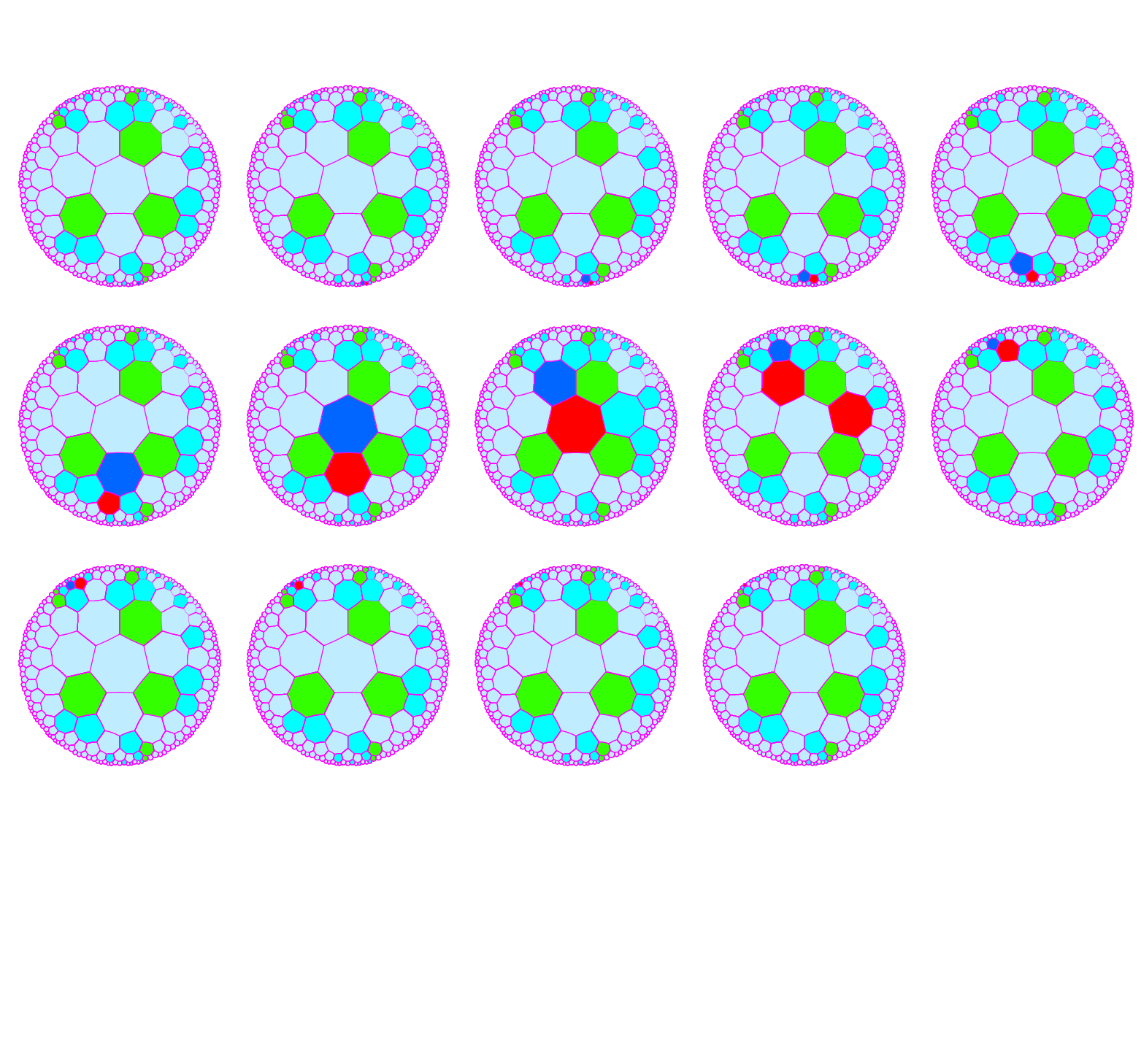,width=340pt}}
\vtop{
\ligne{\hfill
\PlacerEn {-346pt} {0pt} \box110
}
\vspace{-95pt}
\begin{fig}\label{fixe_4}
\leurre
The locomotive actively crosses a fixed switch, here from
sector~$4$. Note the attempt to send another locomotive in the wrong direction: eighth
and ninth pictures. This is a consequence of the rotation invariance of the rules.
\end{fig}
}
\vskip-40pt
   We can see that the idle configuration of the fixed switch looks like the 
idle configuration of a crossing. There is here one path less and the idle cell
is green. Note that in order that the green idle cell remains idle, it has two blue
neighbours: this fixes a configuration which is not disturbed by a nearby passage of
the locomotive.

   Also note that, as in the previous figures, in order that the front of the locomotive
should be distinguished from blue or green cells of the blue track, it is represented in
a darker colour. In the rules, as well as in the computer program to check them,
there is no distinction between the blue colour of the front
of the locomotive and the blue colour for the cells of a blue track. The same holds
for the green colour of the front of the locomotive and the green cells.

\vskip-5pt
\setbox110=\hbox{\epsfig{file=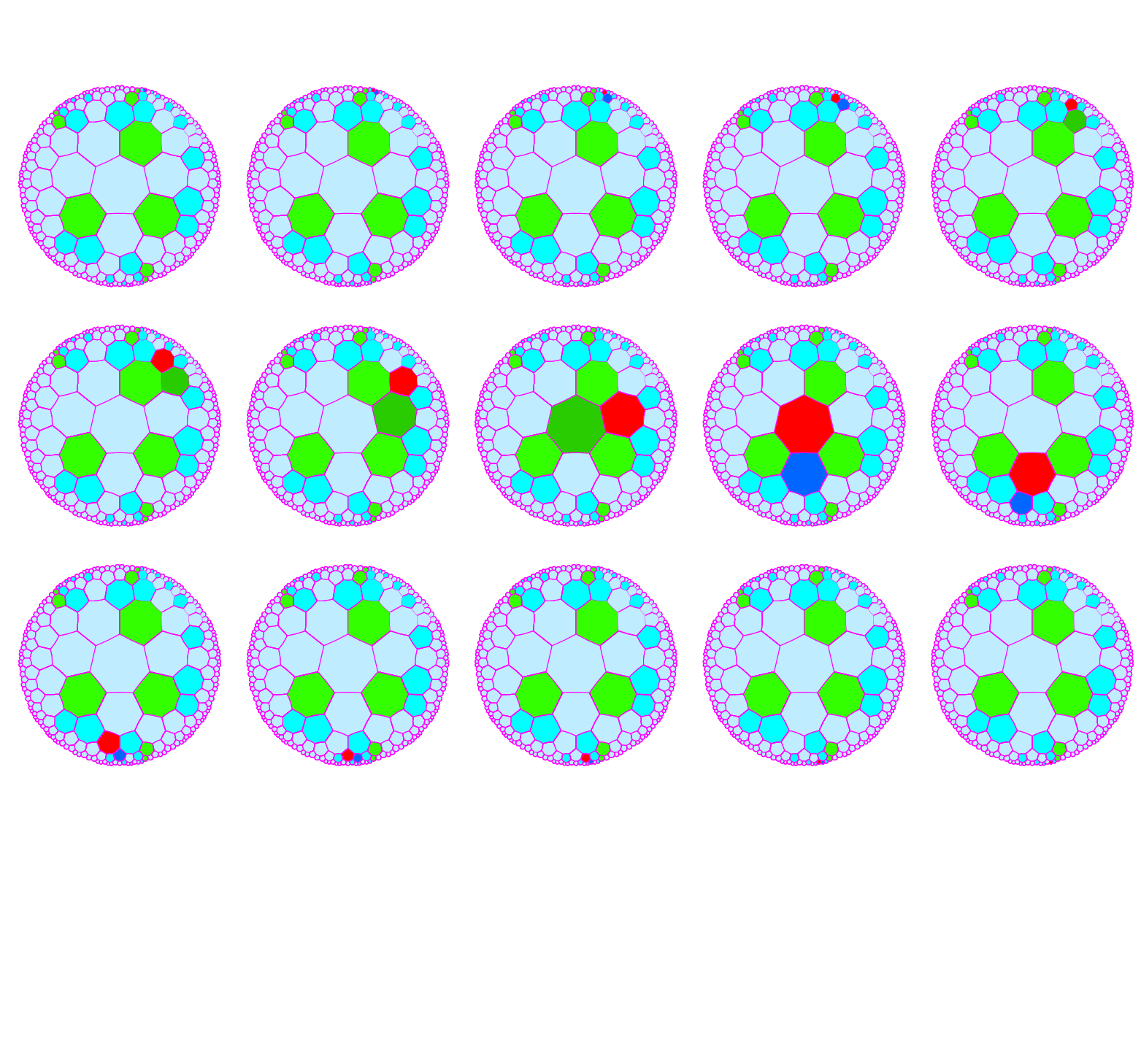,width=340pt}}
\vtop{
\ligne{\hfill
\PlacerEn {-346pt} {0pt} \box110
}
\vspace{-85pt}
\begin{fig}\label{fixe_7}
\leurre
The locomotive passively crosses a fixed switch from the non-selected track, here 
sector~$7$.
\end{fig}
}

\vspace{-50pt}
\subsubsection{Memory switches}

   The idle configuration of a memory switch is very different from the idle configurations
of a crossing or of a fixed switch. Schematically, the intersection at a crossing or at
a fixed switch can be called open while the intersection at a memory or a flip-flop
switch would be called closed.
 
   The memory switch implements the change of the selected path by displacing an
obstacle from the Frost cell of the proper track in the non-selected path to the
first cell of the proper track in the selected cell. The look of an obstacle can
clearly be seen in Fig.~\ref{memorisant_4} and in Fig.~\ref{memorisant_7}.

   In all the situations of this switch, the same basic motion of the locomotive
on the proper track is observed. During the active crossing and the passive one through 
the selected path, the cells which are not on a proper track remain unchanged. 
During the passive crossing through the non-selected path, the basic motion occurs
on the proper track and, at one time, the obstacle on the proper track is removed
and it is placed onto the proper track of the selected path on the next time.
Define by 1(1) and 1(7) the coordinates of the first cell on the proper track of
the selected and non-selected paths respectively, 0~being the coordinate of the centre
of the intersection. The configuration of the memory switch is characterized 
by the blue cell in 2(1) and the fact that the centre has five green neighbours except
at the time when the obstacle on the proper track of the non-selected path is removed,
see Fig.~\ref{memorisant_7}. Also, at the following time, the obstacle which was before
on~1(7) is now placed on~1(1). 

\vskip-10pt
\setbox110=\hbox{\epsfig{file=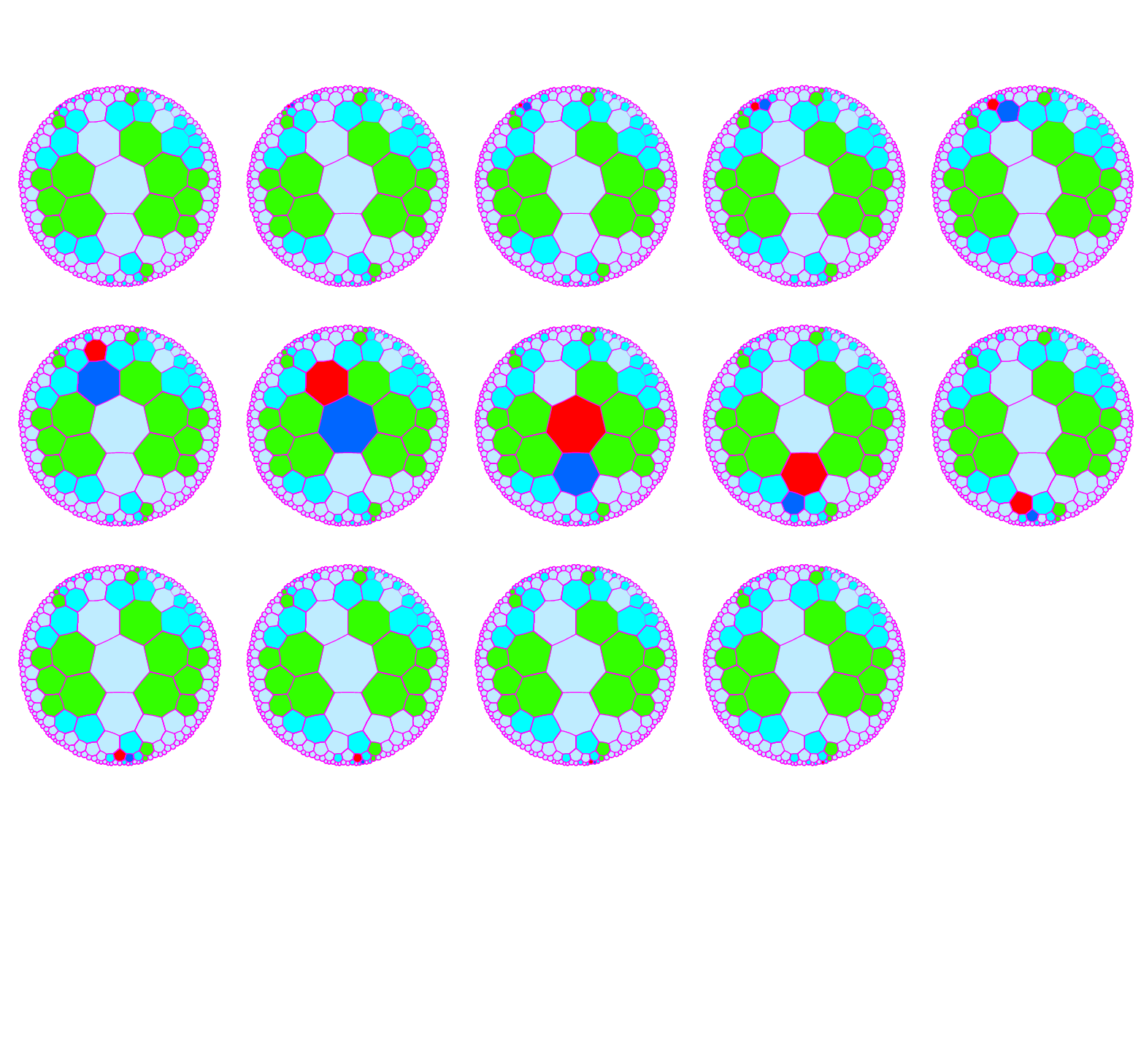,width=340pt}}
\vtop{
\ligne{\hfill
\PlacerEn {-346pt} {0pt} \box110
}
\vspace{-85pt}
\begin{fig}\label{memorisant_1}
\leurre
The locomotive passively crosses a memory switch from the selected track,
here in sector~$1$. 
\end{fig}
}

\vskip-25pt
\setbox110=\hbox{\epsfig{file=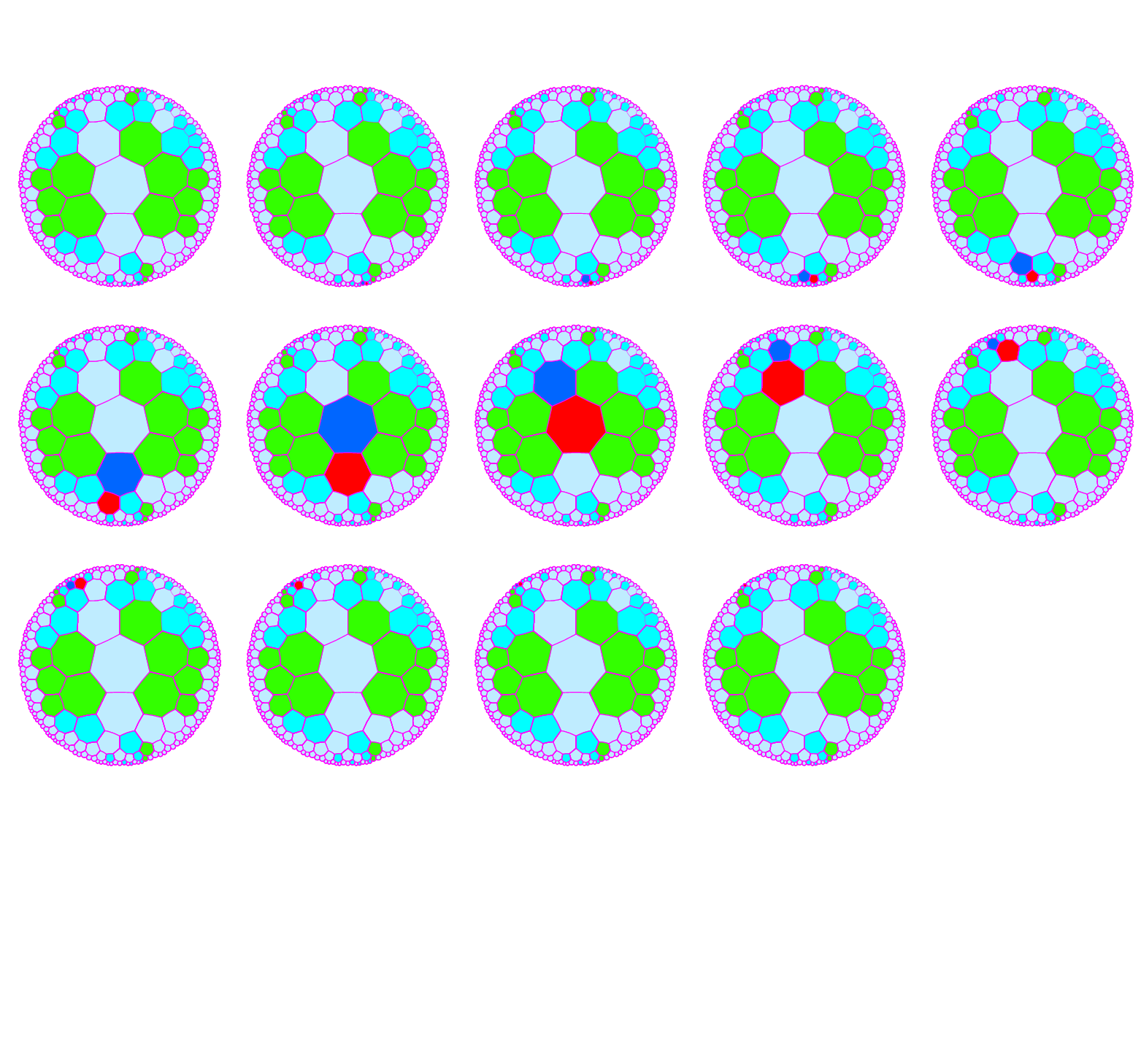,width=340pt}}
\vtop{
\ligne{\hfill
\PlacerEn {-346pt} {0pt} \box110
}
\vspace{-85pt}
\begin{fig}\label{memorisant_4}
\leurre
The locomotive actively crosses a memory switch, here from sector~$4$.
\end{fig}
}

\vskip -50pt
However, during the crossing, the blue cell~2(1) remains
unchanged. Note that this cell is a milestone of the path whose first cell is~1(1).
It is used in this way in the motion of the locomotive when it goes out from this path
or when it enters it.

   On the last picture of Fig.~\ref{memorisant_7}, we can see the idle configuration
of a memory switch whose selected path is the right-hand side one. As the idle configuration
of a memory switch is not symmetric, in particular, due to the situation in sector~5,
assuming the origin at the centre, we have to check that rules can be devised for
such a memory switch. Moreover, the new rules have to be compatible with the ones
already established for a memory switch whose selected track is the left-hand side one.
 
\vskip-5pt
\setbox110=\hbox{\epsfig{file=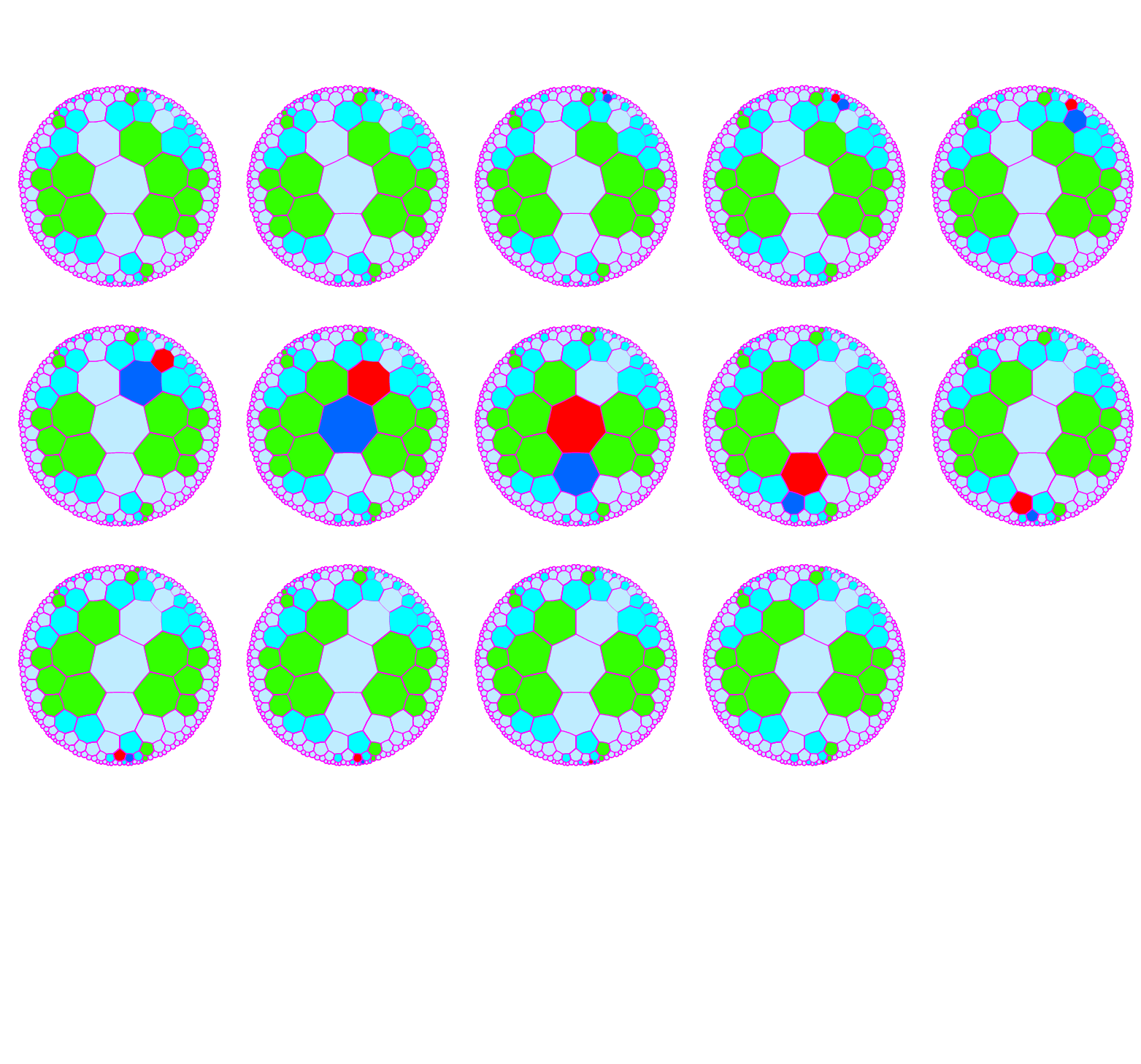,width=340pt}}
\vtop{
\ligne{\hfill
\PlacerEn {-346pt} {0pt} \box110
}
\vspace{-85pt}
\begin{fig}\label{memorisant_7}
\leurre
The locomotive passively crosses a memory switch from the non-selected track, here in
sector~$7$. 
Note the change of the selected track when the rear of the
locomotive leaves the first cell of proper track of the non-selected path.
\end{fig}
}

\vskip -30pt
These rules have been designed and they have been checked by a computer program.
In the Appendix, Fig.~\ref{memorisantd_7}, Fig.~\ref{memorisantd_4} and
Fig.~\ref{memorisantd_1} correspond to a passive crossing through a right-hand side
selected track, to an active crossing and to a passive crossing through a left-hand side
non-selected track respectively.

\subsubsection{Flip-flop switches}

   The idle configuration of a flip-flop switch looks like that of a memory switch.
The main difference comes from the cell~2(1) which is red. In order to keep the red state
during a crossing of the switch by the locomotive, a group of four cells decorates
the cell~2(1), two of them being neighbours of the cell~2(1).

   As we have to study the active passage only, we refer the reader to two figures:
Fig.~\ref{bascule_g} and Fig.~\ref{bascule_d}. In the first figure, the selected path
is the left-hand side one, in the second figure, it is the right-hand side one.
The figures show the change from one idle configuration to the other in both cases,
hence the flip-flop action of the switch. The configuration around the centre is
the same as in a memory switch. This is a difference with the situation 
in~\cite{mmsyBristol,mmsyPPL,mmsyENTCS} where the respective configurations around the
centre are the same.

\vskip-25pt
\setbox110=\hbox{\epsfig{file=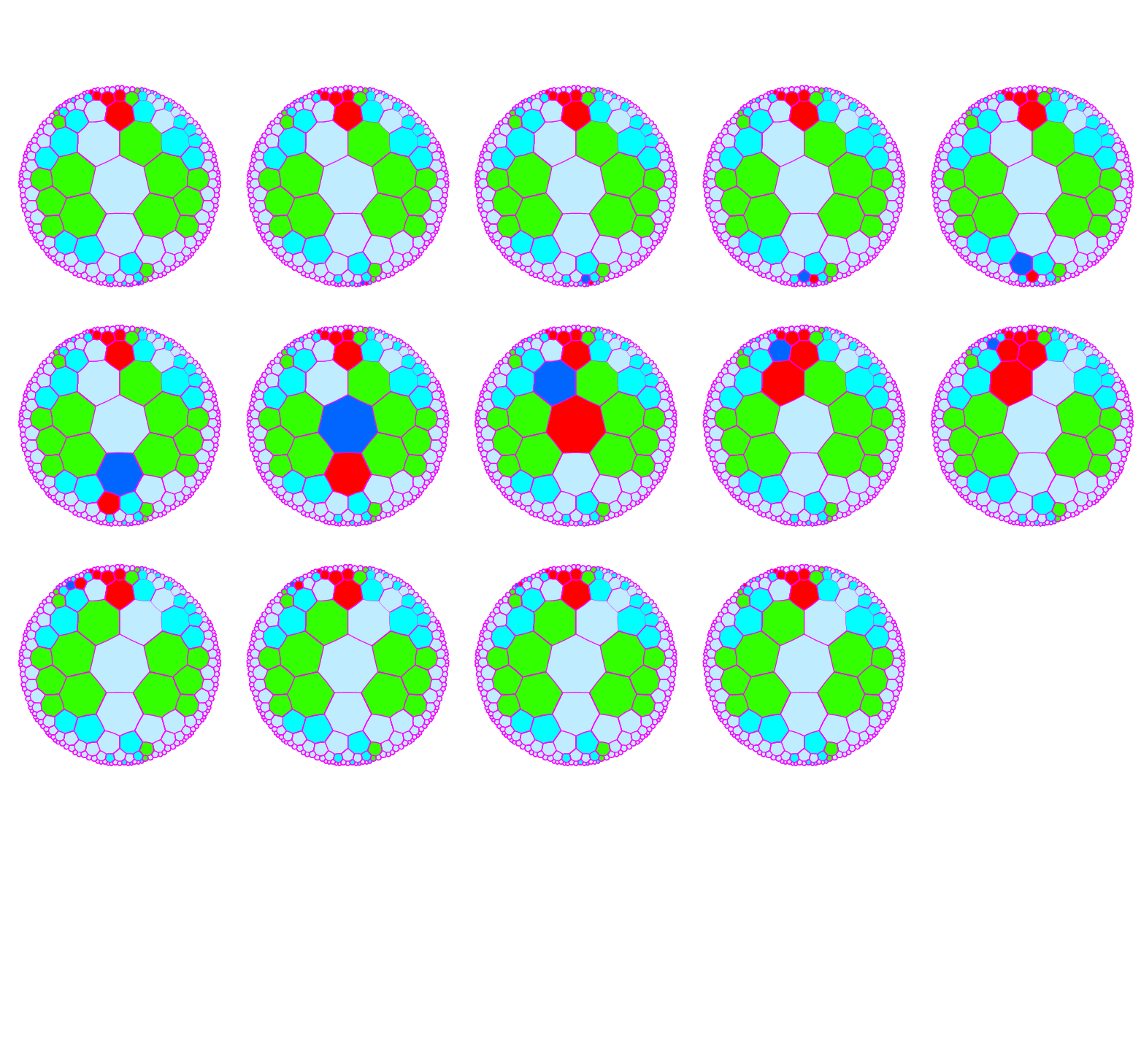,width=340pt}}
\vtop{
\ligne{\hfill
\PlacerEn {-346pt} {0pt} \box110
}
\vspace{-85pt}
\begin{fig}\label{bascule_g}
\leurre
The locomotive actively crosses a left-hand side flip-flop switch: here, it comes
from the path in sector~$4$. Note the change of the selected path when the rear of the
locomotive leaves the first cell of the proper track of the non-selected path.
\end{fig}
}

\vskip-25pt
\setbox110=\hbox{\epsfig{file=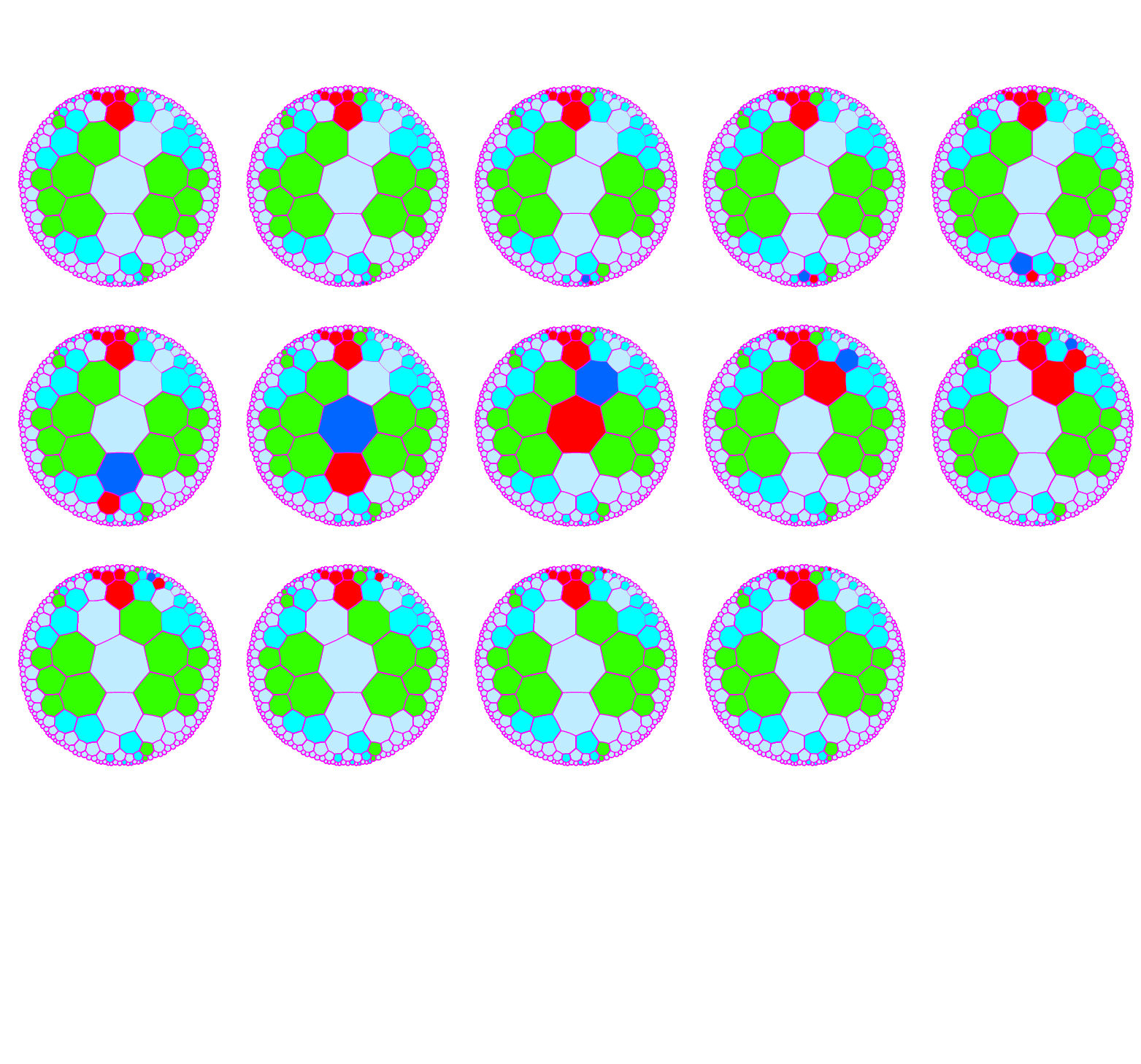,width=340pt}}
\vtop{
\ligne{\hfill
\PlacerEn {-346pt} {0pt} \box110
}
\vspace{-85pt}
\begin{fig}\label{bascule_d}
\leurre
The locomotive actively crosses a right-hand side flip-flop switch: here, it comes
from the track in sector~$4$. Note the change of the selected path when the rear of the
locomotive leaves the first cell of the proper track of the non-selected path. 
\end{fig}
}

\vskip -30pt
   As a final remark about the switches, it can be noticed that the surrounding
of the centre of the intersection and that of the cell~2(1) allow to identify
each case without any ambiguity. What must be distinguished is distinguished, in particular
the change of selection which is different between the memory and the flip-flop switch.
The change does not occur at the same time, mainly because the change occurs during a passive
crossing for the memory switch and during an active one for the memory switch. Also, 
what can be shared in the case of the memory and the flip-flop switch is shared,
in particular the active passage until the centre of the switch is reached.

\section{The rules}

\def\displayrule #1 #2 #3 #4 #5 #6 #7 {%
\hbox{\ttix\hbox to 15pt{#1\hfill}\hbox to 15pt{#2\hfill}\hbox to 15pt{#3\hfill}%
\hbox to 15pt{#4\hfill}\hbox to 15pt{#5\hfill}\hbox to 15pt{#6\hfill}%
\hbox to 15pt{#7\hfill}
}
}

The set of rules of the automaton is displayed in Table~\ref{table_regles}.
Here, we indicate how the rules where computed by a computer program and 
how the rules are represented and dispatched in the table.

\subsection{The format of the rules and rotation invariance}

   The format of the rules is exactly the same as the format described 
in~\cite{mmsyENTCS}. For the sake of self-containedness, we summarize the corresponding
information here.

   Each cell has a  numbering of its sides from 1~to~7. The central cell excepted,
for the other cells, side~1 is the side shared with the father. In the central
cell, side~1 is a side fixed once and for all. Once this side is fixed, all the 
others are also fixed: the numbers of the side increase when counter-clockwise
turning around the tile which supports the cell. Note that the sons of a 
black node in the Fibonacci tree are given by sides~4 and~5 and that those
of a white node are given by sides~3, 4 and~5. 

   As indicated in Subsection~3.2, the rules are denoted as follows:
\vskip 5pt
\ligne{\hfill
$\underline{\eta_0}\eta_1\eta_2\eta_3\eta_4\eta_5\eta_6\eta_7
\underline{\eta_0^1}$, 
\hfill}
\vskip 5pt
\noindent
where $\eta_0$ is the current state of the cell,
$\eta_i$, $i\in\{1..7\}$, is the current states of the neighbour~$i$ of the cell
and $\eta_0^1$ is its new state. We 
define $\eta_0\eta_1\eta_2\eta_3\eta_4\eta_5\eta_6\eta_7$
as the {\bf context} of the rule. it is the biggest proper prefix of the rule as
a word.

The whole set of the rules is given in Table~\ref{table_regles}, in the Appendix.
\vskip 5pt
   To define the features of the computer program, it can be useful to devise
a test in order to check the rotation invariance of the automaton. The notion
of rotated form can be used for that purpose in the following way.

   Consider a rule~$(\eta)$ of the cellular automaton which we represent by
the word $\underline{\eta_0}\eta_1\eta_2\eta_3\eta_4\eta_5\eta_6\eta_7
\underline{\eta_0^1}$, as above indicated. Its rotated forms are defined
by $\underline{\eta_0}\eta_{\pi(1)}\eta_{\pi(2)}\eta_{\pi(3)}%
\eta_{\pi(4)}\eta_{\pi(5)}\eta_{\pi(6)}\eta_{\pi(7)}%
\underline{\eta_0^1}$, where $\pi$ is a circular permutation on $1..7$. We
can see that the difference between the rotated forms of a rule and the rule
itself lies in their contexts: they can be deduced from each other by a suitable
circular permutation on $\{1..7\}$. There are 7~rotated forms for each rule. 
Now, as we consider the contexts as words, we can order them lexicographically. 
And so, there is a {\bf minimal} rotated form for each rule~$(\eta)$: that 
whose context is the minimum of the contexts of the rotated forms of~$(\eta)$ 
with respect to the lexicographical order which is a linear order. 
Denote min$(\eta)$ the minimal rotated form of the rule~$(\eta)$. Remember
that the current state of the cell is the first letter of the context.

   Now, it is plain that we have the following property:

\begin{lem}\label{minim}
The set of rules of a cellular automaton on the heptagrid is invariant by
rotation if and only if for any pair of rules $(\eta)$~and~$(\epsilon)$ of the 
automaton,
if 
{\rm min}$(\eta)$ and {\rm min}$(\epsilon)$ 
have equal contexts then $\eta_0^1=\epsilon_0^1$.
\end{lem}

\subsection{The program}   
   
   The computer program was written in~$ADA$.

   The program uploads the initial configuration of the crossings and of
the switches from a file and puts the corresponding information into a 
table~0. In this table, each row represents a cell. The 
entries of the row indicate the coordinates of the neighbours of the cell as 
well as the states of the cell and of its neighbours. The program also contains 
a copy of table~0 with no state in the cells which we call 
table~1. The set of rules is in a file under an appropriate format, close to 
the one which was depicted in Subsection~5.1. 

    During the construction of the set of rules, the program works as follows.
    When we run the program, it reads the file of the rules which, initially
contains the rule $\underline{\tt W}${\tt WWWWWWW}$\underline{\tt W}$ which
says that a cell in the quiescent state whose neighbours are all 
\ifnum 1=0 {
\subsection{The program and the table of the rules}   

\ligne{\hfill}

\def\lapetiteregle #1 #2 #3 #4 #5 #6 #7 #8 #9 {%
\setbox211=\hbox{$\underline{\hbox{\ttvi#1}}$}
\setbox212=\hbox{$\underline{\hbox{\ttvi#9}}$}
\hskip-15pt\hbox{\ttvi \box211#2#3#4#5#6#7#8\box212}
}
\newcount\reglenum\reglenum=1
\setbox120=
\vtop{\leftskip 0pt\parindent 0pt
\baselineskip 7pt
\hsize=70pt
crossing:
\vskip 5pt\rmviii
track~1:
\vskip 1pt
{\ttv\obeylines
\newcount\reglenum\reglenum=1 
\the\reglenum{}: \lapetiteregle {W} {W} {W} {W} {W} {W} {W} {W} {W} \vskip 0pt 
\global\advance\reglenum by 1 
\the\reglenum{}: \lapetiteregle {B} {B} {W} {B2} {B} {W} {W} {B2} {B} \vskip 0pt 
\global\advance\reglenum by 1 
\the\reglenum{}: \lapetiteregle {B} {B} {B2} {W} {W} {B} {W} {W} {B} \vskip 0pt 
\global\advance\reglenum by 1 
\the\reglenum{}: \lapetiteregle {W} {B} {B2} {B} {W} {W} {W} {W} {W} \vskip 0pt 
\global\advance\reglenum by 1 
\the\reglenum{}: \lapetiteregle {W} {B} {W} {W} {W} {W} {W} {B} {W} \vskip 0pt 
\global\advance\reglenum by 1 
\the\reglenum{}: \lapetiteregle {B} {B} {W} {W} {W} {G} {W} {W} {G} \vskip 0pt 
\global\advance\reglenum by 1 
\the\reglenum{}: \lapetiteregle {W} {W} {B} {B} {W} {W} {W} {W} {W} \vskip 0pt 
\global\advance\reglenum by 1 
\the\reglenum{}: \lapetiteregle {W} {B} {W} {W} {W} {W} {W} {W} {W} \vskip 0pt 
\global\advance\reglenum by 1 
\the\reglenum{}: \lapetiteregle {W} {B} {W} {W} {W} {W} {W} {G} {W} \vskip 0pt 
\global\advance\reglenum by 1 
\the\reglenum{}: \lapetiteregle {G} {B} {W} {W} {W} {R} {W} {W} {R} \vskip 0pt 
\global\advance\reglenum by 1 
\the\reglenum{}: \lapetiteregle {W} {G} {W} {W} {W} {W} {W} {W} {W} \vskip 0pt 
\global\advance\reglenum by 1 
\the\reglenum{}: \lapetiteregle {W} {G} {W} {W} {W} {W} {W} {R} {W} \vskip 0pt 
\global\advance\reglenum by 1 
\the\reglenum{}: \lapetiteregle {R} {G} {W} {W} {W} {B} {W} {W} {B} \vskip 0pt 
\global\advance\reglenum by 1 
\the\reglenum{}: \lapetiteregle {W} {B} {B} {W} {W} {W} {W} {B2} {W} \vskip 0pt 
\global\advance\reglenum by 1 
\the\reglenum{}: \lapetiteregle {W} {W} {B} {G} {W} {W} {W} {W} {W} \vskip 0pt 
\global\advance\reglenum by 1 
\the\reglenum{}: \lapetiteregle {W} {W} {G} {R} {W} {W} {W} {W} {W} \vskip 0pt 
\global\advance\reglenum by 1 
\the\reglenum{}: \lapetiteregle {W} {W} {W} {W} {W} {B} {W} {W} {W} \vskip 0pt 
\global\advance\reglenum by 1 
\the\reglenum{}: \lapetiteregle {B2} {B} {W} {W} {W} {B} {W} {B} {B2} \vskip 0pt 
\global\advance\reglenum by 1 
\the\reglenum{}: \lapetiteregle {W} {B2} {W} {W} {W} {W} {W} {W} {W} \vskip 0pt 
\global\advance\reglenum by 1 
\the\reglenum{}: \lapetiteregle {W} {B2} {W} {W} {W} {W} {W} {B} {W} \vskip 0pt 
\global\advance\reglenum by 1 
\the\reglenum{}: \lapetiteregle {B} {B2} {W} {W} {W} {B} {W} {W} {B} \vskip 0pt 
\global\advance\reglenum by 1 
\the\reglenum{}: \lapetiteregle {B} {B} {W} {W} {W} {B} {W} {W} {B} \vskip 0pt 
\global\advance\reglenum by 1 
\the\reglenum{}: \lapetiteregle {W} {B} {B} {W} {W} {W} {W} {W} {W} \vskip 0pt 
\global\advance\reglenum by 1 
\the\reglenum{}: \lapetiteregle {W} {B} {W} {W} {W} {W} {W} {B2} {W} \vskip 0pt 
\global\advance\reglenum by 1 
\the\reglenum{}: \lapetiteregle {B} {B} {B2} {W} {W} {G} {W} {W} {G} \vskip 0pt 
\global\advance\reglenum by 1 
\the\reglenum{}: \lapetiteregle {W} {R} {W} {W} {W} {W} {W} {W} {W} \vskip 0pt 
\global\advance\reglenum by 1 
\the\reglenum{}: \lapetiteregle {W} {R} {W} {W} {W} {W} {W} {B} {W} \vskip 0pt 
\global\advance\reglenum by 1 
\the\reglenum{}: \lapetiteregle {B} {R} {W} {W} {W} {B} {W} {W} {B} \vskip 0pt 
\global\advance\reglenum by 1 
\the\reglenum{}: \lapetiteregle {W} {W} {R} {B} {W} {W} {W} {W} {W} \vskip 0pt 
\global\advance\reglenum by 1 
\the\reglenum{}: \lapetiteregle {B} {G} {W} {B2} {B} {W} {W} {B2} {G} \vskip 0pt 
\global\advance\reglenum by 1 
\the\reglenum{}: \lapetiteregle {G} {B} {B2} {W} {W} {R} {W} {W} {R} \vskip 0pt 
\global\advance\reglenum by 1 
\the\reglenum{}: \lapetiteregle {W} {G} {B2} {B} {W} {W} {W} {W} {W} \vskip 0pt 
\global\advance\reglenum by 1 
\the\reglenum{}: \lapetiteregle {W} {B} {G} {W} {W} {W} {W} {B2} {W} \vskip 0pt 
\global\advance\reglenum by 1 
\the\reglenum{}: \lapetiteregle {B2} {B} {W} {W} {W} {B} {W} {G} {B2} \vskip 0pt 
\global\advance\reglenum by 1 
\the\reglenum{}: \lapetiteregle {G} {R} {W} {B2} {B} {W} {W} {B2} {R} \vskip 0pt 
\global\advance\reglenum by 1 
\the\reglenum{}: \lapetiteregle {R} {G} {B2} {W} {W} {B} {W} {W} {B} \vskip 0pt 
\global\advance\reglenum by 1 
\the\reglenum{}: \lapetiteregle {W} {R} {B2} {B} {W} {W} {W} {W} {W} \vskip 0pt 
\global\advance\reglenum by 1 
\the\reglenum{}: \lapetiteregle {W} {G} {R} {W} {W} {W} {W} {B2} {W} \vskip 0pt 
\global\advance\reglenum by 1 
\the\reglenum{}: \lapetiteregle {B2} {G} {W} {W} {W} {B} {W} {B} {B2} \vskip 0pt 
\global\advance\reglenum by 1 
\the\reglenum{}: \lapetiteregle {B} {G} {B2} {W} {W} {B} {W} {W} {G} \vskip 0pt 
\global\advance\reglenum by 1 
\the\reglenum{}: \lapetiteregle {W} {G} {B} {W} {W} {W} {W} {W} {W} \vskip 0pt 
\global\advance\reglenum by 1 
\the\reglenum{}: \lapetiteregle {W} {G} {W} {W} {W} {W} {W} {B2} {W} \vskip 0pt 
\global\advance\reglenum by 1 
\the\reglenum{}: \lapetiteregle {B2} {G} {W} {W} {W} {B} {W} {R} {B2} \vskip 0pt 
\global\advance\reglenum by 1 
\the\reglenum{}: \lapetiteregle {R} {B} {W} {B2} {G} {W} {W} {B2} {B} \vskip 0pt 
\global\advance\reglenum by 1 
\the\reglenum{}: \lapetiteregle {B} {R} {B2} {W} {W} {B} {W} {W} {B} \vskip 0pt 
\global\advance\reglenum by 1 
\the\reglenum{}: \lapetiteregle {W} {R} {B} {W} {W} {W} {W} {B2} {W} \vskip 0pt 
\global\advance\reglenum by 1 
\the\reglenum{}: \lapetiteregle {B2} {R} {W} {W} {W} {B} {W} {G} {B2} \vskip 0pt 
\global\advance\reglenum by 1 
\the\reglenum{}: \lapetiteregle {G} {R} {B2} {W} {W} {B} {W} {W} {R} \vskip 0pt 
\global\advance\reglenum by 1 
\the\reglenum{}: \lapetiteregle {W} {G} {W} {W} {W} {W} {W} {B} {W} \vskip 0pt 
\global\advance\reglenum by 1 
\the\reglenum{}: \lapetiteregle {B} {G} {W} {W} {W} {B} {W} {W} {G} \vskip 0pt 
\global\advance\reglenum by 1 
\the\reglenum{}: \lapetiteregle {W} {R} {G} {W} {W} {W} {W} {W} {W} \vskip 0pt 
\global\advance\reglenum by 1 
\the\reglenum{}: \lapetiteregle {W} {W} {G} {B} {W} {W} {W} {W} {W} \vskip 0pt 
\global\advance\reglenum by 1 
\the\reglenum{}: \lapetiteregle {W} {R} {W} {W} {W} {W} {W} {B2} {W} \vskip 0pt 
\global\advance\reglenum by 1 
\the\reglenum{}: \lapetiteregle {B2} {R} {W} {W} {W} {B} {W} {B} {B2} \vskip 0pt 
\global\advance\reglenum by 1 
\the\reglenum{}: \lapetiteregle {B} {B} {W} {B2} {R} {W} {W} {B2} {B} \vskip 0pt 
\global\advance\reglenum by 1 
\the\reglenum{}: \lapetiteregle {B2} {B} {W} {W} {W} {B} {W} {R} {B2} \vskip 0pt 
\global\advance\reglenum by 1 
\the\reglenum{}: \lapetiteregle {R} {B} {B2} {W} {W} {G} {W} {W} {B} \vskip 0pt 
\global\advance\reglenum by 1 
\the\reglenum{}: \lapetiteregle {W} {R} {W} {W} {W} {W} {W} {G} {W} \vskip 0pt 
\global\advance\reglenum by 1 
\the\reglenum{}: \lapetiteregle {G} {R} {W} {W} {W} {B} {W} {W} {R} \vskip 0pt 
\global\advance\reglenum by 1 
\the\reglenum{}: \lapetiteregle {W} {B} {R} {W} {W} {W} {W} {W} {W} \vskip 0pt 
\global\advance\reglenum by 1 
\the\reglenum{}: \lapetiteregle {W} {W} {R} {G} {W} {W} {W} {W} {W} \vskip 0pt 
\global\advance\reglenum by 1 
\the\reglenum{}: \lapetiteregle {B} {B} {B2} {W} {W} {R} {W} {W} {B} \vskip 0pt 
\global\advance\reglenum by 1 
\the\reglenum{}: \lapetiteregle {W} {B} {W} {W} {W} {W} {W} {R} {W} \vskip 0pt 
\global\advance\reglenum by 1 
\the\reglenum{}: \lapetiteregle {R} {B} {W} {W} {W} {G} {W} {W} {B} \vskip 0pt 
\global\advance\reglenum by 1 
\the\reglenum{}: \lapetiteregle {W} {W} {B} {R} {W} {W} {W} {W} {W} \vskip 0pt 
\global\advance\reglenum by 1 
\the\reglenum{}: \lapetiteregle {B} {B} {W} {W} {W} {R} {W} {W} {B} \vskip 0pt 
\global\advance\reglenum by 1 
}
\vskip 2pt
track 4:
\vskip 1pt
{\ttv\obeylines
\the\reglenum{}: \lapetiteregle {B} {B} {W} {B2} {G} {W} {W} {B2} {G} \vskip 0pt 
\global\advance\reglenum by 1 
\the\reglenum{}: \lapetiteregle {W} {B} {G} {W} {W} {W} {W} {W} {W} \vskip 0pt 
\global\advance\reglenum by 1 
\the\reglenum{}: \lapetiteregle {G} {B} {W} {B2} {R} {W} {W} {B2} {R} \vskip 0pt 
\global\advance\reglenum by 1 
}
}
\setbox122=
\vtop{\leftskip 0pt\parindent 0pt
\baselineskip 7pt
\hsize=70pt
\rmviii
crossing:
\vskip 5pt
track~4:
\vskip 1pt
{\ttv\obeylines
\the\reglenum{}: \lapetiteregle {W} {G} {B} {W} {W} {W} {W} {B2} {W} \vskip 0pt 
\global\advance\reglenum by 1 
\the\reglenum{}: \lapetiteregle {W} {G} {R} {W} {W} {W} {W} {W} {W} \vskip 0pt 
\global\advance\reglenum by 1 
\the\reglenum{}: \lapetiteregle {R} {G} {W} {B2} {B} {W} {W} {B2} {B} \vskip 0pt 
\global\advance\reglenum by 1 
\the\reglenum{}: \lapetiteregle {W} {R} {G} {W} {W} {W} {W} {B2} {W} \vskip 0pt 
\global\advance\reglenum by 1 
\the\reglenum{}: \lapetiteregle {W} {R} {B} {W} {W} {W} {W} {W} {W} \vskip 0pt 
\global\advance\reglenum by 1 
\the\reglenum{}: \lapetiteregle {B} {R} {W} {B2} {B} {W} {W} {B2} {B} \vskip 0pt 
\global\advance\reglenum by 1 
\the\reglenum{}: \lapetiteregle {W} {B} {R} {W} {W} {W} {W} {B2} {W} \vskip 0pt 
\global\advance\reglenum by 1 
\the\reglenum{}: \lapetiteregle {B} {B2} {W} {W} {W} {G} {W} {W} {G} \vskip 0pt 
\global\advance\reglenum by 1 
\the\reglenum{}: \lapetiteregle {W} {B} {B2} {G} {W} {W} {W} {W} {W} \vskip 0pt 
\global\advance\reglenum by 1 
}
\vskip 5pt\rmviii
track~7:
\vskip 1pt
{\ttv\obeylines
\the\reglenum{}: \lapetiteregle {B2} {B} {W} {W} {W} {G} {W} {B} {G2} \vskip 0pt 
\global\advance\reglenum by 1 
\the\reglenum{}: \lapetiteregle {W} {B2} {W} {W} {W} {W} {W} {G} {W} \vskip 0pt 
\global\advance\reglenum by 1 
\the\reglenum{}: \lapetiteregle {G} {B2} {W} {W} {W} {R} {W} {W} {R} \vskip 0pt 
\global\advance\reglenum by 1 
\the\reglenum{}: \lapetiteregle {B} {B} {W} {B2} {B} {W} {W} {G2} {G2} \vskip 0pt 
\global\advance\reglenum by 1 
\the\reglenum{}: \lapetiteregle {B} {B} {G2} {W} {W} {B} {W} {W} {B} \vskip 0pt 
\global\advance\reglenum by 1 
\the\reglenum{}: \lapetiteregle {W} {B} {G2} {R} {W} {W} {W} {W} {W} \vskip 0pt 
\global\advance\reglenum by 1 
\the\reglenum{}: \lapetiteregle {W} {B} {W} {W} {W} {W} {W} {G2} {W} \vskip 0pt 
\global\advance\reglenum by 1 
\the\reglenum{}: \lapetiteregle {G2} {B} {W} {W} {W} {R} {W} {B} {R} \vskip 0pt 
\global\advance\reglenum by 1 
\the\reglenum{}: \lapetiteregle {W} {G2} {W} {W} {W} {W} {W} {W} {W} \vskip 0pt 
\global\advance\reglenum by 1 
\the\reglenum{}: \lapetiteregle {W} {G2} {W} {W} {W} {W} {W} {R} {W} \vskip 0pt 
\global\advance\reglenum by 1 
\the\reglenum{}: \lapetiteregle {R} {G2} {W} {W} {W} {B} {W} {W} {B} \vskip 0pt 
\global\advance\reglenum by 1 
\the\reglenum{}: \lapetiteregle {G2} {B} {W} {B2} {B} {W} {W} {R} {R} \vskip 0pt 
\global\advance\reglenum by 1 
\the\reglenum{}: \lapetiteregle {B} {G2} {R} {W} {W} {B} {W} {W} {B} \vskip 0pt 
\global\advance\reglenum by 1 
\the\reglenum{}: \lapetiteregle {W} {B} {R} {B} {W} {W} {W} {W} {W} \vskip 0pt 
\global\advance\reglenum by 1 
\the\reglenum{}: \lapetiteregle {W} {G2} {B} {W} {W} {W} {W} {B2} {W} \vskip 0pt 
\global\advance\reglenum by 1 
\the\reglenum{}: \lapetiteregle {B2} {G2} {W} {W} {W} {B} {W} {B} {G2} \vskip 0pt 
\global\advance\reglenum by 1 
\the\reglenum{}: \lapetiteregle {B} {G2} {B2} {W} {W} {B} {W} {W} {B} \vskip 0pt 
\global\advance\reglenum by 1 
\the\reglenum{}: \lapetiteregle {W} {G2} {B} {W} {W} {W} {W} {W} {W} \vskip 0pt 
\global\advance\reglenum by 1 
\the\reglenum{}: \lapetiteregle {R} {G2} {W} {W} {W} {B} {W} {B} {B2} \vskip 0pt 
\global\advance\reglenum by 1 
\the\reglenum{}: \lapetiteregle {R} {B} {W} {G2} {B} {W} {W} {B2} {B} \vskip 0pt 
\global\advance\reglenum by 1 
\the\reglenum{}: \lapetiteregle {W} {R} {B} {W} {W} {W} {W} {G2} {W} \vskip 0pt 
\global\advance\reglenum by 1 
\the\reglenum{}: \lapetiteregle {G2} {R} {W} {W} {W} {B} {W} {B} {R} \vskip 0pt 
\global\advance\reglenum by 1 
\the\reglenum{}: \lapetiteregle {W} {G2} {W} {W} {W} {W} {W} {B} {W} \vskip 0pt 
\global\advance\reglenum by 1 
\the\reglenum{}: \lapetiteregle {B} {G2} {W} {W} {W} {B} {W} {W} {G} \vskip 0pt 
\global\advance\reglenum by 1 
\the\reglenum{}: \lapetiteregle {B} {R} {G2} {W} {W} {B} {W} {W} {B} \vskip 0pt 
\global\advance\reglenum by 1 
\the\reglenum{}: \lapetiteregle {W} {B} {G2} {B} {W} {W} {W} {W} {W} \vskip 0pt 
\global\advance\reglenum by 1 
\the\reglenum{}: \lapetiteregle {B} {B} {W} {R} {B} {W} {W} {B2} {B} \vskip 0pt 
\global\advance\reglenum by 1 
\the\reglenum{}: \lapetiteregle {W} {B} {B} {W} {W} {W} {W} {R} {W} \vskip 0pt 
\global\advance\reglenum by 1 
\the\reglenum{}: \lapetiteregle {R} {B} {W} {W} {W} {G} {W} {B} {B2} \vskip 0pt 
\global\advance\reglenum by 1 
\the\reglenum{}: \lapetiteregle {B} {B} {R} {W} {W} {B} {W} {W} {B} \vskip 0pt 
\global\advance\reglenum by 1 
\the\reglenum{}: \lapetiteregle {W} {B} {R} {G} {W} {W} {W} {W} {W} \vskip 0pt 
\global\advance\reglenum by 1 
\the\reglenum{}: \lapetiteregle {B2} {B} {W} {W} {W} {R} {W} {B} {B2} \vskip 0pt 
\global\advance\reglenum by 1 
\the\reglenum{}: \lapetiteregle {W} {B2} {W} {W} {W} {W} {W} {R} {W} \vskip 0pt 
\global\advance\reglenum by 1 
\the\reglenum{}: \lapetiteregle {R} {B2} {W} {W} {W} {G} {W} {W} {B} \vskip 0pt 
\global\advance\reglenum by 1 
\the\reglenum{}: \lapetiteregle {W} {B} {B2} {R} {W} {W} {W} {W} {W} \vskip 0pt 
\global\advance\reglenum by 1 
\the\reglenum{}: \lapetiteregle {B} {B2} {W} {W} {W} {R} {W} {W} {B} \vskip 0pt 
\global\advance\reglenum by 1 
\the\reglenum{}: \lapetiteregle {B} {B} {W} {G2} {B} {W} {W} {B2} {G2} \vskip 0pt 
\global\advance\reglenum by 1 
\the\reglenum{}: \lapetiteregle {W} {B} {B} {W} {W} {W} {W} {G2} {W} \vskip 0pt 
\global\advance\reglenum by 1 
}
\vskip 5pt\rmviii
track~3:
\vskip 1pt
{\ttv\obeylines
\the\reglenum{}: \lapetiteregle {G2} {B} {W} {R} {B} {W} {W} {B2} {R} \vskip 0pt 
\global\advance\reglenum by 1 
\the\reglenum{}: \lapetiteregle {W} {G2} {B} {W} {W} {W} {W} {R} {W} \vskip 0pt 
\global\advance\reglenum by 1 
\the\reglenum{}: \lapetiteregle {W} {G2} {W} {W} {W} {W} {W} {B2} {W} \vskip 0pt 
\global\advance\reglenum by 1 
\the\reglenum{}: \lapetiteregle {R} {B} {W} {B2} {B} {W} {W} {G2} {B} \vskip 0pt 
\global\advance\reglenum by 1 
\the\reglenum{}: \lapetiteregle {W} {R} {W} {W} {W} {W} {W} {G2} {W} \vskip 0pt 
\global\advance\reglenum by 1 
\the\reglenum{}: \lapetiteregle {B} {B} {W} {B2} {B} {W} {W} {R} {B} \vskip 0pt 
\global\advance\reglenum by 1 
\the\reglenum{}: \lapetiteregle {W} {B} {W} {W} {W} {B2} {B2} {B} {W} \vskip 0pt 
\global\advance\reglenum by 1 
\the\reglenum{}: \lapetiteregle {B} {B} {W} {W} {B} {W} {W} {B2} {B} \vskip 0pt 
\global\advance\reglenum by 1 
}
\vskip 5pt\rm
fixed switch:
\vskip 2pt\rmviii
track~1:
\vskip 1pt
{\ttv\obeylines
\the\reglenum{}: \lapetiteregle {W} {W} {W} {W} {W} {W} {W} {B2} {W} \vskip 0pt 
\global\advance\reglenum by 1 
\the\reglenum{}: \lapetiteregle {B2} {W} {W} {W} {W} {W} {B2} {B2} {B2} \vskip 0pt 
\global\advance\reglenum by 1 
\the\reglenum{}: \lapetiteregle {W} {B2} {W} {W} {W} {W} {W} {B2} {W} \vskip 0pt 
\global\advance\reglenum by 1 
\the\reglenum{}: \lapetiteregle {B} {B} {W} {B2} {W} {B} {B2} {W} {B} \vskip 0pt 
\global\advance\reglenum by 1 
\the\reglenum{}: \lapetiteregle {B2} {B} {W} {B2} {B2} {W} {W} {W} {B2} \vskip 0pt 
\global\advance\reglenum by 1 
\the\reglenum{}: \lapetiteregle {W} {B} {B2} {W} {W} {W} {W} {B} {W} \vskip 0pt 
\global\advance\reglenum by 1 
\the\reglenum{}: \lapetiteregle {B} {B} {W} {W} {W} {B} {W} {B2} {B} \vskip 0pt 
\global\advance\reglenum by 1 
\the\reglenum{}: \lapetiteregle {B2} {B2} {B2} {W} {W} {W} {W} {W} {B2} \vskip 0pt 
\global\advance\reglenum by 1 
\the\reglenum{}: \lapetiteregle {W} {B2} {B2} {W} {W} {W} {W} {W} {W} \vskip 0pt 
\global\advance\reglenum by 1 
}
}
\setbox124=
\vtop{\leftskip 0pt\parindent 0pt
\baselineskip 7pt
\hsize=70pt
\rmviii
fixed switch:
\vskip 1pt
track~1:
\vskip 1pt
{\ttv\obeylines
\the\reglenum{}: \lapetiteregle {W} {W} {B2} {W} {W} {W} {W} {W} {W} \vskip 0pt 
\global\advance\reglenum by 1 
\the\reglenum{}: \lapetiteregle {W} {B} {B} {B2} {W} {W} {W} {W} {W} \vskip 0pt 
\global\advance\reglenum by 1 
\the\reglenum{}: \lapetiteregle {B2} {W} {B} {B} {W} {B2} {B2} {W} {B2} \vskip 0pt 
\global\advance\reglenum by 1 
\the\reglenum{}: \lapetiteregle {W} {W} {B2} {B2} {W} {W} {W} {W} {W} \vskip 0pt 
\global\advance\reglenum by 1 
\the\reglenum{}: \lapetiteregle {W} {B2} {B} {B} {W} {W} {W} {B2} {W} \vskip 0pt 
\global\advance\reglenum by 1 
\the\reglenum{}: \lapetiteregle {B2} {B2} {W} {W} {W} {W} {W} {B2} {B2} \vskip 0pt 
\global\advance\reglenum by 1 
\the\reglenum{}: \lapetiteregle {B2} {W} {B2} {B2} {W} {W} {W} {W} {B2} \vskip 0pt 
\global\advance\reglenum by 1 
\the\reglenum{}: \lapetiteregle {B} {G} {W} {W} {B} {W} {W} {B2} {G} \vskip 0pt 
\global\advance\reglenum by 1 
\the\reglenum{}: \lapetiteregle {G} {R} {W} {W} {B} {W} {W} {B2} {R} \vskip 0pt 
\global\advance\reglenum by 1 
\the\reglenum{}: \lapetiteregle {W} {G} {W} {W} {W} {B2} {B2} {B} {W} \vskip 0pt 
\global\advance\reglenum by 1 
\the\reglenum{}: \lapetiteregle {B} {G} {W} {B2} {W} {B} {B2} {W} {G} \vskip 0pt 
\global\advance\reglenum by 1 
\the\reglenum{}: \lapetiteregle {W} {G} {B} {B2} {W} {W} {W} {W} {W} \vskip 0pt 
\global\advance\reglenum by 1 
\the\reglenum{}: \lapetiteregle {R} {B} {W} {W} {G} {W} {W} {B2} {B} \vskip 0pt 
\global\advance\reglenum by 1 
\the\reglenum{}: \lapetiteregle {W} {R} {W} {W} {W} {B2} {B2} {G} {W} \vskip 0pt 
\global\advance\reglenum by 1 
\the\reglenum{}: \lapetiteregle {G} {R} {W} {B2} {W} {B} {B2} {W} {R} \vskip 0pt 
\global\advance\reglenum by 1 
\the\reglenum{}: \lapetiteregle {B2} {G} {W} {B2} {B2} {W} {W} {W} {B2} \vskip 0pt 
\global\advance\reglenum by 1 
\the\reglenum{}: \lapetiteregle {W} {G} {B2} {W} {W} {W} {W} {B} {W} \vskip 0pt 
\global\advance\reglenum by 1 
\the\reglenum{}: \lapetiteregle {B} {G} {W} {W} {W} {B} {W} {B2} {G} \vskip 0pt 
\global\advance\reglenum by 1 
\the\reglenum{}: \lapetiteregle {W} {R} {G} {B2} {W} {W} {W} {W} {W} \vskip 0pt 
\global\advance\reglenum by 1 
\the\reglenum{}: \lapetiteregle {B2} {W} {G} {B} {W} {B2} {B2} {W} {B2} \vskip 0pt 
\global\advance\reglenum by 1 
\the\reglenum{}: \lapetiteregle {B} {B} {W} {W} {R} {W} {W} {B2} {B} \vskip 0pt 
\global\advance\reglenum by 1 
\the\reglenum{}: \lapetiteregle {W} {B} {W} {W} {W} {B2} {B2} {R} {W} \vskip 0pt 
\global\advance\reglenum by 1 
\the\reglenum{}: \lapetiteregle {R} {B} {W} {B2} {W} {G} {B2} {W} {B} \vskip 0pt 
\global\advance\reglenum by 1 
\the\reglenum{}: \lapetiteregle {B2} {R} {W} {B2} {B2} {W} {W} {W} {B2} \vskip 0pt 
\global\advance\reglenum by 1 
\the\reglenum{}: \lapetiteregle {W} {R} {B2} {W} {W} {W} {W} {G} {W} \vskip 0pt 
\global\advance\reglenum by 1 
\the\reglenum{}: \lapetiteregle {G} {R} {W} {W} {W} {B} {W} {B2} {R} \vskip 0pt 
\global\advance\reglenum by 1 
\the\reglenum{}: \lapetiteregle {W} {B} {R} {B2} {W} {W} {W} {W} {W} \vskip 0pt 
\global\advance\reglenum by 1 
\the\reglenum{}: \lapetiteregle {B2} {W} {R} {G} {W} {B2} {B2} {W} {B2} \vskip 0pt 
\global\advance\reglenum by 1 
\the\reglenum{}: \lapetiteregle {W} {B2} {G} {B} {W} {W} {W} {B2} {W} \vskip 0pt 
\global\advance\reglenum by 1 
\the\reglenum{}: \lapetiteregle {B} {B} {W} {B2} {W} {R} {B2} {W} {B} \vskip 0pt 
\global\advance\reglenum by 1 
\the\reglenum{}: \lapetiteregle {W} {B} {B2} {W} {W} {W} {W} {R} {W} \vskip 0pt 
\global\advance\reglenum by 1 
\the\reglenum{}: \lapetiteregle {R} {B} {W} {W} {W} {G} {W} {B2} {B} \vskip 0pt 
\global\advance\reglenum by 1 
\the\reglenum{}: \lapetiteregle {B2} {W} {B} {R} {W} {B2} {B2} {W} {B2} \vskip 0pt 
\global\advance\reglenum by 1 
\the\reglenum{}: \lapetiteregle {W} {B2} {R} {G} {W} {W} {W} {B2} {W} \vskip 0pt 
\global\advance\reglenum by 1 
\the\reglenum{}: \lapetiteregle {B} {B} {W} {W} {W} {R} {W} {B2} {B} \vskip 0pt 
\global\advance\reglenum by 1 
\the\reglenum{}: \lapetiteregle {W} {B2} {B} {R} {W} {W} {W} {B2} {W} \vskip 0pt 
\global\advance\reglenum by 1 
\the\reglenum{}: \lapetiteregle {B} {B} {W} {W} {W} {G} {W} {B2} {G} \vskip 0pt 
\global\advance\reglenum by 1 
\the\reglenum{}: \lapetiteregle {W} {B2} {B} {G} {W} {W} {W} {B2} {W} \vskip 0pt 
\global\advance\reglenum by 1 
}
\vskip 3pt\rmviii
track~4:
\vskip 1pt
{\ttv\obeylines
\the\reglenum{}: \lapetiteregle {B} {B} {W} {B2} {W} {G} {B2} {W} {G} \vskip 0pt 
\global\advance\reglenum by 1 
\the\reglenum{}: \lapetiteregle {W} {B} {B2} {W} {W} {W} {W} {G} {W} \vskip 0pt 
\global\advance\reglenum by 1 
\the\reglenum{}: \lapetiteregle {G} {B} {W} {W} {W} {R} {W} {B2} {R} \vskip 0pt 
\global\advance\reglenum by 1 
\the\reglenum{}: \lapetiteregle {B2} {W} {B} {G} {W} {B2} {B2} {W} {B2} \vskip 0pt 
\global\advance\reglenum by 1 
\the\reglenum{}: \lapetiteregle {W} {B2} {G} {R} {W} {W} {W} {B2} {W} \vskip 0pt 
\global\advance\reglenum by 1 
\the\reglenum{}: \lapetiteregle {B} {B} {W} {W} {G} {W} {W} {B2} {G} \vskip 0pt 
\global\advance\reglenum by 1 
\the\reglenum{}: \lapetiteregle {W} {B} {W} {W} {W} {B2} {B2} {G} {W} \vskip 0pt 
\global\advance\reglenum by 1 
\the\reglenum{}: \lapetiteregle {G} {B} {W} {B2} {W} {R} {B2} {W} {R} \vskip 0pt 
\global\advance\reglenum by 1 
\the\reglenum{}: \lapetiteregle {W} {G} {B2} {W} {W} {W} {W} {R} {W} \vskip 0pt 
\global\advance\reglenum by 1 
\the\reglenum{}: \lapetiteregle {R} {G} {W} {W} {W} {B} {W} {B2} {B} \vskip 0pt 
\global\advance\reglenum by 1 
\the\reglenum{}: \lapetiteregle {W} {B} {G} {B2} {W} {W} {W} {W} {W} \vskip 0pt 
\global\advance\reglenum by 1 
\the\reglenum{}: \lapetiteregle {B2} {W} {G} {R} {W} {B2} {B2} {W} {B2} \vskip 0pt 
\global\advance\reglenum by 1 
\the\reglenum{}: \lapetiteregle {W} {B2} {R} {B} {W} {W} {W} {B2} {W} \vskip 0pt 
\global\advance\reglenum by 1 
\the\reglenum{}: \lapetiteregle {G} {B} {W} {W} {R} {W} {W} {B2} {R} \vskip 0pt 
\global\advance\reglenum by 1 
\the\reglenum{}: \lapetiteregle {W} {G} {W} {W} {W} {B2} {B2} {R} {W} \vskip 0pt 
\global\advance\reglenum by 1 
\the\reglenum{}: \lapetiteregle {R} {G} {W} {B2} {W} {B} {B2} {W} {B} \vskip 0pt 
\global\advance\reglenum by 1 
\the\reglenum{}: \lapetiteregle {W} {R} {B2} {W} {W} {W} {W} {B} {W} \vskip 0pt 
\global\advance\reglenum by 1 
\the\reglenum{}: \lapetiteregle {B} {R} {W} {W} {W} {B} {W} {B2} {B} \vskip 0pt 
\global\advance\reglenum by 1 
\the\reglenum{}: \lapetiteregle {W} {G} {R} {B2} {W} {W} {W} {W} {W} \vskip 0pt 
\global\advance\reglenum by 1 
\the\reglenum{}: \lapetiteregle {B2} {W} {R} {B} {W} {B2} {B2} {W} {B2} \vskip 0pt 
\global\advance\reglenum by 1 
\the\reglenum{}: \lapetiteregle {R} {G} {W} {W} {B} {W} {W} {B2} {B} \vskip 0pt 
\global\advance\reglenum by 1 
\the\reglenum{}: \lapetiteregle {W} {R} {W} {W} {W} {B2} {B2} {B} {W} \vskip 0pt 
\global\advance\reglenum by 1 
\the\reglenum{}: \lapetiteregle {B} {R} {W} {B2} {W} {B} {B2} {W} {B} \vskip 0pt 
\global\advance\reglenum by 1 
\the\reglenum{}: \lapetiteregle {W} {R} {B} {B2} {W} {W} {W} {W} {W} \vskip 0pt 
\global\advance\reglenum by 1 
\the\reglenum{}: \lapetiteregle {B} {R} {W} {W} {B} {W} {W} {B2} {B} \vskip 0pt 
\global\advance\reglenum by 1 
}
\vskip 3pt\rmviii
track~7:
\vskip 1pt
{\ttv\obeylines
\the\reglenum{}: \lapetiteregle {B} {B} {W} {W} {B} {W} {W} {G2} {G2} \vskip 0pt 
\global\advance\reglenum by 1 
\the\reglenum{}: \lapetiteregle {G2} {B} {W} {W} {B} {W} {W} {R} {R} \vskip 0pt 
\global\advance\reglenum by 1 
\the\reglenum{}: \lapetiteregle {W} {G2} {W} {W} {W} {B2} {B2} {B} {W} \vskip 0pt 
\global\advance\reglenum by 1 
\the\reglenum{}: \lapetiteregle {B} {G2} {W} {B2} {W} {B} {B2} {W} {G} \vskip 0pt 
\global\advance\reglenum by 1 
\the\reglenum{}: \lapetiteregle {W} {G2} {B} {B2} {W} {W} {W} {W} {W} \vskip 0pt 
\global\advance\reglenum by 1 
}
}
\setbox126=
\vtop{\leftskip 0pt\parindent 0pt
\baselineskip 7pt
\hsize=70pt
\rmviii
\rm
memory switch:
\vskip 2pt\rmviii
track~1:
\vskip 1pt
{\ttv\obeylines
\the\reglenum{}: \lapetiteregle {B} {B} {W} {B2} {B} {B2} {W} {B2} {B} \vskip 0pt 
\global\advance\reglenum by 1 
\the\reglenum{}: \lapetiteregle {B} {B} {B2} {B2} {W} {B} {W} {W} {B} \vskip 0pt 
\global\advance\reglenum by 1 
\the\reglenum{}: \lapetiteregle {B2} {B} {B2} {B} {W} {B2} {B2} {W} {B2} \vskip 0pt 
\global\advance\reglenum by 1 
\the\reglenum{}: \lapetiteregle {W} {B} {B2} {B2} {W} {W} {W} {B} {W} \vskip 0pt 
\global\advance\reglenum by 1 
\the\reglenum{}: \lapetiteregle {B2} {B} {W} {W} {B2} {B2} {W} {B} {B2} \vskip 0pt 
\global\advance\reglenum by 1 
\the\reglenum{}: \lapetiteregle {B} {B} {B2} {W} {W} {B} {W} {B2} {B} \vskip 0pt 
\global\advance\reglenum by 1 
\the\reglenum{}: \lapetiteregle {W} {B} {B2} {B2} {W} {W} {W} {W} {W} \vskip 0pt 
\global\advance\reglenum by 1 
\the\reglenum{}: \lapetiteregle {B2} {B} {B} {W} {B2} {B2} {W} {W} {B2} \vskip 0pt 
\global\advance\reglenum by 1 
\the\reglenum{}: \lapetiteregle {W} {B} {B2} {W} {W} {W} {W} {B2} {W} \vskip 0pt 
\global\advance\reglenum by 1 
\the\reglenum{}: \lapetiteregle {B2} {B} {W} {W} {W} {B} {B2} {B} {B2} \vskip 0pt 
\global\advance\reglenum by 1 
\the\reglenum{}: \lapetiteregle {B} {B2} {W} {W} {W} {B} {W} {B2} {B} \vskip 0pt 
\global\advance\reglenum by 1 
\the\reglenum{}: \lapetiteregle {B} {B} {B2} {B2} {W} {G} {W} {W} {G} \vskip 0pt 
\global\advance\reglenum by 1 
\the\reglenum{}: \lapetiteregle {W} {B} {B2} {B2} {W} {W} {W} {G} {W} \vskip 0pt 
\global\advance\reglenum by 1 
\the\reglenum{}: \lapetiteregle {B} {G} {W} {B2} {B} {B2} {W} {B2} {G} \vskip 0pt 
\global\advance\reglenum by 1 
\the\reglenum{}: \lapetiteregle {G} {B} {B2} {B2} {W} {R} {W} {W} {R} \vskip 0pt 
\global\advance\reglenum by 1 
\the\reglenum{}: \lapetiteregle {B2} {G} {B2} {B} {W} {B2} {B2} {W} {B2} \vskip 0pt 
\global\advance\reglenum by 1 
\the\reglenum{}: \lapetiteregle {W} {G} {B2} {B2} {W} {W} {W} {R} {W} \vskip 0pt 
\global\advance\reglenum by 1 
\the\reglenum{}: \lapetiteregle {B2} {B} {W} {W} {W} {B} {B2} {G} {B2} \vskip 0pt 
\global\advance\reglenum by 1 
\the\reglenum{}: \lapetiteregle {G} {R} {W} {B2} {B} {B2} {W} {B2} {R} \vskip 0pt 
\global\advance\reglenum by 1 
\the\reglenum{}: \lapetiteregle {R} {G} {B2} {B2} {W} {B} {W} {W} {B} \vskip 0pt 
\global\advance\reglenum by 1 
\the\reglenum{}: \lapetiteregle {B2} {R} {B2} {B} {W} {B2} {B2} {W} {B2} \vskip 0pt 
\global\advance\reglenum by 1 
\the\reglenum{}: \lapetiteregle {W} {R} {B2} {B2} {W} {W} {W} {B} {W} \vskip 0pt 
\global\advance\reglenum by 1 
\the\reglenum{}: \lapetiteregle {B2} {G} {W} {W} {B2} {B2} {W} {B} {B2} \vskip 0pt 
\global\advance\reglenum by 1 
\the\reglenum{}: \lapetiteregle {B} {G} {B2} {W} {W} {B} {W} {B2} {G} \vskip 0pt 
\global\advance\reglenum by 1 
\the\reglenum{}: \lapetiteregle {B2} {G} {B} {W} {B2} {B2} {W} {W} {B2} \vskip 0pt 
\global\advance\reglenum by 1 
\the\reglenum{}: \lapetiteregle {W} {G} {B2} {W} {W} {W} {W} {B2} {W} \vskip 0pt 
\global\advance\reglenum by 1 
\the\reglenum{}: \lapetiteregle {B2} {G} {W} {W} {W} {B} {B2} {R} {B2} \vskip 0pt 
\global\advance\reglenum by 1 
\the\reglenum{}: \lapetiteregle {R} {B} {W} {B2} {G} {B2} {W} {B2} {B} \vskip 0pt 
\global\advance\reglenum by 1 
\the\reglenum{}: \lapetiteregle {B} {R} {B2} {B2} {W} {B} {W} {W} {B} \vskip 0pt 
\global\advance\reglenum by 1 
\the\reglenum{}: \lapetiteregle {B2} {R} {W} {W} {B2} {B2} {W} {G} {B2} \vskip 0pt 
\global\advance\reglenum by 1 
\the\reglenum{}: \lapetiteregle {G} {R} {B2} {W} {W} {B} {W} {B2} {R} \vskip 0pt 
\global\advance\reglenum by 1 
\the\reglenum{}: \lapetiteregle {W} {G} {B2} {B2} {W} {W} {W} {W} {W} \vskip 0pt 
\global\advance\reglenum by 1 
\the\reglenum{}: \lapetiteregle {B2} {R} {G} {W} {B2} {B2} {W} {W} {B2} \vskip 0pt 
\global\advance\reglenum by 1 
\the\reglenum{}: \lapetiteregle {W} {R} {B2} {W} {W} {W} {W} {B2} {W} \vskip 0pt 
\global\advance\reglenum by 1 
\the\reglenum{}: \lapetiteregle {B2} {R} {W} {W} {W} {B} {B2} {B} {B2} \vskip 0pt 
\global\advance\reglenum by 1 
\the\reglenum{}: \lapetiteregle {B} {B} {W} {B2} {R} {B2} {W} {B2} {B} \vskip 0pt 
\global\advance\reglenum by 1 
\the\reglenum{}: \lapetiteregle {B2} {B} {W} {W} {B2} {B2} {W} {R} {B2} \vskip 0pt 
\global\advance\reglenum by 1 
\the\reglenum{}: \lapetiteregle {R} {B} {B2} {W} {W} {G} {W} {B2} {B} \vskip 0pt 
\global\advance\reglenum by 1 
\the\reglenum{}: \lapetiteregle {W} {R} {B2} {B2} {W} {W} {W} {W} {W} \vskip 0pt 
\global\advance\reglenum by 1 
\the\reglenum{}: \lapetiteregle {B2} {B} {R} {W} {B2} {B2} {W} {W} {B2} \vskip 0pt 
\global\advance\reglenum by 1 
\the\reglenum{}: \lapetiteregle {B} {B} {B2} {W} {W} {R} {W} {B2} {B} \vskip 0pt 
\global\advance\reglenum by 1 
}
\vskip 3pt\rmviii
track~4:
\vskip 1pt
{\ttv\obeylines
\the\reglenum{}: \lapetiteregle {B} {B} {B2} {W} {W} {G} {W} {B2} {G} \vskip 0pt 
\global\advance\reglenum by 1 
\the\reglenum{}: \lapetiteregle {B} {B} {W} {B2} {G} {B2} {W} {B2} {G} \vskip 0pt 
\global\advance\reglenum by 1 
\the\reglenum{}: \lapetiteregle {B2} {B} {W} {W} {B2} {B2} {W} {G} {B2} \vskip 0pt 
\global\advance\reglenum by 1 
\the\reglenum{}: \lapetiteregle {G} {B} {B2} {W} {W} {R} {W} {B2} {R} \vskip 0pt 
\global\advance\reglenum by 1 
\the\reglenum{}: \lapetiteregle {B2} {B} {G} {W} {B2} {B2} {W} {W} {B2} \vskip 0pt 
\global\advance\reglenum by 1 
\the\reglenum{}: \lapetiteregle {G} {B} {W} {B2} {R} {B2} {W} {B2} {R} \vskip 0pt 
\global\advance\reglenum by 1 
\the\reglenum{}: \lapetiteregle {B} {G} {B2} {B2} {W} {B} {W} {W} {G} \vskip 0pt 
\global\advance\reglenum by 1 
\the\reglenum{}: \lapetiteregle {B2} {G} {W} {W} {B2} {B2} {W} {R} {B2} \vskip 0pt 
\global\advance\reglenum by 1 
\the\reglenum{}: \lapetiteregle {R} {G} {B2} {W} {W} {B} {W} {B2} {B} \vskip 0pt 
\global\advance\reglenum by 1 
\the\reglenum{}: \lapetiteregle {B2} {G} {R} {W} {B2} {B2} {W} {W} {B2} \vskip 0pt 
\global\advance\reglenum by 1 
\the\reglenum{}: \lapetiteregle {B2} {G} {W} {W} {W} {B} {B2} {B} {B2} \vskip 0pt 
\global\advance\reglenum by 1 
\the\reglenum{}: \lapetiteregle {R} {G} {W} {B2} {B} {B2} {W} {B2} {B} \vskip 0pt 
\global\advance\reglenum by 1 
\the\reglenum{}: \lapetiteregle {G} {R} {B2} {B2} {W} {B} {W} {W} {R} \vskip 0pt 
\global\advance\reglenum by 1 
\the\reglenum{}: \lapetiteregle {W} {G} {B2} {B2} {W} {W} {W} {B} {W} \vskip 0pt 
\global\advance\reglenum by 1 
\the\reglenum{}: \lapetiteregle {B2} {R} {W} {W} {B2} {B2} {W} {B} {B2} \vskip 0pt 
\global\advance\reglenum by 1 
\the\reglenum{}: \lapetiteregle {B} {R} {B2} {W} {W} {B} {W} {B2} {B} \vskip 0pt 
\global\advance\reglenum by 1 
\the\reglenum{}: \lapetiteregle {B2} {R} {B} {W} {B2} {B2} {W} {W} {B2} \vskip 0pt 
\global\advance\reglenum by 1 
\the\reglenum{}: \lapetiteregle {B2} {R} {W} {W} {W} {B} {B2} {G} {B2} \vskip 0pt 
\global\advance\reglenum by 1 
\the\reglenum{}: \lapetiteregle {B} {R} {W} {B2} {B} {B2} {W} {B2} {B} \vskip 0pt 
\global\advance\reglenum by 1 
\the\reglenum{}: \lapetiteregle {R} {B} {B2} {B2} {W} {G} {W} {W} {B} \vskip 0pt 
\global\advance\reglenum by 1 
\the\reglenum{}: \lapetiteregle {W} {R} {B2} {B2} {W} {W} {W} {G} {W} \vskip 0pt 
\global\advance\reglenum by 1 
\the\reglenum{}: \lapetiteregle {B2} {B} {W} {W} {W} {B} {B2} {R} {B2} \vskip 0pt 
\global\advance\reglenum by 1 
\the\reglenum{}: \lapetiteregle {B} {B} {B2} {B2} {W} {R} {W} {W} {B} \vskip 0pt 
\global\advance\reglenum by 1 
\the\reglenum{}: \lapetiteregle {W} {B} {B2} {B2} {W} {W} {W} {R} {W} \vskip 0pt 
\global\advance\reglenum by 1 
}
\vskip 3pt\rmviii
track~7:
\vskip 1pt
{\ttv\obeylines
\the\reglenum{}: \lapetiteregle {B} {B2} {W} {W} {W} {G} {W} {B2} {G} \vskip 0pt 
\global\advance\reglenum by 1 
\the\reglenum{}: \lapetiteregle {B2} {B} {B2} {G} {W} {B2} {B2} {W} {B2} \vskip 0pt 
\global\advance\reglenum by 1 
}
}
\setbox130=
\vtop{\leftskip 0pt\parindent 0pt
\baselineskip 7pt
\hsize=70pt
\rmviii
memory switch:
\vskip 2pt
track~7:
\vskip 1pt
{\ttv\obeylines
\the\reglenum{}: \lapetiteregle {B2} {B} {W} {W} {W} {G} {B2} {B} {G2} \vskip 0pt 
\global\advance\reglenum by 1 
\the\reglenum{}: \lapetiteregle {G} {B2} {W} {W} {W} {R} {W} {B2} {R} \vskip 0pt 
\global\advance\reglenum by 1 
\the\reglenum{}: \lapetiteregle {B} {B} {W} {B2} {B} {B2} {W} {G2} {G2} \vskip 0pt 
\global\advance\reglenum by 1 
\the\reglenum{}: \lapetiteregle {B} {B} {G2} {B2} {W} {B} {W} {W} {B} \vskip 0pt 
\global\advance\reglenum by 1 
\the\reglenum{}: \lapetiteregle {B2} {B} {G2} {R} {W} {B2} {B2} {W} {B2} \vskip 0pt 
\global\advance\reglenum by 1 
\the\reglenum{}: \lapetiteregle {W} {B} {B2} {W} {W} {W} {W} {G2} {W} \vskip 0pt 
\global\advance\reglenum by 1 
\the\reglenum{}: \lapetiteregle {G2} {B} {W} {W} {W} {R} {B2} {B} {R} \vskip 0pt 
\global\advance\reglenum by 1 
\the\reglenum{}: \lapetiteregle {R} {G2} {W} {W} {W} {B} {W} {B2} {B} \vskip 0pt 
\global\advance\reglenum by 1 
\the\reglenum{}: \lapetiteregle {G2} {B} {W} {B2} {B} {B2} {W} {R} {R} \vskip 0pt 
\global\advance\reglenum by 1 
\the\reglenum{}: \lapetiteregle {B} {G2} {R} {B2} {W} {B} {W} {W} {B2} \vskip 0pt 
\global\advance\reglenum by 1 
\the\reglenum{}: \lapetiteregle {B2} {B} {R} {B} {W} {B2} {B2} {W} {B2} \vskip 0pt 
\global\advance\reglenum by 1 
\the\reglenum{}: \lapetiteregle {B2} {G2} {W} {W} {B2} {B2} {W} {B} {B2} \vskip 0pt 
\global\advance\reglenum by 1 
\the\reglenum{}: \lapetiteregle {B} {G2} {B2} {W} {W} {B} {W} {B2} {G} \vskip 0pt 
\global\advance\reglenum by 1 
\the\reglenum{}: \lapetiteregle {B2} {G2} {B} {W} {B2} {B2} {W} {W} {B2} \vskip 0pt 
\global\advance\reglenum by 1 
\the\reglenum{}: \lapetiteregle {W} {G2} {B2} {W} {W} {W} {W} {R} {W} \vskip 0pt 
\global\advance\reglenum by 1 
\the\reglenum{}: \lapetiteregle {R} {G2} {W} {W} {W} {B} {B2} {B} {B} \vskip 0pt 
\global\advance\reglenum by 1 
\the\reglenum{}: \lapetiteregle {R} {B2} {W} {B2} {G} {B2} {W} {B} {B} \vskip 0pt 
\global\advance\reglenum by 1 
\the\reglenum{}: \lapetiteregle {B2} {R} {B} {B2} {W} {B} {W} {W} {B2} \vskip 0pt 
\global\advance\reglenum by 1 
\the\reglenum{}: \lapetiteregle {B2} {B2} {B} {B} {W} {B2} {B2} {W} {B2} \vskip 0pt 
\global\advance\reglenum by 1 
\the\reglenum{}: \lapetiteregle {W} {B2} {B2} {B2} {W} {W} {W} {B} {W} \vskip 0pt 
\global\advance\reglenum by 1 
\the\reglenum{}: \lapetiteregle {W} {W} {B2} {B} {W} {W} {W} {W} {W} \vskip 0pt 
\global\advance\reglenum by 1 
\the\reglenum{}: \lapetiteregle {B} {R} {W} {W} {W} {B} {B2} {B2} {B} \vskip 0pt 
\global\advance\reglenum by 1 
\the\reglenum{}: \lapetiteregle {B} {B2} {W} {B2} {R} {B2} {W} {B} {B} \vskip 0pt 
\global\advance\reglenum by 1 
\the\reglenum{}: \lapetiteregle {B2} {B} {B} {B2} {W} {B} {W} {W} {B2} \vskip 0pt 
\global\advance\reglenum by 1 
\the\reglenum{}: \lapetiteregle {B} {B} {W} {W} {W} {B} {B2} {B2} {B} \vskip 0pt 
\global\advance\reglenum by 1 
\the\reglenum{}: \lapetiteregle {B} {B2} {W} {B2} {B} {B2} {W} {B} {B} \vskip 0pt 
\global\advance\reglenum by 1 
}
\vskip 5pt\rm
memory switch L
\vskip 3pt\rmviii
track~7:
\vskip 1pt
{\ttv\obeylines
\the\reglenum{}: \lapetiteregle {B2} {B2} {B} {G} {W} {B2} {B2} {W} {B2} \vskip 0pt 
\global\advance\reglenum by 1 
\the\reglenum{}: \lapetiteregle {B} {B} {W} {W} {W} {G} {B2} {B2} {G} \vskip 0pt 
\global\advance\reglenum by 1 
\the\reglenum{}: \lapetiteregle {B} {B2} {W} {B2} {B} {B2} {W} {G} {G} \vskip 0pt 
\global\advance\reglenum by 1 
\the\reglenum{}: \lapetiteregle {B2} {B} {G} {B2} {W} {B} {W} {W} {B2} \vskip 0pt 
\global\advance\reglenum by 1 
\the\reglenum{}: \lapetiteregle {B2} {B2} {G} {R} {W} {B2} {B2} {W} {B2} \vskip 0pt 
\global\advance\reglenum by 1 
\the\reglenum{}: \lapetiteregle {G} {B} {W} {W} {W} {R} {B2} {B2} {R} \vskip 0pt 
\global\advance\reglenum by 1 
\the\reglenum{}: \lapetiteregle {G} {B2} {W} {B2} {B} {B2} {W} {R} {R} \vskip 0pt 
\global\advance\reglenum by 1 
\the\reglenum{}: \lapetiteregle {B2} {G} {R} {B2} {W} {B} {W} {W} {B2} \vskip 0pt 
\global\advance\reglenum by 1 
\the\reglenum{}: \lapetiteregle {B2} {B2} {R} {B} {W} {B2} {B2} {W} {B2} \vskip 0pt 
\global\advance\reglenum by 1 
\the\reglenum{}: \lapetiteregle {R} {G} {W} {W} {W} {B} {B2} {B2} {B} \vskip 0pt 
\global\advance\reglenum by 1 
}
\vskip 3pt\rmviii
track~4:
\vskip 1pt
{\ttv\obeylines
\the\reglenum{}: \lapetiteregle {B} {B2} {W} {B2} {G} {B2} {W} {B} {G} \vskip 0pt 
\global\advance\reglenum by 1 
\the\reglenum{}: \lapetiteregle {G} {B2} {W} {B2} {R} {B2} {W} {B} {R} \vskip 0pt 
\global\advance\reglenum by 1 
\the\reglenum{}: \lapetiteregle {B2} {G} {B} {B2} {W} {B} {W} {W} {B2} \vskip 0pt 
\global\advance\reglenum by 1 
\the\reglenum{}: \lapetiteregle {B} {G} {W} {W} {W} {B} {B2} {B2} {G} \vskip 0pt 
\global\advance\reglenum by 1 
\the\reglenum{}: \lapetiteregle {R} {B2} {W} {B2} {B} {B2} {W} {G} {B} \vskip 0pt 
\global\advance\reglenum by 1 
\the\reglenum{}: \lapetiteregle {B2} {R} {G} {B2} {W} {B} {W} {W} {B2} \vskip 0pt 
\global\advance\reglenum by 1 
\the\reglenum{}: \lapetiteregle {B2} {B2} {G} {B} {W} {B2} {B2} {W} {B2} \vskip 0pt 
\global\advance\reglenum by 1 
\the\reglenum{}: \lapetiteregle {G} {R} {W} {W} {W} {B} {B2} {B2} {R} \vskip 0pt 
\global\advance\reglenum by 1 
\the\reglenum{}: \lapetiteregle {B} {B2} {W} {B2} {B} {B2} {W} {R} {B} \vskip 0pt 
\global\advance\reglenum by 1 
\the\reglenum{}: \lapetiteregle {B2} {B} {R} {B2} {W} {B} {W} {W} {B2} \vskip 0pt 
\global\advance\reglenum by 1 
\the\reglenum{}: \lapetiteregle {B2} {B2} {R} {G} {W} {B2} {B2} {W} {B2} \vskip 0pt 
\global\advance\reglenum by 1 
\the\reglenum{}: \lapetiteregle {R} {B} {W} {W} {W} {G} {B2} {B2} {B} \vskip 0pt 
\global\advance\reglenum by 1 
\the\reglenum{}: \lapetiteregle {B2} {B2} {B} {R} {W} {B2} {B2} {W} {B2} \vskip 0pt 
\global\advance\reglenum by 1 
\the\reglenum{}: \lapetiteregle {B} {B} {W} {W} {W} {R} {B2} {B2} {B} \vskip 0pt 
\global\advance\reglenum by 1 
}
\vskip 3pt\rmviii
track~1:
\vskip 1pt
{\ttv\obeylines
\the\reglenum{}: \lapetiteregle {B2} {B} {B} {B2} {W} {G} {W} {W} {G2} \vskip 0pt 
\global\advance\reglenum by 1 
\the\reglenum{}: \lapetiteregle {W} {B2} {B2} {B2} {W} {W} {W} {G} {W} \vskip 0pt 
\global\advance\reglenum by 1 
\the\reglenum{}: \lapetiteregle {W} {W} {B2} {G} {W} {W} {W} {W} {W} \vskip 0pt 
\global\advance\reglenum by 1 
\the\reglenum{}: \lapetiteregle {B} {G2} {W} {B2} {B} {B2} {W} {B} {G2} \vskip 0pt 
\global\advance\reglenum by 1 
\the\reglenum{}: \lapetiteregle {G2} {B} {B} {B2} {W} {R} {W} {W} {R} \vskip 0pt 
\global\advance\reglenum by 1 
\the\reglenum{}: \lapetiteregle {B2} {G2} {B} {B} {W} {B2} {B2} {W} {B2} \vskip 0pt 
\global\advance\reglenum by 1 
\the\reglenum{}: \lapetiteregle {W} {G2} {B2} {B2} {W} {W} {W} {R} {W} \vskip 0pt 
\global\advance\reglenum by 1 
\the\reglenum{}: \lapetiteregle {W} {B} {G2} {W} {W} {W} {W} {B2} {W} \vskip 0pt 
\global\advance\reglenum by 1 
\the\reglenum{}: \lapetiteregle {W} {W} {G2} {R} {W} {W} {W} {W} {W} \vskip 0pt 
\global\advance\reglenum by 1 
\the\reglenum{}: \lapetiteregle {B} {B} {W} {W} {W} {B} {B2} {G2} {B} \vskip 0pt 
\global\advance\reglenum by 1 
\the\reglenum{}: \lapetiteregle {G2} {R} {W} {B2} {B} {B2} {W} {B} {R} \vskip 0pt 
\global\advance\reglenum by 1 
\the\reglenum{}: \lapetiteregle {R} {G2} {B} {B2} {W} {B} {W} {W} {B} \vskip 0pt 
\global\advance\reglenum by 1 
\the\reglenum{}: \lapetiteregle {B2} {R} {B} {B} {W} {B2} {B2} {W} {B2} \vskip 0pt 
\global\advance\reglenum by 1 
\the\reglenum{}: \lapetiteregle {W} {G2} {R} {W} {W} {W} {W} {B2} {W} \vskip 0pt 
\global\advance\reglenum by 1 
\the\reglenum{}: \lapetiteregle {W} {G2} {B2} {W} {W} {W} {W} {B} {W} \vskip 0pt 
\global\advance\reglenum by 1 
\the\reglenum{}: \lapetiteregle {B} {G2} {W} {W} {W} {B} {B2} {R} {B2} \vskip 0pt 
\global\advance\reglenum by 1 
}
}
\setbox132=
\vtop{\leftskip 0pt\parindent 0pt
\baselineskip 7pt
\hsize=70pt
\rmviii
\rm
flip-flop switch:
\vskip 3pt\rmviii
track~4:
\vskip 1pt
{\ttv\obeylines
\the\reglenum{}: \lapetiteregle {B} {B} {B2} {G2} {W} {B} {G2} {W} {B} \vskip 0pt 
\global\advance\reglenum by 1 
\the\reglenum{}: \lapetiteregle {G2} {B} {B2} {B} {W} {B2} {B2} {W} {G2} \vskip 0pt 
\global\advance\reglenum by 1 
\the\reglenum{}: \lapetiteregle {W} {B} {G2} {B2} {W} {W} {W} {B} {W} \vskip 0pt 
\global\advance\reglenum by 1 
\the\reglenum{}: \lapetiteregle {B} {B} {W} {W} {W} {B} {W} {G2} {B} \vskip 0pt 
\global\advance\reglenum by 1 
\the\reglenum{}: \lapetiteregle {W} {G2} {B} {B} {W} {W} {W} {B2} {W} \vskip 0pt 
\global\advance\reglenum by 1 
\the\reglenum{}: \lapetiteregle {B2} {G2} {W} {W} {W} {W} {W} {B2} {B2} \vskip 0pt 
\global\advance\reglenum by 1 
\the\reglenum{}: \lapetiteregle {B2} {W} {G2} {B2} {W} {W} {W} {W} {B2} \vskip 0pt 
\global\advance\reglenum by 1 
\the\reglenum{}: \lapetiteregle {W} {B} {B} {G2} {W} {W} {W} {W} {W} \vskip 0pt 
\global\advance\reglenum by 1 
\the\reglenum{}: \lapetiteregle {G2} {W} {B} {B} {W} {B2} {B2} {W} {G2} \vskip 0pt 
\global\advance\reglenum by 1 
\the\reglenum{}: \lapetiteregle {W} {W} {G2} {B2} {W} {W} {W} {W} {W} \vskip 0pt 
\global\advance\reglenum by 1 
\the\reglenum{}: \lapetiteregle {W} {B} {W} {W} {W} {W} {G2} {B2} {W} \vskip 0pt 
\global\advance\reglenum by 1 
\the\reglenum{}: \lapetiteregle {W} {W} {W} {W} {W} {W} {W} {G2} {W} \vskip 0pt 
\global\advance\reglenum by 1 
\the\reglenum{}: \lapetiteregle {B2} {B} {W} {G2} {W} {B} {G2} {B} {B2} \vskip 0pt 
\global\advance\reglenum by 1 
\the\reglenum{}: \lapetiteregle {G2} {B2} {W} {W} {W} {B2} {B2} {W} {G2} \vskip 0pt 
\global\advance\reglenum by 1 
\the\reglenum{}: \lapetiteregle {W} {B2} {G2} {B2} {W} {W} {W} {B} {W} \vskip 0pt 
\global\advance\reglenum by 1 
\the\reglenum{}: \lapetiteregle {B} {B2} {W} {W} {W} {B} {W} {G2} {B} \vskip 0pt 
\global\advance\reglenum by 1 
\the\reglenum{}: \lapetiteregle {B} {G} {B2} {G2} {W} {B} {G2} {W} {G} \vskip 0pt 
\global\advance\reglenum by 1 
\the\reglenum{}: \lapetiteregle {W} {G} {B} {G2} {W} {W} {W} {W} {W} \vskip 0pt 
\global\advance\reglenum by 1 
\the\reglenum{}: \lapetiteregle {W} {G} {W} {W} {W} {W} {G2} {B2} {W} \vskip 0pt 
\global\advance\reglenum by 1 
\the\reglenum{}: \lapetiteregle {B2} {G} {W} {G2} {W} {B} {G2} {B} {B2} \vskip 0pt 
\global\advance\reglenum by 1 
\the\reglenum{}: \lapetiteregle {G} {R} {B2} {G2} {W} {B} {G2} {W} {R} \vskip 0pt 
\global\advance\reglenum by 1 
\the\reglenum{}: \lapetiteregle {G2} {G} {B2} {B} {W} {B2} {B2} {W} {G2} \vskip 0pt 
\global\advance\reglenum by 1 
\the\reglenum{}: \lapetiteregle {W} {G} {G2} {B2} {W} {W} {W} {B} {W} \vskip 0pt 
\global\advance\reglenum by 1 
\the\reglenum{}: \lapetiteregle {B} {G} {W} {W} {W} {B} {W} {G2} {G} \vskip 0pt 
\global\advance\reglenum by 1 
\the\reglenum{}: \lapetiteregle {W} {R} {G} {G2} {W} {W} {W} {W} {W} \vskip 0pt 
\global\advance\reglenum by 1 
\the\reglenum{}: \lapetiteregle {G2} {W} {G} {B} {W} {B2} {B2} {W} {G2} \vskip 0pt 
\global\advance\reglenum by 1 
\the\reglenum{}: \lapetiteregle {W} {R} {W} {W} {W} {W} {G2} {B2} {W} \vskip 0pt 
\global\advance\reglenum by 1 
\the\reglenum{}: \lapetiteregle {B2} {R} {W} {G2} {W} {B} {G2} {G} {B2} \vskip 0pt 
\global\advance\reglenum by 1 
\the\reglenum{}: \lapetiteregle {R} {B} {B2} {G2} {W} {G} {G2} {W} {B2} \vskip 0pt 
\global\advance\reglenum by 1 
\the\reglenum{}: \lapetiteregle {G2} {R} {B2} {B} {W} {B2} {B2} {W} {G2} \vskip 0pt 
\global\advance\reglenum by 1 
\the\reglenum{}: \lapetiteregle {W} {R} {G2} {B2} {W} {W} {W} {G} {W} \vskip 0pt 
\global\advance\reglenum by 1 
\the\reglenum{}: \lapetiteregle {G} {R} {W} {W} {W} {B} {W} {G2} {R} \vskip 0pt 
\global\advance\reglenum by 1 
\the\reglenum{}: \lapetiteregle {W} {B} {R} {G2} {W} {W} {W} {W} {W} \vskip 0pt 
\global\advance\reglenum by 1 
\the\reglenum{}: \lapetiteregle {G2} {W} {R} {G} {W} {B2} {B2} {W} {G2} \vskip 0pt 
\global\advance\reglenum by 1 
\the\reglenum{}: \lapetiteregle {W} {G2} {G} {B} {W} {W} {W} {B2} {W} \vskip 0pt 
\global\advance\reglenum by 1 
\the\reglenum{}: \lapetiteregle {B2} {B} {W} {G2} {W} {B} {G2} {R} {B} \vskip 0pt 
\global\advance\reglenum by 1 
\the\reglenum{}: \lapetiteregle {B} {B2} {W} {W} {B} {W} {W} {B} {B} \vskip 0pt 
\global\advance\reglenum by 1 
\the\reglenum{}: \lapetiteregle {B2} {B} {B} {G2} {W} {R} {G2} {W} {B2} \vskip 0pt 
\global\advance\reglenum by 1 
\the\reglenum{}: \lapetiteregle {G2} {B2} {B} {B} {W} {B2} {B2} {W} {G2} \vskip 0pt 
\global\advance\reglenum by 1 
\the\reglenum{}: \lapetiteregle {W} {B2} {G2} {B2} {W} {W} {W} {R} {W} \vskip 0pt 
\global\advance\reglenum by 1 
\the\reglenum{}: \lapetiteregle {R} {B2} {W} {W} {W} {G} {W} {G2} {B} \vskip 0pt 
\global\advance\reglenum by 1 
\the\reglenum{}: \lapetiteregle {W} {B} {B2} {G2} {W} {W} {W} {W} {W} \vskip 0pt 
\global\advance\reglenum by 1 
\the\reglenum{}: \lapetiteregle {G2} {W} {B2} {R} {W} {B2} {B2} {W} {G2} \vskip 0pt 
\global\advance\reglenum by 1 
\the\reglenum{}: \lapetiteregle {W} {G2} {R} {G} {W} {W} {W} {B2} {W} \vskip 0pt 
\global\advance\reglenum by 1 
\the\reglenum{}: \lapetiteregle {W} {B} {W} {W} {W} {W} {G2} {B} {W} \vskip 0pt 
\global\advance\reglenum by 1 
\the\reglenum{}: \lapetiteregle {B} {B} {W} {G2} {W} {B} {G2} {B2} {B} \vskip 0pt 
\global\advance\reglenum by 1 
\the\reglenum{}: \lapetiteregle {G2} {B} {W} {W} {W} {B2} {B2} {W} {G2} \vskip 0pt 
\global\advance\reglenum by 1 
\the\reglenum{}: \lapetiteregle {B2} {B} {B} {G2} {W} {B} {G2} {W} {B2} \vskip 0pt 
\global\advance\reglenum by 1 
\the\reglenum{}: \lapetiteregle {B} {B2} {W} {W} {W} {R} {W} {G2} {B} \vskip 0pt 
\global\advance\reglenum by 1 
\the\reglenum{}: \lapetiteregle {G2} {W} {B2} {B} {W} {B2} {B2} {W} {G2} \vskip 0pt 
\global\advance\reglenum by 1 
\the\reglenum{}: \lapetiteregle {W} {G2} {B} {R} {W} {W} {W} {B2} {W} \vskip 0pt 
\global\advance\reglenum by 1 
}
\vskip 5pt
flip-flop switch L:
\vskip 2pt\rmviii
track~4:
\vskip 1pt
{\ttv\obeylines
\the\reglenum{}: \lapetiteregle {B} {B2} {W} {W} {G} {W} {W} {B} {G} \vskip 0pt 
\global\advance\reglenum by 1 
\the\reglenum{}: \lapetiteregle {G} {B2} {W} {W} {R} {W} {W} {B} {R} \vskip 0pt 
\global\advance\reglenum by 1 
\the\reglenum{}: \lapetiteregle {B2} {G} {B} {G2} {W} {B} {G2} {W} {B2} \vskip 0pt 
\global\advance\reglenum by 1 
\the\reglenum{}: \lapetiteregle {W} {G} {B2} {G2} {W} {W} {W} {W} {W} \vskip 0pt 
\global\advance\reglenum by 1 
\the\reglenum{}: \lapetiteregle {W} {G} {W} {W} {W} {W} {G2} {B} {W} \vskip 0pt 
\global\advance\reglenum by 1 
\the\reglenum{}: \lapetiteregle {B} {G} {W} {G2} {W} {B} {G2} {B2} {G} \vskip 0pt 
\global\advance\reglenum by 1 
\the\reglenum{}: \lapetiteregle {R} {B2} {W} {W} {B} {W} {W} {G} {B} \vskip 0pt 
\global\advance\reglenum by 1 
\the\reglenum{}: \lapetiteregle {B2} {R} {G} {G2} {W} {B} {G2} {W} {B2} \vskip 0pt 
\global\advance\reglenum by 1 
\the\reglenum{}: \lapetiteregle {G2} {B2} {G} {B} {W} {B2} {B2} {W} {G2} \vskip 0pt 
\global\advance\reglenum by 1 
\the\reglenum{}: \lapetiteregle {W} {R} {B2} {G2} {W} {W} {W} {W} {W} \vskip 0pt 
\global\advance\reglenum by 1 
\the\reglenum{}: \lapetiteregle {W} {R} {W} {W} {W} {W} {G2} {G} {W} \vskip 0pt 
\global\advance\reglenum by 1 
\the\reglenum{}: \lapetiteregle {G} {R} {W} {G2} {W} {B} {G2} {B2} {R} \vskip 0pt 
\global\advance\reglenum by 1 
\the\reglenum{}: \lapetiteregle {G2} {G} {W} {W} {W} {B2} {B2} {W} {G2} \vskip 0pt 
\global\advance\reglenum by 1 
\the\reglenum{}: \lapetiteregle {B} {B2} {W} {W} {B} {W} {W} {R} {B} \vskip 0pt 
\global\advance\reglenum by 1 
\the\reglenum{}: \lapetiteregle {B2} {B} {R} {G2} {W} {B} {G2} {W} {B} \vskip 0pt 
\global\advance\reglenum by 1 
\the\reglenum{}: \lapetiteregle {G2} {B2} {R} {G} {W} {B2} {B2} {W} {G2} \vskip 0pt 
\global\advance\reglenum by 1 
\the\reglenum{}: \lapetiteregle {W} {B} {W} {W} {W} {W} {G2} {R} {W} \vskip 0pt 
\global\advance\reglenum by 1 
\the\reglenum{}: \lapetiteregle {R} {B} {W} {G2} {W} {G} {G2} {B2} {B2} \vskip 0pt 
\global\advance\reglenum by 1 
}
}
\setbox134=
\vtop{\leftskip 0pt\parindent 0pt
\baselineskip 7pt
\hsize=70pt
{\ttv\obeylines
\the\reglenum{}: \lapetiteregle {G2} {R} {W} {W} {W} {B2} {B2} {W} {G2} \vskip 0pt 
\global\advance\reglenum by 1 
\the\reglenum{}: \lapetiteregle {G2} {B} {B2} {R} {W} {B2} {B2} {W} {G2} \vskip 0pt 
\global\advance\reglenum by 1 
\the\reglenum{}: \lapetiteregle {B2} {B} {W} {G2} {W} {R} {G2} {B} {B2} \vskip 0pt 
\global\advance\reglenum by 1 
}
\vskip 5pt\rm
tracks alone:
\vskip 1pt
{\ttv\obeylines
\the\reglenum{}: \lapetiteregle {W} {W} {B} {B} {B} {W} {W} {B} {W} \vskip 0pt 
\global\advance\reglenum by 1 
\the\reglenum{}: \lapetiteregle {W} {W} {B} {W} {W} {W} {W} {B} {W} \vskip 0pt 
\global\advance\reglenum by 1 
\the\reglenum{}: \lapetiteregle {B} {W} {W} {W} {W} {B} {W} {B} {B} \vskip 0pt 
\global\advance\reglenum by 1 
\the\reglenum{}: \lapetiteregle {B} {W} {B} {W} {W} {W} {W} {B} {B} \vskip 0pt 
\global\advance\reglenum by 1 
\the\reglenum{}: \lapetiteregle {W} {B} {B} {B} {W} {W} {W} {W} {W} \vskip 0pt 
\global\advance\reglenum by 1 
\the\reglenum{}: \lapetiteregle {B} {W} {B} {W} {W} {W} {B} {W} {B} \vskip 0pt 
\global\advance\reglenum by 1 
\the\reglenum{}: \lapetiteregle {W} {W} {B} {B} {B} {B} {B} {W} {W} \vskip 0pt 
\global\advance\reglenum by 1 
\the\reglenum{}: \lapetiteregle {W} {W} {W} {B} {B} {W} {B} {B} {W} \vskip 0pt 
\global\advance\reglenum by 1 
\the\reglenum{}: \lapetiteregle {B} {W} {W} {B} {W} {W} {W} {B} {B} \vskip 0pt 
\global\advance\reglenum by 1 
\the\reglenum{}: \lapetiteregle {W} {W} {B} {B} {B} {B} {B} {B} {W} \vskip 0pt 
\global\advance\reglenum by 1 
\the\reglenum{}: \lapetiteregle {B} {W} {W} {B} {W} {B} {W} {W} {B} \vskip 0pt 
\global\advance\reglenum by 1 
\the\reglenum{}: \lapetiteregle {B} {B} {W} {W} {B} {W} {W} {W} {B} \vskip 0pt 
\global\advance\reglenum by 1 
\the\reglenum{}: \lapetiteregle {W} {B} {B} {W} {W} {W} {W} {B} {W} \vskip 0pt 
\global\advance\reglenum by 1 
\the\reglenum{}: \lapetiteregle {B} {B} {W} {B} {W} {W} {W} {W} {B} \vskip 0pt 
\global\advance\reglenum by 1 
\the\reglenum{}: \lapetiteregle {W} {W} {B} {W} {W} {W} {W} {W} {W} \vskip 0pt 
\global\advance\reglenum by 1 
\the\reglenum{}: \lapetiteregle {B} {W} {W} {B} {W} {G} {W} {W} {G} \vskip 0pt 
\global\advance\reglenum by 1 
\the\reglenum{}: \lapetiteregle {W} {B} {B} {W} {W} {W} {W} {G} {W} \vskip 0pt 
\global\advance\reglenum by 1 
\the\reglenum{}: \lapetiteregle {W} {W} {B} {B} {B} {W} {W} {G} {W} \vskip 0pt 
\global\advance\reglenum by 1 
\the\reglenum{}: \lapetiteregle {W} {W} {G} {W} {W} {W} {W} {B} {W} \vskip 0pt 
\global\advance\reglenum by 1 
\the\reglenum{}: \lapetiteregle {W} {W} {W} {B} {B} {W} {B} {G} {W} \vskip 0pt 
\global\advance\reglenum by 1 
\the\reglenum{}: \lapetiteregle {G} {W} {W} {B} {W} {R} {W} {W} {R} \vskip 0pt 
\global\advance\reglenum by 1 
\the\reglenum{}: \lapetiteregle {B} {G} {W} {W} {B} {W} {W} {W} {G} \vskip 0pt 
\global\advance\reglenum by 1 
\the\reglenum{}: \lapetiteregle {W} {G} {B} {W} {W} {W} {W} {R} {W} \vskip 0pt 
\global\advance\reglenum by 1 
\the\reglenum{}: \lapetiteregle {W} {W} {B} {B} {B} {W} {W} {R} {W} \vskip 0pt 
\global\advance\reglenum by 1 
\the\reglenum{}: \lapetiteregle {W} {W} {R} {W} {W} {W} {W} {B} {W} \vskip 0pt 
\global\advance\reglenum by 1 
\the\reglenum{}: \lapetiteregle {W} {W} {W} {B} {B} {W} {G} {R} {W} \vskip 0pt 
\global\advance\reglenum by 1 
\the\reglenum{}: \lapetiteregle {W} {W} {B} {B} {B} {B} {B} {G} {W} \vskip 0pt 
\global\advance\reglenum by 1 
\the\reglenum{}: \lapetiteregle {R} {W} {W} {G} {W} {B} {W} {W} {B} \vskip 0pt 
\global\advance\reglenum by 1 
\the\reglenum{}: \lapetiteregle {G} {R} {W} {W} {B} {W} {W} {W} {R} \vskip 0pt 
\global\advance\reglenum by 1 
\the\reglenum{}: \lapetiteregle {W} {R} {G} {W} {W} {W} {W} {B} {W} \vskip 0pt 
\global\advance\reglenum by 1 
\the\reglenum{}: \lapetiteregle {B} {G} {W} {B} {W} {W} {W} {W} {G} \vskip 0pt 
\global\advance\reglenum by 1 
\the\reglenum{}: \lapetiteregle {W} {W} {G} {W} {W} {W} {W} {W} {W} \vskip 0pt 
\global\advance\reglenum by 1 
\the\reglenum{}: \lapetiteregle {W} {W} {W} {B} {B} {W} {R} {B} {W} \vskip 0pt 
\global\advance\reglenum by 1 
\the\reglenum{}: \lapetiteregle {W} {W} {B} {B} {B} {B} {G} {R} {W} \vskip 0pt 
\global\advance\reglenum by 1 
\the\reglenum{}: \lapetiteregle {B} {W} {B} {W} {W} {W} {W} {G} {G} \vskip 0pt 
\global\advance\reglenum by 1 
\the\reglenum{}: \lapetiteregle {B} {W} {W} {R} {W} {B} {W} {W} {B} \vskip 0pt 
\global\advance\reglenum by 1 
\the\reglenum{}: \lapetiteregle {R} {B} {W} {W} {G} {W} {W} {W} {B} \vskip 0pt 
\global\advance\reglenum by 1 
\the\reglenum{}: \lapetiteregle {W} {B} {R} {W} {W} {W} {W} {B} {W} \vskip 0pt 
\global\advance\reglenum by 1 
\the\reglenum{}: \lapetiteregle {G} {R} {W} {B} {W} {W} {W} {W} {R} \vskip 0pt 
\global\advance\reglenum by 1 
\the\reglenum{}: \lapetiteregle {W} {W} {R} {W} {W} {W} {W} {W} {W} \vskip 0pt 
\global\advance\reglenum by 1 
\the\reglenum{}: \lapetiteregle {W} {W} {B} {B} {B} {G} {R} {B} {W} \vskip 0pt 
\global\advance\reglenum by 1 
\the\reglenum{}: \lapetiteregle {G} {W} {B} {W} {W} {W} {W} {R} {R} \vskip 0pt 
\global\advance\reglenum by 1 
\the\reglenum{}: \lapetiteregle {B} {B} {W} {W} {R} {W} {W} {W} {B} \vskip 0pt 
\global\advance\reglenum by 1 
\the\reglenum{}: \lapetiteregle {R} {B} {W} {G} {W} {W} {W} {W} {B} \vskip 0pt 
\global\advance\reglenum by 1 
\the\reglenum{}: \lapetiteregle {W} {W} {B} {B} {G} {R} {B} {B} {W} \vskip 0pt 
\global\advance\reglenum by 1 
\the\reglenum{}: \lapetiteregle {R} {W} {G} {W} {W} {W} {W} {B} {B} \vskip 0pt 
\global\advance\reglenum by 1 
\the\reglenum{}: \lapetiteregle {B} {B} {W} {R} {W} {W} {W} {W} {B} \vskip 0pt 
\global\advance\reglenum by 1 
\the\reglenum{}: \lapetiteregle {B} {W} {B} {W} {W} {W} {G} {W} {G} \vskip 0pt 
\global\advance\reglenum by 1 
\the\reglenum{}: \lapetiteregle {W} {W} {B} {G} {R} {B} {B} {B} {W} \vskip 0pt 
\global\advance\reglenum by 1 
\the\reglenum{}: \lapetiteregle {B} {W} {R} {W} {W} {W} {W} {B} {B} \vskip 0pt 
\global\advance\reglenum by 1 
\the\reglenum{}: \lapetiteregle {W} {W} {W} {B} {G} {W} {B} {B} {W} \vskip 0pt 
\global\advance\reglenum by 1 
\the\reglenum{}: \lapetiteregle {B} {W} {W} {B} {W} {W} {W} {G} {G} \vskip 0pt 
\global\advance\reglenum by 1 
\the\reglenum{}: \lapetiteregle {G} {W} {B} {W} {W} {W} {R} {W} {R} \vskip 0pt 
\global\advance\reglenum by 1 
\the\reglenum{}: \lapetiteregle {W} {W} {G} {R} {B} {B} {B} {B} {W} \vskip 0pt 
\global\advance\reglenum by 1 
\the\reglenum{}: \lapetiteregle {W} {W} {B} {B} {B} {B} {G} {W} {W} \vskip 0pt 
\global\advance\reglenum by 1 
\the\reglenum{}: \lapetiteregle {W} {W} {W} {G} {R} {W} {B} {B} {W} \vskip 0pt 
\global\advance\reglenum by 1 
\the\reglenum{}: \lapetiteregle {G} {W} {W} {B} {W} {W} {W} {R} {R} \vskip 0pt 
\global\advance\reglenum by 1 
\the\reglenum{}: \lapetiteregle {R} {W} {G} {W} {W} {W} {B} {W} {B} \vskip 0pt 
\global\advance\reglenum by 1 
\the\reglenum{}: \lapetiteregle {W} {W} {R} {B} {B} {B} {B} {B} {W} \vskip 0pt 
\global\advance\reglenum by 1 
\the\reglenum{}: \lapetiteregle {W} {W} {B} {B} {B} {G} {R} {W} {W} \vskip 0pt 
\global\advance\reglenum by 1 
\the\reglenum{}: \lapetiteregle {W} {W} {W} {R} {B} {W} {B} {B} {W} \vskip 0pt 
\global\advance\reglenum by 1 
\the\reglenum{}: \lapetiteregle {R} {W} {W} {G} {W} {W} {W} {B} {B} \vskip 0pt 
\global\advance\reglenum by 1 
\the\reglenum{}: \lapetiteregle {B} {W} {R} {W} {W} {W} {B} {W} {B} \vskip 0pt 
\global\advance\reglenum by 1 
\the\reglenum{}: \lapetiteregle {W} {W} {B} {B} {G} {R} {B} {W} {W} \vskip 0pt 
\global\advance\reglenum by 1 
\the\reglenum{}: \lapetiteregle {B} {W} {W} {R} {W} {W} {W} {B} {B} \vskip 0pt 
\global\advance\reglenum by 1 
\the\reglenum{}: \lapetiteregle {W} {W} {B} {G} {R} {B} {B} {W} {W} \vskip 0pt 
\global\advance\reglenum by 1 
\the\reglenum{}: \lapetiteregle {W} {W} {B} {B} {G} {W} {W} {B} {W} \vskip 0pt 
\global\advance\reglenum by 1 
\the\reglenum{}: \lapetiteregle {W} {W} {G} {R} {B} {B} {B} {W} {W} \vskip 0pt 
\global\advance\reglenum by 1 
\the\reglenum{}: \lapetiteregle {W} {W} {B} {G} {R} {W} {W} {B} {W} \vskip 0pt 
\global\advance\reglenum by 1 
\the\reglenum{}: \lapetiteregle {B} {W} {W} {W} {W} {B} {W} {G} {G} \vskip 0pt 
\global\advance\reglenum by 1 
\the\reglenum{}: \lapetiteregle {W} {G} {B} {B} {W} {W} {W} {W} {W} \vskip 0pt 
\global\advance\reglenum by 1 
}
}
\setbox136=
\vtop{\leftskip 0pt\parindent 0pt
\baselineskip 7pt
\hsize=70pt
\rmviii
tracks alone:
\vskip 1pt
{\ttv\obeylines
\the\reglenum{}: \lapetiteregle {W} {W} {R} {B} {B} {B} {B} {W} {W} \vskip 0pt 
\global\advance\reglenum by 1 
\the\reglenum{}: \lapetiteregle {W} {W} {G} {R} {B} {W} {W} {B} {W} \vskip 0pt 
\global\advance\reglenum by 1 
\the\reglenum{}: \lapetiteregle {W} {W} {B} {W} {W} {W} {W} {G} {W} \vskip 0pt 
\global\advance\reglenum by 1 
\the\reglenum{}: \lapetiteregle {G} {W} {W} {W} {W} {B} {W} {R} {R} \vskip 0pt 
\global\advance\reglenum by 1 
\the\reglenum{}: \lapetiteregle {W} {R} {G} {B} {W} {W} {W} {W} {W} \vskip 0pt 
\global\advance\reglenum by 1 
\the\reglenum{}: \lapetiteregle {W} {W} {R} {B} {B} {W} {W} {B} {W} \vskip 0pt 
\global\advance\reglenum by 1 
\the\reglenum{}: \lapetiteregle {W} {W} {B} {W} {W} {W} {W} {R} {W} \vskip 0pt 
\global\advance\reglenum by 1 
\the\reglenum{}: \lapetiteregle {R} {W} {W} {W} {W} {G} {W} {B} {B} \vskip 0pt 
\global\advance\reglenum by 1 
\the\reglenum{}: \lapetiteregle {B} {W} {W} {W} {W} {R} {W} {B} {B} \vskip 0pt 
\global\advance\reglenum by 1 
\the\reglenum{}: \lapetiteregle {W} {B} {B} {R} {W} {W} {W} {W} {W} \vskip 0pt 
\global\advance\reglenum by 1 
\the\reglenum{}: \lapetiteregle {B} {W} {W} {W} {W} {G} {W} {B} {G} \vskip 0pt 
\global\advance\reglenum by 1 
\the\reglenum{}: \lapetiteregle {W} {B} {B} {G} {W} {W} {W} {W} {W} \vskip 0pt 
\global\advance\reglenum by 1 
\the\reglenum{}: \lapetiteregle {W} {W} {G} {B} {B} {W} {W} {B} {W} \vskip 0pt 
\global\advance\reglenum by 1 
\the\reglenum{}: \lapetiteregle {G} {W} {W} {W} {W} {R} {W} {B} {R} \vskip 0pt 
\global\advance\reglenum by 1 
\the\reglenum{}: \lapetiteregle {B} {W} {G} {W} {W} {W} {W} {B} {G} \vskip 0pt 
\global\advance\reglenum by 1 
\the\reglenum{}: \lapetiteregle {W} {B} {G} {R} {W} {W} {W} {W} {W} \vskip 0pt 
\global\advance\reglenum by 1 
\the\reglenum{}: \lapetiteregle {W} {W} {R} {G} {B} {W} {W} {B} {W} \vskip 0pt 
\global\advance\reglenum by 1 
\the\reglenum{}: \lapetiteregle {R} {W} {W} {W} {W} {B} {W} {G} {B} \vskip 0pt 
\global\advance\reglenum by 1 
\the\reglenum{}: \lapetiteregle {G} {W} {R} {W} {W} {W} {W} {B} {R} \vskip 0pt 
\global\advance\reglenum by 1 
\the\reglenum{}: \lapetiteregle {W} {G} {R} {B} {W} {W} {W} {W} {W} \vskip 0pt 
\global\advance\reglenum by 1 
\the\reglenum{}: \lapetiteregle {B} {W} {G} {W} {W} {W} {B} {W} {G} \vskip 0pt 
\global\advance\reglenum by 1 
\the\reglenum{}: \lapetiteregle {W} {W} {B} {R} {G} {W} {W} {B} {W} \vskip 0pt 
\global\advance\reglenum by 1 
\the\reglenum{}: \lapetiteregle {B} {W} {W} {W} {W} {B} {W} {R} {B} \vskip 0pt 
\global\advance\reglenum by 1 
\the\reglenum{}: \lapetiteregle {R} {W} {B} {W} {W} {W} {W} {G} {B} \vskip 0pt 
\global\advance\reglenum by 1 
\the\reglenum{}: \lapetiteregle {W} {R} {B} {B} {W} {W} {W} {W} {W} \vskip 0pt 
\global\advance\reglenum by 1 
\the\reglenum{}: \lapetiteregle {G} {W} {R} {W} {W} {W} {B} {W} {R} \vskip 0pt 
\global\advance\reglenum by 1 
\the\reglenum{}: \lapetiteregle {W} {W} {G} {B} {B} {B} {B} {W} {W} \vskip 0pt 
\global\advance\reglenum by 1 
\the\reglenum{}: \lapetiteregle {W} {W} {B} {B} {R} {W} {W} {B} {W} \vskip 0pt 
\global\advance\reglenum by 1 
\the\reglenum{}: \lapetiteregle {B} {W} {B} {W} {W} {W} {W} {R} {B} \vskip 0pt 
\global\advance\reglenum by 1 
\the\reglenum{}: \lapetiteregle {R} {W} {B} {W} {W} {W} {G} {W} {B} \vskip 0pt 
\global\advance\reglenum by 1 
\the\reglenum{}: \lapetiteregle {W} {W} {R} {G} {B} {B} {B} {W} {W} \vskip 0pt 
\global\advance\reglenum by 1 
\the\reglenum{}: \lapetiteregle {B} {W} {B} {W} {W} {W} {R} {W} {B} \vskip 0pt 
\global\advance\reglenum by 1 
\the\reglenum{}: \lapetiteregle {W} {W} {B} {R} {G} {B} {B} {W} {W} \vskip 0pt 
\global\advance\reglenum by 1 
\the\reglenum{}: \lapetiteregle {W} {W} {B} {B} {R} {G} {B} {W} {W} \vskip 0pt 
\global\advance\reglenum by 1 
\the\reglenum{}: \lapetiteregle {B} {W} {W} {G} {W} {W} {W} {B} {G} \vskip 0pt 
\global\advance\reglenum by 1 
\the\reglenum{}: \lapetiteregle {W} {W} {B} {B} {B} {R} {G} {W} {W} \vskip 0pt 
\global\advance\reglenum by 1 
\the\reglenum{}: \lapetiteregle {W} {W} {W} {G} {B} {W} {B} {B} {W} \vskip 0pt 
\global\advance\reglenum by 1 
\the\reglenum{}: \lapetiteregle {G} {W} {W} {R} {W} {W} {W} {B} {R} \vskip 0pt 
\global\advance\reglenum by 1 
\the\reglenum{}: \lapetiteregle {W} {W} {B} {B} {B} {B} {R} {W} {W} \vskip 0pt 
\global\advance\reglenum by 1 
\the\reglenum{}: \lapetiteregle {W} {W} {W} {R} {G} {W} {B} {B} {W} \vskip 0pt 
\global\advance\reglenum by 1 
\the\reglenum{}: \lapetiteregle {R} {W} {W} {B} {W} {W} {W} {G} {B} \vskip 0pt 
\global\advance\reglenum by 1 
\the\reglenum{}: \lapetiteregle {W} {W} {G} {B} {B} {B} {B} {B} {W} \vskip 0pt 
\global\advance\reglenum by 1 
\the\reglenum{}: \lapetiteregle {W} {W} {W} {B} {R} {W} {B} {B} {W} \vskip 0pt 
\global\advance\reglenum by 1 
\the\reglenum{}: \lapetiteregle {B} {W} {W} {B} {W} {W} {W} {R} {B} \vskip 0pt 
\global\advance\reglenum by 1 
\the\reglenum{}: \lapetiteregle {W} {W} {R} {G} {B} {B} {B} {B} {W} \vskip 0pt 
\global\advance\reglenum by 1 
\the\reglenum{}: \lapetiteregle {W} {W} {B} {R} {G} {B} {B} {B} {W} \vskip 0pt 
\global\advance\reglenum by 1 
\the\reglenum{}: \lapetiteregle {W} {W} {B} {B} {R} {G} {B} {B} {W} \vskip 0pt 
\global\advance\reglenum by 1 
\the\reglenum{}: \lapetiteregle {B} {B} {W} {G} {W} {W} {W} {W} {G} \vskip 0pt 
\global\advance\reglenum by 1 
\the\reglenum{}: \lapetiteregle {W} {W} {B} {B} {B} {R} {G} {B} {W} \vskip 0pt 
\global\advance\reglenum by 1 
\the\reglenum{}: \lapetiteregle {B} {B} {W} {W} {G} {W} {W} {W} {G} \vskip 0pt 
\global\advance\reglenum by 1 
\the\reglenum{}: \lapetiteregle {G} {B} {W} {R} {W} {W} {W} {W} {R} \vskip 0pt 
\global\advance\reglenum by 1 
\the\reglenum{}: \lapetiteregle {W} {W} {W} {B} {B} {W} {G} {B} {W} \vskip 0pt 
\global\advance\reglenum by 1 
\the\reglenum{}: \lapetiteregle {W} {W} {B} {B} {B} {B} {R} {G} {W} \vskip 0pt 
\global\advance\reglenum by 1 
\the\reglenum{}: \lapetiteregle {B} {W} {W} {G} {W} {B} {W} {W} {G} \vskip 0pt 
\global\advance\reglenum by 1 
\the\reglenum{}: \lapetiteregle {G} {B} {W} {W} {R} {W} {W} {W} {R} \vskip 0pt 
\global\advance\reglenum by 1 
\the\reglenum{}: \lapetiteregle {W} {B} {G} {W} {W} {W} {W} {B} {W} \vskip 0pt 
\global\advance\reglenum by 1 
\the\reglenum{}: \lapetiteregle {R} {G} {W} {B} {W} {W} {W} {W} {B} \vskip 0pt 
\global\advance\reglenum by 1 
\the\reglenum{}: \lapetiteregle {W} {W} {W} {B} {B} {W} {R} {G} {W} \vskip 0pt 
\global\advance\reglenum by 1 
\the\reglenum{}: \lapetiteregle {W} {W} {B} {B} {B} {B} {B} {R} {W} \vskip 0pt 
\global\advance\reglenum by 1 
\the\reglenum{}: \lapetiteregle {G} {W} {W} {R} {W} {B} {W} {W} {R} \vskip 0pt 
\global\advance\reglenum by 1 
\the\reglenum{}: \lapetiteregle {R} {G} {W} {W} {B} {W} {W} {W} {B} \vskip 0pt 
\global\advance\reglenum by 1 
\the\reglenum{}: \lapetiteregle {W} {G} {R} {W} {W} {W} {W} {B} {W} \vskip 0pt 
\global\advance\reglenum by 1 
\the\reglenum{}: \lapetiteregle {B} {R} {W} {B} {W} {W} {W} {W} {B} \vskip 0pt 
\global\advance\reglenum by 1 
\the\reglenum{}: \lapetiteregle {W} {W} {W} {B} {B} {W} {B} {R} {W} \vskip 0pt 
\global\advance\reglenum by 1 
\the\reglenum{}: \lapetiteregle {R} {W} {W} {B} {W} {G} {W} {W} {B} \vskip 0pt 
\global\advance\reglenum by 1 
\the\reglenum{}: \lapetiteregle {B} {R} {W} {W} {B} {W} {W} {W} {B} \vskip 0pt 
\global\advance\reglenum by 1 
\the\reglenum{}: \lapetiteregle {W} {R} {B} {W} {W} {W} {W} {G} {W} \vskip 0pt 
\global\advance\reglenum by 1 
\the\reglenum{}: \lapetiteregle {B} {W} {W} {B} {W} {R} {W} {W} {B} \vskip 0pt 
\global\advance\reglenum by 1 
}
}

\begin{tab}\label{table_regles}
The rules of the cellular automaton of Theorem~{\rm\ref{universal}}.
\end{tab}
\vspace{-8pt}
\grostrait
\ligne{\hfill\copy120\hfill\copy122\hfill\copy124\hfill\copy126\hfill}
\ligne{\hfill\copy130\hfill\copy132\hfill\copy134\hfill\copy136\hfill}
\vskip 10pt
\noindent
} \fi
quiescent
remains quiescent. Starting from the central cell the program scans the sectors 
one after the other and in each one, from the root to the last level, level by 
level. 
The program takes the context~$\kappa$ of~$c$ in table~0. Then, 
it compares~$\kappa$ with the contexts of the rules of the file. 
If it finds a match, it copies the new state of the rule at the address 
of~$c$ in table~1, under column~0. If it does not find a match, it asks for a 
new rule which the user writes into the file. To help the user, the program 
indicates the context of the cell. The user enters the new state only. Then the 
program resumes its computation: it reads again table~0 from the initial 
configuration and performs the computation as far as possible. If it can 
compute
the new state of all cells of table~0, it completes table~1 by computing the
new states of the neighbours of each cell. When this task is over, the program
copies table~1 onto table~0: a new step of the computation of the cellular 
automaton can be processed. This cycle is repeated until no new rule is 
required and until the fixed in advance number of steps is reached.

   Now, when a new rule is entered by the user on a cell~$c$, it may happen 
that the new rule is in conflict with the previously entered rules. 
This happens when there is a rule~$\eta$ whose context is a rotated form 
of the context of~$c$, but the state suggested by the user is not the new 
state of the rule. In this case, the program stops with an error message 
which also displays the rule with which the program have found a mismatch. 
If the rule constructed on the context of the cell and the state indicated 
by the user is a rotated form of an already existing rule, it is appended to 
the set of rules.

   When the program can be run without asking a new rule nor indicating
any error, we know that the set of rules is computed.

   The program also contributes to check that the rules are rotation invariant,
using Lemma~\ref{minim} which is very easy to implement: this is left to the
reader. The same property was used in~\cite{mmsyBristol}. In the display of
the rules in Table~\ref{table_regles}, we assume that the minimal rotated 
forms of the rules are pairwise distinct. 
This means that on the seven rotated forms which should be present for each 
rule, we keep only those needed by the tested configurations. We find 1168~rules.
If minimal rotated forms only would be considered, we would find 595 of them.
\vskip 5pt
   There were 545 rules for the universal cellular
automaton on the heptagrid described in~\cite{mmsyENTCS} and~299 ones in the 
case of the pentagrid, see~\cite{mmsyBristol,mmsyPPL}. It is a well known phenomenon that
when the number of states is reduced, that of the rules is significantly increased.
This comes from the fact that the reduction of the number of states entails a higher
complexity of the configurations.

In Table~\ref{table_regles},
the rules are displayed in the format indicated in Subsection~5.1. Also,
the rules are gathered according to the order in which they were entered in the
set of rules during the execution of the program. This order also corresponds
to the different configurations which we described in Section~4 and to the
figures which are displayed in this section as well as figures which are not
displayed but which correspond to the motions described in the section.
The reader may check in the table that for all crossings and switches, the 
motion of the locomotive was checked for all possible arrivals. Also, the
program have checked configurations which are not represented in the figures
of the paper, in particular the cases of a crossing with an arrival from the other
paths, or the reverse motion on a piece of a slip road connecting a vertical with
a horizontal. 
However, in order to get a correct set of
rules, the program had to test these situations too. As the reader can check it
on the table, all the needed tests were performed. Indeed, each test on a type 
of motion induced new rules where at least one of them was not a rotated form 
of a previously obtained rule.  

   This completes the proof of Theorem~\ref{universal}. \cqfd 

\section{Conclusion}

   We are now closer to the minimal number of states in order to get a universal
cellular automaton on the heptagrid. Two cases remain to be studied: 3~states and
2~states. I had a discussion with Donald Knuth on this topic. Don encouraged me to try
to find an analog of the game of life in the hyperbolic plane. Of course, it would be very
interesting to obtain such a result. Unfortunately, the divergence of the lines and rays
in the heptagrid makes it difficult to define collisions. Moreover, the gliders are not
yet known.


   Accordingly, we remain with some hard work ahead.

%

\section*{Appendix}

   As mentioned in the paper, the front of the locomotive is represented with a
darker colour than it is in the rules. We repeat that the automaton has 4~states only.

   The first series of figures, Fig.~\ref{turnA}, Fig.~\ref{turnB}, Fig.~\ref{turnC}
and Fig.~\ref{turnD} illustrate the way a slip road connect a vertical with an isocline.
The rules for the motion which is opposite to that of the figures have been devised and
they were checked by the computer program.

   Then, Fig.~\ref{memorisantd_7}, Fig.~\ref{memorisantd_4} and Fig.~\ref{memorisantd_1}
illustrate the crossing of a memory switch by the locomotive in which the selected path
is on the right-hand side.

   At last and not the least, we give the rules of the automaton in Table~\ref{table_regles}.
As mentioned, in the table, the rules are displayed according to the test performed
on the configurations of the crossings, the switches and the paths. This is indicated
in the table were rules for a given situation have been gathered together. However, we
have to keep in mind that all rules can be simultaneously applied to the set of all cells.

\setbox110=\hbox{\epsfig{file=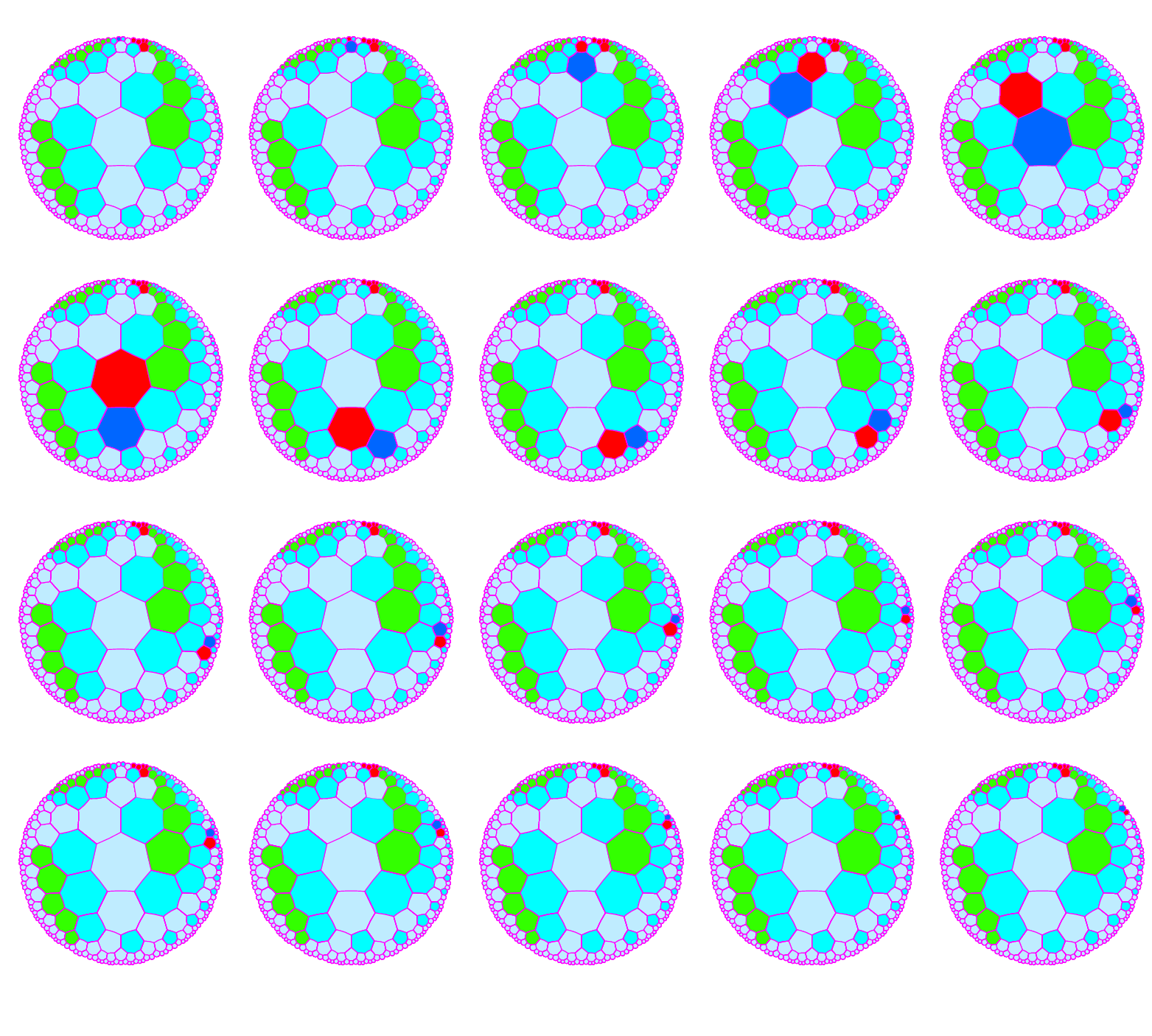,width=300pt}}
\vtop{
\ligne{\hfill
\PlacerEn {-345pt} {0pt} \box110
}
\vspace{-15pt}
\begin{fig}\label{turnA}
The motion of the locomotive along a turn at an angle of a quadrangle. Here, 
the locomotive goes down along a vertical and then turns to right to follow an isocline.
\end{fig}
}

\vskip -20pt
\setbox110=\hbox{\epsfig{file=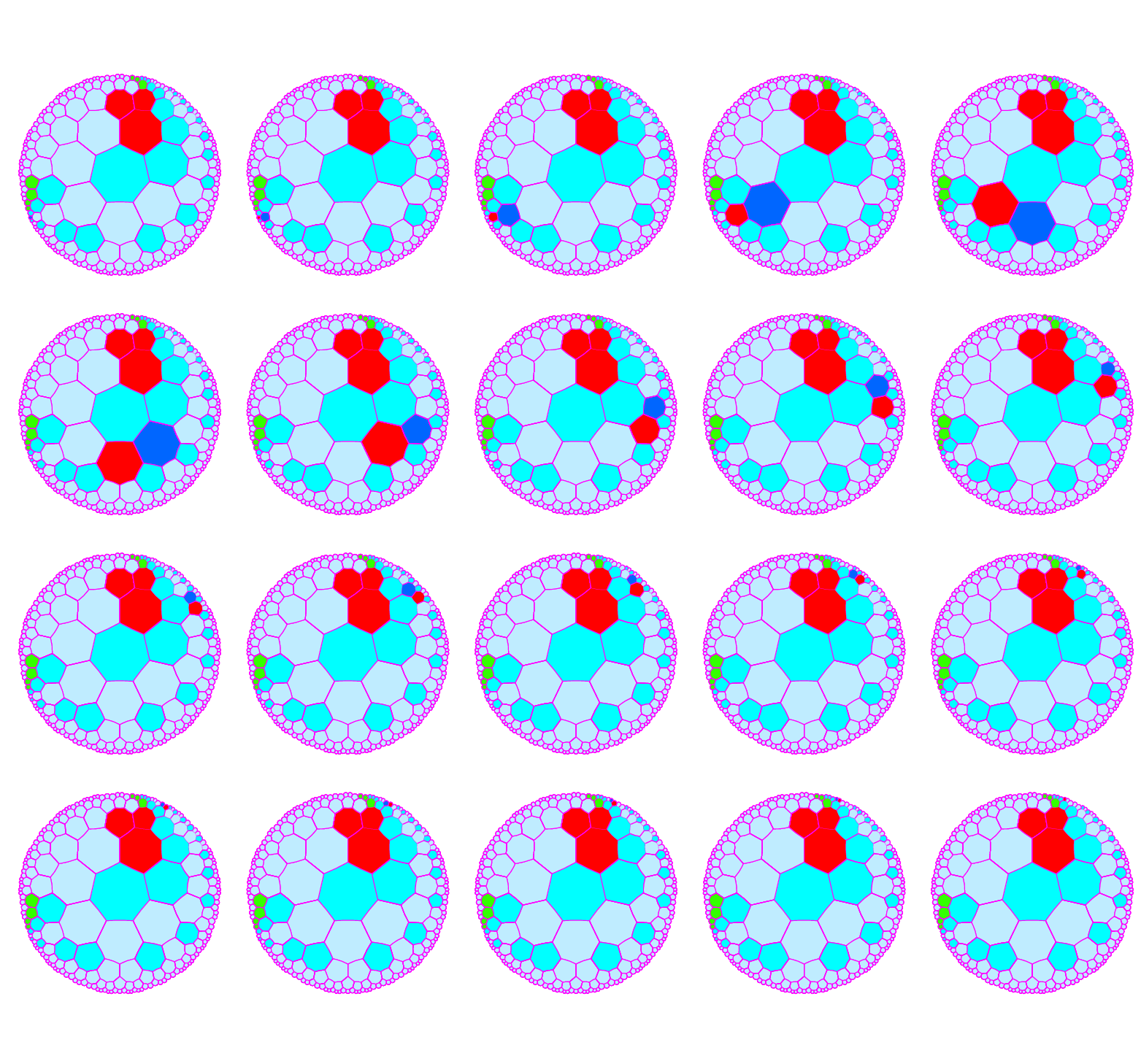,width=300pt}}
\vtop{
\ligne{\hfill
\PlacerEn {-325pt} {0pt} \box110
}
\vspace{-25pt}
\begin{fig}\label{turnB}
Here, the locomotive goes up along a vertical and turns to right to follow an isocline.
\end{fig}
}
\vskip 10pt
\setbox110=\hbox{\epsfig{file=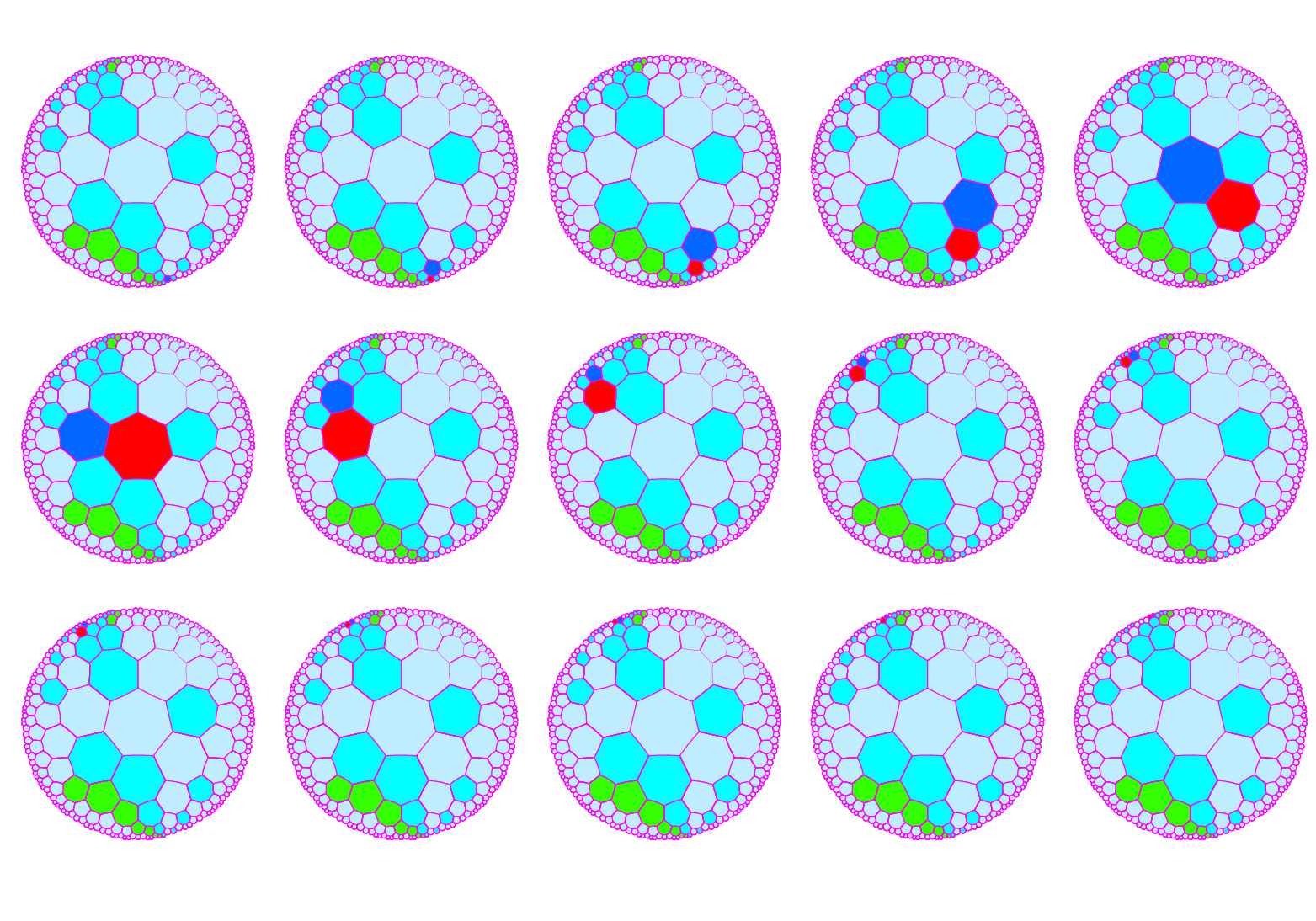,width=300pt}}
\vtop{
\ligne{\hfill
\PlacerEn {-345pt} {0pt} \box110
}
\vspace{-15pt}
\begin{fig}\label{turnC}
Here, the locomotive goes up along a vertical and turns to left to follow an isocline. 
\end{fig}
}
\vskip 10pt
\setbox110=\hbox{\epsfig{file=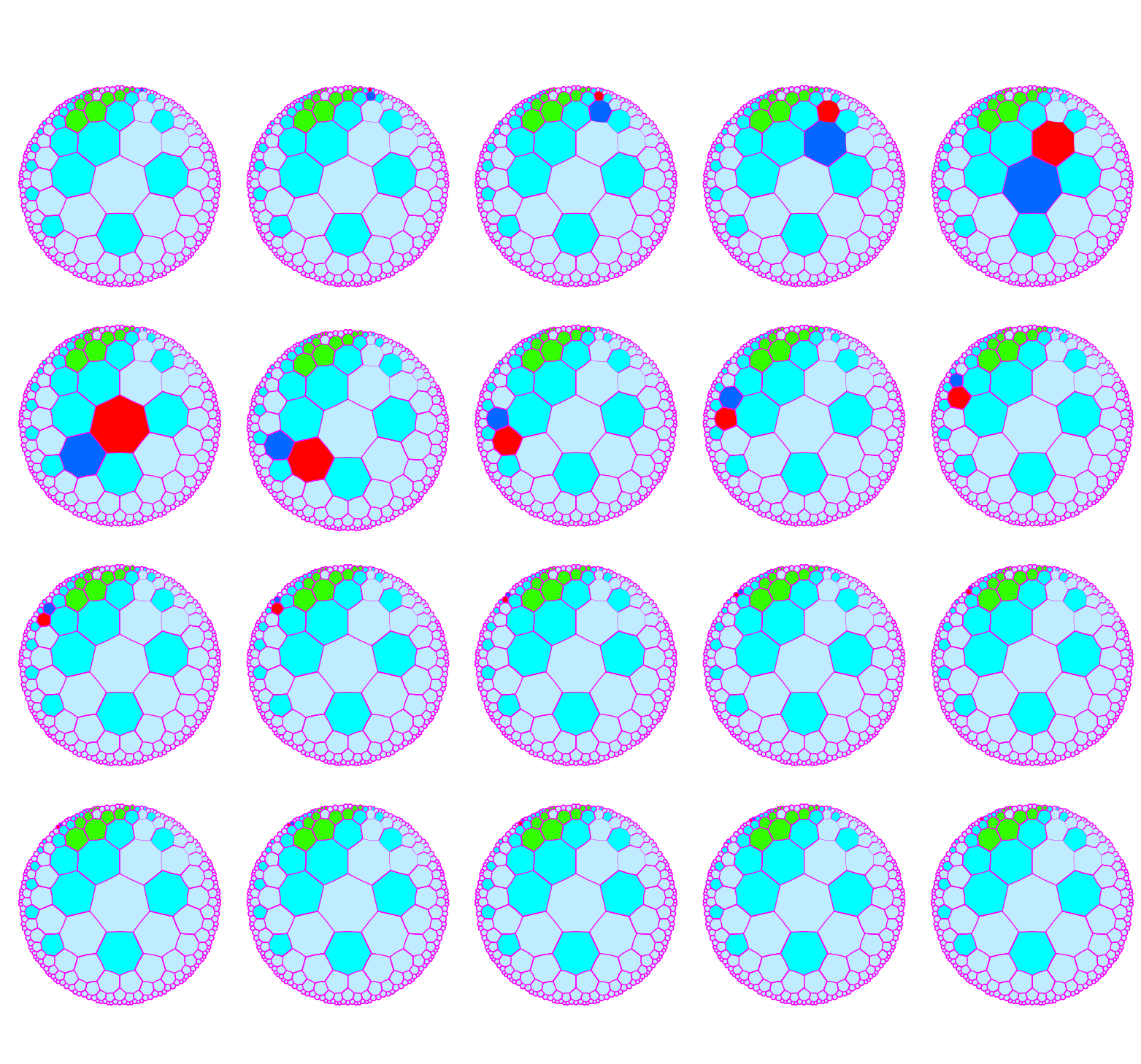,width=300pt}}
\vtop{
\ligne{\hfill
\PlacerEn {-345pt} {0pt} \box110
}
\vspace{-15pt}
\begin{fig}\label{turnD}
Now, the locomotive goes down along a vertical and turns to left to follow an isocline.
\end{fig}
}
\vskip 10pt
\vskip-5pt
\setbox110=\hbox{\epsfig{file=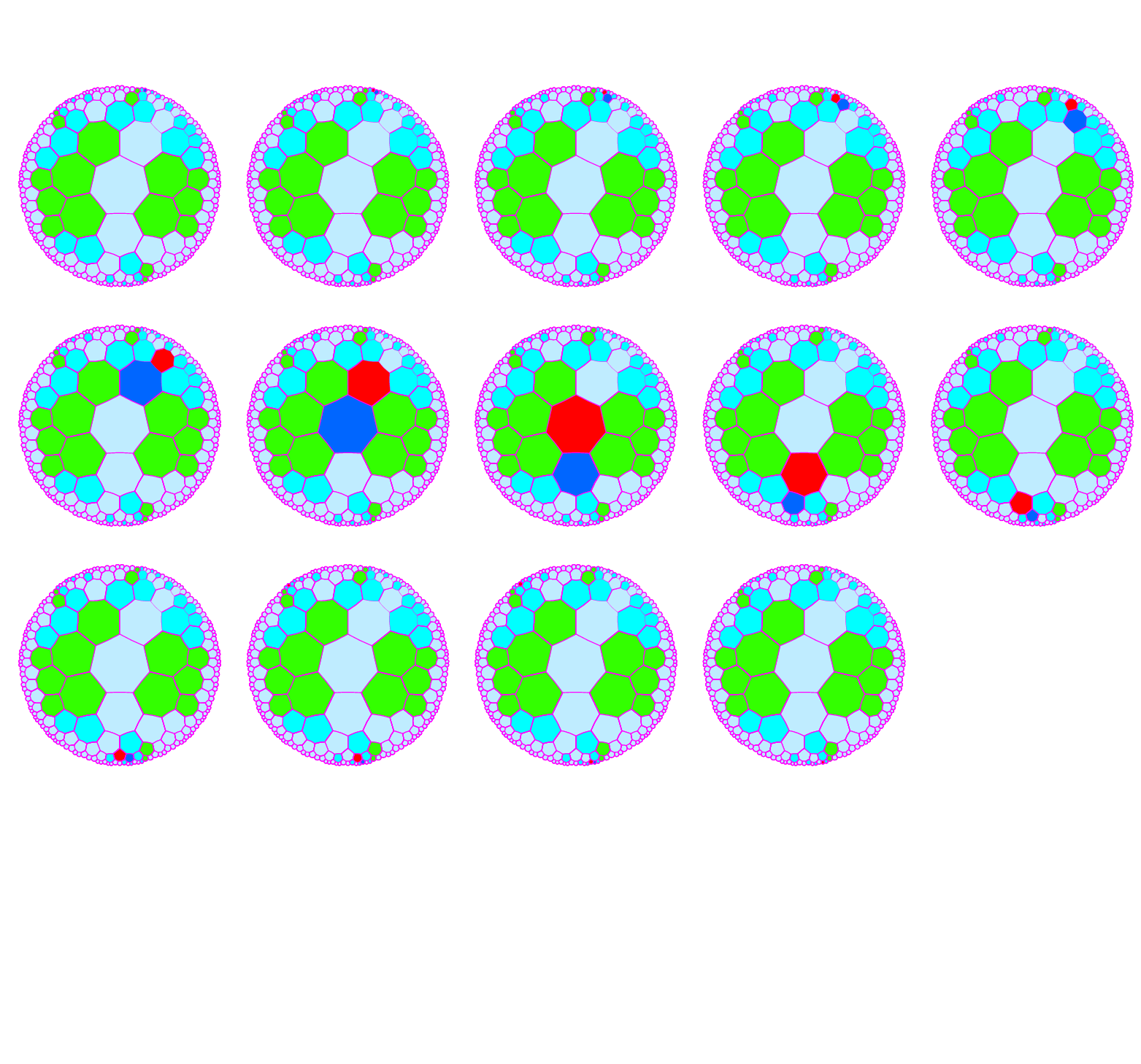,width=340pt}}
\vtop{
\ligne{\hfill
\PlacerEn {-346pt} {0pt} \box110
}
\vspace{-85pt}
\begin{fig}\label{memorisantd_7}
\leurre
The locomotive passively crosses a right-hand side memory switch from the selected track,
here in sector~$7$. 
\end{fig}
}
\vskip 10pt
\vskip-5pt
\setbox110=\hbox{\epsfig{file=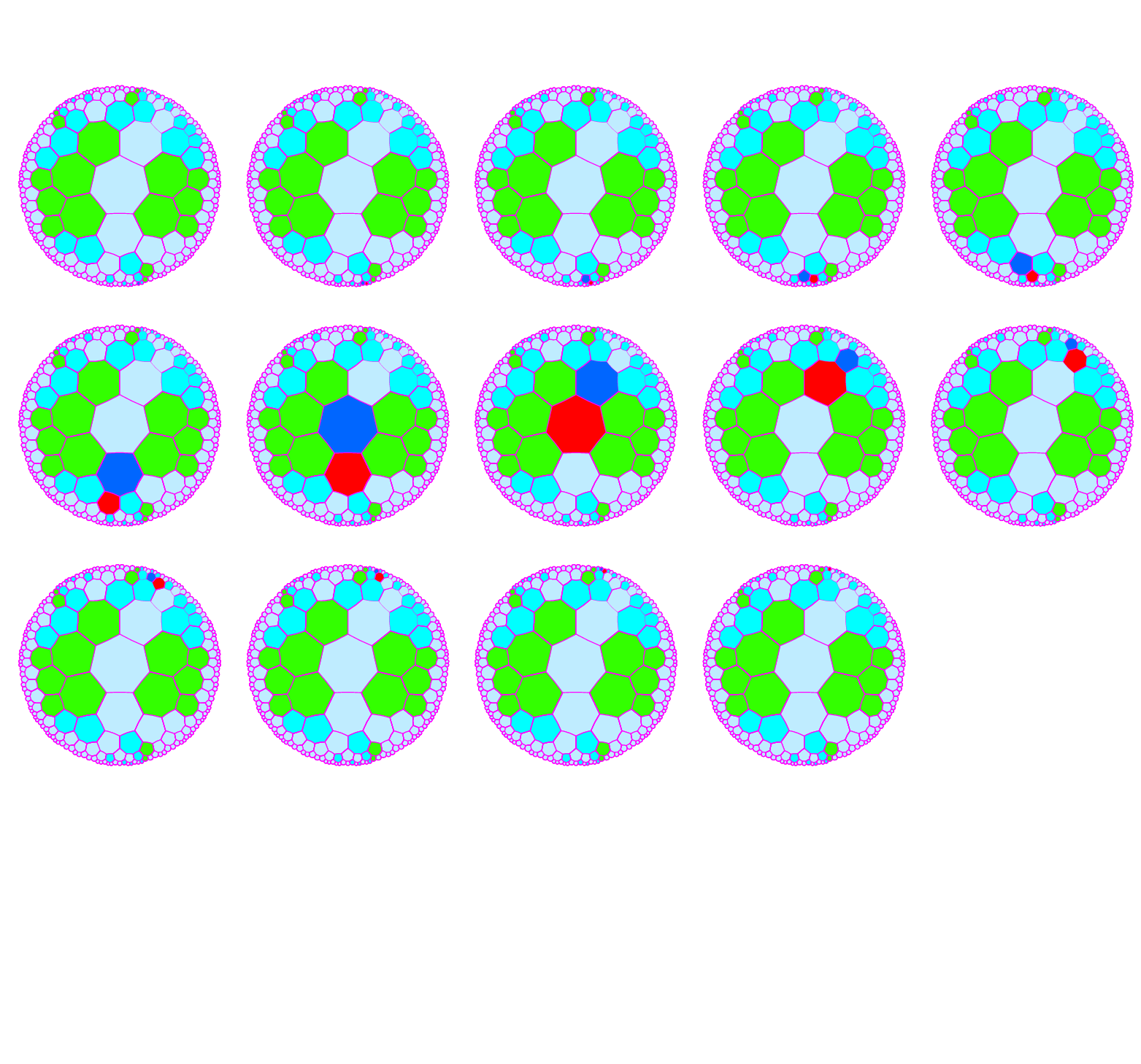,width=340pt}}
\vtop{
\ligne{\hfill
\PlacerEn {-346pt} {0pt} \box110
}
\vspace{-85pt}
\begin{fig}\label{memorisantd_4}
\leurre
The locomotive actively crosses a right-hand side memory switch, here from sector~$4$.
\end{fig}
}
\vskip-5pt
\setbox110=\hbox{\epsfig{file=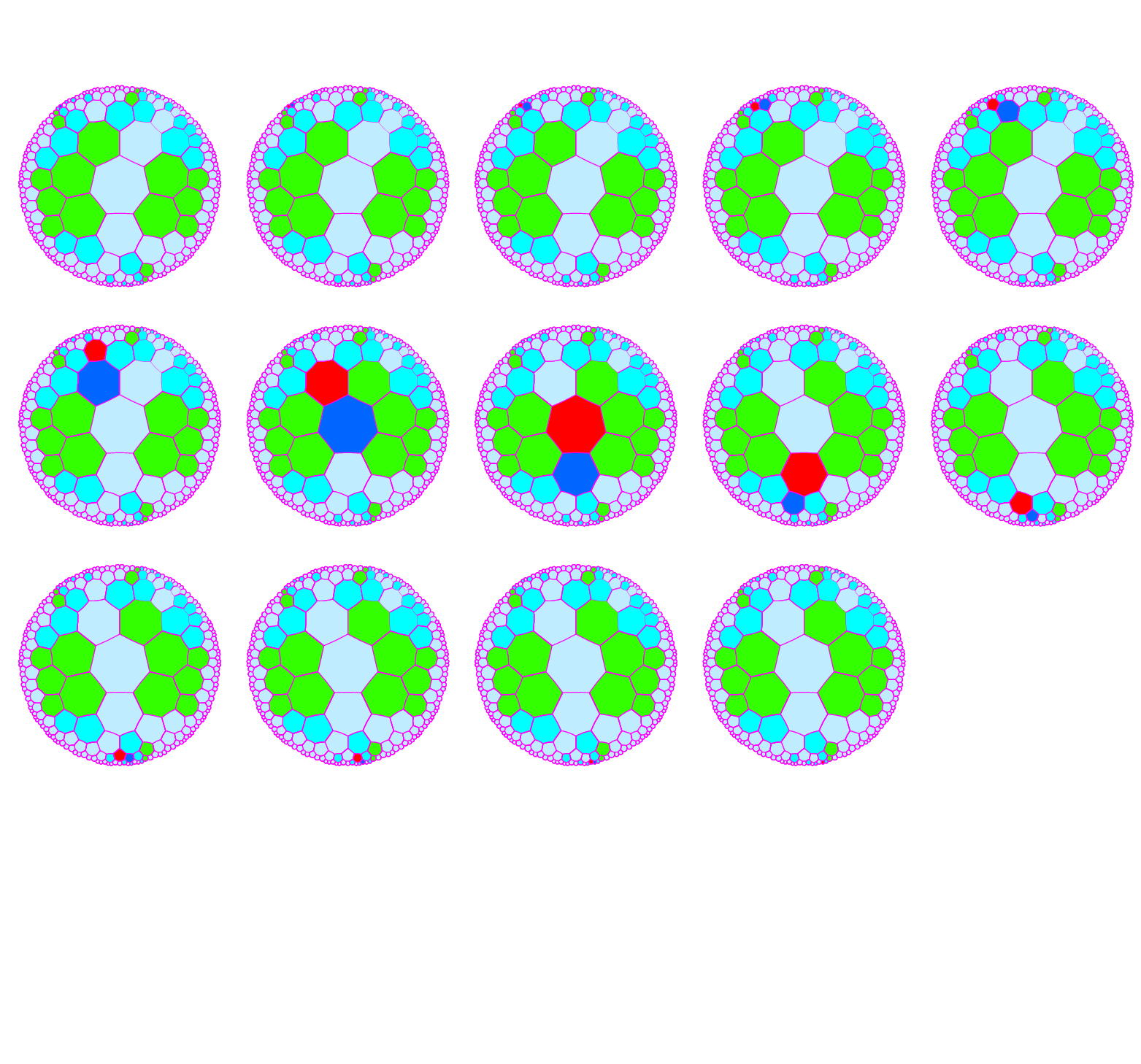,width=340pt}}
\vtop{
\ligne{\hfill
\PlacerEn {-346pt} {0pt} \box110
}
\vspace{-85pt}
\begin{fig}\label{memorisantd_1}
\leurre
The locomotive passively crosses a memory switch from the non-selected track, here in
sector~$1$. 
Note the change of the selected track when the rear of the
locomotive leaves the first cell of the non-selected track.
\end{fig}
}

\def\lapetiteregle #1 #2 #3 #4 #5 #6 #7 #8 #9 {%
\setbox211=\hbox{$\underline{\hbox{\ttvi#1}}$}
\setbox212=\hbox{$\underline{\hbox{\ttvi#9}}$}
\hskip-15pt\hbox{\ttvi \box211#2#3#4#5#6#7#8\box212}
}
\newcount\reglenum\reglenum=1

\begin{tab}\label{table_regles}
The rules of the universal automaton on the heptagrid with four states.
\end{tab}
\vspace{-16pt}
\grostrait

\setbox120=\vtop{\leftskip 0pt\parindent 0pt
\baselineskip 7pt
\hsize=70pt
crossing:
\vskip 5pt\rmviii
path~1:
\vskip 1pt
{\ttv\obeylines
\newcount\reglenum\reglenum=1 
\the\reglenum{}: \lapetiteregle {W} {W} {W} {W} {W} {W} {W} {W} {W} \vskip 0pt
\global\advance\reglenum by 1
\the\reglenum{}: \lapetiteregle {W} {W} {W} {G} {W} {W} {W} {G} {W} \vskip 0pt
\global\advance\reglenum by 1
\the\reglenum{}: \lapetiteregle {W} {W} {G} {B} {W} {B} {B} {W} {W} \vskip 0pt
\global\advance\reglenum by 1
\the\reglenum{}: \lapetiteregle {B} {W} {G} {B} {W} {W} {B} {W} {B} \vskip 0pt
\global\advance\reglenum by 1
\the\reglenum{}: \lapetiteregle {W} {W} {B} {W} {W} {B} {B} {B} {B} \vskip 0pt
\global\advance\reglenum by 1
\the\reglenum{}: \lapetiteregle {B} {W} {W} {B} {R} {B} {W} {W} {B} \vskip 0pt
\global\advance\reglenum by 1
\the\reglenum{}: \lapetiteregle {W} {B} {B} {B} {W} {W} {W} {W} {W} \vskip 0pt
\global\advance\reglenum by 1
\the\reglenum{}: \lapetiteregle {W} {B} {W} {W} {W} {W} {W} {B} {W} \vskip 0pt
\global\advance\reglenum by 1
\the\reglenum{}: \lapetiteregle {B} {W} {B} {W} {W} {W} {W} {W} {B} \vskip 0pt
\global\advance\reglenum by 1
\the\reglenum{}: \lapetiteregle {W} {W} {B} {W} {W} {W} {W} {B} {W} \vskip 0pt
\global\advance\reglenum by 1
\the\reglenum{}: \lapetiteregle {B} {W} {W} {W} {W} {W} {W} {B} {B} \vskip 0pt
\global\advance\reglenum by 1
\the\reglenum{}: \lapetiteregle {B} {B} {W} {B} {W} {W} {B} {R} {R} \vskip 0pt
\global\advance\reglenum by 1
\the\reglenum{}: \lapetiteregle {R} {B} {B} {B} {W} {B} {W} {B} {W} \vskip 0pt
\global\advance\reglenum by 1
\the\reglenum{}: \lapetiteregle {B} {B} {R} {W} {W} {B} {W} {W} {B} \vskip 0pt
\global\advance\reglenum by 1
\the\reglenum{}: \lapetiteregle {W} {W} {B} {B} {W} {W} {W} {W} {W} \vskip 0pt
\global\advance\reglenum by 1
\the\reglenum{}: \lapetiteregle {W} {B} {W} {W} {W} {W} {W} {W} {W} \vskip 0pt
\global\advance\reglenum by 1
\the\reglenum{}: \lapetiteregle {W} {W} {B} {W} {W} {W} {W} {W} {W} \vskip 0pt
\global\advance\reglenum by 1
\the\reglenum{}: \lapetiteregle {W} {B} {B} {W} {W} {W} {W} {W} {W} \vskip 0pt
\global\advance\reglenum by 1
\the\reglenum{}: \lapetiteregle {B} {R} {B} {W} {W} {W} {W} {W} {B} \vskip 0pt
\global\advance\reglenum by 1
\the\reglenum{}: \lapetiteregle {W} {R} {B} {W} {W} {W} {W} {B} {W} \vskip 0pt
\global\advance\reglenum by 1
\the\reglenum{}: \lapetiteregle {B} {R} {W} {W} {W} {W} {W} {W} {B} \vskip 0pt
\global\advance\reglenum by 1
\the\reglenum{}: \lapetiteregle {W} {B} {R} {B} {W} {W} {W} {W} {W} \vskip 0pt
\global\advance\reglenum by 1
\the\reglenum{}: \lapetiteregle {B} {B} {W} {W} {W} {B} {W} {W} {B} \vskip 0pt
\global\advance\reglenum by 1
\the\reglenum{}: \lapetiteregle {W} {W} {W} {W} {W} {B} {W} {G} {W} \vskip 0pt
\global\advance\reglenum by 1
\the\reglenum{}: \lapetiteregle {B} {W} {W} {B} {B} {W} {W} {W} {B} \vskip 0pt
\global\advance\reglenum by 1
\the\reglenum{}: \lapetiteregle {B} {W} {W} {W} {W} {W} {W} {W} {B} \vskip 0pt
\global\advance\reglenum by 1
\the\reglenum{}: \lapetiteregle {B} {B} {B} {B} {W} {W} {W} {W} {B} \vskip 0pt
\global\advance\reglenum by 1
\the\reglenum{}: \lapetiteregle {G} {W} {W} {W} {W} {B} {W} {W} {G} \vskip 0pt
\global\advance\reglenum by 1
\the\reglenum{}: \lapetiteregle {W} {G} {W} {B} {W} {W} {B} {W} {W} \vskip 0pt
\global\advance\reglenum by 1
\the\reglenum{}: \lapetiteregle {W} {G} {W} {B} {W} {B} {W} {B} {W} \vskip 0pt
\global\advance\reglenum by 1
\the\reglenum{}: \lapetiteregle {B} {G} {W} {W} {W} {B} {W} {W} {B} \vskip 0pt
\global\advance\reglenum by 1
\the\reglenum{}: \lapetiteregle {W} {W} {W} {W} {W} {W} {W} {B} {W} \vskip 0pt
\global\advance\reglenum by 1
\the\reglenum{}: \lapetiteregle {W} {B} {W} {B} {W} {W} {B} {W} {W} \vskip 0pt 
\global\advance\reglenum by 1 
}
}

\setbox122=\vtop{\leftskip 0pt\parindent 0pt
\baselineskip 7pt
\hsize=70pt
{\ttv\obeylines
\the\reglenum{}: \lapetiteregle {W} {B} {W} {B} {W} {B} {W} {B} {W} \vskip 0pt 
\global\advance\reglenum by 1 
\the\reglenum{}: \lapetiteregle {W} {B} {W} {B} {W} {W} {W} {W} {W} \vskip 0pt 
\global\advance\reglenum by 1 
\the\reglenum{}: \lapetiteregle {W} {W} {G} {B} {W} {B} {W} {W} {W} \vskip 0pt 
\global\advance\reglenum by 1 
\the\reglenum{}: \lapetiteregle {B} {W} {G} {B} {W} {W} {W} {W} {B} \vskip 0pt 
\global\advance\reglenum by 1 
\the\reglenum{}: \lapetiteregle {W} {W} {B} {W} {W} {B} {W} {B} {W} \vskip 0pt 
\global\advance\reglenum by 1 
\the\reglenum{}: \lapetiteregle {B} {W} {W} {W} {W} {B} {W} {W} {B} \vskip 0pt 
\global\advance\reglenum by 1 
\the\reglenum{}: \lapetiteregle {W} {W} {W} {W} {W} {W} {B} {W} {W} \vskip 0pt 
\global\advance\reglenum by 1 
\the\reglenum{}: \lapetiteregle {W} {W} {W} {B} {W} {W} {W} {W} {W} \vskip 0pt 
\global\advance\reglenum by 1 
\the\reglenum{}: \lapetiteregle {W} {W} {W} {B} {W} {B} {W} {G} {W} \vskip 0pt 
\global\advance\reglenum by 1 
\the\reglenum{}: \lapetiteregle {G} {W} {W} {W} {W} {B} {B} {W} {G} \vskip 0pt 
\global\advance\reglenum by 1 
\the\reglenum{}: \lapetiteregle {B} {G} {W} {W} {W} {B} {W} {B} {B} \vskip 0pt 
\global\advance\reglenum by 1 
\the\reglenum{}: \lapetiteregle {W} {W} {G} {B} {B} {B} {W} {W} {B} \vskip 0pt 
\global\advance\reglenum by 1 
\the\reglenum{}: \lapetiteregle {B} {W} {G} {B} {W} {W} {W} {B} {B} \vskip 0pt 
\global\advance\reglenum by 1 
\the\reglenum{}: \lapetiteregle {B} {W} {B} {W} {W} {B} {R} {B} {R} \vskip 0pt 
\global\advance\reglenum by 1 
\the\reglenum{}: \lapetiteregle {B} {W} {B} {R} {W} {B} {W} {W} {B} \vskip 0pt 
\global\advance\reglenum by 1 
\the\reglenum{}: \lapetiteregle {B} {B} {B} {W} {W} {W} {W} {W} {B} \vskip 0pt 
\global\advance\reglenum by 1 
\the\reglenum{}: \lapetiteregle {W} {B} {B} {W} {W} {W} {W} {B} {W} \vskip 0pt 
\global\advance\reglenum by 1 
\the\reglenum{}: \lapetiteregle {B} {B} {W} {W} {W} {W} {W} {R} {B} \vskip 0pt 
\global\advance\reglenum by 1 
\the\reglenum{}: \lapetiteregle {R} {B} {B} {B} {W} {W} {B} {W} {W} \vskip 0pt 
\global\advance\reglenum by 1 
\the\reglenum{}: \lapetiteregle {W} {B} {R} {B} {W} {B} {W} {B} {W} \vskip 0pt 
\global\advance\reglenum by 1 
\the\reglenum{}: \lapetiteregle {B} {B} {W} {W} {W} {B} {W} {B} {B} \vskip 0pt 
\global\advance\reglenum by 1 
\the\reglenum{}: \lapetiteregle {W} {R} {B} {W} {W} {W} {W} {W} {W} \vskip 0pt 
\global\advance\reglenum by 1 
\the\reglenum{}: \lapetiteregle {W} {R} {W} {W} {W} {W} {W} {B} {W} \vskip 0pt 
\global\advance\reglenum by 1 
\the\reglenum{}: \lapetiteregle {B} {W} {R} {W} {W} {W} {W} {W} {B} \vskip 0pt 
\global\advance\reglenum by 1 
\the\reglenum{}: \lapetiteregle {W} {B} {W} {G} {W} {W} {W} {G} {B} \vskip 0pt 
\global\advance\reglenum by 1 
\the\reglenum{}: \lapetiteregle {B} {W} {G} {B} {R} {B} {W} {W} {R} \vskip 0pt 
\global\advance\reglenum by 1 
\the\reglenum{}: \lapetiteregle {B} {B} {G} {B} {W} {W} {W} {R} {B} \vskip 0pt 
\global\advance\reglenum by 1 
\the\reglenum{}: \lapetiteregle {R} {B} {B} {W} {W} {B} {W} {B} {W} \vskip 0pt 
\global\advance\reglenum by 1 
\the\reglenum{}: \lapetiteregle {B} {B} {R} {W} {W} {B} {B} {B} {B} \vskip 0pt 
\global\advance\reglenum by 1 
\the\reglenum{}: \lapetiteregle {W} {B} {R} {B} {W} {W} {B} {W} {W} \vskip 0pt 
\global\advance\reglenum by 1 
\the\reglenum{}: \lapetiteregle {W} {W} {B} {W} {W} {B} {W} {G} {W} \vskip 0pt 
\global\advance\reglenum by 1 
\the\reglenum{}: \lapetiteregle {B} {W} {B} {B} {B} {W} {W} {W} {B} \vskip 0pt 
\global\advance\reglenum by 1 
\the\reglenum{}: \lapetiteregle {G} {W} {W} {W} {W} {B} {B} {B} {G} \vskip 0pt 
\global\advance\reglenum by 1 
\the\reglenum{}: \lapetiteregle {B} {R} {W} {G} {W} {W} {W} {G} {R} \vskip 0pt 
\global\advance\reglenum by 1 
\the\reglenum{}: \lapetiteregle {R} {B} {G} {B} {W} {B} {W} {W} {W} \vskip 0pt 
\global\advance\reglenum by 1 
\the\reglenum{}: \lapetiteregle {B} {R} {G} {B} {W} {W} {W} {W} {B} \vskip 0pt 
\global\advance\reglenum by 1 
}
}

\setbox124=\vtop{\leftskip 0pt\parindent 0pt
\baselineskip 7pt
\hsize=70pt
{\ttv\obeylines
\the\reglenum{}: \lapetiteregle {W} {R} {B} {W} {W} {B} {W} {B} {W} \vskip 0pt 
\global\advance\reglenum by 1 
\the\reglenum{}: \lapetiteregle {B} {R} {W} {W} {W} {B} {W} {W} {B} \vskip 0pt 
\global\advance\reglenum by 1 
\the\reglenum{}: \lapetiteregle {W} {B} {R} {W} {W} {B} {W} {G} {W} \vskip 0pt 
\global\advance\reglenum by 1 
\the\reglenum{}: \lapetiteregle {B} {W} {R} {B} {B} {W} {W} {W} {B} \vskip 0pt 
\global\advance\reglenum by 1 
\the\reglenum{}: \lapetiteregle {G} {B} {W} {W} {W} {B} {B} {W} {G} \vskip 0pt 
\global\advance\reglenum by 1 
\the\reglenum{}: \lapetiteregle {W} {B} {G} {B} {W} {B} {W} {W} {B} \vskip 0pt 
\global\advance\reglenum by 1 
\the\reglenum{}: \lapetiteregle {W} {B} {W} {W} {W} {W} {B} {W} {W} \vskip 0pt 
\global\advance\reglenum by 1 
\the\reglenum{}: \lapetiteregle {W} {B} {W} {B} {W} {B} {W} {G} {W} \vskip 0pt 
\global\advance\reglenum by 1 
\the\reglenum{}: \lapetiteregle {G} {B} {W} {W} {W} {B} {B} {R} {G} \vskip 0pt 
\global\advance\reglenum by 1 
\the\reglenum{}: \lapetiteregle {R} {W} {W} {G} {B} {W} {W} {G} {W} \vskip 0pt 
\global\advance\reglenum by 1 
\the\reglenum{}: \lapetiteregle {W} {R} {G} {B} {W} {B} {W} {W} {W} \vskip 0pt 
\global\advance\reglenum by 1 
\the\reglenum{}: \lapetiteregle {W} {W} {B} {B} {W} {B} {W} {B} {W} \vskip 0pt 
\global\advance\reglenum by 1 
\the\reglenum{}: \lapetiteregle {B} {W} {W} {W} {W} {B} {B} {B} {B} \vskip 0pt 
\global\advance\reglenum by 1 
\the\reglenum{}: \lapetiteregle {W} {R} {W} {W} {W} {B} {W} {G} {W} \vskip 0pt 
\global\advance\reglenum by 1 
\the\reglenum{}: \lapetiteregle {G} {R} {W} {W} {W} {B} {B} {B} {G} \vskip 0pt 
\global\advance\reglenum by 1 
\the\reglenum{}: \lapetiteregle {B} {R} {G} {B} {W} {B} {W} {W} {R} \vskip 0pt 
\global\advance\reglenum by 1 
\the\reglenum{}: \lapetiteregle {B} {B} {G} {B} {W} {W} {W} {W} {B} \vskip 0pt 
\global\advance\reglenum by 1 
\the\reglenum{}: \lapetiteregle {W} {B} {B} {W} {W} {B} {W} {B} {B} \vskip 0pt 
\global\advance\reglenum by 1 
\the\reglenum{}: \lapetiteregle {W} {R} {B} {W} {W} {W} {B} {W} {W} \vskip 0pt 
\global\advance\reglenum by 1 
\the\reglenum{}: \lapetiteregle {W} {R} {W} {B} {W} {B} {W} {G} {W} \vskip 0pt 
\global\advance\reglenum by 1 
\the\reglenum{}: \lapetiteregle {G} {R} {W} {W} {W} {B} {B} {W} {G} \vskip 0pt 
\global\advance\reglenum by 1 
\the\reglenum{}: \lapetiteregle {W} {W} {W} {G} {R} {W} {W} {G} {W} \vskip 0pt 
\global\advance\reglenum by 1 
\the\reglenum{}: \lapetiteregle {G} {W} {W} {W} {W} {B} {B} {R} {G} \vskip 0pt 
\global\advance\reglenum by 1 
\the\reglenum{}: \lapetiteregle {R} {W} {G} {B} {B} {B} {W} {W} {W} \vskip 0pt 
\global\advance\reglenum by 1 
\the\reglenum{}: \lapetiteregle {B} {R} {G} {B} {W} {W} {W} {B} {B} \vskip 0pt 
\global\advance\reglenum by 1 
\the\reglenum{}: \lapetiteregle {B} {R} {B} {W} {W} {B} {W} {B} {R} \vskip 0pt 
\global\advance\reglenum by 1 
\the\reglenum{}: \lapetiteregle {B} {R} {B} {W} {W} {B} {W} {W} {B} \vskip 0pt 
\global\advance\reglenum by 1 
\the\reglenum{}: \lapetiteregle {B} {B} {W} {W} {W} {W} {W} {W} {B} \vskip 0pt 
\global\advance\reglenum by 1 
\the\reglenum{}: \lapetiteregle {W} {B} {B} {B} {W} {W} {B} {W} {B} \vskip 0pt 
\global\advance\reglenum by 1 
\the\reglenum{}: \lapetiteregle {W} {W} {R} {W} {W} {W} {B} {W} {W} \vskip 0pt 
\global\advance\reglenum by 1 
\the\reglenum{}: \lapetiteregle {W} {W} {R} {B} {W} {W} {W} {W} {W} \vskip 0pt 
\global\advance\reglenum by 1 
\the\reglenum{}: \lapetiteregle {W} {W} {G} {B} {R} {B} {W} {W} {W} \vskip 0pt 
\global\advance\reglenum by 1 
\the\reglenum{}: \lapetiteregle {B} {W} {G} {B} {W} {W} {W} {R} {B} \vskip 0pt 
\global\advance\reglenum by 1 
\the\reglenum{}: \lapetiteregle {R} {W} {B} {W} {W} {B} {B} {B} {W} \vskip 0pt 
\global\advance\reglenum by 1 
\the\reglenum{}: \lapetiteregle {B} {W} {R} {B} {W} {B} {W} {W} {B} \vskip 0pt 
\global\advance\reglenum by 1 
\the\reglenum{}: \lapetiteregle {B} {R} {W} {W} {W} {W} {W} {B} {B} \vskip 0pt 
\global\advance\reglenum by 1 
}
}

\setbox126=\vtop{\leftskip 0pt\parindent 0pt
\baselineskip 7pt
\hsize=70pt
{\ttv\obeylines
\the\reglenum{}: \lapetiteregle {B} {B} {R} {B} {W} {W} {B} {W} {R} \vskip 0pt 
\global\advance\reglenum by 1 
\the\reglenum{}: \lapetiteregle {W} {B} {B} {B} {W} {B} {W} {B} {B} \vskip 0pt 
\global\advance\reglenum by 1 
\the\reglenum{}: \lapetiteregle {W} {W} {B} {W} {W} {B} {R} {B} {W} \vskip 0pt 
\global\advance\reglenum by 1 
\the\reglenum{}: \lapetiteregle {B} {W} {W} {R} {B} {B} {W} {W} {B} \vskip 0pt 
\global\advance\reglenum by 1 
\the\reglenum{}: \lapetiteregle {B} {W} {W} {W} {W} {W} {W} {R} {B} \vskip 0pt 
\global\advance\reglenum by 1 
\the\reglenum{}: \lapetiteregle {R} {B} {W} {B} {W} {W} {B} {B} {W} \vskip 0pt 
\global\advance\reglenum by 1 
\the\reglenum{}: \lapetiteregle {B} {B} {R} {B} {W} {B} {W} {B} {R} \vskip 0pt 
\global\advance\reglenum by 1 
\the\reglenum{}: \lapetiteregle {B} {B} {B} {W} {W} {B} {W} {W} {B} \vskip 0pt 
\global\advance\reglenum by 1 
\the\reglenum{}: \lapetiteregle {B} {B} {R} {W} {W} {W} {W} {W} {B} \vskip 0pt 
\global\advance\reglenum by 1 
}
\vskip 3pt\rmviii
same path,\vskip 0pt
other\vskip 2pt
conditions:
\vskip 1pt
{\ttv\obeylines
\the\reglenum{}: \lapetiteregle {W} {B} {W} {W} {W} {G} {B} {W} {W} \vskip 0pt 
\global\advance\reglenum by 1 
\the\reglenum{}: \lapetiteregle {B} {W} {W} {W} {W} {B} {G} {W} {B} \vskip 0pt 
\global\advance\reglenum by 1 
\the\reglenum{}: \lapetiteregle {W} {W} {B} {G} {W} {W} {W} {W} {W} \vskip 0pt 
\global\advance\reglenum by 1 
\the\reglenum{}: \lapetiteregle {G} {W} {B} {B} {G} {W} {W} {W} {G} \vskip 0pt 
\global\advance\reglenum by 1 
\the\reglenum{}: \lapetiteregle {W} {W} {W} {W} {B} {W} {G} {W} {W} \vskip 0pt 
\global\advance\reglenum by 1 
\the\reglenum{}: \lapetiteregle {B} {W} {G} {B} {G} {W} {W} {W} {B} \vskip 0pt 
\global\advance\reglenum by 1 
\the\reglenum{}: \lapetiteregle {W} {W} {W} {B} {G} {W} {W} {W} {W} \vskip 0pt 
\global\advance\reglenum by 1 
\the\reglenum{}: \lapetiteregle {G} {B} {B} {B} {W} {W} {W} {W} {G} \vskip 0pt 
\global\advance\reglenum by 1 
\the\reglenum{}: \lapetiteregle {G} {W} {W} {W} {W} {W} {B} {B} {G} \vskip 0pt 
\global\advance\reglenum by 1 
\the\reglenum{}: \lapetiteregle {W} {G} {W} {W} {B} {W} {B} {W} {W} \vskip 0pt 
\global\advance\reglenum by 1 
\the\reglenum{}: \lapetiteregle {B} {B} {G} {W} {W} {W} {B} {G} {B} \vskip 0pt 
\global\advance\reglenum by 1 
\the\reglenum{}: \lapetiteregle {G} {B} {B} {B} {G} {W} {W} {W} {G} \vskip 0pt 
\global\advance\reglenum by 1 
\the\reglenum{}: \lapetiteregle {B} {G} {B} {W} {W} {W} {B} {G} {B} \vskip 0pt 
\global\advance\reglenum by 1 
\the\reglenum{}: \lapetiteregle {G} {G} {B} {W} {W} {W} {W} {W} {G} \vskip 0pt 
\global\advance\reglenum by 1 
\the\reglenum{}: \lapetiteregle {B} {W} {W} {W} {B} {B} {G} {W} {B} \vskip 0pt 
\global\advance\reglenum by 1 
\the\reglenum{}: \lapetiteregle {B} {W} {W} {B} {R} {B} {G} {W} {B} \vskip 0pt 
\global\advance\reglenum by 1 
\the\reglenum{}: \lapetiteregle {W} {B} {W} {W} {W} {G} {B} {B} {B} \vskip 0pt 
\global\advance\reglenum by 1 
\the\reglenum{}: \lapetiteregle {B} {W} {B} {R} {W} {B} {G} {W} {B} \vskip 0pt 
\global\advance\reglenum by 1 
\the\reglenum{}: \lapetiteregle {B} {W} {G} {B} {G} {W} {W} {B} {B} \vskip 0pt 
\global\advance\reglenum by 1 
\the\reglenum{}: \lapetiteregle {B} {B} {W} {W} {W} {G} {B} {R} {R} \vskip 0pt 
\global\advance\reglenum by 1 
\the\reglenum{}: \lapetiteregle {W} {B} {B} {G} {W} {W} {W} {W} {W} \vskip 0pt 
\global\advance\reglenum by 1 
\the\reglenum{}: \lapetiteregle {W} {B} {W} {W} {B} {W} {G} {W} {W} \vskip 0pt 
\global\advance\reglenum by 1 
\the\reglenum{}: \lapetiteregle {G} {B} {W} {W} {W} {W} {B} {B} {G} \vskip 0pt 
\global\advance\reglenum by 1 
}
}

\setbox128=\vtop{\leftskip 0pt\parindent 0pt
\baselineskip 7pt
\hsize=70pt
{\ttv\obeylines
\the\reglenum{}: \lapetiteregle {B} {B} {G} {B} {G} {W} {W} {R} {B} \vskip 0pt 
\global\advance\reglenum by 1 
\the\reglenum{}: \lapetiteregle {R} {B} {W} {W} {B} {G} {B} {W} {W} \vskip 0pt 
\global\advance\reglenum by 1 
\the\reglenum{}: \lapetiteregle {B} {R} {W} {W} {W} {B} {G} {W} {B} \vskip 0pt 
\global\advance\reglenum by 1 
\the\reglenum{}: \lapetiteregle {W} {R} {B} {G} {W} {W} {W} {W} {W} \vskip 0pt 
\global\advance\reglenum by 1 
\the\reglenum{}: \lapetiteregle {W} {R} {W} {W} {B} {W} {G} {B} {W} \vskip 0pt 
\global\advance\reglenum by 1 
\the\reglenum{}: \lapetiteregle {G} {R} {B} {W} {W} {W} {B} {B} {G} \vskip 0pt 
\global\advance\reglenum by 1 
\the\reglenum{}: \lapetiteregle {W} {G} {B} {W} {B} {W} {B} {W} {W} \vskip 0pt 
\global\advance\reglenum by 1 
\the\reglenum{}: \lapetiteregle {B} {R} {G} {B} {G} {W} {W} {W} {B} \vskip 0pt 
\global\advance\reglenum by 1 
\the\reglenum{}: \lapetiteregle {W} {B} {W} {W} {R} {G} {B} {W} {W} \vskip 0pt 
\global\advance\reglenum by 1 
\the\reglenum{}: \lapetiteregle {W} {W} {W} {W} {B} {W} {G} {R} {W} \vskip 0pt 
\global\advance\reglenum by 1 
\the\reglenum{}: \lapetiteregle {B} {B} {G} {B} {G} {W} {W} {W} {B} \vskip 0pt 
\global\advance\reglenum by 1 
\the\reglenum{}: \lapetiteregle {W} {W} {B} {B} {G} {W} {W} {W} {W} \vskip 0pt 
\global\advance\reglenum by 1 
\the\reglenum{}: \lapetiteregle {G} {W} {R} {W} {W} {W} {B} {B} {G} \vskip 0pt 
\global\advance\reglenum by 1 
\the\reglenum{}: \lapetiteregle {W} {G} {R} {W} {B} {W} {B} {W} {W} \vskip 0pt 
\global\advance\reglenum by 1 
\the\reglenum{}: \lapetiteregle {W} {B} {W} {W} {B} {G} {B} {W} {B} \vskip 0pt 
\global\advance\reglenum by 1 
\the\reglenum{}: \lapetiteregle {W} {W} {W} {W} {B} {W} {G} {B} {W} \vskip 0pt 
\global\advance\reglenum by 1 
\the\reglenum{}: \lapetiteregle {W} {W} {R} {B} {G} {W} {W} {W} {W} \vskip 0pt 
\global\advance\reglenum by 1 
\the\reglenum{}: \lapetiteregle {G} {W} {B} {W} {W} {W} {B} {B} {G} \vskip 0pt 
\global\advance\reglenum by 1 
\the\reglenum{}: \lapetiteregle {B} {B} {W} {W} {R} {G} {B} {W} {R} \vskip 0pt 
\global\advance\reglenum by 1 
\the\reglenum{}: \lapetiteregle {W} {B} {W} {W} {B} {W} {G} {R} {W} \vskip 0pt 
\global\advance\reglenum by 1 
\the\reglenum{}: \lapetiteregle {G} {B} {R} {W} {W} {W} {B} {B} {G} \vskip 0pt 
\global\advance\reglenum by 1 
\the\reglenum{}: \lapetiteregle {R} {B} {W} {W} {W} {G} {B} {B} {W} \vskip 0pt 
\global\advance\reglenum by 1 
\the\reglenum{}: \lapetiteregle {B} {R} {B} {W} {W} {B} {G} {W} {B} \vskip 0pt 
\global\advance\reglenum by 1 
\the\reglenum{}: \lapetiteregle {W} {R} {W} {W} {B} {W} {G} {W} {W} \vskip 0pt 
\global\advance\reglenum by 1 
\the\reglenum{}: \lapetiteregle {G} {R} {W} {W} {W} {W} {B} {B} {G} \vskip 0pt 
\global\advance\reglenum by 1 
\the\reglenum{}: \lapetiteregle {B} {R} {G} {B} {G} {W} {W} {B} {B} \vskip 0pt 
\global\advance\reglenum by 1 
\the\reglenum{}: \lapetiteregle {W} {B} {W} {W} {W} {G} {B} {R} {W} \vskip 0pt 
\global\advance\reglenum by 1 
\the\reglenum{}: \lapetiteregle {B} {W} {R} {B} {W} {B} {G} {W} {B} \vskip 0pt 
\global\advance\reglenum by 1 
\the\reglenum{}: \lapetiteregle {B} {W} {G} {B} {G} {W} {W} {R} {B} \vskip 0pt 
\global\advance\reglenum by 1 
\the\reglenum{}: \lapetiteregle {B} {W} {W} {R} {B} {B} {G} {W} {B} \vskip 0pt 
\global\advance\reglenum by 1 
\the\reglenum{}: \lapetiteregle {B} {W} {W} {W} {R} {B} {G} {W} {B} \vskip 0pt 
\global\advance\reglenum by 1 
}
\vskip 5pt\rmviii
path~7:
\vskip 1pt
{\ttv\obeylines
\the\reglenum{}: \lapetiteregle {W} {G} {W} {B} {W} {B} {B} {B} {G} \vskip 0pt 
\global\advance\reglenum by 1 
\the\reglenum{}: \lapetiteregle {B} {G} {W} {B} {R} {B} {W} {B} {B} \vskip 0pt 
\global\advance\reglenum by 1 
\the\reglenum{}: \lapetiteregle {G} {W} {W} {W} {G} {B} {B} {W} {G} \vskip 0pt 
\global\advance\reglenum by 1 
}
}

\ligne{\box120\hfill
\box122\hfill
\box124\hfill
\box126\hfill
\box128\hfill
}

\setbox120=\vtop{\leftskip 0pt\parindent 0pt
\baselineskip 7pt
\hsize=70pt
{\ttv\obeylines
\the\reglenum{}: \lapetiteregle {W} {G} {W} {B} {W} {W} {B} {G} {G} \vskip 0pt 
\global\advance\reglenum by 1 
\the\reglenum{}: \lapetiteregle {G} {G} {W} {B} {W} {B} {R} {B} {R} \vskip 0pt 
\global\advance\reglenum by 1 
\the\reglenum{}: \lapetiteregle {B} {G} {G} {R} {W} {B} {W} {B} {B} \vskip 0pt 
\global\advance\reglenum by 1 
\the\reglenum{}: \lapetiteregle {B} {G} {W} {W} {W} {W} {W} {W} {B} \vskip 0pt 
\global\advance\reglenum by 1 
\the\reglenum{}: \lapetiteregle {W} {G} {B} {W} {W} {W} {W} {B} {W} \vskip 0pt 
\global\advance\reglenum by 1 
\the\reglenum{}: \lapetiteregle {B} {G} {W} {W} {W} {W} {W} {R} {B} \vskip 0pt 
\global\advance\reglenum by 1 
\the\reglenum{}: \lapetiteregle {R} {B} {G} {B} {W} {W} {B} {W} {W} \vskip 0pt 
\global\advance\reglenum by 1 
\the\reglenum{}: \lapetiteregle {W} {W} {W} {B} {W} {B} {G} {G} {G} \vskip 0pt 
\global\advance\reglenum by 1 
\the\reglenum{}: \lapetiteregle {B} {W} {W} {W} {W} {W} {W} {G} {B} \vskip 0pt 
\global\advance\reglenum by 1 
\the\reglenum{}: \lapetiteregle {G} {W} {W} {G} {R} {B} {B} {W} {G} \vskip 0pt 
\global\advance\reglenum by 1 
\the\reglenum{}: \lapetiteregle {G} {G} {W} {B} {W} {W} {B} {R} {R} \vskip 0pt 
\global\advance\reglenum by 1 
\the\reglenum{}: \lapetiteregle {R} {G} {G} {B} {W} {B} {W} {B} {W} \vskip 0pt 
\global\advance\reglenum by 1 
\the\reglenum{}: \lapetiteregle {B} {G} {R} {W} {W} {B} {W} {B} {B} \vskip 0pt 
\global\advance\reglenum by 1 
\the\reglenum{}: \lapetiteregle {W} {G} {B} {W} {W} {W} {W} {W} {W} \vskip 0pt 
\global\advance\reglenum by 1 
\the\reglenum{}: \lapetiteregle {W} {G} {W} {W} {W} {W} {W} {B} {W} \vskip 0pt 
\global\advance\reglenum by 1 
\the\reglenum{}: \lapetiteregle {B} {R} {G} {W} {W} {W} {W} {W} {B} \vskip 0pt 
\global\advance\reglenum by 1 
\the\reglenum{}: \lapetiteregle {W} {W} {W} {G} {W} {W} {G} {G} {G} \vskip 0pt 
\global\advance\reglenum by 1 
\the\reglenum{}: \lapetiteregle {W} {W} {W} {W} {W} {W} {B} {G} {W} \vskip 0pt 
\global\advance\reglenum by 1 
\the\reglenum{}: \lapetiteregle {G} {W} {W} {B} {W} {B} {R} {G} {R} \vskip 0pt 
\global\advance\reglenum by 1 
\the\reglenum{}: \lapetiteregle {G} {W} {G} {R} {W} {B} {B} {W} {G} \vskip 0pt 
\global\advance\reglenum by 1 
\the\reglenum{}: \lapetiteregle {R} {G} {G} {B} {W} {W} {B} {W} {W} \vskip 0pt 
\global\advance\reglenum by 1 
\the\reglenum{}: \lapetiteregle {W} {G} {R} {B} {W} {B} {W} {B} {W} \vskip 0pt 
\global\advance\reglenum by 1 
\the\reglenum{}: \lapetiteregle {G} {W} {W} {G} {W} {W} {R} {G} {R} \vskip 0pt 
\global\advance\reglenum by 1 
\the\reglenum{}: \lapetiteregle {W} {G} {G} {B} {W} {B} {W} {W} {W} \vskip 0pt 
\global\advance\reglenum by 1 
\the\reglenum{}: \lapetiteregle {W} {G} {W} {W} {W} {B} {W} {G} {G} \vskip 0pt 
\global\advance\reglenum by 1 
\the\reglenum{}: \lapetiteregle {G} {G} {W} {W} {W} {B} {B} {W} {G} \vskip 0pt 
\global\advance\reglenum by 1 
\the\reglenum{}: \lapetiteregle {W} {G} {W} {W} {W} {W} {B} {R} {W} \vskip 0pt 
\global\advance\reglenum by 1 
\the\reglenum{}: \lapetiteregle {R} {G} {W} {B} {W} {B} {W} {G} {W} \vskip 0pt 
\global\advance\reglenum by 1 
\the\reglenum{}: \lapetiteregle {G} {G} {R} {W} {W} {B} {B} {W} {G} \vskip 0pt 
\global\advance\reglenum by 1 
\the\reglenum{}: \lapetiteregle {W} {G} {R} {B} {W} {W} {B} {W} {W} \vskip 0pt 
\global\advance\reglenum by 1 
\the\reglenum{}: \lapetiteregle {R} {W} {G} {G} {W} {W} {W} {G} {W} \vskip 0pt 
\global\advance\reglenum by 1 
\the\reglenum{}: \lapetiteregle {W} {R} {G} {B} {W} {B} {W} {G} {W} \vskip 0pt 
\global\advance\reglenum by 1 
\the\reglenum{}: \lapetiteregle {G} {R} {W} {W} {W} {B} {W} {G} {R} \vskip 0pt 
\global\advance\reglenum by 1 
\the\reglenum{}: \lapetiteregle {W} {G} {W} {B} {W} {W} {W} {W} {W} \vskip 0pt 
\global\advance\reglenum by 1 
\the\reglenum{}: \lapetiteregle {G} {R} {G} {W} {W} {B} {B} {W} {G} \vskip 0pt 
\global\advance\reglenum by 1 
\the\reglenum{}: \lapetiteregle {W} {G} {G} {B} {W} {W} {B} {W} {G} \vskip 0pt 
\global\advance\reglenum by 1 
\the\reglenum{}: \lapetiteregle {W} {R} {W} {W} {W} {W} {B} {W} {W} \vskip 0pt 
\global\advance\reglenum by 1 
\the\reglenum{}: \lapetiteregle {W} {W} {R} {G} {W} {W} {W} {G} {W} \vskip 0pt 
\global\advance\reglenum by 1 
\the\reglenum{}: \lapetiteregle {W} {W} {G} {B} {W} {B} {W} {R} {W} \vskip 0pt 
\global\advance\reglenum by 1 
\the\reglenum{}: \lapetiteregle {R} {W} {W} {W} {W} {B} {G} {G} {W} \vskip 0pt 
\global\advance\reglenum by 1 
\the\reglenum{}: \lapetiteregle {W} {R} {W} {B} {W} {W} {W} {W} {W} \vskip 0pt 
\global\advance\reglenum by 1 
\the\reglenum{}: \lapetiteregle {B} {R} {W} {W} {W} {W} {W} {G} {B} \vskip 0pt 
\global\advance\reglenum by 1 
\the\reglenum{}: \lapetiteregle {G} {W} {R} {G} {W} {B} {B} {W} {G} \vskip 0pt 
\global\advance\reglenum by 1 
\the\reglenum{}: \lapetiteregle {G} {G} {R} {B} {W} {W} {B} {W} {R} \vskip 0pt 
\global\advance\reglenum by 1 
\the\reglenum{}: \lapetiteregle {W} {G} {G} {B} {W} {B} {W} {B} {G} \vskip 0pt 
\global\advance\reglenum by 1 
\the\reglenum{}: \lapetiteregle {B} {W} {G} {W} {W} {W} {W} {W} {B} \vskip 0pt 
\global\advance\reglenum by 1 
\the\reglenum{}: \lapetiteregle {W} {W} {W} {W} {W} {B} {R} {G} {W} \vskip 0pt 
\global\advance\reglenum by 1 
\the\reglenum{}: \lapetiteregle {G} {W} {W} {R} {G} {B} {B} {W} {G} \vskip 0pt 
\global\advance\reglenum by 1 
\the\reglenum{}: \lapetiteregle {R} {G} {W} {B} {W} {W} {B} {G} {W} \vskip 0pt 
\global\advance\reglenum by 1 
\the\reglenum{}: \lapetiteregle {G} {G} {R} {B} {W} {B} {W} {B} {R} \vskip 0pt 
\global\advance\reglenum by 1 
\the\reglenum{}: \lapetiteregle {B} {G} {G} {W} {W} {B} {W} {B} {B} \vskip 0pt 
\global\advance\reglenum by 1 
\the\reglenum{}: \lapetiteregle {B} {G} {R} {W} {W} {W} {W} {W} {B} \vskip 0pt 
\global\advance\reglenum by 1 
\the\reglenum{}: \lapetiteregle {W} {B} {G} {B} {W} {W} {B} {W} {B} \vskip 0pt 
\global\advance\reglenum by 1 
}
\vskip 3pt\rmviii
same path,\vskip 0pt
other\vskip 2pt
conditions:
\vskip 1pt
{\ttv\obeylines
\the\reglenum{}: \lapetiteregle {B} {B} {G} {W} {B} {R} {B} {G} {B} \vskip 0pt 
\global\advance\reglenum by 1 
\the\reglenum{}: \lapetiteregle {B} {G} {B} {R} {W} {W} {B} {G} {B} \vskip 0pt 
\global\advance\reglenum by 1 
\the\reglenum{}: \lapetiteregle {G} {W} {W} {W} {W} {G} {B} {B} {G} \vskip 0pt 
\global\advance\reglenum by 1 
\the\reglenum{}: \lapetiteregle {B} {B} {G} {G} {R} {W} {B} {G} {B} \vskip 0pt 
\global\advance\reglenum by 1 
\the\reglenum{}: \lapetiteregle {G} {W} {W} {W} {G} {R} {B} {B} {G} \vskip 0pt 
\global\advance\reglenum by 1 
\the\reglenum{}: \lapetiteregle {W} {G} {W} {W} {B} {W} {B} {G} {G} \vskip 0pt 
\global\advance\reglenum by 1 
\the\reglenum{}: \lapetiteregle {B} {B} {G} {R} {W} {W} {B} {G} {B} \vskip 0pt 
\global\advance\reglenum by 1 
\the\reglenum{}: \lapetiteregle {G} {W} {W} {G} {R} {W} {B} {B} {G} \vskip 0pt 
\global\advance\reglenum by 1 
\the\reglenum{}: \lapetiteregle {G} {G} {W} {W} {B} {W} {B} {R} {R} \vskip 0pt 
\global\advance\reglenum by 1 
\the\reglenum{}: \lapetiteregle {W} {B} {W} {W} {G} {G} {B} {W} {W} \vskip 0pt 
\global\advance\reglenum by 1 
\the\reglenum{}: \lapetiteregle {W} {W} {W} {W} {B} {W} {G} {G} {G} \vskip 0pt 
\global\advance\reglenum by 1 
\the\reglenum{}: \lapetiteregle {G} {W} {G} {R} {W} {W} {B} {B} {G} \vskip 0pt 
\global\advance\reglenum by 1 
\the\reglenum{}: \lapetiteregle {R} {G} {G} {W} {B} {W} {B} {W} {W} \vskip 0pt 
\global\advance\reglenum by 1 
\the\reglenum{}: \lapetiteregle {W} {B} {W} {G} {R} {G} {B} {W} {W} \vskip 0pt 
\global\advance\reglenum by 1 
\the\reglenum{}: \lapetiteregle {W} {W} {B} {G} {W} {W} {W} {G} {W} \vskip 0pt 
\global\advance\reglenum by 1 
\the\reglenum{}: \lapetiteregle {G} {W} {W} {W} {B} {W} {G} {R} {R} \vskip 0pt 
\global\advance\reglenum by 1 
\the\reglenum{}: \lapetiteregle {W} {G} {B} {W} {W} {B} {W} {G} {G} \vskip 0pt 
\global\advance\reglenum by 1 
\the\reglenum{}: \lapetiteregle {W} {B} {W} {R} {W} {G} {B} {W} {W} \vskip 0pt 
\global\advance\reglenum by 1 
\the\reglenum{}: \lapetiteregle {W} {W} {B} {G} {W} {W} {W} {R} {W} \vskip 0pt 
\global\advance\reglenum by 1 
\the\reglenum{}: \lapetiteregle {R} {W} {W} {W} {B} {G} {G} {W} {W} \vskip 0pt 
\global\advance\reglenum by 1 
\the\reglenum{}: \lapetiteregle {G} {R} {B} {W} {W} {B} {W} {G} {R} \vskip 0pt 
\global\advance\reglenum by 1 
\the\reglenum{}: \lapetiteregle {W} {W} {W} {W} {B} {G} {G} {W} {G} \vskip 0pt 
\global\advance\reglenum by 1 
\the\reglenum{}: \lapetiteregle {G} {W} {B} {W} {W} {B} {R} {G} {R} \vskip 0pt 
\global\advance\reglenum by 1 
\the\reglenum{}: \lapetiteregle {B} {G} {R} {W} {W} {B} {G} {B} {B} \vskip 0pt 
\global\advance\reglenum by 1 
\the\reglenum{}: \lapetiteregle {W} {B} {W} {G} {W} {G} {B} {W} {W} \vskip 0pt 
\global\advance\reglenum by 1 
\the\reglenum{}: \lapetiteregle {G} {W} {W} {W} {B} {R} {G} {W} {R} \vskip 0pt 
\global\advance\reglenum by 1 
\the\reglenum{}: \lapetiteregle {R} {G} {B} {W} {W} {B} {W} {G} {W} \vskip 0pt 
\global\advance\reglenum by 1 
\the\reglenum{}: \lapetiteregle {W} {B} {W} {R} {G} {G} {B} {W} {W} \vskip 0pt 
\global\advance\reglenum by 1 
\the\reglenum{}: \lapetiteregle {R} {W} {W} {W} {B} {W} {G} {G} {W} \vskip 0pt 
\global\advance\reglenum by 1 
}
}

\setbox122=\vtop{\leftskip 0pt\parindent 0pt
\baselineskip 7pt
\hsize=70pt
{\ttv\obeylines
\the\reglenum{}: \lapetiteregle {G} {W} {G} {W} {W} {W} {B} {B} {G} \vskip 0pt 
\global\advance\reglenum by 1 
\the\reglenum{}: \lapetiteregle {W} {G} {G} {W} {B} {W} {B} {W} {G} \vskip 0pt 
\global\advance\reglenum by 1 
\the\reglenum{}: \lapetiteregle {G} {W} {R} {G} {W} {W} {B} {B} {G} \vskip 0pt 
\global\advance\reglenum by 1 
\the\reglenum{}: \lapetiteregle {G} {G} {R} {W} {B} {W} {B} {W} {R} \vskip 0pt 
\global\advance\reglenum by 1 
\the\reglenum{}: \lapetiteregle {G} {W} {W} {R} {G} {W} {B} {B} {G} \vskip 0pt 
\global\advance\reglenum by 1 
\the\reglenum{}: \lapetiteregle {R} {G} {W} {W} {B} {W} {B} {G} {W} \vskip 0pt 
\global\advance\reglenum by 1 
\the\reglenum{}: \lapetiteregle {G} {W} {W} {W} {R} {G} {B} {B} {G} \vskip 0pt 
\global\advance\reglenum by 1 
\the\reglenum{}: \lapetiteregle {W} {G} {W} {W} {B} {W} {B} {R} {W} \vskip 0pt 
\global\advance\reglenum by 1 
\the\reglenum{}: \lapetiteregle {B} {B} {G} {G} {W} {W} {B} {G} {B} \vskip 0pt 
\global\advance\reglenum by 1 
\the\reglenum{}: \lapetiteregle {G} {W} {W} {W} {W} {R} {B} {B} {G} \vskip 0pt 
\global\advance\reglenum by 1 
\the\reglenum{}: \lapetiteregle {B} {B} {G} {R} {B} {W} {B} {G} {B} \vskip 0pt 
\global\advance\reglenum by 1 
}
\vskip 5pt\rmviii
path~4:
\vskip 1pt
{\ttv\obeylines
\the\reglenum{}: \lapetiteregle {W} {W} {W} {G} {B} {W} {W} {G} {B} \vskip 0pt 
\global\advance\reglenum by 1 
\the\reglenum{}: \lapetiteregle {W} {W} {B} {W} {W} {W} {B} {W} {W} \vskip 0pt 
\global\advance\reglenum by 1 
\the\reglenum{}: \lapetiteregle {B} {W} {W} {G} {R} {W} {W} {G} {R} \vskip 0pt 
\global\advance\reglenum by 1 
\the\reglenum{}: \lapetiteregle {W} {B} {W} {W} {W} {B} {W} {G} {W} \vskip 0pt 
\global\advance\reglenum by 1 
\the\reglenum{}: \lapetiteregle {W} {B} {R} {W} {W} {W} {B} {W} {W} \vskip 0pt 
\global\advance\reglenum by 1 
\the\reglenum{}: \lapetiteregle {R} {B} {W} {G} {W} {W} {W} {G} {W} \vskip 0pt 
\global\advance\reglenum by 1 
\the\reglenum{}: \lapetiteregle {W} {R} {B} {W} {W} {B} {W} {G} {W} \vskip 0pt 
\global\advance\reglenum by 1 
\the\reglenum{}: \lapetiteregle {W} {R} {W} {G} {W} {W} {W} {G} {W} \vskip 0pt 
\global\advance\reglenum by 1 
\the\reglenum{}: \lapetiteregle {W} {W} {R} {W} {W} {B} {W} {G} {W} \vskip 0pt 
\global\advance\reglenum by 1 
}
\vskip 5pt\rmviii
path~3:
\vskip 1pt
{\ttv\obeylines
\the\reglenum{}: \lapetiteregle {W} {W} {W} {W} {W} {B} {G} {G} {G} \vskip 0pt 
\global\advance\reglenum by 1 
\the\reglenum{}: \lapetiteregle {W} {W} {G} {G} {W} {W} {W} {G} {G} \vskip 0pt 
\global\advance\reglenum by 1 
\the\reglenum{}: \lapetiteregle {W} {W} {G} {B} {W} {B} {W} {G} {W} \vskip 0pt 
\global\advance\reglenum by 1 
\the\reglenum{}: \lapetiteregle {G} {W} {W} {W} {W} {B} {R} {G} {R} \vskip 0pt 
\global\advance\reglenum by 1 
\the\reglenum{}: \lapetiteregle {G} {W} {R} {G} {W} {W} {W} {G} {R} \vskip 0pt 
\global\advance\reglenum by 1 
\the\reglenum{}: \lapetiteregle {W} {G} {G} {B} {W} {B} {W} {R} {W} \vskip 0pt 
\global\advance\reglenum by 1 
\the\reglenum{}: \lapetiteregle {R} {G} {W} {W} {W} {B} {W} {G} {W} \vskip 0pt 
\global\advance\reglenum by 1 
\the\reglenum{}: \lapetiteregle {W} {G} {W} {W} {W} {W} {B} {W} {W} \vskip 0pt 
\global\advance\reglenum by 1 
\the\reglenum{}: \lapetiteregle {W} {G} {W} {B} {W} {B} {W} {G} {G} \vskip 0pt 
\global\advance\reglenum by 1 
\the\reglenum{}: \lapetiteregle {R} {W} {W} {G} {W} {W} {G} {G} {W} \vskip 0pt 
\global\advance\reglenum by 1 
\the\reglenum{}: \lapetiteregle {W} {R} {W} {W} {W} {W} {B} {G} {W} \vskip 0pt 
\global\advance\reglenum by 1 
\the\reglenum{}: \lapetiteregle {G} {R} {W} {B} {W} {B} {W} {G} {R} \vskip 0pt 
\global\advance\reglenum by 1 
\the\reglenum{}: \lapetiteregle {W} {W} {W} {G} {W} {W} {R} {G} {W} \vskip 0pt 
\global\advance\reglenum by 1 
\the\reglenum{}: \lapetiteregle {W} {W} {W} {W} {W} {W} {B} {R} {W} \vskip 0pt 
\global\advance\reglenum by 1 
\the\reglenum{}: \lapetiteregle {R} {W} {W} {B} {W} {B} {G} {G} {W} \vskip 0pt 
\global\advance\reglenum by 1 
\the\reglenum{}: \lapetiteregle {W} {W} {W} {B} {W} {B} {R} {G} {W} \vskip 0pt 
\global\advance\reglenum by 1 
\the\reglenum{}: \lapetiteregle {G} {W} {W} {W} {R} {B} {B} {W} {G} \vskip 0pt 
\global\advance\reglenum by 1 
\the\reglenum{}: \lapetiteregle {W} {G} {W} {B} {W} {W} {B} {R} {W} \vskip 0pt 
\global\advance\reglenum by 1 
\the\reglenum{}: \lapetiteregle {R} {G} {W} {B} {W} {B} {B} {B} {W} \vskip 0pt 
\global\advance\reglenum by 1 
\the\reglenum{}: \lapetiteregle {B} {G} {R} {B} {W} {B} {W} {B} {B} \vskip 0pt 
\global\advance\reglenum by 1 
}
\vskip 5pt
{\sc switches}
\vskip 5pt
fixed switch
\vskip 5pt\rmviii
path~1:
\vskip 1pt
{\ttv\obeylines
\vskip 1pt
\the\reglenum{}: \lapetiteregle {W} {W} {W} {G} {W} {G} {W} {G} {W} \vskip 0pt 
\global\advance\reglenum by 1 
\the\reglenum{}: \lapetiteregle {W} {W} {W} {W} {W} {W} {W} {G} {W} \vskip 0pt 
\global\advance\reglenum by 1 
\the\reglenum{}: \lapetiteregle {W} {G} {W} {W} {W} {W} {W} {W} {W} \vskip 0pt 
\global\advance\reglenum by 1 
\the\reglenum{}: \lapetiteregle {B} {G} {W} {W} {W} {W} {W} {B} {B} \vskip 0pt 
\global\advance\reglenum by 1 
\the\reglenum{}: \lapetiteregle {W} {B} {W} {G} {W} {G} {W} {G} {B} \vskip 0pt 
\global\advance\reglenum by 1 
\the\reglenum{}: \lapetiteregle {W} {W} {B} {W} {W} {W} {W} {G} {W} \vskip 0pt 
\global\advance\reglenum by 1 
\the\reglenum{}: \lapetiteregle {B} {R} {W} {G} {W} {G} {W} {G} {R} \vskip 0pt 
\global\advance\reglenum by 1 
\the\reglenum{}: \lapetiteregle {W} {B} {R} {W} {W} {W} {W} {G} {W} \vskip 0pt 
\global\advance\reglenum by 1 
\the\reglenum{}: \lapetiteregle {W} {B} {G} {B} {W} {B} {W} {G} {B} \vskip 0pt 
\global\advance\reglenum by 1 
\the\reglenum{}: \lapetiteregle {R} {W} {W} {G} {B} {G} {B} {G} {W} \vskip 0pt 
\global\advance\reglenum by 1 
\the\reglenum{}: \lapetiteregle {W} {R} {W} {W} {W} {W} {W} {G} {W} \vskip 0pt 
\global\advance\reglenum by 1 
\the\reglenum{}: \lapetiteregle {B} {R} {G} {B} {W} {B} {W} {G} {R} \vskip 0pt 
\global\advance\reglenum by 1 
\the\reglenum{}: \lapetiteregle {G} {R} {B} {W} {W} {B} {B} {B} {G} \vskip 0pt 
\global\advance\reglenum by 1 
\the\reglenum{}: \lapetiteregle {W} {G} {B} {B} {W} {W} {W} {W} {W} \vskip 0pt 
\global\advance\reglenum by 1 
\the\reglenum{}: \lapetiteregle {G} {R} {B} {W} {W} {B} {B} {W} {G} \vskip 0pt 
\global\advance\reglenum by 1 
\the\reglenum{}: \lapetiteregle {W} {G} {B} {B} {W} {W} {B} {W} {W} \vskip 0pt 
\global\advance\reglenum by 1 
\the\reglenum{}: \lapetiteregle {W} {W} {W} {G} {R} {G} {R} {G} {W} \vskip 0pt 
\global\advance\reglenum by 1 
\the\reglenum{}: \lapetiteregle {R} {W} {G} {B} {B} {B} {W} {G} {W} \vskip 0pt 
\global\advance\reglenum by 1 
\the\reglenum{}: \lapetiteregle {G} {W} {R} {W} {W} {B} {B} {R} {G} \vskip 0pt 
\global\advance\reglenum by 1 
\the\reglenum{}: \lapetiteregle {W} {G} {R} {B} {W} {W} {W} {W} {W} \vskip 0pt 
\global\advance\reglenum by 1 
\the\reglenum{}: \lapetiteregle {R} {W} {G} {B} {W} {B} {W} {G} {W} \vskip 0pt 
\global\advance\reglenum by 1 
\the\reglenum{}: \lapetiteregle {G} {W} {R} {W} {W} {B} {B} {W} {G} \vskip 0pt 
\global\advance\reglenum by 1 
\the\reglenum{}: \lapetiteregle {W} {W} {G} {B} {R} {B} {W} {G} {W} \vskip 0pt 
\global\advance\reglenum by 1 
}
\vskip 3pt\rmviii
same path,\vskip 0pt
other\vskip 2pt
conditions:
\vskip 1pt
{\ttv\obeylines
\the\reglenum{}: \lapetiteregle {W} {W} {W} {W} {W} {W} {G} {W} {W} \vskip 0pt 
\global\advance\reglenum by 1 
\the\reglenum{}: \lapetiteregle {W} {G} {W} {B} {G} {W} {W} {W} {W} \vskip 0pt 
\global\advance\reglenum by 1 
\the\reglenum{}: \lapetiteregle {W} {B} {W} {W} {W} {B} {W} {W} {W} \vskip 0pt 
\global\advance\reglenum by 1 
\the\reglenum{}: \lapetiteregle {W} {G} {W} {G} {B} {W} {B} {W} {W} \vskip 0pt 
\global\advance\reglenum by 1 
\the\reglenum{}: \lapetiteregle {W} {B} {W} {W} {W} {W} {G} {W} {W} \vskip 0pt 
\global\advance\reglenum by 1 
\the\reglenum{}: \lapetiteregle {W} {R} {W} {W} {W} {W} {G} {B} {W} \vskip 0pt 
\global\advance\reglenum by 1 
\the\reglenum{}: \lapetiteregle {W} {G} {B} {G} {B} {W} {B} {W} {B} \vskip 0pt 
\global\advance\reglenum by 1 
\the\reglenum{}: \lapetiteregle {W} {W} {W} {W} {W} {W} {G} {R} {W} \vskip 0pt 
\global\advance\reglenum by 1 
\the\reglenum{}: \lapetiteregle {W} {G} {B} {B} {G} {W} {W} {W} {W} \vskip 0pt 
\global\advance\reglenum by 1 
}
}

\setbox124=\vtop{\leftskip 0pt\parindent 0pt
\baselineskip 7pt
\hsize=70pt
{\ttv\obeylines
\the\reglenum{}: \lapetiteregle {G} {W} {R} {B} {W} {W} {B} {B} {G} \vskip 0pt 
\global\advance\reglenum by 1 
\the\reglenum{}: \lapetiteregle {B} {G} {R} {G} {B} {W} {B} {W} {R} \vskip 0pt 
\global\advance\reglenum by 1 
\the\reglenum{}: \lapetiteregle {W} {W} {W} {W} {W} {W} {G} {B} {W} \vskip 0pt 
\global\advance\reglenum by 1 
\the\reglenum{}: \lapetiteregle {W} {G} {R} {B} {G} {W} {W} {W} {W} \vskip 0pt 
\global\advance\reglenum by 1 
\the\reglenum{}: \lapetiteregle {W} {B} {W} {W} {W} {W} {G} {R} {W} \vskip 0pt 
\global\advance\reglenum by 1 
\the\reglenum{}: \lapetiteregle {G} {B} {R} {B} {W} {W} {B} {B} {G} \vskip 0pt 
\global\advance\reglenum by 1 
\the\reglenum{}: \lapetiteregle {W} {R} {W} {W} {W} {W} {G} {W} {W} \vskip 0pt 
\global\advance\reglenum by 1 
\the\reglenum{}: \lapetiteregle {G} {R} {W} {R} {W} {W} {B} {B} {G} \vskip 0pt 
\global\advance\reglenum by 1 
\the\reglenum{}: \lapetiteregle {R} {G} {W} {G} {B} {W} {B} {W} {W} \vskip 0pt 
\global\advance\reglenum by 1 
}
\vskip 5pt\rmviii
path~4:
\vskip 1pt
{\ttv\obeylines
\the\reglenum{}: \lapetiteregle {W} {W} {G} {B} {B} {B} {W} {G} {B} \vskip 0pt 
\global\advance\reglenum by 1 
\the\reglenum{}: \lapetiteregle {W} {W} {W} {G} {B} {G} {W} {G} {B} \vskip 0pt 
\global\advance\reglenum by 1 
\the\reglenum{}: \lapetiteregle {B} {W} {G} {B} {R} {B} {W} {G} {R} \vskip 0pt 
\global\advance\reglenum by 1 
\the\reglenum{}: \lapetiteregle {G} {W} {B} {W} {W} {B} {B} {W} {G} \vskip 0pt 
\global\advance\reglenum by 1 
\the\reglenum{}: \lapetiteregle {B} {W} {W} {G} {R} {G} {W} {G} {R} \vskip 0pt 
\global\advance\reglenum by 1 
\the\reglenum{}: \lapetiteregle {W} {B} {W} {W} {W} {W} {W} {G} {W} \vskip 0pt 
\global\advance\reglenum by 1 
\the\reglenum{}: \lapetiteregle {R} {B} {G} {B} {W} {B} {W} {G} {W} \vskip 0pt 
\global\advance\reglenum by 1 
\the\reglenum{}: \lapetiteregle {G} {B} {R} {W} {W} {B} {B} {W} {G} \vskip 0pt 
\global\advance\reglenum by 1 
\the\reglenum{}: \lapetiteregle {R} {B} {W} {G} {W} {G} {B} {G} {W} \vskip 0pt 
\global\advance\reglenum by 1 
\the\reglenum{}: \lapetiteregle {W} {R} {B} {W} {W} {W} {W} {G} {W} \vskip 0pt 
\global\advance\reglenum by 1 
\the\reglenum{}: \lapetiteregle {W} {R} {W} {G} {W} {G} {R} {G} {W} \vskip 0pt 
\global\advance\reglenum by 1 
\the\reglenum{}: \lapetiteregle {W} {W} {R} {W} {W} {W} {W} {G} {W} \vskip 0pt 
\global\advance\reglenum by 1 
}
\vskip 3pt\rmviii
same path,\vskip 0pt
other\vskip 2pt
conditions:
\vskip 1pt
{\ttv\obeylines
\the\reglenum{}: \lapetiteregle {W} {G} {B} {W} {B} {W} {G} {W} {W} \vskip 0pt 
\global\advance\reglenum by 1 
\the\reglenum{}: \lapetiteregle {G} {W} {W} {W} {B} {B} {W} {W} {G} \vskip 0pt 
\global\advance\reglenum by 1 
\the\reglenum{}: \lapetiteregle {B} {G} {B} {W} {W} {W} {W} {W} {B} \vskip 0pt 
\global\advance\reglenum by 1 
\the\reglenum{}: \lapetiteregle {W} {W} {G} {W} {G} {B} {W} {G} {B} \vskip 0pt 
\global\advance\reglenum by 1 
\the\reglenum{}: \lapetiteregle {W} {G} {B} {W} {B} {W} {G} {B} {B} \vskip 0pt 
\global\advance\reglenum by 1 
\the\reglenum{}: \lapetiteregle {W} {G} {B} {R} {W} {W} {W} {W} {W} \vskip 0pt 
\global\advance\reglenum by 1 
\the\reglenum{}: \lapetiteregle {G} {W} {W} {W} {B} {B} {W} {B} {G} \vskip 0pt 
\global\advance\reglenum by 1 
\the\reglenum{}: \lapetiteregle {B} {W} {G} {W} {G} {R} {W} {G} {R} \vskip 0pt 
\global\advance\reglenum by 1 
\the\reglenum{}: \lapetiteregle {B} {G} {B} {W} {B} {W} {G} {R} {R} \vskip 0pt 
\global\advance\reglenum by 1 
\the\reglenum{}: \lapetiteregle {W} {G} {R} {W} {W} {W} {W} {W} {W} \vskip 0pt 
\global\advance\reglenum by 1 
\the\reglenum{}: \lapetiteregle {W} {B} {B} {G} {W} {W} {W} {G} {W} \vskip 0pt 
\global\advance\reglenum by 1 
\the\reglenum{}: \lapetiteregle {G} {B} {W} {W} {B} {B} {B} {R} {G} \vskip 0pt 
\global\advance\reglenum by 1 
\the\reglenum{}: \lapetiteregle {R} {B} {G} {B} {G} {W} {W} {G} {W} \vskip 0pt 
\global\advance\reglenum by 1 
\the\reglenum{}: \lapetiteregle {R} {G} {B} {B} {B} {W} {G} {W} {W} \vskip 0pt 
\global\advance\reglenum by 1 
\the\reglenum{}: \lapetiteregle {W} {R} {B} {G} {W} {W} {W} {G} {W} \vskip 0pt 
\global\advance\reglenum by 1 
\the\reglenum{}: \lapetiteregle {G} {R} {W} {W} {B} {B} {R} {W} {G} \vskip 0pt 
\global\advance\reglenum by 1 
\the\reglenum{}: \lapetiteregle {B} {G} {B} {W} {W} {W} {W} {R} {B} \vskip 0pt 
\global\advance\reglenum by 1 
\the\reglenum{}: \lapetiteregle {W} {R} {G} {R} {G} {W} {W} {G} {W} \vskip 0pt 
\global\advance\reglenum by 1 
\the\reglenum{}: \lapetiteregle {W} {G} {B} {R} {B} {W} {G} {W} {W} \vskip 0pt 
\global\advance\reglenum by 1 
\the\reglenum{}: \lapetiteregle {W} {W} {G} {W} {G} {W} {W} {G} {W} \vskip 0pt 
\global\advance\reglenum by 1 
\the\reglenum{}: \lapetiteregle {G} {W} {W} {W} {B} {B} {G} {W} {G} \vskip 0pt 
\global\advance\reglenum by 1 
\the\reglenum{}: \lapetiteregle {W} {W} {G} {G} {G} {W} {W} {G} {G} \vskip 0pt 
\global\advance\reglenum by 1 
\the\reglenum{}: \lapetiteregle {W} {G} {B} {W} {B} {W} {G} {G} {B} \vskip 0pt 
\global\advance\reglenum by 1 
\the\reglenum{}: \lapetiteregle {G} {W} {W} {W} {B} {B} {R} {G} {G} \vskip 0pt 
\global\advance\reglenum by 1 
\the\reglenum{}: \lapetiteregle {G} {W} {G} {R} {G} {W} {W} {G} {R} \vskip 0pt 
\global\advance\reglenum by 1 
\the\reglenum{}: \lapetiteregle {G} {B} {W} {W} {B} {B} {W} {R} {G} \vskip 0pt 
\global\advance\reglenum by 1 
\the\reglenum{}: \lapetiteregle {R} {B} {G} {W} {G} {W} {W} {G} {W} \vskip 0pt 
\global\advance\reglenum by 1 
\the\reglenum{}: \lapetiteregle {G} {R} {W} {W} {B} {B} {W} {W} {G} \vskip 0pt 
\global\advance\reglenum by 1 
\the\reglenum{}: \lapetiteregle {W} {R} {G} {W} {G} {W} {W} {G} {W} \vskip 0pt 
\global\advance\reglenum by 1 
\the\reglenum{}: \lapetiteregle {W} {G} {B} {B} {B} {W} {G} {W} {B} \vskip 0pt 
\global\advance\reglenum by 1 
\the\reglenum{}: \lapetiteregle {B} {G} {B} {R} {B} {W} {G} {W} {R} \vskip 0pt 
\global\advance\reglenum by 1 
\the\reglenum{}: \lapetiteregle {G} {B} {W} {W} {B} {B} {W} {W} {G} \vskip 0pt 
\global\advance\reglenum by 1 
\the\reglenum{}: \lapetiteregle {W} {B} {G} {W} {G} {W} {W} {G} {B} \vskip 0pt 
\global\advance\reglenum by 1 
\the\reglenum{}: \lapetiteregle {R} {G} {B} {W} {B} {W} {G} {B} {W} \vskip 0pt 
\global\advance\reglenum by 1 
\the\reglenum{}: \lapetiteregle {G} {R} {W} {W} {B} {B} {W} {B} {G} \vskip 0pt 
\global\advance\reglenum by 1 
\the\reglenum{}: \lapetiteregle {B} {R} {G} {W} {G} {W} {W} {G} {R} \vskip 0pt 
\global\advance\reglenum by 1 
\the\reglenum{}: \lapetiteregle {W} {G} {B} {W} {B} {W} {G} {R} {W} \vskip 0pt 
\global\advance\reglenum by 1 
\the\reglenum{}: \lapetiteregle {G} {W} {W} {W} {B} {B} {B} {R} {G} \vskip 0pt 
\global\advance\reglenum by 1 
\the\reglenum{}: \lapetiteregle {R} {W} {G} {B} {G} {B} {W} {G} {W} \vskip 0pt 
\global\advance\reglenum by 1 
}
\vskip 5pt\rmviii
path~7:
\vskip 1pt
{\ttv\obeylines
\the\reglenum{}: \lapetiteregle {W} {W} {G} {B} {W} {B} {G} {G} {G} \vskip 0pt 
\global\advance\reglenum by 1 
\the\reglenum{}: \lapetiteregle {W} {W} {W} {G} {W} {G} {G} {G} {G} \vskip 0pt 
\global\advance\reglenum by 1 
\the\reglenum{}: \lapetiteregle {G} {W} {W} {W} {W} {B} {B} {G} {G} \vskip 0pt 
\global\advance\reglenum by 1 
\the\reglenum{}: \lapetiteregle {G} {W} {G} {B} {W} {B} {R} {G} {R} \vskip 0pt 
\global\advance\reglenum by 1 
\the\reglenum{}: \lapetiteregle {B} {G} {G} {B} {W} {W} {W} {W} {B} \vskip 0pt 
\global\advance\reglenum by 1 
\the\reglenum{}: \lapetiteregle {G} {W} {W} {G} {W} {G} {R} {G} {R} \vskip 0pt 
\global\advance\reglenum by 1 
\the\reglenum{}: \lapetiteregle {W} {G} {W} {W} {W} {W} {W} {G} {W} \vskip 0pt 
\global\advance\reglenum by 1 
\the\reglenum{}: \lapetiteregle {W} {G} {G} {B} {W} {B} {W} {G} {B} \vskip 0pt 
\global\advance\reglenum by 1 
\the\reglenum{}: \lapetiteregle {G} {G} {W} {W} {W} {B} {B} {R} {G} \vskip 0pt 
\global\advance\reglenum by 1 
\the\reglenum{}: \lapetiteregle {R} {G} {G} {B} {W} {B} {W} {G} {W} \vskip 0pt 
\global\advance\reglenum by 1 
\the\reglenum{}: \lapetiteregle {R} {W} {W} {G} {B} {G} {W} {G} {W} \vskip 0pt 
\global\advance\reglenum by 1 
\the\reglenum{}: \lapetiteregle {W} {W} {W} {G} {R} {G} {W} {G} {W} \vskip 0pt 
\global\advance\reglenum by 1 
}
\vskip 3pt\rmviii
same path,\vskip 0pt
other\vskip 2pt
conditions:
\vskip 1pt
{\ttv\obeylines
\the\reglenum{}: \lapetiteregle {W} {G} {W} {G} {B} {W} {B} {G} {G} \vskip 0pt 
\global\advance\reglenum by 1 
\the\reglenum{}: \lapetiteregle {G} {G} {W} {G} {B} {W} {B} {R} {R} \vskip 0pt 
\global\advance\reglenum by 1 
}
}

\setbox126=\vtop{\leftskip 0pt\parindent 0pt
\baselineskip 7pt
\hsize=70pt
{\ttv\obeylines
\the\reglenum{}: \lapetiteregle {W} {W} {W} {W} {W} {W} {G} {G} {W} \vskip 0pt 
\global\advance\reglenum by 1 
\the\reglenum{}: \lapetiteregle {R} {G} {G} {G} {B} {W} {B} {W} {W} \vskip 0pt 
\global\advance\reglenum by 1 
\the\reglenum{}: \lapetiteregle {W} {G} {R} {G} {B} {W} {B} {W} {W} \vskip 0pt 
\global\advance\reglenum by 1 
}
\vskip 5pt
left-hand\vskip 2pt
side memory\vskip 2pt
switch
\vskip 5pt\rmviii
path~1:
\vskip 1pt
{\ttv\obeylines
\the\reglenum{}: \lapetiteregle {W} {W} {G} {G} {W} {G} {G} {G} {W} \vskip 0pt 
\global\advance\reglenum by 1 
\the\reglenum{}: \lapetiteregle {W} {W} {G} {B} {W} {B} {B} {G} {W} \vskip 0pt 
\global\advance\reglenum by 1 
\the\reglenum{}: \lapetiteregle {B} {W} {G} {B} {G} {G} {B} {W} {B} \vskip 0pt 
\global\advance\reglenum by 1 
\the\reglenum{}: \lapetiteregle {W} {W} {B} {B} {W} {B} {B} {B} {B} \vskip 0pt 
\global\advance\reglenum by 1 
\the\reglenum{}: \lapetiteregle {B} {W} {W} {B} {R} {B} {B} {B} {B} \vskip 0pt 
\global\advance\reglenum by 1 
\the\reglenum{}: \lapetiteregle {G} {B} {B} {B} {W} {W} {W} {G} {G} \vskip 0pt 
\global\advance\reglenum by 1 
\the\reglenum{}: \lapetiteregle {G} {B} {G} {W} {W} {W} {W} {B} {G} \vskip 0pt 
\global\advance\reglenum by 1 
\the\reglenum{}: \lapetiteregle {B} {W} {B} {G} {W} {W} {W} {W} {B} \vskip 0pt 
\global\advance\reglenum by 1 
\the\reglenum{}: \lapetiteregle {B} {B} {R} {W} {W} {B} {W} {B} {B} \vskip 0pt 
\global\advance\reglenum by 1 
\the\reglenum{}: \lapetiteregle {W} {G} {G} {W} {W} {W} {W} {W} {W} \vskip 0pt 
\global\advance\reglenum by 1 
\the\reglenum{}: \lapetiteregle {W} {B} {G} {W} {W} {W} {W} {W} {W} \vskip 0pt 
\global\advance\reglenum by 1 
\the\reglenum{}: \lapetiteregle {G} {W} {W} {B} {B} {G} {G} {G} {G} \vskip 0pt 
\global\advance\reglenum by 1 
\the\reglenum{}: \lapetiteregle {B} {G} {W} {B} {B} {W} {W} {B} {B} \vskip 0pt 
\global\advance\reglenum by 1 
\the\reglenum{}: \lapetiteregle {B} {G} {B} {W} {W} {W} {W} {G} {B} \vskip 0pt 
\global\advance\reglenum by 1 
\the\reglenum{}: \lapetiteregle {G} {G} {B} {W} {W} {W} {W} {G} {G} \vskip 0pt 
\global\advance\reglenum by 1 
\the\reglenum{}: \lapetiteregle {G} {W} {G} {G} {G} {B} {B} {W} {G} \vskip 0pt 
\global\advance\reglenum by 1 
\the\reglenum{}: \lapetiteregle {G} {G} {G} {G} {W} {W} {W} {G} {G} \vskip 0pt 
\global\advance\reglenum by 1 
\the\reglenum{}: \lapetiteregle {G} {G} {G} {W} {W} {W} {W} {B} {G} \vskip 0pt 
\global\advance\reglenum by 1 
\the\reglenum{}: \lapetiteregle {B} {G} {G} {W} {W} {W} {W} {B} {B} \vskip 0pt 
\global\advance\reglenum by 1 
\the\reglenum{}: \lapetiteregle {G} {W} {W} {W} {W} {G} {G} {G} {G} \vskip 0pt 
\global\advance\reglenum by 1 
\the\reglenum{}: \lapetiteregle {G} {G} {W} {W} {W} {W} {W} {G} {G} \vskip 0pt 
\global\advance\reglenum by 1 
\the\reglenum{}: \lapetiteregle {G} {W} {G} {G} {G} {B} {B} {G} {G} \vskip 0pt 
\global\advance\reglenum by 1 
\the\reglenum{}: \lapetiteregle {B} {G} {G} {B} {W} {W} {W} {B} {B} \vskip 0pt 
\global\advance\reglenum by 1 
\the\reglenum{}: \lapetiteregle {B} {G} {B} {W} {W} {W} {W} {B} {B} \vskip 0pt 
\global\advance\reglenum by 1 
\the\reglenum{}: \lapetiteregle {B} {G} {G} {W} {W} {W} {B} {B} {B} \vskip 0pt 
\global\advance\reglenum by 1 
\the\reglenum{}: \lapetiteregle {G} {W} {G} {B} {W} {B} {B} {W} {G} \vskip 0pt 
\global\advance\reglenum by 1 
\the\reglenum{}: \lapetiteregle {B} {G} {G} {B} {B} {B} {B} {W} {B} \vskip 0pt 
\global\advance\reglenum by 1 
\the\reglenum{}: \lapetiteregle {W} {G} {B} {B} {W} {B} {W} {B} {W} \vskip 0pt 
\global\advance\reglenum by 1 
\the\reglenum{}: \lapetiteregle {B} {G} {W} {W} {W} {B} {G} {B} {B} \vskip 0pt 
\global\advance\reglenum by 1 
\the\reglenum{}: \lapetiteregle {B} {B} {B} {W} {W} {W} {W} {B} {B} \vskip 0pt 
\global\advance\reglenum by 1 
\the\reglenum{}: \lapetiteregle {B} {W} {B} {B} {W} {W} {W} {W} {B} \vskip 0pt 
\global\advance\reglenum by 1 
\the\reglenum{}: \lapetiteregle {B} {B} {W} {W} {W} {B} {W} {G} {B} \vskip 0pt 
\global\advance\reglenum by 1 
\the\reglenum{}: \lapetiteregle {W} {W} {G} {B} {B} {B} {B} {G} {B} \vskip 0pt 
\global\advance\reglenum by 1 
\the\reglenum{}: \lapetiteregle {B} {W} {G} {B} {G} {G} {B} {B} {B} \vskip 0pt 
\global\advance\reglenum by 1 
\the\reglenum{}: \lapetiteregle {B} {W} {B} {B} {W} {B} {R} {B} {R} \vskip 0pt 
\global\advance\reglenum by 1 
\the\reglenum{}: \lapetiteregle {B} {W} {B} {R} {W} {B} {B} {B} {B} \vskip 0pt 
\global\advance\reglenum by 1 
\the\reglenum{}: \lapetiteregle {B} {B} {B} {G} {W} {W} {W} {W} {B} \vskip 0pt 
\global\advance\reglenum by 1 
\the\reglenum{}: \lapetiteregle {W} {B} {G} {G} {W} {G} {G} {G} {B} \vskip 0pt 
\global\advance\reglenum by 1 
\the\reglenum{}: \lapetiteregle {B} {W} {G} {B} {R} {B} {B} {G} {R} \vskip 0pt 
\global\advance\reglenum by 1 
\the\reglenum{}: \lapetiteregle {B} {B} {G} {B} {G} {G} {B} {R} {B} \vskip 0pt 
\global\advance\reglenum by 1 
\the\reglenum{}: \lapetiteregle {B} {R} {B} {G} {W} {W} {W} {W} {B} \vskip 0pt 
\global\advance\reglenum by 1 
\the\reglenum{}: \lapetiteregle {G} {W} {B} {B} {B} {G} {G} {G} {G} \vskip 0pt 
\global\advance\reglenum by 1 
\the\reglenum{}: \lapetiteregle {B} {G} {B} {B} {B} {W} {W} {B} {B} \vskip 0pt 
\global\advance\reglenum by 1 
\the\reglenum{}: \lapetiteregle {G} {W} {G} {B} {W} {B} {B} {B} {G} \vskip 0pt 
\global\advance\reglenum by 1 
\the\reglenum{}: \lapetiteregle {B} {R} {G} {G} {W} {G} {G} {G} {R} \vskip 0pt 
\global\advance\reglenum by 1 
\the\reglenum{}: \lapetiteregle {R} {B} {G} {B} {W} {B} {B} {G} {W} \vskip 0pt 
\global\advance\reglenum by 1 
\the\reglenum{}: \lapetiteregle {B} {R} {G} {B} {G} {G} {B} {W} {B} \vskip 0pt 
\global\advance\reglenum by 1 
\the\reglenum{}: \lapetiteregle {W} {R} {B} {B} {W} {B} {W} {B} {W} \vskip 0pt 
\global\advance\reglenum by 1 
\the\reglenum{}: \lapetiteregle {B} {R} {W} {W} {W} {B} {B} {B} {B} \vskip 0pt 
\global\advance\reglenum by 1 
\the\reglenum{}: \lapetiteregle {G} {B} {R} {B} {B} {G} {G} {G} {G} \vskip 0pt 
\global\advance\reglenum by 1 
\the\reglenum{}: \lapetiteregle {B} {G} {R} {B} {B} {W} {W} {B} {B} \vskip 0pt 
\global\advance\reglenum by 1 
\the\reglenum{}: \lapetiteregle {G} {B} {G} {G} {G} {B} {B} {W} {G} \vskip 0pt 
\global\advance\reglenum by 1 
\the\reglenum{}: \lapetiteregle {G} {B} {W} {W} {W} {G} {G} {G} {G} \vskip 0pt 
\global\advance\reglenum by 1 
\the\reglenum{}: \lapetiteregle {G} {B} {G} {G} {G} {B} {B} {G} {G} \vskip 0pt 
\global\advance\reglenum by 1 
\the\reglenum{}: \lapetiteregle {G} {B} {G} {B} {W} {B} {B} {R} {G} \vskip 0pt 
\global\advance\reglenum by 1 
\the\reglenum{}: \lapetiteregle {R} {W} {G} {G} {B} {G} {G} {G} {W} \vskip 0pt 
\global\advance\reglenum by 1 
\the\reglenum{}: \lapetiteregle {W} {R} {G} {B} {W} {B} {B} {G} {W} \vskip 0pt 
\global\advance\reglenum by 1 
\the\reglenum{}: \lapetiteregle {G} {R} {W} {B} {B} {G} {G} {G} {G} \vskip 0pt 
\global\advance\reglenum by 1 
\the\reglenum{}: \lapetiteregle {G} {R} {G} {G} {G} {B} {B} {B} {G} \vskip 0pt 
\global\advance\reglenum by 1 
\the\reglenum{}: \lapetiteregle {G} {R} {B} {W} {W} {G} {G} {G} {G} \vskip 0pt 
\global\advance\reglenum by 1 
\the\reglenum{}: \lapetiteregle {G} {R} {G} {G} {G} {B} {B} {G} {G} \vskip 0pt 
\global\advance\reglenum by 1 
\the\reglenum{}: \lapetiteregle {G} {R} {G} {B} {W} {B} {B} {W} {G} \vskip 0pt 
\global\advance\reglenum by 1 
\the\reglenum{}: \lapetiteregle {W} {W} {G} {G} {R} {G} {G} {G} {W} \vskip 0pt 
\global\advance\reglenum by 1 
\the\reglenum{}: \lapetiteregle {G} {W} {G} {G} {G} {B} {B} {R} {G} \vskip 0pt 
\global\advance\reglenum by 1 
\the\reglenum{}: \lapetiteregle {G} {W} {R} {W} {W} {G} {G} {G} {G} \vskip 0pt 
\global\advance\reglenum by 1 
}
\vskip 3pt\rmviii
same path,\vskip 0pt
other\vskip 2pt
conditions:
\vskip 1pt
{\ttv\obeylines
\the\reglenum{}: \lapetiteregle {W} {B} {B} {G} {W} {G} {B} {W} {W} \vskip 0pt 
\global\advance\reglenum by 1 
\the\reglenum{}: \lapetiteregle {B} {W} {B} {G} {W} {W} {B} {G} {B} \vskip 0pt 
\global\advance\reglenum by 1 
\the\reglenum{}: \lapetiteregle {B} {B} {B} {B} {G} {W} {W} {W} {B} \vskip 0pt 
\global\advance\reglenum by 1 
\the\reglenum{}: \lapetiteregle {G} {W} {B} {B} {G} {G} {G} {W} {G} \vskip 0pt 
\global\advance\reglenum by 1 
\the\reglenum{}: \lapetiteregle {G} {G} {G} {W} {W} {W} {G} {G} {G} \vskip 0pt 
\global\advance\reglenum by 1 
\the\reglenum{}: \lapetiteregle {W} {G} {W} {W} {W} {B} {W} {W} {W} \vskip 0pt 
\global\advance\reglenum by 1 
\the\reglenum{}: \lapetiteregle {G} {W} {W} {G} {B} {W} {B} {B} {G} \vskip 0pt 
\global\advance\reglenum by 1 
}
}

\setbox128=\vtop{\leftskip 0pt\parindent 0pt
\baselineskip 7pt
\hsize=70pt
{\ttv\obeylines
\the\reglenum{}: \lapetiteregle {G} {G} {W} {G} {G} {G} {B} {B} {G} \vskip 0pt 
\global\advance\reglenum by 1 
\the\reglenum{}: \lapetiteregle {B} {B} {G} {G} {W} {W} {W} {B} {B} \vskip 0pt 
\global\advance\reglenum by 1 
\the\reglenum{}: \lapetiteregle {G} {B} {B} {B} {G} {W} {W} {G} {G} \vskip 0pt 
\global\advance\reglenum by 1 
\the\reglenum{}: \lapetiteregle {B} {W} {W} {W} {B} {B} {G} {B} {B} \vskip 0pt 
\global\advance\reglenum by 1 
\the\reglenum{}: \lapetiteregle {W} {B} {B} {G} {W} {G} {B} {B} {B} \vskip 0pt 
\global\advance\reglenum by 1 
\the\reglenum{}: \lapetiteregle {B} {B} {B} {G} {W} {G} {B} {R} {R} \vskip 0pt 
\global\advance\reglenum by 1 
\the\reglenum{}: \lapetiteregle {B} {B} {B} {G} {W} {W} {B} {G} {B} \vskip 0pt 
\global\advance\reglenum by 1 
\the\reglenum{}: \lapetiteregle {G} {B} {B} {B} {G} {G} {G} {W} {G} \vskip 0pt 
\global\advance\reglenum by 1 
\the\reglenum{}: \lapetiteregle {G} {B} {W} {G} {B} {W} {B} {B} {G} \vskip 0pt 
\global\advance\reglenum by 1 
\the\reglenum{}: \lapetiteregle {R} {B} {B} {G} {B} {G} {B} {W} {W} \vskip 0pt 
\global\advance\reglenum by 1 
\the\reglenum{}: \lapetiteregle {B} {R} {B} {G} {W} {W} {B} {G} {B} \vskip 0pt 
\global\advance\reglenum by 1 
\the\reglenum{}: \lapetiteregle {G} {R} {B} {B} {G} {G} {G} {B} {G} \vskip 0pt 
\global\advance\reglenum by 1 
\the\reglenum{}: \lapetiteregle {G} {R} {B} {G} {B} {W} {B} {B} {G} \vskip 0pt 
\global\advance\reglenum by 1 
\the\reglenum{}: \lapetiteregle {G} {G} {B} {G} {G} {G} {B} {B} {G} \vskip 0pt 
\global\advance\reglenum by 1 
\the\reglenum{}: \lapetiteregle {W} {B} {B} {G} {R} {G} {B} {W} {W} \vskip 0pt 
\global\advance\reglenum by 1 
\the\reglenum{}: \lapetiteregle {G} {W} {B} {B} {G} {G} {G} {R} {G} \vskip 0pt 
\global\advance\reglenum by 1 
\the\reglenum{}: \lapetiteregle {G} {W} {R} {G} {B} {W} {B} {B} {G} \vskip 0pt 
\global\advance\reglenum by 1 
\the\reglenum{}: \lapetiteregle {G} {G} {R} {G} {G} {G} {B} {B} {G} \vskip 0pt 
\global\advance\reglenum by 1 
\the\reglenum{}: \lapetiteregle {W} {B} {B} {G} {B} {G} {B} {W} {B} \vskip 0pt 
\global\advance\reglenum by 1 
\the\reglenum{}: \lapetiteregle {G} {W} {B} {B} {G} {G} {G} {B} {G} \vskip 0pt 
\global\advance\reglenum by 1 
\the\reglenum{}: \lapetiteregle {G} {W} {B} {G} {B} {W} {B} {B} {G} \vskip 0pt 
\global\advance\reglenum by 1 
\the\reglenum{}: \lapetiteregle {B} {B} {B} {G} {R} {G} {B} {W} {R} \vskip 0pt 
\global\advance\reglenum by 1 
\the\reglenum{}: \lapetiteregle {G} {B} {B} {B} {G} {G} {G} {R} {G} \vskip 0pt 
\global\advance\reglenum by 1 
\the\reglenum{}: \lapetiteregle {G} {B} {R} {G} {B} {W} {B} {B} {G} \vskip 0pt 
\global\advance\reglenum by 1 
\the\reglenum{}: \lapetiteregle {R} {B} {B} {G} {W} {G} {B} {B} {W} \vskip 0pt 
\global\advance\reglenum by 1 
\the\reglenum{}: \lapetiteregle {G} {R} {B} {B} {G} {G} {G} {W} {G} \vskip 0pt 
\global\advance\reglenum by 1 
\the\reglenum{}: \lapetiteregle {G} {R} {W} {G} {B} {W} {B} {B} {G} \vskip 0pt 
\global\advance\reglenum by 1 
\the\reglenum{}: \lapetiteregle {W} {B} {B} {G} {W} {G} {B} {R} {W} \vskip 0pt 
\global\advance\reglenum by 1 
\the\reglenum{}: \lapetiteregle {B} {W} {W} {W} {R} {B} {G} {B} {B} \vskip 0pt 
\global\advance\reglenum by 1 
}
\vskip 5pt\rmviii
path~4:
\vskip 1pt
{\ttv\obeylines
\the\reglenum{}: \lapetiteregle {W} {W} {G} {G} {B} {G} {G} {G} {B} \vskip 0pt 
\global\advance\reglenum by 1 
\the\reglenum{}: \lapetiteregle {G} {W} {G} {G} {G} {B} {B} {B} {G} \vskip 0pt 
\global\advance\reglenum by 1 
\the\reglenum{}: \lapetiteregle {G} {W} {B} {W} {W} {G} {G} {G} {G} \vskip 0pt 
\global\advance\reglenum by 1 
\the\reglenum{}: \lapetiteregle {B} {W} {G} {G} {R} {G} {G} {G} {R} \vskip 0pt 
\global\advance\reglenum by 1 
\the\reglenum{}: \lapetiteregle {W} {B} {G} {B} {W} {B} {B} {G} {B} \vskip 0pt 
\global\advance\reglenum by 1 
\the\reglenum{}: \lapetiteregle {G} {B} {W} {B} {B} {G} {G} {G} {G} \vskip 0pt 
\global\advance\reglenum by 1 
\the\reglenum{}: \lapetiteregle {G} {B} {G} {G} {G} {B} {B} {R} {G} \vskip 0pt 
\global\advance\reglenum by 1 
\the\reglenum{}: \lapetiteregle {G} {B} {R} {W} {W} {G} {G} {G} {G} \vskip 0pt 
\global\advance\reglenum by 1 
\the\reglenum{}: \lapetiteregle {G} {B} {G} {B} {W} {B} {B} {W} {G} \vskip 0pt 
\global\advance\reglenum by 1 
\the\reglenum{}: \lapetiteregle {R} {B} {G} {G} {W} {G} {G} {G} {W} \vskip 0pt 
\global\advance\reglenum by 1 
\the\reglenum{}: \lapetiteregle {B} {R} {G} {B} {W} {B} {B} {G} {R} \vskip 0pt 
\global\advance\reglenum by 1 
\the\reglenum{}: \lapetiteregle {B} {B} {G} {B} {G} {G} {B} {W} {B} \vskip 0pt 
\global\advance\reglenum by 1 
\the\reglenum{}: \lapetiteregle {B} {B} {W} {W} {W} {B} {B} {B} {B} \vskip 0pt 
\global\advance\reglenum by 1 
\the\reglenum{}: \lapetiteregle {G} {R} {B} {B} {B} {G} {G} {G} {G} \vskip 0pt 
\global\advance\reglenum by 1 
\the\reglenum{}: \lapetiteregle {G} {R} {G} {G} {G} {B} {B} {W} {G} \vskip 0pt 
\global\advance\reglenum by 1 
\the\reglenum{}: \lapetiteregle {G} {R} {W} {W} {W} {G} {G} {G} {G} \vskip 0pt 
\global\advance\reglenum by 1 
\the\reglenum{}: \lapetiteregle {G} {R} {G} {B} {W} {B} {B} {B} {G} \vskip 0pt 
\global\advance\reglenum by 1 
\the\reglenum{}: \lapetiteregle {W} {R} {G} {G} {W} {G} {G} {G} {W} \vskip 0pt 
\global\advance\reglenum by 1 
\the\reglenum{}: \lapetiteregle {R} {W} {G} {B} {B} {B} {B} {G} {W} \vskip 0pt 
\global\advance\reglenum by 1 
\the\reglenum{}: \lapetiteregle {B} {R} {G} {B} {G} {G} {B} {B} {B} \vskip 0pt 
\global\advance\reglenum by 1 
\the\reglenum{}: \lapetiteregle {B} {R} {B} {B} {W} {B} {W} {B} {R} \vskip 0pt 
\global\advance\reglenum by 1 
\the\reglenum{}: \lapetiteregle {B} {R} {B} {W} {W} {B} {B} {B} {B} \vskip 0pt 
\global\advance\reglenum by 1 
\the\reglenum{}: \lapetiteregle {G} {W} {R} {B} {B} {G} {G} {G} {G} \vskip 0pt 
\global\advance\reglenum by 1 
\the\reglenum{}: \lapetiteregle {G} {W} {G} {B} {W} {B} {B} {R} {G} \vskip 0pt 
\global\advance\reglenum by 1 
\the\reglenum{}: \lapetiteregle {W} {G} {G} {G} {W} {G} {G} {W} {W} \vskip 0pt 
\global\advance\reglenum by 1 
\the\reglenum{}: \lapetiteregle {W} {W} {G} {B} {R} {B} {B} {G} {W} \vskip 0pt 
\global\advance\reglenum by 1 
\the\reglenum{}: \lapetiteregle {B} {W} {G} {B} {G} {G} {B} {R} {B} \vskip 0pt 
\global\advance\reglenum by 1 
\the\reglenum{}: \lapetiteregle {R} {W} {B} {B} {W} {B} {B} {B} {W} \vskip 0pt 
\global\advance\reglenum by 1 
\the\reglenum{}: \lapetiteregle {B} {W} {R} {B} {W} {B} {B} {B} {B} \vskip 0pt 
\global\advance\reglenum by 1 
\the\reglenum{}: \lapetiteregle {W} {W} {B} {B} {W} {B} {R} {B} {W} \vskip 0pt 
\global\advance\reglenum by 1 
\the\reglenum{}: \lapetiteregle {B} {W} {W} {R} {B} {B} {B} {B} {B} \vskip 0pt 
\global\advance\reglenum by 1 
\the\reglenum{}: \lapetiteregle {B} {B} {B} {W} {W} {B} {W} {B} {B} \vskip 0pt 
\global\advance\reglenum by 1 
}
\vskip 3pt\rmviii
same path,\vskip 0pt
other\vskip 2pt
conditions:
\vskip 1pt
{\ttv\obeylines
\the\reglenum{}: \lapetiteregle {G} {G} {W} {B} {B} {B} {G} {G} {G} \vskip 0pt 
\global\advance\reglenum by 1 
\the\reglenum{}: \lapetiteregle {B} {G} {B} {B} {G} {W} {W} {B} {B} \vskip 0pt 
\global\advance\reglenum by 1 
\the\reglenum{}: \lapetiteregle {G} {G} {G} {B} {W} {W} {W} {W} {G} \vskip 0pt 
\global\advance\reglenum by 1 
\the\reglenum{}: \lapetiteregle {B} {B} {G} {G} {W} {W} {W} {W} {B} \vskip 0pt 
\global\advance\reglenum by 1 
\the\reglenum{}: \lapetiteregle {G} {W} {W} {W} {G} {G} {G} {W} {G} \vskip 0pt 
\global\advance\reglenum by 1 
\the\reglenum{}: \lapetiteregle {W} {W} {G} {G} {G} {B} {G} {G} {B} \vskip 0pt 
\global\advance\reglenum by 1 
\the\reglenum{}: \lapetiteregle {G} {G} {B} {R} {B} {B} {G} {G} {G} \vskip 0pt 
\global\advance\reglenum by 1 
\the\reglenum{}: \lapetiteregle {B} {G} {R} {B} {G} {W} {W} {B} {B} \vskip 0pt 
\global\advance\reglenum by 1 
\the\reglenum{}: \lapetiteregle {G} {W} {W} {W} {G} {G} {G} {B} {G} \vskip 0pt 
\global\advance\reglenum by 1 
\the\reglenum{}: \lapetiteregle {B} {W} {G} {G} {G} {R} {G} {G} {R} \vskip 0pt 
\global\advance\reglenum by 1 
\the\reglenum{}: \lapetiteregle {G} {G} {R} {W} {B} {B} {G} {G} {G} \vskip 0pt 
\global\advance\reglenum by 1 
\the\reglenum{}: \lapetiteregle {B} {G} {W} {B} {G} {W} {W} {B} {B} \vskip 0pt 
\global\advance\reglenum by 1 
\the\reglenum{}: \lapetiteregle {G} {B} {W} {W} {G} {G} {G} {R} {G} \vskip 0pt 
\global\advance\reglenum by 1 
\the\reglenum{}: \lapetiteregle {R} {B} {G} {G} {G} {W} {G} {G} {W} \vskip 0pt 
\global\advance\reglenum by 1 
\the\reglenum{}: \lapetiteregle {G} {G} {W} {W} {B} {B} {G} {G} {G} \vskip 0pt 
\global\advance\reglenum by 1 
\the\reglenum{}: \lapetiteregle {G} {R} {W} {W} {G} {G} {G} {W} {G} \vskip 0pt 
\global\advance\reglenum by 1 
\the\reglenum{}: \lapetiteregle {W} {R} {G} {G} {G} {W} {G} {G} {W} \vskip 0pt 
\global\advance\reglenum by 1 
\the\reglenum{}: \lapetiteregle {W} {W} {G} {G} {G} {W} {G} {G} {W} \vskip 0pt 
\global\advance\reglenum by 1 
\the\reglenum{}: \lapetiteregle {G} {B} {W} {W} {G} {G} {G} {W} {G} \vskip 0pt 
\global\advance\reglenum by 1 
}
}

\ligne{\box120\hfill
\box122\hfill
\box124\hfill
\box126\hfill
\box128\hfill
}

\setbox120=\vtop{\leftskip 0pt\parindent 0pt
\baselineskip 7pt
\hsize=70pt
{\ttv\obeylines
\the\reglenum{}: \lapetiteregle {W} {B} {G} {G} {G} {W} {G} {G} {B} \vskip 0pt 
\global\advance\reglenum by 1 
\the\reglenum{}: \lapetiteregle {G} {G} {B} {W} {B} {B} {G} {G} {G} \vskip 0pt 
\global\advance\reglenum by 1 
\the\reglenum{}: \lapetiteregle {G} {R} {W} {W} {G} {G} {G} {B} {G} \vskip 0pt 
\global\advance\reglenum by 1 
\the\reglenum{}: \lapetiteregle {B} {R} {G} {G} {G} {W} {G} {G} {R} \vskip 0pt 
\global\advance\reglenum by 1 
\the\reglenum{}: \lapetiteregle {G} {G} {R} {B} {B} {B} {G} {G} {G} \vskip 0pt 
\global\advance\reglenum by 1 
\the\reglenum{}: \lapetiteregle {G} {W} {W} {W} {G} {G} {G} {R} {G} \vskip 0pt 
\global\advance\reglenum by 1 
\the\reglenum{}: \lapetiteregle {R} {W} {G} {G} {G} {B} {G} {G} {W} \vskip 0pt 
\global\advance\reglenum by 1 
\the\reglenum{}: \lapetiteregle {W} {W} {G} {G} {B} {W} {G} {G} {B} \vskip 0pt 
\global\advance\reglenum by 1 
\the\reglenum{}: \lapetiteregle {G} {G} {B} {G} {B} {B} {G} {G} {G} \vskip 0pt 
\global\advance\reglenum by 1 
\the\reglenum{}: \lapetiteregle {B} {G} {G} {B} {G} {W} {W} {B} {B} \vskip 0pt 
\global\advance\reglenum by 1 
\the\reglenum{}: \lapetiteregle {G} {G} {R} {G} {B} {B} {G} {G} {G} \vskip 0pt 
\global\advance\reglenum by 1 
\the\reglenum{}: \lapetiteregle {G} {G} {W} {G} {B} {B} {G} {G} {G} \vskip 0pt 
\global\advance\reglenum by 1 
}
\vskip 5pt\rmviii
path~7:
\vskip 1pt
{\ttv\obeylines
\the\reglenum{}: \lapetiteregle {W} {G} {B} {B} {W} {B} {B} {B} {B} \vskip 0pt 
\global\advance\reglenum by 1 
\the\reglenum{}: \lapetiteregle {B} {G} {W} {B} {R} {B} {G} {B} {B} \vskip 0pt 
\global\advance\reglenum by 1 
\the\reglenum{}: \lapetiteregle {B} {B} {R} {W} {W} {B} {W} {G} {B} \vskip 0pt 
\global\advance\reglenum by 1 
\the\reglenum{}: \lapetiteregle {G} {W} {G} {B} {B} {B} {B} {W} {B} \vskip 0pt 
\global\advance\reglenum by 1 
\the\reglenum{}: \lapetiteregle {B} {G} {G} {B} {B} {B} {B} {B} {B} \vskip 0pt 
\global\advance\reglenum by 1 
\the\reglenum{}: \lapetiteregle {B} {G} {B} {B} {W} {B} {R} {B} {R} \vskip 0pt 
\global\advance\reglenum by 1 
\the\reglenum{}: \lapetiteregle {B} {G} {B} {R} {W} {B} {G} {B} {B} \vskip 0pt 
\global\advance\reglenum by 1 
\the\reglenum{}: \lapetiteregle {W} {W} {G} {G} {W} {G} {G} {B} {B} \vskip 0pt 
\global\advance\reglenum by 1 
\the\reglenum{}: \lapetiteregle {W} {W} {B} {B} {W} {B} {B} {G} {G} \vskip 0pt 
\global\advance\reglenum by 1 
\the\reglenum{}: \lapetiteregle {B} {W} {B} {B} {G} {G} {B} {W} {B} \vskip 0pt 
\global\advance\reglenum by 1 
\the\reglenum{}: \lapetiteregle {B} {W} {G} {B} {R} {B} {B} {W} {R} \vskip 0pt 
\global\advance\reglenum by 1 
\the\reglenum{}: \lapetiteregle {B} {B} {G} {B} {B} {B} {B} {R} {B} \vskip 0pt 
\global\advance\reglenum by 1 
\the\reglenum{}: \lapetiteregle {B} {B} {R} {W} {W} {B} {G} {B} {B} \vskip 0pt 
\global\advance\reglenum by 1 
\the\reglenum{}: \lapetiteregle {B} {R} {B} {B} {W} {W} {W} {W} {B} \vskip 0pt 
\global\advance\reglenum by 1 
\the\reglenum{}: \lapetiteregle {B} {G} {G} {G} {W} {G} {G} {R} {R} \vskip 0pt 
\global\advance\reglenum by 1 
\the\reglenum{}: \lapetiteregle {G} {B} {R} {B} {W} {B} {B} {G} {G} \vskip 0pt 
\global\advance\reglenum by 1 
\the\reglenum{}: \lapetiteregle {B} {G} {R} {B} {G} {G} {B} {W} {B} \vskip 0pt 
\global\advance\reglenum by 1 
\the\reglenum{}: \lapetiteregle {B} {G} {W} {W} {W} {B} {B} {B} {B} \vskip 0pt 
\global\advance\reglenum by 1 
\the\reglenum{}: \lapetiteregle {G} {B} {G} {B} {B} {G} {G} {G} {G} \vskip 0pt 
\global\advance\reglenum by 1 
\the\reglenum{}: \lapetiteregle {B} {G} {G} {B} {B} {W} {W} {B} {B} \vskip 0pt 
\global\advance\reglenum by 1 
\the\reglenum{}: \lapetiteregle {B} {R} {G} {B} {B} {B} {B} {W} {B} \vskip 0pt 
\global\advance\reglenum by 1 
\the\reglenum{}: \lapetiteregle {B} {R} {W} {W} {W} {B} {G} {B} {B} \vskip 0pt 
\global\advance\reglenum by 1 
\the\reglenum{}: \lapetiteregle {R} {G} {G} {G} {B} {G} {G} {W} {W} \vskip 0pt 
\global\advance\reglenum by 1 
\the\reglenum{}: \lapetiteregle {G} {R} {W} {B} {W} {B} {B} {G} {G} \vskip 0pt 
\global\advance\reglenum by 1 
\the\reglenum{}: \lapetiteregle {B} {G} {W} {B} {G} {G} {B} {W} {B} \vskip 0pt 
\global\advance\reglenum by 1 
\the\reglenum{}: \lapetiteregle {G} {R} {G} {B} {B} {G} {G} {G} {G} \vskip 0pt 
\global\advance\reglenum by 1 
\the\reglenum{}: \lapetiteregle {B} {W} {G} {B} {B} {B} {B} {W} {B} \vskip 0pt 
\global\advance\reglenum by 1 
\the\reglenum{}: \lapetiteregle {B} {W} {W} {W} {W} {B} {G} {B} {B} \vskip 0pt 
\global\advance\reglenum by 1 
\the\reglenum{}: \lapetiteregle {W} {G} {G} {G} {R} {G} {G} {W} {W} \vskip 0pt 
\global\advance\reglenum by 1 
\the\reglenum{}: \lapetiteregle {G} {W} {W} {B} {W} {B} {B} {G} {G} \vskip 0pt 
\global\advance\reglenum by 1 
\the\reglenum{}: \lapetiteregle {G} {W} {G} {B} {B} {G} {G} {G} {G} \vskip 0pt 
\global\advance\reglenum by 1 
}
\vskip 3pt\rmviii
same path,\vskip 0pt
other\vskip 2pt
conditions:
\vskip 1pt
{\ttv\obeylines
\the\reglenum{}: \lapetiteregle {G} {W} {W} {G} {B} {B} {B} {B} {B} \vskip 0pt 
\global\advance\reglenum by 1 
\the\reglenum{}: \lapetiteregle {B} {B} {G} {B} {R} {W} {B} {G} {B} \vskip 0pt 
\global\advance\reglenum by 1 
\the\reglenum{}: \lapetiteregle {W} {B} {B} {G} {W} {B} {B} {W} {G} \vskip 0pt 
\global\advance\reglenum by 1 
\the\reglenum{}: \lapetiteregle {B} {W} {W} {G} {B} {R} {B} {B} {R} \vskip 0pt 
\global\advance\reglenum by 1 
\the\reglenum{}: \lapetiteregle {G} {B} {W} {G} {G} {G} {B} {B} {G} \vskip 0pt 
\global\advance\reglenum by 1 
\the\reglenum{}: \lapetiteregle {B} {B} {B} {R} {W} {W} {B} {G} {B} \vskip 0pt 
\global\advance\reglenum by 1 
\the\reglenum{}: \lapetiteregle {G} {B} {B} {G} {B} {R} {B} {W} {G} \vskip 0pt 
\global\advance\reglenum by 1 
\the\reglenum{}: \lapetiteregle {B} {G} {B} {G} {W} {W} {B} {G} {B} \vskip 0pt 
\global\advance\reglenum by 1 
\the\reglenum{}: \lapetiteregle {G} {G} {B} {B} {G} {G} {G} {B} {G} \vskip 0pt 
\global\advance\reglenum by 1 
\the\reglenum{}: \lapetiteregle {R} {G} {B} {G} {B} {W} {B} {B} {W} \vskip 0pt 
\global\advance\reglenum by 1 
\the\reglenum{}: \lapetiteregle {G} {R} {B} {G} {G} {G} {B} {B} {G} \vskip 0pt 
\global\advance\reglenum by 1 
\the\reglenum{}: \lapetiteregle {B} {B} {R} {W} {W} {W} {B} {G} {B} \vskip 0pt 
\global\advance\reglenum by 1 
\the\reglenum{}: \lapetiteregle {G} {B} {B} {G} {R} {W} {B} {W} {G} \vskip 0pt 
\global\advance\reglenum by 1 
\the\reglenum{}: \lapetiteregle {G} {G} {B} {B} {G} {G} {G} {R} {G} \vskip 0pt 
\global\advance\reglenum by 1 
\the\reglenum{}: \lapetiteregle {W} {G} {R} {G} {B} {W} {B} {B} {W} \vskip 0pt 
\global\advance\reglenum by 1 
\the\reglenum{}: \lapetiteregle {G} {W} {R} {G} {G} {G} {B} {B} {G} \vskip 0pt 
\global\advance\reglenum by 1 
\the\reglenum{}: \lapetiteregle {B} {B} {W} {W} {W} {W} {B} {G} {B} \vskip 0pt 
\global\advance\reglenum by 1 
\the\reglenum{}: \lapetiteregle {G} {B} {B} {G} {W} {W} {B} {W} {G} \vskip 0pt 
\global\advance\reglenum by 1 
\the\reglenum{}: \lapetiteregle {G} {G} {B} {B} {G} {G} {G} {W} {G} \vskip 0pt 
\global\advance\reglenum by 1 
\the\reglenum{}: \lapetiteregle {W} {G} {W} {G} {B} {W} {B} {B} {W} \vskip 0pt 
\global\advance\reglenum by 1 
\the\reglenum{}: \lapetiteregle {G} {W} {W} {G} {G} {G} {B} {B} {G} \vskip 0pt 
\global\advance\reglenum by 1 
}
\vskip 5pt\rm
right-hand\vskip 0pt
side memory\vskip 0pt
switch
\vskip 5pt\rmviii
path~7:
\vskip 1pt
{\ttv\obeylines
\the\reglenum{}: \lapetiteregle {B} {W} {W} {B} {R} {B} {G} {B} {B} \vskip 0pt 
\global\advance\reglenum by 1 
\the\reglenum{}: \lapetiteregle {B} {W} {G} {B} {B} {B} {B} {B} {B} \vskip 0pt 
\global\advance\reglenum by 1 
\the\reglenum{}: \lapetiteregle {B} {W} {B} {R} {W} {B} {G} {B} {B} \vskip 0pt 
\global\advance\reglenum by 1 
\the\reglenum{}: \lapetiteregle {W} {G} {G} {G} {W} {G} {G} {B} {B} \vskip 0pt 
\global\advance\reglenum by 1 
\the\reglenum{}: \lapetiteregle {G} {W} {B} {B} {W} {B} {B} {G} {G} \vskip 0pt 
\global\advance\reglenum by 1 
\the\reglenum{}: \lapetiteregle {B} {G} {B} {B} {G} {G} {B} {W} {B} \vskip 0pt 
\global\advance\reglenum by 1 
}
}

\setbox122=\vtop{\leftskip 0pt\parindent 0pt
\baselineskip 7pt
\hsize=70pt
\rmviii
same path,\vskip 0pt
other\vskip 2pt
conditions:
\vskip 1pt
{\ttv\obeylines
\the\reglenum{}: \lapetiteregle {B} {B} {W} {W} {B} {R} {B} {G} {B} \vskip 0pt 
\global\advance\reglenum by 1 
\the\reglenum{}: \lapetiteregle {W} {G} {W} {G} {B} {B} {B} {B} {B} \vskip 0pt 
\global\advance\reglenum by 1 
\the\reglenum{}: \lapetiteregle {B} {B} {W} {B} {R} {W} {B} {G} {B} \vskip 0pt 
\global\advance\reglenum by 1 
\the\reglenum{}: \lapetiteregle {G} {B} {B} {G} {W} {B} {B} {W} {G} \vskip 0pt 
\global\advance\reglenum by 1 
\the\reglenum{}: \lapetiteregle {B} {G} {W} {G} {B} {R} {B} {B} {R} \vskip 0pt 
\global\advance\reglenum by 1 
\the\reglenum{}: \lapetiteregle {G} {B} {B} {G} {B} {W} {B} {W} {G} \vskip 0pt 
\global\advance\reglenum by 1 
\the\reglenum{}: \lapetiteregle {W} {G} {B} {G} {B} {W} {B} {B} {B} \vskip 0pt 
\global\advance\reglenum by 1 
\the\reglenum{}: \lapetiteregle {G} {W} {B} {G} {G} {G} {B} {B} {G} \vskip 0pt 
\global\advance\reglenum by 1 
\the\reglenum{}: \lapetiteregle {G} {B} {B} {G} {R} {B} {B} {W} {G} \vskip 0pt 
\global\advance\reglenum by 1 
\the\reglenum{}: \lapetiteregle {B} {G} {R} {G} {B} {W} {B} {B} {R} \vskip 0pt 
\global\advance\reglenum by 1 
\the\reglenum{}: \lapetiteregle {G} {B} {R} {G} {G} {G} {B} {B} {G} \vskip 0pt 
\global\advance\reglenum by 1 
\the\reglenum{}: \lapetiteregle {B} {B} {B} {W} {W} {W} {B} {G} {B} \vskip 0pt 
\global\advance\reglenum by 1 
\the\reglenum{}: \lapetiteregle {G} {B} {B} {G} {W} {R} {B} {W} {G} \vskip 0pt 
\global\advance\reglenum by 1 
\the\reglenum{}: \lapetiteregle {R} {G} {W} {G} {B} {B} {B} {B} {W} \vskip 0pt 
\global\advance\reglenum by 1 
\the\reglenum{}: \lapetiteregle {G} {R} {W} {G} {G} {G} {B} {B} {G} \vskip 0pt 
\global\advance\reglenum by 1 
\the\reglenum{}: \lapetiteregle {B} {B} {R} {B} {W} {W} {B} {G} {B} \vskip 0pt 
\global\advance\reglenum by 1 
\the\reglenum{}: \lapetiteregle {W} {G} {W} {G} {B} {R} {B} {B} {W} \vskip 0pt 
\global\advance\reglenum by 1 
\the\reglenum{}: \lapetiteregle {B} {B} {W} {R} {B} {W} {B} {G} {B} \vskip 0pt 
\global\advance\reglenum by 1 
\the\reglenum{}: \lapetiteregle {B} {B} {W} {W} {R} {B} {B} {G} {B} \vskip 0pt 
\global\advance\reglenum by 1 
\the\reglenum{}: \lapetiteregle {B} {G} {B} {B} {W} {W} {B} {G} {B} \vskip 0pt 
\global\advance\reglenum by 1 
}
\vskip 5pt\rmviii
path~4:
\vskip 1pt
{\ttv\obeylines
\the\reglenum{}: \lapetiteregle {W} {G} {G} {G} {B} {G} {G} {W} {B} \vskip 0pt 
\global\advance\reglenum by 1 
\the\reglenum{}: \lapetiteregle {B} {G} {G} {G} {R} {G} {G} {W} {R} \vskip 0pt 
\global\advance\reglenum by 1 
\the\reglenum{}: \lapetiteregle {G} {B} {W} {B} {W} {B} {B} {G} {G} \vskip 0pt 
\global\advance\reglenum by 1 
\the\reglenum{}: \lapetiteregle {R} {G} {G} {G} {W} {G} {G} {B} {W} \vskip 0pt 
\global\advance\reglenum by 1 
\the\reglenum{}: \lapetiteregle {G} {R} {B} {B} {W} {B} {B} {G} {G} \vskip 0pt 
\global\advance\reglenum by 1 
\the\reglenum{}: \lapetiteregle {B} {B} {G} {B} {B} {B} {B} {W} {B} \vskip 0pt 
\global\advance\reglenum by 1 
\the\reglenum{}: \lapetiteregle {B} {B} {W} {W} {W} {B} {G} {B} {B} \vskip 0pt 
\global\advance\reglenum by 1 
\the\reglenum{}: \lapetiteregle {W} {G} {G} {G} {W} {G} {G} {R} {W} \vskip 0pt 
\global\advance\reglenum by 1 
\the\reglenum{}: \lapetiteregle {G} {W} {R} {B} {W} {B} {B} {G} {G} \vskip 0pt 
\global\advance\reglenum by 1 
\the\reglenum{}: \lapetiteregle {B} {R} {G} {B} {B} {B} {B} {B} {B} \vskip 0pt 
\global\advance\reglenum by 1 
\the\reglenum{}: \lapetiteregle {B} {R} {B} {W} {W} {B} {G} {B} {B} \vskip 0pt 
\global\advance\reglenum by 1 
\the\reglenum{}: \lapetiteregle {B} {W} {G} {B} {B} {B} {B} {R} {B} \vskip 0pt 
\global\advance\reglenum by 1 
\the\reglenum{}: \lapetiteregle {B} {W} {R} {B} {W} {B} {G} {B} {B} \vskip 0pt 
\global\advance\reglenum by 1 
\the\reglenum{}: \lapetiteregle {B} {W} {W} {R} {B} {B} {G} {B} {B} \vskip 0pt 
\global\advance\reglenum by 1 
\the\reglenum{}: \lapetiteregle {B} {B} {B} {W} {W} {B} {W} {G} {B} \vskip 0pt 
\global\advance\reglenum by 1 
\the\reglenum{}: \lapetiteregle {W} {W} {G} {G} {W} {B} {G} {G} {B} \vskip 0pt 
\global\advance\reglenum by 1 
}
\vskip 5pt\rmviii
path~1:
\vskip 1pt
{\ttv\obeylines
\the\reglenum{}: \lapetiteregle {B} {G} {W} {B} {R} {B} {B} {B} {B} \vskip 0pt 
\global\advance\reglenum by 1 
\the\reglenum{}: \lapetiteregle {G} {W} {W} {B} {B} {B} {B} {G} {B} \vskip 0pt 
\global\advance\reglenum by 1 
\the\reglenum{}: \lapetiteregle {B} {G} {W} {B} {G} {G} {B} {B} {B} \vskip 0pt 
\global\advance\reglenum by 1 
\the\reglenum{}: \lapetiteregle {B} {G} {B} {R} {W} {B} {B} {B} {B} \vskip 0pt 
\global\advance\reglenum by 1 
\the\reglenum{}: \lapetiteregle {W} {B} {G} {G} {W} {G} {G} {W} {B} \vskip 0pt 
\global\advance\reglenum by 1 
\the\reglenum{}: \lapetiteregle {B} {W} {W} {B} {R} {B} {B} {G} {R} \vskip 0pt 
\global\advance\reglenum by 1 
\the\reglenum{}: \lapetiteregle {B} {B} {W} {B} {G} {G} {B} {R} {B} \vskip 0pt 
\global\advance\reglenum by 1 
\the\reglenum{}: \lapetiteregle {W} {W} {G} {B} {W} {B} {B} {B} {G} \vskip 0pt 
\global\advance\reglenum by 1 
}
\vskip 3pt\rmviii
same path,\vskip 0pt
other\vskip 2pt
conditions:
\vskip 1pt
{\ttv\obeylines
\the\reglenum{}: \lapetiteregle {B} {G} {W} {W} {B} {B} {G} {B} {B} \vskip 0pt 
\global\advance\reglenum by 1 
\the\reglenum{}: \lapetiteregle {G} {B} {B} {G} {W} {W} {B} {B} {B} \vskip 0pt 
\global\advance\reglenum by 1 
\the\reglenum{}: \lapetiteregle {B} {B} {B} {G} {W} {W} {B} {R} {R} \vskip 0pt 
\global\advance\reglenum by 1 
\the\reglenum{}: \lapetiteregle {W} {B} {W} {G} {B} {W} {B} {B} {G} \vskip 0pt 
\global\advance\reglenum by 1 
}
\vskip 5pt\rm
left-hand\vskip 2pt
side flip flop\vskip 0pt
switch
\vskip 5pt\rmviii
path~4:
\vskip 1pt
{\ttv\obeylines
\the\reglenum{}: \lapetiteregle {W} {W} {G} {R} {W} {B} {B} {G} {W} \vskip 0pt 
\global\advance\reglenum by 1 
\the\reglenum{}: \lapetiteregle {R} {W} {G} {B} {B} {R} {R} {W} {R} \vskip 0pt 
\global\advance\reglenum by 1 
\the\reglenum{}: \lapetiteregle {W} {W} {R} {R} {R} {B} {W} {B} {W} \vskip 0pt 
\global\advance\reglenum by 1 
\the\reglenum{}: \lapetiteregle {B} {R} {B} {B} {W} {W} {W} {R} {B} \vskip 0pt 
\global\advance\reglenum by 1 
\the\reglenum{}: \lapetiteregle {R} {R} {B} {W} {W} {W} {W} {R} {R} \vskip 0pt 
\global\advance\reglenum by 1 
\the\reglenum{}: \lapetiteregle {R} {W} {R} {R} {W} {W} {W} {R} {R} \vskip 0pt 
\global\advance\reglenum by 1 
\the\reglenum{}: \lapetiteregle {R} {W} {R} {W} {W} {W} {R} {B} {R} \vskip 0pt 
\global\advance\reglenum by 1 
\the\reglenum{}: \lapetiteregle {B} {W} {R} {R} {W} {W} {W} {W} {B} \vskip 0pt 
\global\advance\reglenum by 1 
\the\reglenum{}: \lapetiteregle {W} {R} {B} {B} {W} {W} {W} {W} {W} \vskip 0pt 
\global\advance\reglenum by 1 
\the\reglenum{}: \lapetiteregle {W} {R} {W} {W} {W} {W} {W} {W} {W} \vskip 0pt 
\global\advance\reglenum by 1 
\the\reglenum{}: \lapetiteregle {W} {R} {R} {W} {W} {W} {W} {W} {W} \vskip 0pt 
\global\advance\reglenum by 1 
\the\reglenum{}: \lapetiteregle {W} {B} {R} {W} {W} {W} {W} {W} {W} \vskip 0pt 
\global\advance\reglenum by 1 
\the\reglenum{}: \lapetiteregle {W} {R} {W} {W} {W} {W} {W} {R} {W} \vskip 0pt 
\global\advance\reglenum by 1 
\the\reglenum{}: \lapetiteregle {R} {B} {R} {W} {W} {W} {W} {W} {R} \vskip 0pt 
\global\advance\reglenum by 1 
\the\reglenum{}: \lapetiteregle {G} {W} {G} {B} {W} {B} {R} {W} {G} \vskip 0pt 
\global\advance\reglenum by 1 
\the\reglenum{}: \lapetiteregle {B} {G} {W} {W} {W} {B} {B} {R} {B} \vskip 0pt 
\global\advance\reglenum by 1 
\the\reglenum{}: \lapetiteregle {B} {B} {W} {W} {W} {B} {W} {R} {B} \vskip 0pt 
\global\advance\reglenum by 1 
\the\reglenum{}: \lapetiteregle {W} {B} {G} {R} {W} {B} {B} {G} {B} \vskip 0pt 
\global\advance\reglenum by 1 
\the\reglenum{}: \lapetiteregle {G} {B} {G} {B} {W} {B} {R} {W} {G} \vskip 0pt 
\global\advance\reglenum by 1 
}
}

\setbox124=\vtop{\leftskip 0pt\parindent 0pt
\baselineskip 7pt
\hsize=70pt
{\ttv\obeylines
\the\reglenum{}: \lapetiteregle {B} {R} {G} {R} {W} {B} {B} {G} {R} \vskip 0pt 
\global\advance\reglenum by 1 
\the\reglenum{}: \lapetiteregle {R} {B} {G} {B} {B} {R} {R} {W} {R} \vskip 0pt 
\global\advance\reglenum by 1 
\the\reglenum{}: \lapetiteregle {W} {B} {R} {R} {R} {B} {W} {B} {B} \vskip 0pt 
\global\advance\reglenum by 1 
\the\reglenum{}: \lapetiteregle {G} {R} {G} {B} {W} {B} {R} {B} {G} \vskip 0pt 
\global\advance\reglenum by 1 
\the\reglenum{}: \lapetiteregle {R} {W} {G} {R} {B} {B} {B} {G} {R} \vskip 0pt 
\global\advance\reglenum by 1 
\the\reglenum{}: \lapetiteregle {R} {R} {G} {B} {B} {R} {R} {B} {R} \vskip 0pt 
\global\advance\reglenum by 1 
\the\reglenum{}: \lapetiteregle {B} {R} {R} {R} {R} {B} {W} {B} {R} \vskip 0pt 
\global\advance\reglenum by 1 
\the\reglenum{}: \lapetiteregle {R} {B} {R} {R} {W} {W} {W} {R} {R} \vskip 0pt 
\global\advance\reglenum by 1 
\the\reglenum{}: \lapetiteregle {R} {B} {R} {W} {W} {W} {R} {B} {R} \vskip 0pt 
\global\advance\reglenum by 1 
\the\reglenum{}: \lapetiteregle {B} {B} {R} {R} {W} {W} {W} {W} {B} \vskip 0pt 
\global\advance\reglenum by 1 
\the\reglenum{}: \lapetiteregle {G} {W} {G} {B} {W} {B} {R} {R} {W} \vskip 0pt 
\global\advance\reglenum by 1 
\the\reglenum{}: \lapetiteregle {W} {R} {G} {G} {W} {G} {G} {W} {W} \vskip 0pt 
\global\advance\reglenum by 1 
\the\reglenum{}: \lapetiteregle {R} {W} {W} {R} {R} {B} {B} {G} {G} \vskip 0pt 
\global\advance\reglenum by 1 
\the\reglenum{}: \lapetiteregle {R} {R} {W} {B} {B} {R} {R} {R} {R} \vskip 0pt 
\global\advance\reglenum by 1 
\the\reglenum{}: \lapetiteregle {R} {R} {R} {R} {R} {B} {B} {B} {W} \vskip 0pt 
\global\advance\reglenum by 1 
\the\reglenum{}: \lapetiteregle {B} {R} {R} {B} {W} {B} {B} {B} {B} \vskip 0pt 
\global\advance\reglenum by 1 
\the\reglenum{}: \lapetiteregle {R} {R} {R} {R} {W} {W} {W} {R} {R} \vskip 0pt 
\global\advance\reglenum by 1 
\the\reglenum{}: \lapetiteregle {R} {R} {R} {W} {W} {W} {R} {B} {R} \vskip 0pt 
\global\advance\reglenum by 1 
\the\reglenum{}: \lapetiteregle {B} {R} {R} {R} {W} {W} {W} {B} {B} \vskip 0pt 
\global\advance\reglenum by 1 
\the\reglenum{}: \lapetiteregle {W} {W} {G} {B} {W} {B} {R} {R} {W} \vskip 0pt 
\global\advance\reglenum by 1 
\the\reglenum{}: \lapetiteregle {B} {W} {W} {W} {W} {B} {B} {R} {B} \vskip 0pt 
\global\advance\reglenum by 1 
\the\reglenum{}: \lapetiteregle {G} {W} {W} {R} {W} {B} {B} {G} {G} \vskip 0pt 
\global\advance\reglenum by 1 
\the\reglenum{}: \lapetiteregle {R} {G} {W} {B} {B} {R} {R} {W} {R} \vskip 0pt 
\global\advance\reglenum by 1 
\the\reglenum{}: \lapetiteregle {W} {G} {R} {R} {R} {B} {R} {B} {W} \vskip 0pt 
\global\advance\reglenum by 1 
\the\reglenum{}: \lapetiteregle {B} {G} {W} {R} {B} {B} {B} {B} {B} \vskip 0pt 
\global\advance\reglenum by 1 
\the\reglenum{}: \lapetiteregle {B} {W} {R} {R} {W} {W} {W} {R} {B} \vskip 0pt 
\global\advance\reglenum by 1 
\the\reglenum{}: \lapetiteregle {W} {W} {G} {B} {W} {B} {R} {G} {W} \vskip 0pt 
\global\advance\reglenum by 1 
}
\vskip 3pt\rmviii
same path,\vskip 0pt
other\vskip 2pt
conditions:
\vskip 1pt
{\ttv\obeylines
\the\reglenum{}: \lapetiteregle {W} {B} {B} {G} {W} {G} {R} {W} {W} \vskip 0pt 
\global\advance\reglenum by 1 
\the\reglenum{}: \lapetiteregle {G} {W} {W} {G} {B} {W} {B} {R} {G} \vskip 0pt 
\global\advance\reglenum by 1 
\the\reglenum{}: \lapetiteregle {R} {W} {G} {B} {G} {R} {R} {W} {R} \vskip 0pt 
\global\advance\reglenum by 1 
\the\reglenum{}: \lapetiteregle {B} {R} {G} {W} {W} {W} {B} {G} {B} \vskip 0pt 
\global\advance\reglenum by 1 
\the\reglenum{}: \lapetiteregle {G} {R} {B} {B} {G} {W} {W} {R} {G} \vskip 0pt 
\global\advance\reglenum by 1 
\the\reglenum{}: \lapetiteregle {R} {R} {G} {W} {W} {W} {W} {R} {R} \vskip 0pt 
\global\advance\reglenum by 1 
\the\reglenum{}: \lapetiteregle {W} {R} {G} {W} {W} {W} {W} {W} {W} \vskip 0pt 
\global\advance\reglenum by 1 
\the\reglenum{}: \lapetiteregle {W} {W} {R} {W} {W} {W} {W} {W} {W} \vskip 0pt 
\global\advance\reglenum by 1 
\the\reglenum{}: \lapetiteregle {W} {B} {B} {G} {B} {G} {R} {W} {B} \vskip 0pt 
\global\advance\reglenum by 1 
\the\reglenum{}: \lapetiteregle {G} {W} {B} {G} {B} {W} {B} {R} {G} \vskip 0pt 
\global\advance\reglenum by 1 
\the\reglenum{}: \lapetiteregle {B} {B} {B} {G} {R} {G} {R} {W} {R} \vskip 0pt 
\global\advance\reglenum by 1 
\the\reglenum{}: \lapetiteregle {G} {B} {R} {G} {B} {W} {B} {R} {G} \vskip 0pt 
\global\advance\reglenum by 1 
\the\reglenum{}: \lapetiteregle {R} {B} {G} {B} {G} {R} {R} {W} {R} \vskip 0pt 
\global\advance\reglenum by 1 
\the\reglenum{}: \lapetiteregle {R} {B} {B} {G} {W} {G} {R} {B} {R} \vskip 0pt 
\global\advance\reglenum by 1 
\the\reglenum{}: \lapetiteregle {G} {R} {W} {G} {B} {W} {B} {R} {W} \vskip 0pt 
\global\advance\reglenum by 1 
\the\reglenum{}: \lapetiteregle {R} {R} {G} {B} {G} {R} {R} {B} {R} \vskip 0pt 
\global\advance\reglenum by 1 
\the\reglenum{}: \lapetiteregle {R} {B} {B} {G} {W} {W} {R} {R} {G} \vskip 0pt 
\global\advance\reglenum by 1 
\the\reglenum{}: \lapetiteregle {B} {R} {R} {B} {W} {B} {G} {B} {B} \vskip 0pt 
\global\advance\reglenum by 1 
\the\reglenum{}: \lapetiteregle {W} {R} {W} {G} {B} {W} {B} {R} {W} \vskip 0pt 
\global\advance\reglenum by 1 
\the\reglenum{}: \lapetiteregle {R} {R} {W} {B} {G} {R} {R} {R} {R} \vskip 0pt 
\global\advance\reglenum by 1 
\the\reglenum{}: \lapetiteregle {B} {G} {W} {R} {B} {B} {G} {B} {B} \vskip 0pt 
\global\advance\reglenum by 1 
\the\reglenum{}: \lapetiteregle {B} {G} {W} {W} {R} {B} {G} {B} {B} \vskip 0pt 
\global\advance\reglenum by 1 
}
\vskip 5pt\rm
right-hand\vskip 0pt
side flip flop\vskip 0pt
switch
\vskip 5pt\rmviii
path~4:
\vskip 1pt
{\ttv\obeylines
\the\reglenum{}: \lapetiteregle {W} {G} {R} {R} {R} {B} {W} {B} {W} \vskip 0pt 
\global\advance\reglenum by 1 
\the\reglenum{}: \lapetiteregle {G} {B} {W} {R} {W} {B} {B} {G} {G} \vskip 0pt 
\global\advance\reglenum by 1 
\the\reglenum{}: \lapetiteregle {W} {B} {G} {B} {W} {B} {R} {G} {B} \vskip 0pt 
\global\advance\reglenum by 1 
\the\reglenum{}: \lapetiteregle {G} {R} {B} {R} {W} {B} {B} {G} {G} \vskip 0pt 
\global\advance\reglenum by 1 
\the\reglenum{}: \lapetiteregle {R} {G} {B} {B} {B} {R} {R} {W} {R} \vskip 0pt 
\global\advance\reglenum by 1 
\the\reglenum{}: \lapetiteregle {B} {R} {G} {B} {W} {B} {R} {G} {R} \vskip 0pt 
\global\advance\reglenum by 1 
\the\reglenum{}: \lapetiteregle {B} {B} {W} {W} {W} {B} {B} {R} {B} \vskip 0pt 
\global\advance\reglenum by 1 
\the\reglenum{}: \lapetiteregle {G} {W} {R} {R} {W} {B} {B} {G} {W} \vskip 0pt 
\global\advance\reglenum by 1 
\the\reglenum{}: \lapetiteregle {R} {G} {R} {B} {B} {R} {R} {W} {R} \vskip 0pt 
\global\advance\reglenum by 1 
\the\reglenum{}: \lapetiteregle {R} {W} {G} {B} {B} {B} {R} {G} {R} \vskip 0pt 
\global\advance\reglenum by 1 
\the\reglenum{}: \lapetiteregle {B} {R} {B} {W} {W} {B} {B} {R} {B} \vskip 0pt 
\global\advance\reglenum by 1 
\the\reglenum{}: \lapetiteregle {W} {W} {G} {G} {W} {G} {G} {R} {W} \vskip 0pt 
\global\advance\reglenum by 1 
\the\reglenum{}: \lapetiteregle {W} {W} {R} {R} {W} {B} {B} {G} {W} \vskip 0pt 
\global\advance\reglenum by 1 
\the\reglenum{}: \lapetiteregle {R} {W} {R} {B} {B} {R} {R} {W} {R} \vskip 0pt 
\global\advance\reglenum by 1 
\the\reglenum{}: \lapetiteregle {R} {W} {G} {B} {R} {B} {R} {W} {G} \vskip 0pt 
\global\advance\reglenum by 1 
\the\reglenum{}: \lapetiteregle {B} {R} {G} {B} {B} {B} {B} {R} {B} \vskip 0pt 
\global\advance\reglenum by 1 
\the\reglenum{}: \lapetiteregle {R} {R} {B} {B} {W} {B} {B} {B} {W} \vskip 0pt 
\global\advance\reglenum by 1 
\the\reglenum{}: \lapetiteregle {B} {R} {R} {B} {W} {B} {B} {R} {B} \vskip 0pt 
\global\advance\reglenum by 1 
\the\reglenum{}: \lapetiteregle {W} {G} {B} {B} {W} {B} {R} {B} {W} \vskip 0pt 
\global\advance\reglenum by 1 
\the\reglenum{}: \lapetiteregle {B} {G} {W} {R} {B} {B} {B} {R} {B} \vskip 0pt 
\global\advance\reglenum by 1 
}
\vskip 3pt\rmviii
same path,\vskip 0pt
other\vskip 2pt
conditions:
\vskip 1pt
{\ttv\obeylines
\the\reglenum{}: \lapetiteregle {G} {B} {B} {G} {W} {W} {R} {W} {G} \vskip 0pt 
\global\advance\reglenum by 1 
\the\reglenum{}: \lapetiteregle {W} {G} {W} {G} {B} {W} {B} {R} {W} \vskip 0pt 
\global\advance\reglenum by 1 
}
}

\setbox126=\vtop{\leftskip 0pt\parindent 0pt
\baselineskip 7pt
\hsize=70pt
{\ttv\obeylines
\the\reglenum{}: \lapetiteregle {R} {G} {W} {B} {G} {R} {R} {W} {R} \vskip 0pt 
\global\advance\reglenum by 1 
\the\reglenum{}: \lapetiteregle {B} {R} {W} {W} {W} {W} {B} {G} {B} \vskip 0pt 
\global\advance\reglenum by 1 
\the\reglenum{}: \lapetiteregle {G} {B} {B} {G} {B} {W} {R} {W} {G} \vskip 0pt 
\global\advance\reglenum by 1 
\the\reglenum{}: \lapetiteregle {W} {G} {B} {G} {B} {W} {B} {R} {B} \vskip 0pt 
\global\advance\reglenum by 1 
\the\reglenum{}: \lapetiteregle {G} {B} {B} {G} {R} {B} {R} {W} {G} \vskip 0pt 
\global\advance\reglenum by 1 
\the\reglenum{}: \lapetiteregle {B} {G} {R} {G} {B} {W} {B} {R} {R} \vskip 0pt 
\global\advance\reglenum by 1 
\the\reglenum{}: \lapetiteregle {R} {G} {B} {B} {G} {R} {R} {W} {R} \vskip 0pt 
\global\advance\reglenum by 1 
\the\reglenum{}: \lapetiteregle {B} {R} {B} {W} {W} {W} {B} {G} {B} \vskip 0pt 
\global\advance\reglenum by 1 
\the\reglenum{}: \lapetiteregle {G} {B} {B} {G} {W} {R} {R} {W} {W} \vskip 0pt 
\global\advance\reglenum by 1 
\the\reglenum{}: \lapetiteregle {R} {G} {W} {G} {B} {B} {B} {R} {R} \vskip 0pt 
\global\advance\reglenum by 1 
\the\reglenum{}: \lapetiteregle {R} {G} {R} {B} {G} {R} {R} {W} {R} \vskip 0pt 
\global\advance\reglenum by 1 
\the\reglenum{}: \lapetiteregle {B} {R} {R} {B} {W} {W} {B} {G} {B} \vskip 0pt 
\global\advance\reglenum by 1 
\the\reglenum{}: \lapetiteregle {W} {B} {B} {G} {W} {R} {R} {W} {W} \vskip 0pt 
\global\advance\reglenum by 1 
\the\reglenum{}: \lapetiteregle {R} {W} {W} {G} {B} {R} {B} {R} {G} \vskip 0pt 
\global\advance\reglenum by 1 
\the\reglenum{}: \lapetiteregle {R} {W} {R} {B} {G} {R} {R} {W} {R} \vskip 0pt 
\global\advance\reglenum by 1 
\the\reglenum{}: \lapetiteregle {B} {R} {R} {R} {B} {W} {B} {G} {B} \vskip 0pt 
\global\advance\reglenum by 1 
\the\reglenum{}: \lapetiteregle {B} {R} {G} {W} {R} {B} {B} {G} {B} \vskip 0pt 
\global\advance\reglenum by 1 
\the\reglenum{}: \lapetiteregle {B} {R} {G} {W} {W} {R} {B} {G} {B} \vskip 0pt 
\global\advance\reglenum by 1 
}
\vskip 5pt
{\sc paths}
\vskip 5pt
vertical
\vskip 5pt\rmviii
from top\vskip 1pt
to bottom~:
\vskip 1pt
{\ttv\obeylines
\the\reglenum{}: \lapetiteregle {B} {B} {G} {G} {G} {B} {W} {W} {B} \vskip 0pt 
\global\advance\reglenum by 1 
\the\reglenum{}: \lapetiteregle {B} {B} {W} {W} {W} {B} {G} {G} {B} \vskip 0pt 
\global\advance\reglenum by 1 
\the\reglenum{}: \lapetiteregle {W} {B} {W} {B} {W} {B} {B} {B} {B} \vskip 0pt 
\global\advance\reglenum by 1 
\the\reglenum{}: \lapetiteregle {B} {B} {W} {B} {R} {B} {G} {G} {B} \vskip 0pt 
\global\advance\reglenum by 1 
\the\reglenum{}: \lapetiteregle {B} {B} {R} {W} {W} {B} {G} {G} {B} \vskip 0pt 
\global\advance\reglenum by 1 
\the\reglenum{}: \lapetiteregle {B} {B} {W} {W} {W} {W} {W} {G} {B} \vskip 0pt 
\global\advance\reglenum by 1 
\the\reglenum{}: \lapetiteregle {G} {B} {B} {G} {W} {W} {W} {G} {G} \vskip 0pt 
\global\advance\reglenum by 1 
\the\reglenum{}: \lapetiteregle {G} {G} {B} {B} {G} {W} {W} {W} {G} \vskip 0pt 
\global\advance\reglenum by 1 
\the\reglenum{}: \lapetiteregle {W} {W} {G} {W} {W} {W} {W} {W} {W} \vskip 0pt 
\global\advance\reglenum by 1 
\the\reglenum{}: \lapetiteregle {G} {G} {B} {B} {W} {W} {W} {W} {G} \vskip 0pt 
\global\advance\reglenum by 1 
\the\reglenum{}: \lapetiteregle {W} {W} {W} {W} {W} {B} {W} {W} {W} \vskip 0pt 
\global\advance\reglenum by 1 
\the\reglenum{}: \lapetiteregle {G} {B} {G} {W} {W} {W} {W} {G} {G} \vskip 0pt 
\global\advance\reglenum by 1 
\the\reglenum{}: \lapetiteregle {B} {B} {G} {W} {W} {W} {W} {W} {B} \vskip 0pt 
\global\advance\reglenum by 1 
\the\reglenum{}: \lapetiteregle {W} {B} {B} {W} {B} {W} {B} {W} {W} \vskip 0pt 
\global\advance\reglenum by 1 
\the\reglenum{}: \lapetiteregle {W} {W} {B} {W} {W} {W} {G} {B} {W} \vskip 0pt 
\global\advance\reglenum by 1 
\the\reglenum{}: \lapetiteregle {B} {W} {W} {W} {B} {W} {W} {W} {B} \vskip 0pt 
\global\advance\reglenum by 1 
\the\reglenum{}: \lapetiteregle {W} {W} {B} {W} {B} {W} {B} {B} {W} \vskip 0pt 
\global\advance\reglenum by 1 
\the\reglenum{}: \lapetiteregle {W} {B} {W} {W} {W} {W} {G} {B} {W} \vskip 0pt 
\global\advance\reglenum by 1 
\the\reglenum{}: \lapetiteregle {B} {B} {W} {G} {B} {W} {W} {W} {B} \vskip 0pt 
\global\advance\reglenum by 1 
\the\reglenum{}: \lapetiteregle {W} {W} {B} {W} {B} {W} {W} {B} {W} \vskip 0pt 
\global\advance\reglenum by 1 
\the\reglenum{}: \lapetiteregle {G} {B} {W} {W} {W} {W} {W} {B} {G} \vskip 0pt 
\global\advance\reglenum by 1 
\the\reglenum{}: \lapetiteregle {B} {W} {W} {W} {B} {B} {W} {W} {B} \vskip 0pt 
\global\advance\reglenum by 1 
\the\reglenum{}: \lapetiteregle {B} {B} {W} {W} {W} {W} {W} {B} {B} \vskip 0pt 
\global\advance\reglenum by 1 
\the\reglenum{}: \lapetiteregle {B} {B} {W} {W} {B} {B} {G} {G} {B} \vskip 0pt 
\global\advance\reglenum by 1 
\the\reglenum{}: \lapetiteregle {W} {B} {W} {B} {W} {W} {B} {B} {B} \vskip 0pt 
\global\advance\reglenum by 1 
\the\reglenum{}: \lapetiteregle {B} {B} {W} {B} {W} {B} {R} {B} {R} \vskip 0pt 
\global\advance\reglenum by 1 
\the\reglenum{}: \lapetiteregle {B} {B} {B} {R} {W} {B} {G} {G} {B} \vskip 0pt 
\global\advance\reglenum by 1 
\the\reglenum{}: \lapetiteregle {B} {B} {G} {G} {G} {B} {W} {B} {B} \vskip 0pt 
\global\advance\reglenum by 1 
\the\reglenum{}: \lapetiteregle {W} {B} {B} {W} {B} {W} {B} {B} {B} \vskip 0pt 
\global\advance\reglenum by 1 
\the\reglenum{}: \lapetiteregle {B} {B} {W} {W} {B} {B} {W} {W} {B} \vskip 0pt 
\global\advance\reglenum by 1 
\the\reglenum{}: \lapetiteregle {B} {B} {G} {G} {G} {B} {B} {R} {B} \vskip 0pt 
\global\advance\reglenum by 1 
\the\reglenum{}: \lapetiteregle {B} {B} {G} {W} {W} {W} {W} {B} {B} \vskip 0pt 
\global\advance\reglenum by 1 
\the\reglenum{}: \lapetiteregle {B} {B} {B} {W} {B} {W} {B} {R} {R} \vskip 0pt 
\global\advance\reglenum by 1 
\the\reglenum{}: \lapetiteregle {B} {B} {W} {W} {B} {W} {W} {W} {B} \vskip 0pt 
\global\advance\reglenum by 1 
\the\reglenum{}: \lapetiteregle {W} {B} {B} {W} {B} {W} {W} {B} {B} \vskip 0pt 
\global\advance\reglenum by 1 
\the\reglenum{}: \lapetiteregle {B} {R} {B} {W} {B} {B} {W} {W} {B} \vskip 0pt 
\global\advance\reglenum by 1 
\the\reglenum{}: \lapetiteregle {B} {B} {G} {G} {G} {B} {R} {W} {B} \vskip 0pt 
\global\advance\reglenum by 1 
\the\reglenum{}: \lapetiteregle {B} {B} {G} {W} {W} {W} {W} {R} {B} \vskip 0pt 
\global\advance\reglenum by 1 
\the\reglenum{}: \lapetiteregle {R} {B} {B} {W} {B} {B} {B} {W} {W} \vskip 0pt 
\global\advance\reglenum by 1 
\the\reglenum{}: \lapetiteregle {B} {R} {W} {W} {B} {W} {W} {B} {B} \vskip 0pt 
\global\advance\reglenum by 1 
\the\reglenum{}: \lapetiteregle {B} {R} {B} {W} {B} {W} {W} {B} {R} \vskip 0pt 
\global\advance\reglenum by 1 
\the\reglenum{}: \lapetiteregle {B} {W} {R} {B} {W} {W} {W} {W} {B} \vskip 0pt 
\global\advance\reglenum by 1 
\the\reglenum{}: \lapetiteregle {W} {B} {B} {W} {B} {R} {B} {W} {W} \vskip 0pt 
\global\advance\reglenum by 1 
\the\reglenum{}: \lapetiteregle {B} {W} {W} {W} {B} {W} {B} {R} {B} \vskip 0pt 
\global\advance\reglenum by 1 
\the\reglenum{}: \lapetiteregle {R} {W} {B} {B} {B} {W} {W} {B} {W} \vskip 0pt 
\global\advance\reglenum by 1 
\the\reglenum{}: \lapetiteregle {B} {W} {W} {R} {W} {W} {W} {W} {B} \vskip 0pt 
\global\advance\reglenum by 1 
\the\reglenum{}: \lapetiteregle {B} {W} {W} {W} {B} {B} {R} {W} {B} \vskip 0pt 
\global\advance\reglenum by 1 
\the\reglenum{}: \lapetiteregle {W} {W} {B} {R} {B} {W} {W} {B} {W} \vskip 0pt 
\global\advance\reglenum by 1 
\the\reglenum{}: \lapetiteregle {B} {B} {W} {G} {B} {W} {W} {B} {B} \vskip 0pt 
\global\advance\reglenum by 1 
\the\reglenum{}: \lapetiteregle {B} {W} {W} {W} {B} {R} {W} {W} {B} \vskip 0pt 
\global\advance\reglenum by 1 
\the\reglenum{}: \lapetiteregle {B} {B} {W} {G} {B} {W} {W} {R} {B} \vskip 0pt 
\global\advance\reglenum by 1 
\the\reglenum{}: \lapetiteregle {R} {B} {B} {W} {B} {W} {B} {W} {W} \vskip 0pt 
\global\advance\reglenum by 1 
}
\vskip 5pt\rmviii
from bottom\vskip 1pt
to top~:
\vskip 1pt
{\ttv\obeylines
%
\the\reglenum{}: \lapetiteregle {B} {W} {W} {W} {B} {R} {B} {W} {B} \vskip 0pt 
\global\advance\reglenum by 1 
\the\reglenum{}: \lapetiteregle {W} {W} {B} {B} {B} {W} {W} {B} {B} \vskip 0pt 
\global\advance\reglenum by 1 
\the\reglenum{}: \lapetiteregle {R} {B} {B} {W} {B} {W} {B} {B} {W} \vskip 0pt 
\global\advance\reglenum by 1 
\the\reglenum{}: \lapetiteregle {B} {W} {B} {R} {B} {W} {W} {B} {R} \vskip 0pt 
\global\advance\reglenum by 1 
\the\reglenum{}: \lapetiteregle {W} {B} {B} {W} {B} {B} {B} {W} {B} \vskip 0pt 
\global\advance\reglenum by 1 
\the\reglenum{}: \lapetiteregle {B} {W} {W} {W} {B} {W} {R} {B} {B} \vskip 0pt 
\global\advance\reglenum by 1 
\the\reglenum{}: \lapetiteregle {W} {B} {B} {W} {B} {W} {B} {R} {W} \vskip 0pt 
\global\advance\reglenum by 1 
}
}

\setbox128=\vtop{\leftskip 0pt\parindent 0pt
\baselineskip 7pt
\hsize=70pt
{\ttv\obeylines
\the\reglenum{}: \lapetiteregle {R} {B} {B} {W} {B} {W} {W} {B} {W} \vskip 0pt 
\global\advance\reglenum by 1 
\the\reglenum{}: \lapetiteregle {B} {W} {W} {B} {W} {W} {W} {W} {B} \vskip 0pt 
\global\advance\reglenum by 1 
\the\reglenum{}: \lapetiteregle {B} {B} {G} {G} {G} {B} {B} {W} {B} \vskip 0pt 
\global\advance\reglenum by 1 
\the\reglenum{}: \lapetiteregle {B} {B} {B} {W} {B} {R} {B} {W} {R} \vskip 0pt 
\global\advance\reglenum by 1 
\the\reglenum{}: \lapetiteregle {B} {B} {W} {W} {B} {W} {W} {R} {B} \vskip 0pt 
\global\advance\reglenum by 1 
\the\reglenum{}: \lapetiteregle {W} {R} {B} {W} {B} {W} {W} {B} {W} \vskip 0pt 
\global\advance\reglenum by 1 
\the\reglenum{}: \lapetiteregle {B} {W} {B} {R} {W} {W} {W} {W} {B} \vskip 0pt 
\global\advance\reglenum by 1 
\the\reglenum{}: \lapetiteregle {B} {B} {G} {G} {G} {B} {R} {B} {B} \vskip 0pt 
\global\advance\reglenum by 1 
\the\reglenum{}: \lapetiteregle {B} {B} {B} {W} {W} {B} {G} {G} {B} \vskip 0pt 
\global\advance\reglenum by 1 
\the\reglenum{}: \lapetiteregle {B} {R} {W} {W} {B} {W} {W} {W} {B} \vskip 0pt 
\global\advance\reglenum by 1 
\the\reglenum{}: \lapetiteregle {B} {B} {G} {G} {G} {B} {W} {R} {B} \vskip 0pt 
\global\advance\reglenum by 1 
\the\reglenum{}: \lapetiteregle {B} {B} {R} {B} {W} {B} {G} {G} {B} \vskip 0pt 
\global\advance\reglenum by 1 
\the\reglenum{}: \lapetiteregle {R} {B} {W} {B} {W} {B} {B} {B} {W} \vskip 0pt 
\global\advance\reglenum by 1 
\the\reglenum{}: \lapetiteregle {B} {B} {W} {R} {B} {B} {G} {G} {B} \vskip 0pt 
\global\advance\reglenum by 1 
\the\reglenum{}: \lapetiteregle {W} {B} {W} {B} {W} {B} {R} {B} {W} \vskip 0pt 
\global\advance\reglenum by 1 
\the\reglenum{}: \lapetiteregle {B} {B} {W} {W} {R} {B} {G} {G} {B} \vskip 0pt 
\global\advance\reglenum by 1 
\the\reglenum{}: \lapetiteregle {W} {B} {W} {B} {W} {W} {B} {R} {W} \vskip 0pt 
\global\advance\reglenum by 1 
}
\vskip 5pt
hairpin
\vskip 5pt\rmviii
from left\vskip 1pt
to right~:
\vskip 1pt
{\ttv\obeylines
\the\reglenum{}: \lapetiteregle {B} {B} {W} {W} {B} {G} {G} {G} {B} \vskip 0pt 
\global\advance\reglenum by 1 
\the\reglenum{}: \lapetiteregle {W} {W} {B} {B} {G} {W} {W} {B} {W} \vskip 0pt 
\global\advance\reglenum by 1 
\the\reglenum{}: \lapetiteregle {G} {W} {B} {B} {W} {W} {W} {W} {G} \vskip 0pt 
\global\advance\reglenum by 1 
\the\reglenum{}: \lapetiteregle {R} {B} {B} {B} {W} {W} {W} {W} {W} \vskip 0pt 
\global\advance\reglenum by 1 
\the\reglenum{}: \lapetiteregle {W} {B} {R} {W} {W} {W} {W} {B} {W} \vskip 0pt 
\global\advance\reglenum by 1 
\the\reglenum{}: \lapetiteregle {G} {B} {B} {G} {W} {W} {W} {W} {G} \vskip 0pt 
\global\advance\reglenum by 1 
\the\reglenum{}: \lapetiteregle {G} {G} {B} {G} {W} {W} {W} {W} {G} \vskip 0pt 
\global\advance\reglenum by 1 
\the\reglenum{}: \lapetiteregle {B} {B} {W} {W} {W} {B} {G} {W} {B} \vskip 0pt 
\global\advance\reglenum by 1 
\the\reglenum{}: \lapetiteregle {B} {B} {W} {B} {B} {G} {G} {G} {B} \vskip 0pt 
\global\advance\reglenum by 1 
\the\reglenum{}: \lapetiteregle {W} {B} {B} {B} {G} {W} {W} {B} {W} \vskip 0pt 
\global\advance\reglenum by 1 
\the\reglenum{}: \lapetiteregle {B} {W} {W} {W} {W} {B} {W} {B} {B} \vskip 0pt 
\global\advance\reglenum by 1 
\the\reglenum{}: \lapetiteregle {B} {B} {B} {R} {B} {G} {G} {G} {B} \vskip 0pt 
\global\advance\reglenum by 1 
\the\reglenum{}: \lapetiteregle {W} {R} {B} {B} {G} {W} {W} {B} {W} \vskip 0pt 
\global\advance\reglenum by 1 
\the\reglenum{}: \lapetiteregle {W} {B} {W} {W} {W} {W} {B} {B} {W} \vskip 0pt 
\global\advance\reglenum by 1 
\the\reglenum{}: \lapetiteregle {B} {B} {R} {W} {B} {G} {G} {G} {B} \vskip 0pt 
\global\advance\reglenum by 1 
\the\reglenum{}: \lapetiteregle {W} {B} {W} {W} {W} {W} {B} {R} {W} \vskip 0pt 
\global\advance\reglenum by 1 
\the\reglenum{}: \lapetiteregle {R} {B} {W} {B} {B} {B} {W} {B} {W} \vskip 0pt 
\global\advance\reglenum by 1 
\the\reglenum{}: \lapetiteregle {W} {B} {W} {B} {R} {B} {W} {B} {W} \vskip 0pt 
\global\advance\reglenum by 1 
\the\reglenum{}: \lapetiteregle {B} {B} {B} {W} {W} {B} {G} {W} {B} \vskip 0pt 
\global\advance\reglenum by 1 
\the\reglenum{}: \lapetiteregle {B} {W} {W} {W} {R} {B} {W} {W} {B} \vskip 0pt 
\global\advance\reglenum by 1 
\the\reglenum{}: \lapetiteregle {R} {B} {W} {B} {W} {B} {W} {B} {W} \vskip 0pt 
\global\advance\reglenum by 1 
\the\reglenum{}: \lapetiteregle {B} {B} {R} {W} {W} {B} {G} {W} {B} \vskip 0pt 
\global\advance\reglenum by 1 
}
\vskip 5pt\rmviii
from right\vskip 1pt
to left~:
\vskip 1pt
{\ttv\obeylines
%
\the\reglenum{}: \lapetiteregle {B} {B} {B} {R} {W} {B} {G} {W} {B} \vskip 0pt 
\global\advance\reglenum by 1 
\the\reglenum{}: \lapetiteregle {W} {B} {W} {B} {B} {B} {W} {B} {B} \vskip 0pt 
\global\advance\reglenum by 1 
\the\reglenum{}: \lapetiteregle {B} {B} {B} {W} {B} {G} {G} {G} {B} \vskip 0pt 
\global\advance\reglenum by 1 
\the\reglenum{}: \lapetiteregle {B} {B} {W} {B} {R} {B} {W} {B} {R} \vskip 0pt 
\global\advance\reglenum by 1 
\the\reglenum{}: \lapetiteregle {B} {B} {R} {B} {B} {G} {G} {G} {B} \vskip 0pt 
\global\advance\reglenum by 1 
\the\reglenum{}: \lapetiteregle {B} {R} {W} {W} {W} {B} {W} {B} {B} \vskip 0pt 
\global\advance\reglenum by 1 
\the\reglenum{}: \lapetiteregle {B} {B} {W} {R} {B} {G} {G} {G} {B} \vskip 0pt 
\global\advance\reglenum by 1 
\the\reglenum{}: \lapetiteregle {B} {W} {W} {W} {W} {B} {W} {R} {B} \vskip 0pt 
\global\advance\reglenum by 1 
}
\vskip 5pt\rmviii
horizontal\vskip 5pt
from left\vskip 1pt
to right~:
\vskip 1pt
{\ttv\obeylines
\the\reglenum{}: \lapetiteregle {B} {W} {B} {G} {G} {B} {W} {W} {B} \vskip 0pt 
\global\advance\reglenum by 1 
\the\reglenum{}: \lapetiteregle {G} {B} {G} {W} {W} {W} {G} {B} {G} \vskip 0pt 
\global\advance\reglenum by 1 
\the\reglenum{}: \lapetiteregle {B} {B} {G} {G} {B} {W} {W} {W} {B} \vskip 0pt 
\global\advance\reglenum by 1 
\the\reglenum{}: \lapetiteregle {B} {B} {B} {G} {G} {B} {W} {W} {B} \vskip 0pt 
\global\advance\reglenum by 1 
\the\reglenum{}: \lapetiteregle {B} {R} {B} {G} {G} {B} {W} {B} {B} \vskip 0pt 
\global\advance\reglenum by 1 
\the\reglenum{}: \lapetiteregle {B} {W} {B} {G} {G} {B} {B} {R} {B} \vskip 0pt 
\global\advance\reglenum by 1 
\the\reglenum{}: \lapetiteregle {B} {B} {G} {G} {B} {W} {W} {B} {B} \vskip 0pt 
\global\advance\reglenum by 1 
\the\reglenum{}: \lapetiteregle {B} {W} {B} {G} {G} {B} {R} {W} {B} \vskip 0pt 
\global\advance\reglenum by 1 
\the\reglenum{}: \lapetiteregle {B} {B} {G} {G} {B} {W} {B} {R} {B} \vskip 0pt 
\global\advance\reglenum by 1 
\the\reglenum{}: \lapetiteregle {R} {B} {B} {B} {B} {W} {B} {W} {W} \vskip 0pt 
\global\advance\reglenum by 1 
\the\reglenum{}: \lapetiteregle {B} {B} {G} {G} {B} {B} {R} {W} {B} \vskip 0pt 
\global\advance\reglenum by 1 
\the\reglenum{}: \lapetiteregle {W} {B} {B} {R} {B} {W} {B} {W} {W} \vskip 0pt 
\global\advance\reglenum by 1 
\the\reglenum{}: \lapetiteregle {B} {B} {G} {G} {B} {R} {W} {W} {B} \vskip 0pt 
\global\advance\reglenum by 1 
}
\vskip 5pt\rmviii
from left\vskip 1pt
to right~:
\vskip 1pt
{\ttv\obeylines
\the\reglenum{}: \lapetiteregle {B} {B} {G} {G} {B} {R} {B} {W} {B} \vskip 0pt 
\global\advance\reglenum by 1 
\the\reglenum{}: \lapetiteregle {W} {B} {B} {B} {B} {W} {B} {W} {B} \vskip 0pt 
\global\advance\reglenum by 1 
\the\reglenum{}: \lapetiteregle {B} {B} {G} {G} {B} {W} {W} {R} {B} \vskip 0pt 
\global\advance\reglenum by 1 
\the\reglenum{}: \lapetiteregle {B} {B} {G} {G} {B} {B} {W} {W} {B} \vskip 0pt 
\global\advance\reglenum by 1 
\the\reglenum{}: \lapetiteregle {B} {B} {G} {G} {B} {W} {R} {B} {B} \vskip 0pt 
\global\advance\reglenum by 1 
\the\reglenum{}: \lapetiteregle {B} {B} {B} {R} {B} {W} {B} {W} {R} \vskip 0pt 
\global\advance\reglenum by 1 
\the\reglenum{}: \lapetiteregle {B} {W} {B} {G} {G} {B} {B} {W} {B} \vskip 0pt 
\global\advance\reglenum by 1 
\the\reglenum{}: \lapetiteregle {B} {W} {B} {G} {G} {B} {R} {B} {B} \vskip 0pt 
\global\advance\reglenum by 1 
\the\reglenum{}: \lapetiteregle {B} {B} {B} {G} {G} {B} {W} {R} {B} \vskip 0pt 
\global\advance\reglenum by 1 
\the\reglenum{}: \lapetiteregle {B} {R} {B} {G} {G} {B} {W} {W} {B} \vskip 0pt 
\global\advance\reglenum by 1 
}
}

\ligne{\box120\hfill
\box122\hfill
\box124\hfill
\box126\hfill
\box128\hfill
}

\setbox120=\vtop{\leftskip 0pt\parindent 0pt
\baselineskip 7pt
\hsize=70pt
isocline
\vskip 5pt\rmviii
from left\vskip 2pt
to right:
\vskip 1pt
{\ttv\obeylines
\the\reglenum{}: \lapetiteregle {G} {B} {B} {G} {W} {G} {B} {B} {G} \vskip 0pt 
\global\advance\reglenum by 1 
\the\reglenum{}: \lapetiteregle {G} {G} {B} {B} {B} {G} {W} {W} {G} \vskip 0pt 
\global\advance\reglenum by 1 
\the\reglenum{}: \lapetiteregle {W} {G} {G} {W} {W} {W} {W} {G} {W} \vskip 0pt 
\global\advance\reglenum by 1 
\the\reglenum{}: \lapetiteregle {G} {G} {W} {W} {G} {B} {B} {B} {G} \vskip 0pt 
\global\advance\reglenum by 1 
\the\reglenum{}: \lapetiteregle {G} {G} {B} {B} {B} {B} {G} {W} {G} \vskip 0pt 
\global\advance\reglenum by 1 
\the\reglenum{}: \lapetiteregle {W} {W} {G} {G} {G} {W} {W} {W} {W} \vskip 0pt 
\global\advance\reglenum by 1 
\the\reglenum{}: \lapetiteregle {W} {G} {W} {W} {W} {G} {G} {G} {W} \vskip 0pt 
\global\advance\reglenum by 1 
\the\reglenum{}: \lapetiteregle {G} {G} {W} {G} {B} {B} {B} {B} {G} \vskip 0pt 
\global\advance\reglenum by 1 
\the\reglenum{}: \lapetiteregle {G} {W} {G} {B} {W} {W} {W} {W} {G} \vskip 0pt 
\global\advance\reglenum by 1 
\the\reglenum{}: \lapetiteregle {G} {W} {W} {W} {W} {W} {W} {G} {G} \vskip 0pt 
\global\advance\reglenum by 1 
\the\reglenum{}: \lapetiteregle {G} {G} {W} {G} {W} {W} {W} {B} {G} \vskip 0pt 
\global\advance\reglenum by 1 
\the\reglenum{}: \lapetiteregle {B} {B} {G} {B} {B} {W} {W} {W} {B} \vskip 0pt 
\global\advance\reglenum by 1 
\the\reglenum{}: \lapetiteregle {B} {B} {G} {B} {R} {B} {W} {W} {B} \vskip 0pt 
\global\advance\reglenum by 1 
\the\reglenum{}: \lapetiteregle {B} {B} {G} {B} {W} {R} {B} {W} {B} \vskip 0pt 
\global\advance\reglenum by 1 
\the\reglenum{}: \lapetiteregle {B} {G} {B} {B} {W} {W} {W} {B} {B} \vskip 0pt 
\global\advance\reglenum by 1 
\the\reglenum{}: \lapetiteregle {B} {B} {G} {B} {W} {W} {R} {B} {B} \vskip 0pt 
\global\advance\reglenum by 1 
\the\reglenum{}: \lapetiteregle {B} {G} {B} {R} {B} {W} {W} {B} {B} \vskip 0pt 
\global\advance\reglenum by 1 
\the\reglenum{}: \lapetiteregle {B} {G} {B} {W} {R} {B} {W} {B} {B} \vskip 0pt 
\global\advance\reglenum by 1 
\the\reglenum{}: \lapetiteregle {B} {G} {B} {W} {W} {R} {B} {B} {B} \vskip 0pt 
\global\advance\reglenum by 1 
\the\reglenum{}: \lapetiteregle {B} {G} {B} {W} {W} {W} {R} {B} {B} \vskip 0pt 
\global\advance\reglenum by 1 
}
\vskip 5pt\rmviii
from right\vskip 0pt
to left:
\vskip 1pt
{\ttv\obeylines
\the\reglenum{}: \lapetiteregle {B} {G} {B} {W} {W} {W} {B} {B} {B} \vskip 0pt 
\global\advance\reglenum by 1 
\the\reglenum{}: \lapetiteregle {B} {G} {B} {W} {W} {B} {R} {B} {B} \vskip 0pt 
\global\advance\reglenum by 1 
\the\reglenum{}: \lapetiteregle {B} {G} {B} {W} {B} {R} {W} {B} {B} \vskip 0pt 
\global\advance\reglenum by 1 
\the\reglenum{}: \lapetiteregle {B} {G} {B} {B} {R} {W} {W} {B} {B} \vskip 0pt 
\global\advance\reglenum by 1 
\the\reglenum{}: \lapetiteregle {B} {B} {G} {B} {W} {W} {W} {B} {B} \vskip 0pt 
\global\advance\reglenum by 1 
\the\reglenum{}: \lapetiteregle {B} {G} {B} {R} {W} {W} {W} {B} {B} \vskip 0pt 
\global\advance\reglenum by 1 
\the\reglenum{}: \lapetiteregle {B} {B} {G} {B} {W} {W} {B} {R} {B} \vskip 0pt 
\global\advance\reglenum by 1 
\the\reglenum{}: \lapetiteregle {B} {B} {G} {B} {W} {B} {R} {W} {B} \vskip 0pt 
\global\advance\reglenum by 1 
\the\reglenum{}: \lapetiteregle {B} {B} {G} {B} {B} {R} {W} {W} {B} \vskip 0pt 
\global\advance\reglenum by 1 
\the\reglenum{}: \lapetiteregle {B} {B} {G} {B} {R} {W} {W} {W} {B} \vskip 0pt 
\global\advance\reglenum by 1 
}
\vskip 5pt
{\sc slip roads}
\vskip 5pt\rmviii
vertical $\downarrow$\vskip 2pt   
to isocline $\rightarrow$
\vskip 1pt
{\ttv\obeylines
\the\reglenum{}: \lapetiteregle {W} {W} {B} {B} {W} {B} {G} {B} {W} \vskip 0pt 
\global\advance\reglenum by 1 
\the\reglenum{}: \lapetiteregle {W} {W} {B} {B} {B} {B} {W} {B} {B} \vskip 0pt 
\global\advance\reglenum by 1 
\the\reglenum{}: \lapetiteregle {B} {W} {B} {W} {B} {R} {B} {B} {R} \vskip 0pt 
\global\advance\reglenum by 1 
\the\reglenum{}: \lapetiteregle {B} {W} {B} {B} {G} {G} {G} {B} {B} \vskip 0pt 
\global\advance\reglenum by 1 
\the\reglenum{}: \lapetiteregle {B} {W} {B} {G} {G} {G} {B} {W} {B} \vskip 0pt 
\global\advance\reglenum by 1 
\the\reglenum{}: \lapetiteregle {B} {B} {W} {R} {R} {R} {W} {R} {B} \vskip 0pt 
\global\advance\reglenum by 1 
\the\reglenum{}: \lapetiteregle {R} {B} {R} {W} {W} {W} {W} {R} {R} \vskip 0pt 
\global\advance\reglenum by 1 
\the\reglenum{}: \lapetiteregle {W} {R} {B} {R} {W} {W} {W} {B} {W} \vskip 0pt 
\global\advance\reglenum by 1 
\the\reglenum{}: \lapetiteregle {B} {B} {R} {W} {W} {W} {W} {G} {B} \vskip 0pt 
\global\advance\reglenum by 1 
\the\reglenum{}: \lapetiteregle {G} {B} {B} {W} {W} {W} {W} {G} {G} \vskip 0pt 
\global\advance\reglenum by 1 
\the\reglenum{}: \lapetiteregle {B} {W} {W} {W} {W} {G} {G} {B} {B} \vskip 0pt 
\global\advance\reglenum by 1 
\the\reglenum{}: \lapetiteregle {W} {B} {W} {B} {B} {W} {W} {W} {W} \vskip 0pt 
\global\advance\reglenum by 1 
\the\reglenum{}: \lapetiteregle {G} {B} {W} {W} {W} {W} {W} {G} {G} \vskip 0pt 
\global\advance\reglenum by 1 
\the\reglenum{}: \lapetiteregle {B} {W} {B} {G} {G} {W} {W} {W} {B} \vskip 0pt 
\global\advance\reglenum by 1 
\the\reglenum{}: \lapetiteregle {G} {B} {G} {W} {W} {W} {W} {W} {G} \vskip 0pt 
\global\advance\reglenum by 1 
\the\reglenum{}: \lapetiteregle {B} {W} {W} {W} {W} {W} {B} {G} {B} \vskip 0pt 
\global\advance\reglenum by 1 
\the\reglenum{}: \lapetiteregle {G} {W} {B} {B} {B} {B} {G} {B} {G} \vskip 0pt 
\global\advance\reglenum by 1 
\the\reglenum{}: \lapetiteregle {B} {W} {G} {G} {G} {W} {B} {W} {B} \vskip 0pt 
\global\advance\reglenum by 1 
\the\reglenum{}: \lapetiteregle {G} {B} {G} {B} {B} {B} {B} {G} {G} \vskip 0pt 
\global\advance\reglenum by 1 
\the\reglenum{}: \lapetiteregle {G} {B} {G} {B} {B} {B} {G} {W} {G} \vskip 0pt 
\global\advance\reglenum by 1 
\the\reglenum{}: \lapetiteregle {W} {B} {G} {G} {G} {R} {B} {B} {W} \vskip 0pt 
\global\advance\reglenum by 1 
\the\reglenum{}: \lapetiteregle {G} {W} {G} {B} {B} {B} {B} {G} {G} \vskip 0pt 
\global\advance\reglenum by 1 
\the\reglenum{}: \lapetiteregle {G} {W} {G} {B} {B} {B} {G} {R} {G} \vskip 0pt 
\global\advance\reglenum by 1 
\the\reglenum{}: \lapetiteregle {R} {W} {G} {G} {R} {R} {R} {B} {R} \vskip 0pt 
\global\advance\reglenum by 1 
\the\reglenum{}: \lapetiteregle {G} {R} {G} {B} {W} {W} {W} {R} {G} \vskip 0pt 
\global\advance\reglenum by 1 
\the\reglenum{}: \lapetiteregle {R} {R} {R} {W} {W} {B} {W} {R} {R} \vskip 0pt 
\global\advance\reglenum by 1 
\the\reglenum{}: \lapetiteregle {W} {B} {B} {B} {W} {B} {G} {B} {B} \vskip 0pt 
\global\advance\reglenum by 1 
\the\reglenum{}: \lapetiteregle {B} {W} {B} {R} {B} {B} {W} {B} {R} \vskip 0pt 
\global\advance\reglenum by 1 
\the\reglenum{}: \lapetiteregle {B} {B} {R} {B} {G} {G} {G} {B} {B} \vskip 0pt 
\global\advance\reglenum by 1 
\the\reglenum{}: \lapetiteregle {B} {B} {B} {G} {G} {G} {B} {W} {B} \vskip 0pt 
\global\advance\reglenum by 1 
\the\reglenum{}: \lapetiteregle {B} {R} {W} {R} {R} {R} {W} {W} {B} \vskip 0pt 
\global\advance\reglenum by 1 
\the\reglenum{}: \lapetiteregle {W} {R} {B} {W} {B} {W} {B} {B} {W} \vskip 0pt 
\global\advance\reglenum by 1 
\the\reglenum{}: \lapetiteregle {G} {B} {G} {W} {W} {G} {G} {B} {G} \vskip 0pt 
\global\advance\reglenum by 1 
\the\reglenum{}: \lapetiteregle {W} {W} {B} {R} {W} {W} {W} {B} {W} \vskip 0pt 
\global\advance\reglenum by 1 
\the\reglenum{}: \lapetiteregle {G} {G} {W} {W} {W} {B} {W} {G} {G} \vskip 0pt 
\global\advance\reglenum by 1 
\the\reglenum{}: \lapetiteregle {B} {W} {B} {W} {W} {G} {G} {B} {B} \vskip 0pt 
\global\advance\reglenum by 1 
\the\reglenum{}: \lapetiteregle {W} {B} {B} {B} {B} {W} {W} {W} {W} \vskip 0pt 
\global\advance\reglenum by 1 
\the\reglenum{}: \lapetiteregle {G} {B} {G} {G} {W} {W} {W} {G} {G} \vskip 0pt 
\global\advance\reglenum by 1 
\the\reglenum{}: \lapetiteregle {B} {W} {G} {G} {G} {W} {R} {B} {B} \vskip 0pt 
\global\advance\reglenum by 1 
\the\reglenum{}: \lapetiteregle {W} {B} {G} {G} {G} {R} {B} {R} {W} \vskip 0pt 
\global\advance\reglenum by 1 
\the\reglenum{}: \lapetiteregle {R} {R} {R} {W} {W} {G} {W} {R} {R} \vskip 0pt 
\global\advance\reglenum by 1 
\the\reglenum{}: \lapetiteregle {B} {R} {B} {B} {W} {B} {G} {B} {R} \vskip 0pt 
\global\advance\reglenum by 1 
\the\reglenum{}: \lapetiteregle {R} {B} {B} {W} {B} {B} {W} {B} {W} \vskip 0pt 
\global\advance\reglenum by 1 
\the\reglenum{}: \lapetiteregle {B} {R} {W} {B} {G} {G} {G} {B} {B} \vskip 0pt 
\global\advance\reglenum by 1 
}
}

\setbox122=\vtop{\leftskip 0pt\parindent 0pt
\baselineskip 7pt
\hsize=70pt
{\ttv\obeylines
\the\reglenum{}: \lapetiteregle {B} {R} {B} {G} {G} {G} {B} {W} {B} \vskip 0pt 
\global\advance\reglenum by 1 
\the\reglenum{}: \lapetiteregle {B} {W} {W} {R} {R} {R} {W} {W} {B} \vskip 0pt 
\global\advance\reglenum by 1 
\the\reglenum{}: \lapetiteregle {B} {B} {R} {W} {W} {G} {G} {B} {B} \vskip 0pt 
\global\advance\reglenum by 1 
\the\reglenum{}: \lapetiteregle {W} {B} {R} {B} {B} {W} {W} {W} {W} \vskip 0pt 
\global\advance\reglenum by 1 
\the\reglenum{}: \lapetiteregle {G} {B} {B} {B} {B} {B} {G} {B} {G} \vskip 0pt 
\global\advance\reglenum by 1 
\the\reglenum{}: \lapetiteregle {B} {B} {G} {G} {G} {W} {W} {R} {B} \vskip 0pt 
\global\advance\reglenum by 1 
\the\reglenum{}: \lapetiteregle {W} {B} {G} {G} {G} {R} {B} {W} {W} \vskip 0pt 
\global\advance\reglenum by 1 
\the\reglenum{}: \lapetiteregle {R} {W} {B} {B} {B} {B} {G} {B} {W} \vskip 0pt 
\global\advance\reglenum by 1 
\the\reglenum{}: \lapetiteregle {W} {R} {B} {W} {B} {B} {W} {B} {W} \vskip 0pt 
\global\advance\reglenum by 1 
\the\reglenum{}: \lapetiteregle {B} {W} {W} {B} {G} {G} {G} {B} {B} \vskip 0pt 
\global\advance\reglenum by 1 
\the\reglenum{}: \lapetiteregle {B} {R} {W} {W} {W} {G} {G} {B} {B} \vskip 0pt 
\global\advance\reglenum by 1 
\the\reglenum{}: \lapetiteregle {B} {R} {B} {G} {G} {G} {B} {B} {B} \vskip 0pt 
\global\advance\reglenum by 1 
\the\reglenum{}: \lapetiteregle {B} {B} {B} {G} {G} {W} {W} {W} {B} \vskip 0pt 
\global\advance\reglenum by 1 
\the\reglenum{}: \lapetiteregle {G} {R} {B} {B} {B} {B} {G} {B} {G} \vskip 0pt 
\global\advance\reglenum by 1 
\the\reglenum{}: \lapetiteregle {B} {R} {G} {G} {G} {W} {W} {W} {B} \vskip 0pt 
\global\advance\reglenum by 1 
\the\reglenum{}: \lapetiteregle {W} {W} {B} {B} {R} {B} {G} {B} {W} \vskip 0pt 
\global\advance\reglenum by 1 
\the\reglenum{}: \lapetiteregle {W} {W} {B} {W} {B} {B} {W} {B} {W} \vskip 0pt 
\global\advance\reglenum by 1 
\the\reglenum{}: \lapetiteregle {B} {W} {B} {G} {G} {G} {B} {R} {B} \vskip 0pt 
\global\advance\reglenum by 1 
\the\reglenum{}: \lapetiteregle {B} {R} {B} {G} {G} {W} {W} {W} {B} \vskip 0pt 
\global\advance\reglenum by 1 
\the\reglenum{}: \lapetiteregle {B} {W} {R} {B} {W} {W} {B} {G} {B} \vskip 0pt 
\global\advance\reglenum by 1 
\the\reglenum{}: \lapetiteregle {B} {W} {G} {G} {G} {W} {W} {W} {B} \vskip 0pt 
\global\advance\reglenum by 1 
\the\reglenum{}: \lapetiteregle {B} {W} {W} {R} {B} {W} {B} {G} {B} \vskip 0pt 
\global\advance\reglenum by 1 
\the\reglenum{}: \lapetiteregle {B} {W} {W} {W} {R} {B} {B} {G} {B} \vskip 0pt 
\global\advance\reglenum by 1 
\the\reglenum{}: \lapetiteregle {B} {W} {W} {W} {W} {R} {B} {G} {B} \vskip 0pt 
\global\advance\reglenum by 1 
}
\vskip 5pt\rmviii
isocline $\leftarrow$\vskip 2pt   
to vertical $\uparrow$\vskip 2pt
\vskip 1pt
{\ttv\obeylines
\the\reglenum{}: \lapetiteregle {B} {G} {B} {B} {R} {W} {B} {G} {B} \vskip 0pt 
\global\advance\reglenum by 1 
\the\reglenum{}: \lapetiteregle {B} {W} {W} {W} {W} {B} {B} {G} {B} \vskip 0pt 
\global\advance\reglenum by 1 
\the\reglenum{}: \lapetiteregle {B} {W} {W} {W} {B} {R} {B} {G} {B} \vskip 0pt 
\global\advance\reglenum by 1 
\the\reglenum{}: \lapetiteregle {B} {W} {W} {B} {R} {W} {B} {G} {B} \vskip 0pt 
\global\advance\reglenum by 1 
\the\reglenum{}: \lapetiteregle {W} {W} {B} {B} {B} {B} {G} {B} {B} \vskip 0pt 
\global\advance\reglenum by 1 
\the\reglenum{}: \lapetiteregle {B} {W} {B} {G} {G} {G} {B} {B} {B} \vskip 0pt 
\global\advance\reglenum by 1 
\the\reglenum{}: \lapetiteregle {B} {W} {B} {R} {W} {W} {B} {G} {B} \vskip 0pt 
\global\advance\reglenum by 1 
\the\reglenum{}: \lapetiteregle {B} {W} {B} {B} {R} {B} {G} {B} {R} \vskip 0pt 
\global\advance\reglenum by 1 
\the\reglenum{}: \lapetiteregle {W} {B} {B} {W} {B} {B} {W} {B} {B} \vskip 0pt 
\global\advance\reglenum by 1 
\the\reglenum{}: \lapetiteregle {B} {B} {W} {W} {W} {G} {G} {B} {B} \vskip 0pt 
\global\advance\reglenum by 1 
\the\reglenum{}: \lapetiteregle {B} {B} {B} {G} {G} {G} {B} {R} {B} \vskip 0pt 
\global\advance\reglenum by 1 
\the\reglenum{}: \lapetiteregle {B} {B} {G} {G} {G} {W} {W} {W} {B} \vskip 0pt 
\global\advance\reglenum by 1 
\the\reglenum{}: \lapetiteregle {R} {B} {B} {B} {W} {B} {G} {B} {W} \vskip 0pt 
\global\advance\reglenum by 1 
\the\reglenum{}: \lapetiteregle {B} {R} {B} {W} {B} {B} {W} {B} {R} \vskip 0pt 
\global\advance\reglenum by 1 
\the\reglenum{}: \lapetiteregle {B} {B} {W} {B} {G} {G} {G} {B} {B} \vskip 0pt 
\global\advance\reglenum by 1 
\the\reglenum{}: \lapetiteregle {B} {R} {B} {W} {W} {G} {G} {B} {B} \vskip 0pt 
\global\advance\reglenum by 1 
\the\reglenum{}: \lapetiteregle {B} {R} {G} {G} {G} {W} {W} {B} {B} \vskip 0pt 
\global\advance\reglenum by 1 
\the\reglenum{}: \lapetiteregle {W} {R} {B} {B} {W} {B} {G} {B} {W} \vskip 0pt 
\global\advance\reglenum by 1 
\the\reglenum{}: \lapetiteregle {R} {W} {B} {B} {B} {B} {W} {B} {W} \vskip 0pt 
\global\advance\reglenum by 1 
\the\reglenum{}: \lapetiteregle {B} {R} {B} {W} {B} {W} {B} {B} {R} \vskip 0pt 
\global\advance\reglenum by 1 
\the\reglenum{}: \lapetiteregle {B} {R} {B} {B} {G} {G} {G} {B} {B} \vskip 0pt 
\global\advance\reglenum by 1 
\the\reglenum{}: \lapetiteregle {B} {B} {W} {R} {R} {R} {W} {W} {B} \vskip 0pt 
\global\advance\reglenum by 1 
\the\reglenum{}: \lapetiteregle {B} {W} {R} {W} {W} {G} {G} {B} {B} \vskip 0pt 
\global\advance\reglenum by 1 
\the\reglenum{}: \lapetiteregle {B} {W} {G} {G} {G} {W} {B} {R} {B} \vskip 0pt 
\global\advance\reglenum by 1 
}
\vskip 5pt\rmviii
vertical $\uparrow$\vskip 2pt          
to isocline $\rightarrow$\vskip 2pt
\vskip 1pt
{\ttv\obeylines
\the\reglenum{}: \lapetiteregle {B} {W} {W} {W} {W} {W} {B} {R} {B} \vskip 0pt 
\global\advance\reglenum by 1 
\the\reglenum{}: \lapetiteregle {W} {B} {R} {R} {W} {W} {W} {W} {W} \vskip 0pt 
\global\advance\reglenum by 1 
\the\reglenum{}: \lapetiteregle {R} {W} {R} {R} {W} {W} {W} {W} {R} \vskip 0pt 
\global\advance\reglenum by 1 
\the\reglenum{}: \lapetiteregle {W} {R} {R} {G} {G} {G} {W} {W} {W} \vskip 0pt 
\global\advance\reglenum by 1 
\the\reglenum{}: \lapetiteregle {G} {W} {G} {B} {W} {W} {W} {G} {G} \vskip 0pt 
\global\advance\reglenum by 1 
\the\reglenum{}: \lapetiteregle {G} {W} {G} {W} {W} {W} {W} {W} {G} \vskip 0pt 
\global\advance\reglenum by 1 
\the\reglenum{}: \lapetiteregle {W} {W} {W} {G} {W} {W} {W} {W} {W} \vskip 0pt 
\global\advance\reglenum by 1 
\the\reglenum{}: \lapetiteregle {W} {B} {W} {B} {B} {B} {B} {W} {B} \vskip 0pt 
\global\advance\reglenum by 1 
\the\reglenum{}: \lapetiteregle {B} {W} {W} {W} {G} {G} {B} {B} {B} \vskip 0pt 
\global\advance\reglenum by 1 
\the\reglenum{}: \lapetiteregle {B} {W} {B} {B} {R} {B} {W} {B} {R} \vskip 0pt 
\global\advance\reglenum by 1 
\the\reglenum{}: \lapetiteregle {B} {W} {B} {W} {W} {W} {W} {B} {B} \vskip 0pt 
\global\advance\reglenum by 1 
\the\reglenum{}: \lapetiteregle {B} {B} {W} {W} {W} {W} {B} {R} {B} \vskip 0pt 
\global\advance\reglenum by 1 
\the\reglenum{}: \lapetiteregle {R} {B} {B} {B} {B} {R} {R} {W} {R} \vskip 0pt 
\global\advance\reglenum by 1 
\the\reglenum{}: \lapetiteregle {B} {R} {B} {W} {W} {W} {W} {B} {B} \vskip 0pt 
\global\advance\reglenum by 1 
\the\reglenum{}: \lapetiteregle {B} {R} {B} {W} {W} {W} {B} {R} {B} \vskip 0pt 
\global\advance\reglenum by 1 
\the\reglenum{}: \lapetiteregle {R} {R} {B} {B} {B} {G} {W} {R} {R} \vskip 0pt 
\global\advance\reglenum by 1 
\the\reglenum{}: \lapetiteregle {G} {R} {B} {B} {B} {B} {G} {W} {G} \vskip 0pt 
\global\advance\reglenum by 1 
\the\reglenum{}: \lapetiteregle {B} {G} {B} {W} {W} {B} {W} {G} {B} \vskip 0pt 
\global\advance\reglenum by 1 
\the\reglenum{}: \lapetiteregle {B} {W} {W} {B} {W} {W} {B} {R} {B} \vskip 0pt 
\global\advance\reglenum by 1 
\the\reglenum{}: \lapetiteregle {B} {B} {W} {B} {R} {B} {B} {W} {R} \vskip 0pt 
\global\advance\reglenum by 1 
\the\reglenum{}: \lapetiteregle {B} {B} {W} {W} {G} {G} {B} {R} {B} \vskip 0pt 
\global\advance\reglenum by 1 
\the\reglenum{}: \lapetiteregle {B} {B} {R} {W} {W} {W} {W} {B} {B} \vskip 0pt 
\global\advance\reglenum by 1 
\the\reglenum{}: \lapetiteregle {B} {W} {W} {R} {B} {W} {B} {R} {B} \vskip 0pt 
\global\advance\reglenum by 1 
\the\reglenum{}: \lapetiteregle {B} {R} {W} {W} {G} {G} {B} {W} {B} \vskip 0pt 
\global\advance\reglenum by 1 
\the\reglenum{}: \lapetiteregle {B} {B} {R} {B} {W} {W} {W} {W} {B} \vskip 0pt 
\global\advance\reglenum by 1 
\the\reglenum{}: \lapetiteregle {B} {W} {W} {W} {R} {B} {B} {R} {B} \vskip 0pt 
\global\advance\reglenum by 1 
\the\reglenum{}: \lapetiteregle {W} {B} {W} {B} {W} {B} {B} {R} {W} \vskip 0pt 
\global\advance\reglenum by 1 
\the\reglenum{}: \lapetiteregle {B} {W} {W} {W} {G} {G} {B} {W} {B} \vskip 0pt 
\global\advance\reglenum by 1 
\the\reglenum{}: \lapetiteregle {B} {R} {W} {B} {W} {W} {W} {W} {B} \vskip 0pt 
\global\advance\reglenum by 1 
\the\reglenum{}: \lapetiteregle {B} {B} {B} {W} {W} {W} {B} {R} {B} \vskip 0pt 
\global\advance\reglenum by 1 
\the\reglenum{}: \lapetiteregle {B} {W} {W} {W} {W} {R} {B} {R} {B} \vskip 0pt 
\global\advance\reglenum by 1 
}
}

\setbox124=\vtop{\leftskip 0pt\parindent 0pt
\baselineskip 7pt
\hsize=70pt
{\ttv\obeylines
\the\reglenum{}: \lapetiteregle {W} {B} {W} {B} {W} {B} {B} {W} {W} \vskip 0pt 
\global\advance\reglenum by 1 
\the\reglenum{}: \lapetiteregle {B} {B} {R} {B} {W} {W} {B} {R} {B} \vskip 0pt 
\global\advance\reglenum by 1 
\the\reglenum{}: \lapetiteregle {B} {B} {W} {R} {B} {W} {B} {R} {B} \vskip 0pt 
\global\advance\reglenum by 1 
\the\reglenum{}: \lapetiteregle {B} {B} {W} {W} {R} {B} {B} {R} {B} \vskip 0pt 
\global\advance\reglenum by 1 
\the\reglenum{}: \lapetiteregle {B} {R} {B} {B} {W} {W} {W} {B} {B} \vskip 0pt 
\global\advance\reglenum by 1 
\the\reglenum{}: \lapetiteregle {B} {B} {W} {W} {W} {R} {B} {R} {B} \vskip 0pt 
\global\advance\reglenum by 1 
\the\reglenum{}: \lapetiteregle {B} {R} {B} {R} {B} {W} {W} {B} {B} \vskip 0pt 
\global\advance\reglenum by 1 
\the\reglenum{}: \lapetiteregle {B} {R} {B} {W} {R} {B} {W} {B} {B} \vskip 0pt 
\global\advance\reglenum by 1 
\the\reglenum{}: \lapetiteregle {B} {R} {B} {W} {W} {R} {B} {B} {B} \vskip 0pt 
\global\advance\reglenum by 1 
\the\reglenum{}: \lapetiteregle {B} {R} {B} {B} {W} {W} {B} {R} {B} \vskip 0pt 
\global\advance\reglenum by 1 
\the\reglenum{}: \lapetiteregle {B} {R} {B} {W} {W} {W} {R} {B} {B} \vskip 0pt 
\global\advance\reglenum by 1 
\the\reglenum{}: \lapetiteregle {B} {R} {B} {R} {B} {W} {B} {R} {B} \vskip 0pt 
\global\advance\reglenum by 1 
}
\vskip 5pt\rmviii
isocline $\leftarrow$\vskip 2pt        
to vertical $\downarrow$\vskip 2pt
\vskip 1pt
{\ttv\obeylines
\the\reglenum{}: \lapetiteregle {B} {R} {B} {W} {B} {R} {B} {R} {B} \vskip 0pt 
\global\advance\reglenum by 1 
\the\reglenum{}: \lapetiteregle {B} {R} {B} {R} {W} {W} {W} {B} {B} \vskip 0pt 
\global\advance\reglenum by 1 
\the\reglenum{}: \lapetiteregle {B} {R} {B} {W} {W} {W} {B} {B} {B} \vskip 0pt 
\global\advance\reglenum by 1 
\the\reglenum{}: \lapetiteregle {B} {R} {B} {B} {R} {W} {B} {R} {B} \vskip 0pt 
\global\advance\reglenum by 1 
\the\reglenum{}: \lapetiteregle {B} {R} {B} {W} {W} {B} {R} {B} {B} \vskip 0pt 
\global\advance\reglenum by 1 
\the\reglenum{}: \lapetiteregle {B} {R} {B} {R} {W} {W} {B} {R} {B} \vskip 0pt 
\global\advance\reglenum by 1 
\the\reglenum{}: \lapetiteregle {B} {R} {B} {W} {B} {R} {W} {B} {B} \vskip 0pt 
\global\advance\reglenum by 1 
\the\reglenum{}: \lapetiteregle {B} {R} {B} {B} {R} {W} {W} {B} {B} \vskip 0pt 
\global\advance\reglenum by 1 
\the\reglenum{}: \lapetiteregle {B} {B} {W} {W} {B} {R} {B} {R} {B} \vskip 0pt 
\global\advance\reglenum by 1 
\the\reglenum{}: \lapetiteregle {B} {B} {W} {B} {R} {W} {B} {R} {B} \vskip 0pt 
\global\advance\reglenum by 1 
\the\reglenum{}: \lapetiteregle {B} {B} {B} {R} {W} {W} {B} {R} {B} \vskip 0pt 
\global\advance\reglenum by 1 
\the\reglenum{}: \lapetiteregle {B} {W} {W} {W} {B} {R} {B} {R} {B} \vskip 0pt 
\global\advance\reglenum by 1 
\the\reglenum{}: \lapetiteregle {B} {B} {W} {B} {W} {W} {W} {W} {B} \vskip 0pt 
\global\advance\reglenum by 1 
\the\reglenum{}: \lapetiteregle {B} {B} {R} {W} {W} {W} {B} {R} {B} \vskip 0pt 
\global\advance\reglenum by 1 
\the\reglenum{}: \lapetiteregle {B} {W} {W} {B} {R} {W} {B} {R} {B} \vskip 0pt 
\global\advance\reglenum by 1 
\the\reglenum{}: \lapetiteregle {B} {B} {W} {B} {W} {B} {B} {R} {R} \vskip 0pt 
\global\advance\reglenum by 1 
\the\reglenum{}: \lapetiteregle {B} {B} {W} {W} {G} {G} {B} {W} {B} \vskip 0pt 
\global\advance\reglenum by 1 
\the\reglenum{}: \lapetiteregle {B} {W} {W} {R} {W} {W} {B} {R} {B} \vskip 0pt 
\global\advance\reglenum by 1 
\the\reglenum{}: \lapetiteregle {R} {B} {W} {B} {B} {B} {B} {W} {W} \vskip 0pt 
\global\advance\reglenum by 1 
\the\reglenum{}: \lapetiteregle {B} {R} {W} {W} {G} {G} {B} {B} {B} \vskip 0pt 
\global\advance\reglenum by 1 
}
\vskip 5pt\rmviii
vertical $\uparrow$\vskip 2pt        
to isocline $\leftarrow$\vskip 2pt
\vskip 1pt
{\ttv\obeylines
\the\reglenum{}: \lapetiteregle {W} {B} {W} {B} {B} {W} {B} {W} {W} \vskip 0pt 
\global\advance\reglenum by 1 
\the\reglenum{}: \lapetiteregle {B} {B} {W} {W} {G} {B} {B} {B} {B} \vskip 0pt 
\global\advance\reglenum by 1 
\the\reglenum{}: \lapetiteregle {W} {B} {W} {W} {W} {W} {G} {G} {W} \vskip 0pt 
\global\advance\reglenum by 1 
\the\reglenum{}: \lapetiteregle {G} {B} {W} {G} {B} {B} {B} {B} {G} \vskip 0pt 
\global\advance\reglenum by 1 
\the\reglenum{}: \lapetiteregle {G} {G} {W} {W} {W} {W} {W} {B} {G} \vskip 0pt 
\global\advance\reglenum by 1 
\the\reglenum{}: \lapetiteregle {W} {W} {W} {W} {W} {B} {W} {B} {W} \vskip 0pt 
\global\advance\reglenum by 1 
\the\reglenum{}: \lapetiteregle {B} {B} {G} {W} {W} {B} {W} {W} {B} \vskip 0pt 
\global\advance\reglenum by 1 
\the\reglenum{}: \lapetiteregle {W} {R} {B} {B} {W} {W} {W} {B} {W} \vskip 0pt 
\global\advance\reglenum by 1 
\the\reglenum{}: \lapetiteregle {W} {W} {B} {B} {W} {W} {W} {B} {W} \vskip 0pt 
\global\advance\reglenum by 1 
\the\reglenum{}: \lapetiteregle {B} {B} {W} {B} {B} {R} {B} {W} {R} \vskip 0pt 
\global\advance\reglenum by 1 
\the\reglenum{}: \lapetiteregle {B} {R} {W} {W} {B} {B} {W} {B} {B} \vskip 0pt 
\global\advance\reglenum by 1 
\the\reglenum{}: \lapetiteregle {W} {B} {R} {B} {B} {W} {B} {W} {W} \vskip 0pt 
\global\advance\reglenum by 1 
\the\reglenum{}: \lapetiteregle {B} {W} {W} {W} {B} {B} {B} {R} {B} \vskip 0pt 
\global\advance\reglenum by 1 
\the\reglenum{}: \lapetiteregle {B} {B} {B} {B} {W} {W} {W} {B} {B} \vskip 0pt 
\global\advance\reglenum by 1 
\the\reglenum{}: \lapetiteregle {B} {B} {B} {B} {W} {W} {B} {R} {B} \vskip 0pt 
\global\advance\reglenum by 1 
\the\reglenum{}: \lapetiteregle {B} {B} {B} {B} {W} {B} {R} {W} {B} \vskip 0pt 
\global\advance\reglenum by 1 
\the\reglenum{}: \lapetiteregle {W} {B} {W} {W} {W} {B} {W} {B} {W} \vskip 0pt 
\global\advance\reglenum by 1 
\the\reglenum{}: \lapetiteregle {W} {W} {B} {B} {R} {B} {W} {B} {W} \vskip 0pt 
\global\advance\reglenum by 1 
}
\vskip 5pt\rmviii
isocline $\rightarrow$\vskip 2pt          
to vertical $\downarrow$\vskip 2pt       
\vskip 1pt
{\ttv\obeylines
\the\reglenum{}: \lapetiteregle {B} {B} {B} {B} {R} {B} {W} {W} {B} \vskip 0pt 
\global\advance\reglenum by 1 
\the\reglenum{}: \lapetiteregle {B} {B} {B} {B} {W} {W} {W} {R} {B} \vskip 0pt 
\global\advance\reglenum by 1 
\the\reglenum{}: \lapetiteregle {B} {B} {B} {B} {W} {R} {B} {W} {B} \vskip 0pt 
\global\advance\reglenum by 1 
\the\reglenum{}: \lapetiteregle {W} {R} {W} {W} {W} {B} {W} {B} {W} \vskip 0pt 
\global\advance\reglenum by 1 
\the\reglenum{}: \lapetiteregle {B} {W} {W} {W} {B} {B} {B} {W} {B} \vskip 0pt 
\global\advance\reglenum by 1 
\the\reglenum{}: \lapetiteregle {B} {B} {B} {B} {W} {W} {R} {B} {B} \vskip 0pt 
\global\advance\reglenum by 1 
\the\reglenum{}: \lapetiteregle {B} {W} {W} {W} {B} {B} {R} {B} {B} \vskip 0pt 
\global\advance\reglenum by 1 
\the\reglenum{}: \lapetiteregle {B} {B} {R} {B} {B} {W} {B} {W} {R} \vskip 0pt 
\global\advance\reglenum by 1 
\the\reglenum{}: \lapetiteregle {B} {B} {W} {W} {B} {B} {W} {R} {B} \vskip 0pt 
\global\advance\reglenum by 1 
\the\reglenum{}: \lapetiteregle {B} {R} {W} {W} {B} {B} {W} {W} {B} \vskip 0pt 
\global\advance\reglenum by 1 
\the\reglenum{}: \lapetiteregle {W} {B} {W} {B} {B} {R} {B} {W} {W} \vskip 0pt 
\global\advance\reglenum by 1 
}
\vskip 5pt\rmviii
vertical $\downarrow$\vskip 2pt           
to isocline $\leftarrow$\vskip 2pt     
\vskip 1pt
{\ttv\obeylines
\the\reglenum{}: \lapetiteregle {B} {W} {W} {B} {G} {G} {B} {B} {B} \vskip 0pt 
\global\advance\reglenum by 1 
\the\reglenum{}: \lapetiteregle {G} {B} {B} {G} {W} {G} {G} {G} {G} \vskip 0pt 
\global\advance\reglenum by 1 
\the\reglenum{}: \lapetiteregle {G} {B} {G} {G} {B} {B} {B} {B} {G} \vskip 0pt 
\global\advance\reglenum by 1 
\the\reglenum{}: \lapetiteregle {W} {G} {G} {W} {G} {G} {G} {G} {W} \vskip 0pt 
\global\advance\reglenum by 1 
\the\reglenum{}: \lapetiteregle {G} {G} {W} {G} {B} {B} {B} {G} {G} \vskip 0pt 
\global\advance\reglenum by 1 
\the\reglenum{}: \lapetiteregle {G} {G} {G} {G} {B} {B} {B} {B} {G} \vskip 0pt 
\global\advance\reglenum by 1 
\the\reglenum{}: \lapetiteregle {W} {W} {G} {W} {W} {W} {W} {G} {W} \vskip 0pt 
\global\advance\reglenum by 1 
\the\reglenum{}: \lapetiteregle {G} {W} {G} {W} {W} {W} {W} {G} {G} \vskip 0pt 
\global\advance\reglenum by 1 
\the\reglenum{}: \lapetiteregle {B} {B} {B} {G} {B} {W} {W} {W} {B} \vskip 0pt 
\global\advance\reglenum by 1 
\the\reglenum{}: \lapetiteregle {W} {W} {B} {W} {B} {B} {B} {B} {B} \vskip 0pt 
\global\advance\reglenum by 1 
}
}

\setbox126=\vtop{\leftskip 0pt\parindent 0pt
\baselineskip 7pt
\hsize=70pt
{\ttv\obeylines
\the\reglenum{}: \lapetiteregle {B} {W} {B} {B} {G} {G} {B} {B} {B} \vskip 0pt 
\global\advance\reglenum by 1 
\the\reglenum{}: \lapetiteregle {B} {B} {R} {B} {G} {G} {B} {B} {B} \vskip 0pt 
\global\advance\reglenum by 1 
\the\reglenum{}: \lapetiteregle {B} {R} {W} {B} {G} {G} {B} {B} {B} \vskip 0pt 
\global\advance\reglenum by 1 
\the\reglenum{}: \lapetiteregle {B} {W} {B} {B} {W} {W} {B} {R} {B} \vskip 0pt 
\global\advance\reglenum by 1 
\the\reglenum{}: \lapetiteregle {B} {W} {B} {B} {W} {B} {R} {W} {B} \vskip 0pt 
\global\advance\reglenum by 1 
\the\reglenum{}: \lapetiteregle {B} {W} {B} {B} {B} {R} {W} {W} {B} \vskip 0pt 
\global\advance\reglenum by 1 
\the\reglenum{}: \lapetiteregle {B} {B} {B} {G} {B} {W} {W} {B} {B} \vskip 0pt 
\global\advance\reglenum by 1 
}
\vskip 5pt\rmviii
isocline $\rightarrow$\vskip 2pt        
vertical $\uparrow$\vskip 2pt           
\vskip 1pt
{\ttv\obeylines
\the\reglenum{}: \lapetiteregle {B} {B} {B} {G} {B} {W} {R} {B} {B} \vskip 0pt 
\global\advance\reglenum by 1 
\the\reglenum{}: \lapetiteregle {B} {W} {B} {B} {R} {B} {W} {W} {B} \vskip 0pt 
\global\advance\reglenum by 1 
\the\reglenum{}: \lapetiteregle {B} {B} {B} {G} {B} {W} {W} {R} {B} \vskip 0pt 
\global\advance\reglenum by 1 
\the\reglenum{}: \lapetiteregle {B} {W} {B} {B} {W} {R} {B} {W} {B} \vskip 0pt 
\global\advance\reglenum by 1 
\the\reglenum{}: \lapetiteregle {B} {W} {B} {B} {W} {W} {R} {B} {B} \vskip 0pt 
\global\advance\reglenum by 1 
\the\reglenum{}: \lapetiteregle {B} {B} {W} {B} {G} {G} {B} {B} {B} \vskip 0pt 
\global\advance\reglenum by 1 
\the\reglenum{}: \lapetiteregle {B} {R} {B} {B} {G} {G} {B} {B} {B} \vskip 0pt 
\global\advance\reglenum by 1 
\the\reglenum{}: \lapetiteregle {B} {W} {R} {B} {G} {G} {B} {B} {B} \vskip 0pt 
\global\advance\reglenum by 1 
\the\reglenum{}: \lapetiteregle {R} {W} {B} {W} {B} {B} {B} {B} {W} \vskip 0pt 
\global\advance\reglenum by 1 
}
\vskip 5pt\rmviii
vertical\vskip 2pt
slip road $\downarrow$\vskip 2pt           
\vskip 1pt
{\ttv\obeylines
\the\reglenum{}: \lapetiteregle {B} {W} {B} {R} {B} {W} {W} {W} {B} \vskip 0pt 
\global\advance\reglenum by 1 
\the\reglenum{}: \lapetiteregle {W} {B} {W} {B} {W} {B} {W} {W} {W} \vskip 0pt 
\global\advance\reglenum by 1 
\the\reglenum{}: \lapetiteregle {B} {B} {W} {W} {B} {R} {B} {W} {R} \vskip 0pt 
\global\advance\reglenum by 1 
\the\reglenum{}: \lapetiteregle {B} {B} {R} {W} {B} {W} {W} {W} {B} \vskip 0pt 
\global\advance\reglenum by 1 
\the\reglenum{}: \lapetiteregle {B} {W} {W} {W} {W} {R} {B} {W} {B} \vskip 0pt 
\global\advance\reglenum by 1 
\the\reglenum{}: \lapetiteregle {W} {B} {W} {W} {W} {W} {W} {R} {W} \vskip 0pt 
\global\advance\reglenum by 1 
\the\reglenum{}: \lapetiteregle {R} {B} {W} {W} {B} {W} {B} {B} {W} \vskip 0pt 
\global\advance\reglenum by 1 
\the\reglenum{}: \lapetiteregle {B} {W} {W} {W} {W} {W} {R} {B} {B} \vskip 0pt 
\global\advance\reglenum by 1 
\the\reglenum{}: \lapetiteregle {W} {B} {W} {W} {B} {W} {B} {R} {W} \vskip 0pt 
\global\advance\reglenum by 1 
\the\reglenum{}: \lapetiteregle {W} {B} {W} {W} {B} {W} {B} {W} {W} \vskip 0pt 
\global\advance\reglenum by 1 
\the\reglenum{}: \lapetiteregle {B} {W} {W} {W} {W} {W} {R} {W} {B} \vskip 0pt 
\global\advance\reglenum by 1 
\the\reglenum{}: \lapetiteregle {W} {B} {B} {B} {W} {B} {W} {W} {B} \vskip 0pt 
\global\advance\reglenum by 1 
}
\vskip 5pt\rmviii
vertical\vskip 2pt
slip road $\uparrow$\vskip 2pt           
\vskip 1pt
{\ttv\obeylines
\the\reglenum{}: \lapetiteregle {B} {W} {W} {B} {R} {W} {W} {W} {B} \vskip 0pt 
\global\advance\reglenum by 1 
\the\reglenum{}: \lapetiteregle {R} {B} {B} {B} {W} {B} {W} {W} {W} \vskip 0pt 
\global\advance\reglenum by 1 
\the\reglenum{}: \lapetiteregle {B} {W} {W} {W} {W} {W} {B} {W} {B} \vskip 0pt 
\global\advance\reglenum by 1 
\the\reglenum{}: \lapetiteregle {W} {B} {R} {B} {W} {B} {W} {W} {W} \vskip 0pt 
\global\advance\reglenum by 1 
\the\reglenum{}: \lapetiteregle {W} {B} {W} {W} {B} {W} {B} {B} {B} \vskip 0pt 
\global\advance\reglenum by 1 
\the\reglenum{}: \lapetiteregle {B} {W} {W} {W} {W} {B} {R} {W} {B} \vskip 0pt 
\global\advance\reglenum by 1 
\the\reglenum{}: \lapetiteregle {B} {B} {W} {W} {B} {W} {B} {R} {R} \vskip 0pt 
\global\advance\reglenum by 1 
}
}

\setbox128=\vtop{\leftskip 0pt\parindent 0pt
\baselineskip 7pt
\hsize=70pt
{\ttv\obeylines
\ligne{\hfill}
}
}

\ligne{\box120\hfill
\box122\hfill
\box124\hfill
\box126\hfill
\box128\hfill
}

\end{document}